\documentclass[twocolumn]{aastex}

\usepackage{amsmath}
\usepackage{xcolor}
\usepackage{footmisc}

\usepackage{longtable}
\usepackage{lineno}

\received{--- 2021}
\revised{--- 2021}
\accepted{--- 2021}
\submitjournal{ApJ}

\shorttitle{Sh 2-190}
\shortauthors{Sinha et al.}

\begin{document}

\title{Photometric variability  of the pre-main sequence stars towards the Sh 2-190 region}

\correspondingauthor{Tirthendu Sinha}
\email{tirthendu@aries.res.in}

\author[0000-0001-5508-6575]{Tirthendu Sinha}
\affil{Aryabhatta Research Institute of Observational Sciences (ARIES),
Manora Peak, Nainital 263 002, India}
\affil{Kumaun University, Nainital 263 002, India}

\author[0000-0001-5731-3057]{Saurabh Sharma}
\affil{Aryabhatta Research Institute of Observational Sciences (ARIES),
Manora Peak, Nainital 263 002, India}

\author{Neelam Panwar}
\affil{Aryabhatta Research Institute of Observational Sciences (ARIES),
Manora Peak, Nainital 263 002, India}

\author{N. Matsunaga}
\affil{Department of Astronomy, The University of Tokyo
7-3-1 Hongo, Bunkyo-ku, Tokyo 113-0033, Japan}

\author{K. Ogura}
\affil{Kokugakuin University, Higashi, Shibuya-ku, Tokyo 150-8440, Japan}

\author{N. Kobayashi}
\affil{Institute of Astronomy, University of Tokyo, 2-21-1 Osawa, Mitaka, Tokyo 181-0015, Japan}

\author{R. K. Yadav}
\affil{National Astronomical Research Institute of Thailand, Chiang Mai 50200, Thailand}

\author{A. Ghosh}
\affil{Aryabhatta Research Institute of Observational Sciences (ARIES),
Manora Peak, Nainital 263 002, India}

\author{R. Pandey}
\affil{Aryabhatta Research Institute of Observational Sciences (ARIES),
Manora Peak, Nainital 263 002, India}

\author{P. S. Bisht}
\affil{Department of Physics, SSJ University, Almora 263 601, India}



\begin{abstract}
We present the results from our time-series imaging  data
taken  with the 1.3m Devasthal fast optical telescope and 0.81m Tenagara telescope
in  $V$, $R_{c}$, $I_{c}$ bands covering an area of $\sim18^\prime.4\times 18^\prime.4$
towards the star-forming region Sh 2-190. This photometric data helped us to explore the nature of the variability of 
pre-main sequence (PMS) stars. 
We have identified 85 PMS variables, i.e., 37  Class\,{\sc ii} and 48  Class\,{\sc iii} sources. 
Forty-five of the PMS variables are showing periodicity in their light curves. 
We show that the stars with thicker discs and envelopes rotate slower and exhibit 
larger photometric variations compared to their disc-less counterparts. 
This result suggests that rotation of the PMS stars is regulated by the presence of circumstellar discs. 
We also found that the period of the stars show a decreasing trend with increasing mass in the range of $\sim$0.5-2.5 M$_\odot$. Our result indicates that most of the variability in Class\,{\sc ii} sources is ascribed to the presence of thick disc, while the presence of cool spots on the stellar surface causes the brightness variation in Class\,{\sc iii} sources. X-ray activities in the PMS stars were found to be at the saturation level reported for the main sequence (MS) stars. The younger counterparts of the PMS variables are showing less X-ray activity hinting towards a less significant role of a stellar disc in X-ray generation.
\end{abstract}




\section{Introduction}

Circumstellar discs are an integral part of pre-main sequence (PMS) stars and are potential sites 
for planet formation \citep{2002astro.ph.10520H}. Depending on their evolutionary stages PMS stars are 
classified as Class\,{\sc i}, Class\,{\sc ii} and Class\,{\sc iii} objects. Class\,{\sc i} objects are deeply 
embedded protostars with in-falling material from the envelope and circumstellar discs through which material 
is accreted  onto  the  growing  star.  Class\,{\sc ii} objects possess thick circumstellar disc and 
residual envelope and are found to accrete material from the disc through magnetic channels connecting 
the disc and the stars, while the Class\,{\sc iii} objects are yet to reach the main-sequence (MS) 
with depleted or no accretion discs \citep{1994ApJ...434..614G,2016A&A...586A..44C}. 
Based on their mass, PMS stars are also  classified as T Tauri stars (TTSs) ($\lesssim$ 3 M$_\odot$) and Herbig Ae/Be stars (3-10 M$_\odot$). On the basis of the strength of H$\alpha$ emission, TTSs are further 
classified into classical TTSs (CTTSs; equivalent width (EW)$>$10\AA) 
and weak-line TTSs (WTTSs; EW$\leq$10\AA) \citep{1988cels.book.....H,1989AJ.....97.1451S}. 
CTTSs and WTTSs more or less resemble Class\,{\sc ii} and  Class\,{\sc iii} objects, respectively. 
Brightness variability is one of the distinguishing features of stars in their TTS phase \citep{1945ApJ...102..168J}. 
These variations are attributed to a combination of physical processes operating at and near their surfaces \citep{2014AJ....147...82C}. 
WTTSs tend to display periodic light curves (LCs) due to the presence of an asymmetric distribution of cool or 
dark spots which modulate the observed luminosity of the stars during its rotation  \citep{2008A&A...479..827G,2009A&A...502..883R}. 
On the other hand CTTSs display more complex signatures of different time-scales categorized as stochastic events \citep[e.g.,][]{2008MNRAS.391.1913R,2011MNRAS.410.2725S}, occasional fading/brightening \citep{2010ApJS..191..389C,2018ApJ...852...56G} 
and periodic/semi-periodic \citep{1994AJ....108.1906H} variability. The first category is caused by time variable accretion from the circumstellar disc onto the surface of the star where the accretion zones or hot spots are non-uniformly distributed \citep{2007prpl.conf..297H,2014AJ....147...82C}. In addition, CTTSs occasionally exhibit short (1-5 days, called dippers) and/or long (weeks to years, called faders) term extinction events caused when the star is occulted by disc components at or near the disc truncation radius \citep{2014AJ....147...82C,2010A&A...519A..88A,2013A&A...557A..77B,2017ApJ...840...23L,2018ApJ...852...56G}.
Like WTTSs, CTTSs can also show periodic/semi-periodic brightness variation due to either presence of hot and/or cool spots on their 
surface or any disc wrap which periodically occults the central star \citep{2007A&A...463.1017B,2015A&A...577A..11M}. 
Most of the variability in Herbig Ae/Be stars, the intermediate-mass counterpart of CTTSs, are caused due to obscuration 
by circumstellar dust \citep{2007prpl.conf..297H,2011BlgAJ..15...65S}. 
Studying variability properties of PMS stars can lead to a better understanding of the physical processes happening 
in their evolution and imposing constraints on the stellar evolutionary models.
Recent work on PMS variability can be found, e.g.,  in  \citet{2016MNRAS.462.2396F,2018A&A...619A..41T,2019MNRAS.482..658X}, but still we are not able to construct a complete paradigm for the stars and disc evolution.

In this study, we have identified PMS variables, both periodic and non-periodic, in the Sh 2-190 star-forming region and
investigate the correlations between their physical properties 
(period, amplitude, age, mass, IR-excess, accretion rate, X-ray activities, etc.). 
Then we discuss the physical processes responsible for the early evolution of the central star and the disc.
We organize this paper into six sections.
Following the introduction section, we present an overview of the Sh 2-190 star-forming region in Section \ref{sect:over}.
We describe the observation and data reduction in Section \ref{sect:obs}.
The identification of PMS variables and derivation of their physical parameters are presented in Section \ref{sect:Result}.
Physical processes responsible for the PMS variability are discussed in Section \ref{PMS}.
We conclude our results in Section \ref{conclusion}.

\begin{figure}
\centering
\includegraphics[width=0.48\textwidth]{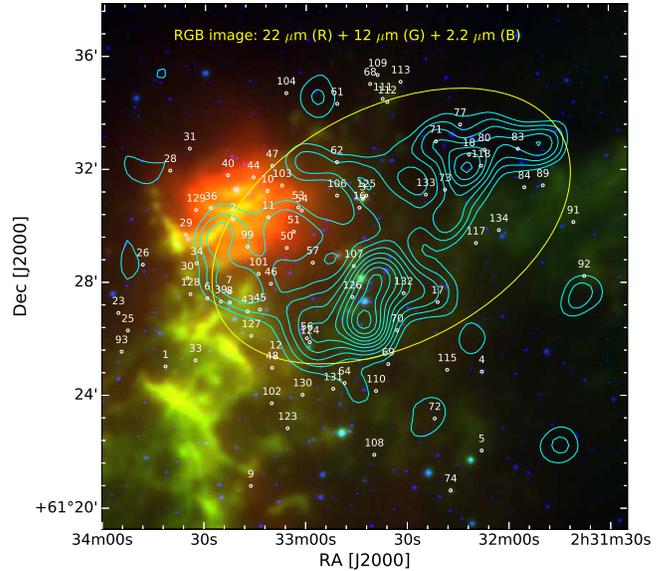}
\caption{ Color-composite image obtained by using the 2.2 $\mu$m $K_{s}$ (blue, 2MASS), 12 $\mu$m (green, $WISE$) and  22 $\mu$m (red, $WISE$)
images of the $\sim18^\prime.4\times 18^\prime.4$ FOV around the Sh 2-190 H\,{\sc ii} region.
The cyan contours represent the $K_{s}$ band stellar surface density distribution. 
The  yellow ellipse represents the extent of the cluster IC 1805.
Identified PMS variable stars are encircled  along with their identification numbers (see text for details).} 
 \label{fig:image}
\end{figure}

\section{Overview of Sh 2-190}
\label{sect:over}

Sh 2-190 is a young \citep[age $\sim$ 2.3 Myr,][]{2017MNRAS.468.2684P} star forming region located at a moderate distance \citep[2.0 kpc,][]{2013A&A...554A...3S} consisting the central cluster IC 1805 ($\alpha_{2000}$ = 02$^{h}$32$^{m}$42$^{s}$, $\delta_{2000}$ = +61$^\circ$27$^{m}$00$^{s}$) surrounded by the giant molecular cloud W4 in the Perseus spiral arm of the galaxy.  
There are nine O-type stars in IC 1805 cluster, which are considered to be the triggering
source of star-formation activities in this region \citep{2017ApJS..230....3S}. 
The infrared color-composite image of the studied region of the Sh 2-190 is shown in Figure \ref{fig:image}.
Heated dust grains (22 $\micron$ emission, red) can be seen adjacent to the stellar overdensity region.
This warm dust is also surrounded by 12 $\mu$m emission (green). $WISE$  12 $\mu$m band covers the prominent
PAH features at 11.3 $\mu$m, indicative of photon dominant region (PDR)
under the influence of feedback from massive stars \citep[see e.g.,][]{2004ApJ...613..986P}.
This fact denotes that we are dealing with a site showing signatures of recent star-formation activities.
Naturally, this region is also known to host many young stellar objects (YSOs) \citep[e.g.,][]{2017ApJS..230....3S,2017MNRAS.468.2684P}.
All of these features make this cluster an ideal site to look for PMS variable stars and to study their properties.

\begin{table*}
\centering  
\caption{Observation log for Sh 2-190 region.}
 \label{obs_log}
\begin{tabular}{@{}c@{   }   c@{       }  c@{ }c@{ }c@{}}
  \hline
Date of observation  & Telescope & N$\times$Exp.(s) &  Number of nights &  Filter\\
  \hline

30.09.2012-20.03.2013    & 0.81m Tenagara  &   230$\times$120, 240$\times$60, 246$\times$100 &82, 81, 82     &$I_{c}$,$R_{c}$,$V$   \\
                         & Telescope       &                                                 &               &   \\
28.06.2018-16.03.2019    & ZTF       &   107$\times$30,  107$\times$30                 &107, 107    &$zg$,$zr$       \\
06.12.2018               & DFOT       &   05$\times$120,               05$\times$300   & 01      &$I_{c}$    ,$V$   \\
09.01.2019               & DFOT       &   04$\times$120,               04$\times$300   & 01      &$I_{c}$    ,$V$   \\
01.02.2019               & DFOT       &   01$\times$120, 01$\times$180                 & 01      &$I_{c}$,$R_{c}$       \\
02.02.2019               & DFOT       &   11$\times$120, 11$\times$180, 11$\times$300  & 01      &$I_{c}$,$R_{c}$,$V$   \\
03.02.2019               & DFOT       &   09$\times$120, 09$\times$180, 09$\times$300  & 01     &$I_{c}$,$R_{c}$,$V$   \\
05.02.2019               & DFOT       &   02$\times$120, 02$\times$180, 02$\times$300  & 01      &$I_{c}$,$R_{c}$,$V$   \\
09.03.2019               & DFOT       &   03$\times$120, 03$\times$180, 01$\times$300  & 01      &$I_{c}$,$R_{c}$,$V$   \\
10.03.2019               & DFOT       &   03$\times$120, 03$\times$180, 01$\times$300  & 01      &$I_{c}$,$R_{c}$,$V$   \\
26.10.2019               & DFOT       &                  05$\times$180, 05$\times$300  & 01      &    $R_{c}$,$V$   \\
27.10.2019               & DFOT       &                  03$\times$180, 03$\times$300  & 01      &    $R_{c}$,$V$   \\
28.10.2019               & DFOT       &                  08$\times$180, 08$\times$300  & 01      &    $R_{c}$,$V$   \\
01.11.2019               & DFOT       &   25$\times$120, 25$\times$180, 25$\times$300  & 01      &$I_{c}$,$R_{c}$,$V$   \\
01.12.2019               & DFOT       &   27$\times$120, 27$\times$180, 27$\times$300  & 01      &$I_{c}$,$R_{c}$,$V$   \\
25.12.2019               & DFOT       &   03$\times$120, 04$\times$180, 03$\times$300  & 01      &$I_{c}$,$R_{c}$,$V$   \\
18.02.2020               & DFOT       &   04$\times$120, 04$\times$180, 04$\times$300  & 01      &$I_{c}$,$R_{c}$,$V$   \\  \hline       
\end{tabular}
\end{table*}

\section{Observation and data reduction}
\label{sect:obs}
\subsection{Optical photometric data}

Optical photometric observations of Sh 2-190  were taken in $V$, $R_{c}$, $I_{c}$ bands during September 2012 to
March 2013 with 0.81 m Tenagara Automated Telescope (South Arizona) and from December 2018 to
February 2020 using the 1.3 m Devasthal Fast Optical Telescope (DFOT, India). 
DFOT has a 2048 $\times$ 2048 pixel square CCD which covers $18^{\prime}.4 \times 18^{\prime}.4$ field of view (FOV), whereas 
Tenagara telescope covers $14^{\prime}.6 \times 14^{\prime}.6$ FOV with a 1024 $\times$ 1024 pixel square CCD. 
The FITS files of $18^{\prime}.4 \times 18^{\prime}.4$ FOV in $zg$ and $zr$ 
bands were also downloaded from Zwicky Transient Facility (ZTF) archive \citep{2019PASP..131a8003M} giving time-series images from June 2018 to March 2019. 
In total, the object was observed on 96, 94, 94, 107 and 107 nights 
in $V$, $R_{c}$, $I_{c}$, $zg$ and $zr$ bands, respectively, 
and we use 354, 344, 327, 107 and 107 frames of the five bands in the following analysis.  
Details of observation are given in Table \ref{obs_log}. 
We have used standard data reduction procedures for the image cleaning, photometry and astrometry \citep[for details, see ][]{2020MNRAS.493..267S}.  
The photometric detection from different telescopes were cross-matched using their astrometry within a search radius of 1 arcsec. 
The instrumental magnitudes in $V$, $R_{c}$, $I_{c}$ bands were converted to standard $V$, $R_{c}$ and $I_{c}$ magnitudes, also $zg$ and $zr$ magnitudes were converted to standard $R_{c}$, $I_{c}$ magnitudes using the already published photometry of stars in the same region \citep{2017ApJS..230....3S}. 

\subsection{Archival Photometric data }

For our analyses, we have also used the following archival data covering various wavelengths : \\

(i) NIR $JHKs$ photometric data were taken from  the 2MASS All-Sky Point Source Catalog
\citep{2006AJ....131.1163S,2003yCat.2246....0C}. \\

(ii) Spitzer-IRAC observations at 3.6, 4.5, 5.8 and 8.0 $\mu$m were taken from the GLIMPSE360 Catalog and Archive
\citep{2004ApJS..154....1W}. \\

(iii) MIR data at  3.4, 4.6, 12 and  22 $\mu$m were taken from  the Wide-field Infrared Survey Explorer (WISE) All-sky Survey
Data release \citep{2010AJ....140.1868W}. \\

(iv) X-ray data were taken from `The massive Star-Forming Regions Omnibus X-ray Catalog' \citep{2014yCat..22130001T}.

\section{Results and Analysis}
\label{sect:Result}

\subsection{Structure of the cluster}
\label{sect:structure}

To study the stellar surface density distribution of the Sh 2-190 region,
we generated surface density contours for a sample of stars taken from the 2MASS
All-Sky Point Source Catalog,
covering $18^{\prime}.4 \times 18^{\prime}.4$ FOV around this region \citep[for details, see ][]{2020MNRAS.493..267S}.   
These density contours are plotted in Figure \ref{fig:image} as cyan curves.
The lowest contour is 1$\sigma$ above the mean stellar density (i.e., 9.4$\pm$2.8 stars/arcmin$^2$)
and the step size is equal to 1$\sigma$ (2.8 stars/arcmin$^2$). Stellar density enhancements at 
($\alpha_{2000}$, $\delta_{2000}$) $\sim$ ($02^h32^m37^s.2$, $+61^\circ30^\prime23^{\prime \prime}.7$)
and $\sim$ ($02^h32^m22^s.5$, $+61^\circ31^\prime56^{\prime \prime}$)
can easily be seen from the contours. 
With these two peaks the shape of the cluster looks like an ellipse (shown with a yellow ellipse in Figure \ref{fig:image}) with a major 
and minor axis of 7 arcmin and 4.5 arcmin, respectively. On the basis of the radial density profile, \citet{2017MNRAS.468.2684P} have estimated the cluster 
radius as 7 arcmin which is comparable with the present estimate.

\begin{table*}
\centering
\caption{ Sample of stars identified as members in the Sh 2-190.
The complete table is available in the electronic form only.\label{PMT} }
\begin{tabular}{@{}c@{ }c@{ }c@{ }c@{ }c@{ }c@{ }c@{ }c@{ }c@{}c@{}c@{}c@{}}
\hline
ID & $\alpha_{(2000)}$ & $\delta_{(2000)}$ & $V$ & $R_{c}$ & $I_{c}$ &$\mu_\alpha$&$\mu_\delta$ &  Parallax& Probability& $G$ & $G_{BP}-G_{RP}$ \\
& {\rm $(degrees)$} & {\rm $(degrees) $} &  (mag) &  (mag) & (mag) &   (mas/yr)& (mas/yr)& (mas) & (\%) & (mag) & (mag) \\
\hline
 M1    &  38.300182   &  61.354942  & 15.046   $\pm$  0.004  & 14.516  $\pm$  0.008  & 13.968  $\pm$  0.007  & -0.66  $\pm$  0.03  & -0.85$\pm$  0.05  &  0.39$\pm$  0.03   & 97  & 14.744  &  1.182 \\
    M2 &  38.396862   &  61.390205  & 18.705   $\pm$  0.001  & 17.689  $\pm$  0.011  & 16.615  $\pm$  0.001  & -0.60  $\pm$  0.15  & -0.50$\pm$  0.19  &  0.48$\pm$  0.12   & 95  & 17.779  &  2.120 \\ 
    M3 &  38.112164   &  61.312714  & 19.969   $\pm$  0.004  & 19.005  $\pm$  0.005  & 18.074  $\pm$  0.001  & -1.12  $\pm$  0.36  & -0.49$\pm$  0.46  &  0.34$\pm$  0.30   & 84  & 19.218  &  1.683 \\ 
    M4 &  38.189682   &  61.441990  & 18.045   $\pm$  0.037  & 16.868  $\pm$  0.005  & 15.791  $\pm$  0.010  & -0.72  $\pm$  0.09  & -0.93$\pm$  0.13  &  0.41$\pm$  0.07   & 96  & 16.928  &  2.262 \\
    
\hline
\end{tabular}
\\
\hspace{-10.9 cm} $V$, $R_{c}$ and $I_{c}$ data taken from \citet{2017ApJS..230....3S}. 
\end{table*}

\subsection{Membership}
\label{membership}

We have estimated the membership probability of stars for their association with Sh 2-190 using the method described in \citet{1998A&AS..133..387B}. This method has been extensively used recently \citep [cf. ][] {2020MNRAS.493..267S,2020MNRAS.498.2309S, 2020ApJ...896...29K, 2020ApJ...891...81P}. For this, we have used $Gaia$ DR2 proper motion (PM) data of stars located within $18^{\prime}.4 \times 18^{\prime}.4$ FOV in the Sh 2-190 region. We first construct the frequency distributions of cluster stars ($\phi^\nu_c$) and field stars ($\phi^\nu_f$) using the equations 3 and 4 of \citet{1998A&AS..133..387B}.
The input parameters such as the center of Proper motion ($\mu_\alpha$cos($\delta$) = -0.75 mas yr$^{-1}$, $\mu_\delta$ = -0.70 mas yr$^{-1}$), its dispersion for the cluster stars ($\sigma_c$, $\sim$0.06 mas yr$^{-1}$) and field proper motion center ($\mu_{xf}$ = -0.003 mas yr$^{-1}$, $\mu_{yf}$ = -0.536 mas yr$^{-1}$), field intrinsic proper motion dispersion ($\sigma_{xf}$ = 3.35 mas yr$^{-1}$, $\sigma_{yf}$ = 3.21 mas yr$^{-1}$) are estimated 
similarly as have been discussed in our earlier work \citep{2020MNRAS.493..267S}.  
The membership probability (ratio of the distribution of cluster stars with all the stars) for the $i^{th}$ star is then estimated using the equation (\ref{eq1}).

\begin{equation}
P_\mu(i) = {{n_c\times\phi^\nu_c(i)}\over{n_c\times\phi^\nu_c(i)+n_f\times\phi^\nu_f(i)}}
\label{eq1}
\end{equation}

where $n_c$ (=0.12) and $n_f$ (=0.88) are the normalized numbers of stars for the cluster
and field regions ($n_c$+$n_f$ = 1), respectively.
 
\begin{figure}
\centering
\includegraphics[width=0.45\textwidth, angle= 0]{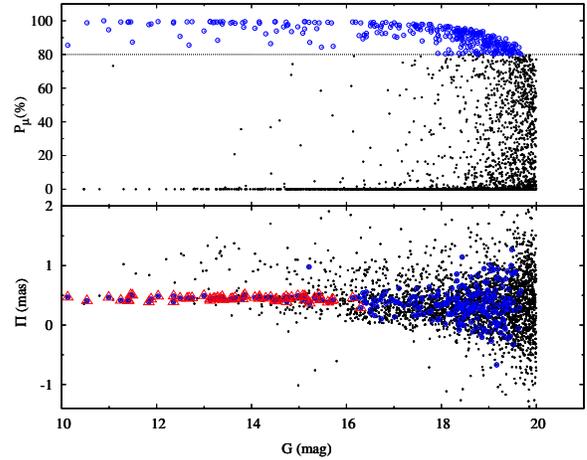}
\caption{
Membership probability (P$_\mu$) and parallax ($\Pi$) as a function of $G$ magnitude for all the stars (black dots). 
The probable member stars (P$_\mu\geq$ 80\%) are shown by blue circles in both sub-panels.
Red triangles indicate  probable member stars with parallax errors better than 0.05 mas.}
 \label{fig: VPD}
\end{figure}

The membership probability derived as above and parallax
are plotted as a function of $G$ magnitude in the top and bottom panels of Figure \ref{fig: VPD}.
 As can be seen a high membership probability (P$_\mu \geqslant$ 80\%) extends down to $G$ $\sim$ 20 mag.
The bottom panel of Figure \ref{fig: VPD} displays the parallax of the
same stars as a function of $G$ magnitude. Except few outliers, most of the stars with high membership
probability (P$_\mu \geqslant$ 80\%) follow a tight distribution.
We estimated the membership probability for 4551 stars and
found 308 stars as cluster members (P$_\mu \geqslant$ 80\%).
The details of these members are given in Table \ref{PMT}.  

\subsection{Distance and Reddening}
\label{distance and age of cluster}

We adopted the distances of the 64 identified member stars having parallax values with high accuracy
(i.e. $error < 0.05$ mas, shown as red triangles in the bottom panel of Figure \ref{fig: VPD}) from \citet{2018AJ....156...58B}. The average distance of these 
stars is $2.1\pm0.2$ kpc, comparable to that obtained by
\citet[][]{2013A&A...554A...3S} for IC 1805 (i.e., 2.0 kpc). 
We adopted the extinction value A$_{V} =$ 2.46 mag for this cluster \citep{2013A&A...554A...3S}.

\begin{table*}
\centering
\tiny
\caption{Sample of the identified PMS variables and their physical parameters as derived from CMD and/or SED. The complete table 
is available in the electronic form only. \label{Variables}} 
\begin{tabular}{@{}c@{  }c@{  }c@{  }c@{  }c@{ }c@{ }c@{ }c@{ }c@{ }c@{ }c@{ }c@{ }}
\hline
 ID  &  $\alpha_{(2000)}$ & $\delta_{(2000)}$ &  $V$   &  $R_{c}$   &  $I_{c}$  &  Period &  Amplitude  &  $\Delta(I_{c}-K_{s})$ &  $[3.6]-[4.5]$  & Age$\dagger$ \\
 {}  &  (degrees)         &  (degrees)        &  (mag) &  (mag)     &  (mag)    &  (days) &     (mag)   &   (mag)                &      (mag)      &    (Myr)     \\
   
\hline
    V1   &  38.422897   &  61.417152  & 16.038$\pm$0.007  & 15.218$\pm$0.009  & 14.511$\pm$0.007  &  1.922$\pm$0.005   &   0.1$\pm$0.1   &  0.52$\pm$0.03   &  0.04$\pm$0.01  &    8.3$\pm$0.7      \\
    V2   &  38.340141   &  61.504135  & 16.342$\pm$0.012  & 15.312$\pm$0.006  & 14.419$\pm$0.010  &  1.123$\pm$0.002   &   0.1$\pm$0.1   &  0.05$\pm$0.02   &  0.05$\pm$0.01  &    1.0$\pm$0.2       \\
    V3   &  38.179836   &  61.516354  & 15.808$\pm$0.003  & 14.836$\pm$0.010  & 13.974$\pm$0.009  &     ---        &   1.0$\pm$0.1   &  0.71$\pm$0.02   &  0.47$\pm$0.01  &    1.2$\pm$0.2       \\
    V4   &  38.033024   &  61.414288  & 16.102$\pm$0.002  & 15.211$\pm$0.008  & 14.358$\pm$0.010  &     ---        &   1.3$\pm$0.1   &  1.38$\pm$0.02   &  0.41$\pm$0.01  &    2.8$\pm$0.6       \\
\hline
    ID   & Mass$\dagger$ & Data points* &  $\chi^{2}*$ &  Age*    &  Mass*       &  Disc mass*  &  Disc accretion rate* & Log ($L_X/L_{bol}$)  &  Classification & \\
         &  (M$_\odot$)  &     {}       &      {}      &   (Myr)  &  (M$_\odot$) &  (M$_\odot$) & (M$_\odot$~$yr^{-1}$) &                      &                 & \\

\hline
    V1     &   1.8$\pm$0.0 &     10  &     6.8   &    5.5  $\pm$     2.1   &    2.7    $\pm$   0.6   &  4.7E-04   $\pm$   3.7E-03   &  1.7E-10  $\pm$  8.0E-10  &  -3.59     &      WTT  \\  
    V2     &   2.5$\pm$0.1 &     10  &     3.6   &    2.1  $\pm$     1.4   &    2.8    $\pm$   0.4   &  2.0E-03   $\pm$   7.7E-03   &  3.0E-09  $\pm$  3.8E-08  &  -3.74     &      WTT  \\ 
    V3     &   3.0$\pm$0.1 &     14  &    12.4   &    1.8  $\pm$     1.3   &    4.0    $\pm$   1.3   &  1.9E-02   $\pm$   3.3E-02   &  6.2E-08  $\pm$  2.2E-07  &  -4.19     &      CTT  \\ 
    V4     &   2.4$\pm$0.1 &     14  &    14.1   &    3.8  $\pm$     2.2   &    3.6    $\pm$   0.8   &  1.1E-03   $\pm$   6.7E-03   &  1.0E-08  $\pm$  6.2E-08  &  -3.84     &      CTT  \\ 
\hline

\end{tabular}
\\
\vspace{0.1cm}
\hspace{-8.5 cm} $\dagger$ : Parameters derived from CMD ; *  Parameters derived from SED
\end{table*}

\subsection{Identification of  variables}
\label{variable identification}

\begin{figure}
\centering
\includegraphics[width=0.48\columnwidth, angle= 0]{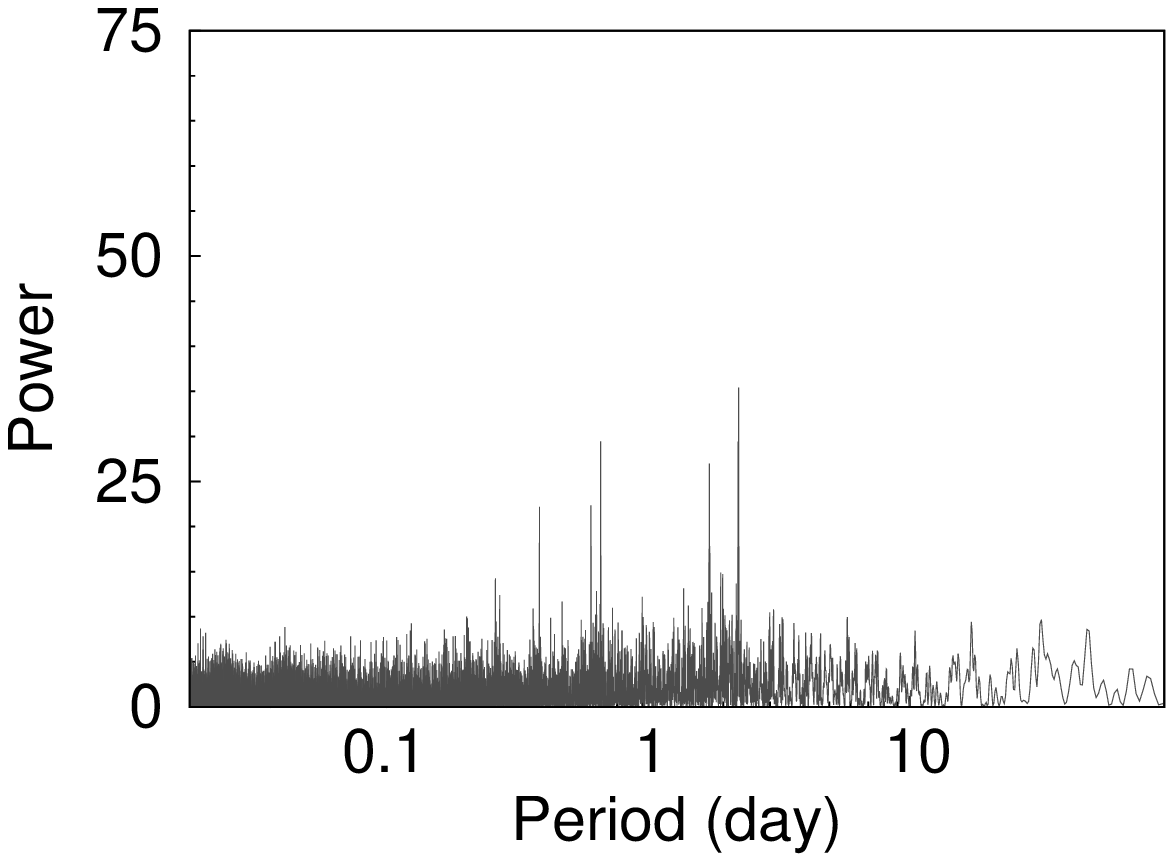}
\includegraphics[width=0.48\columnwidth, angle= 0]{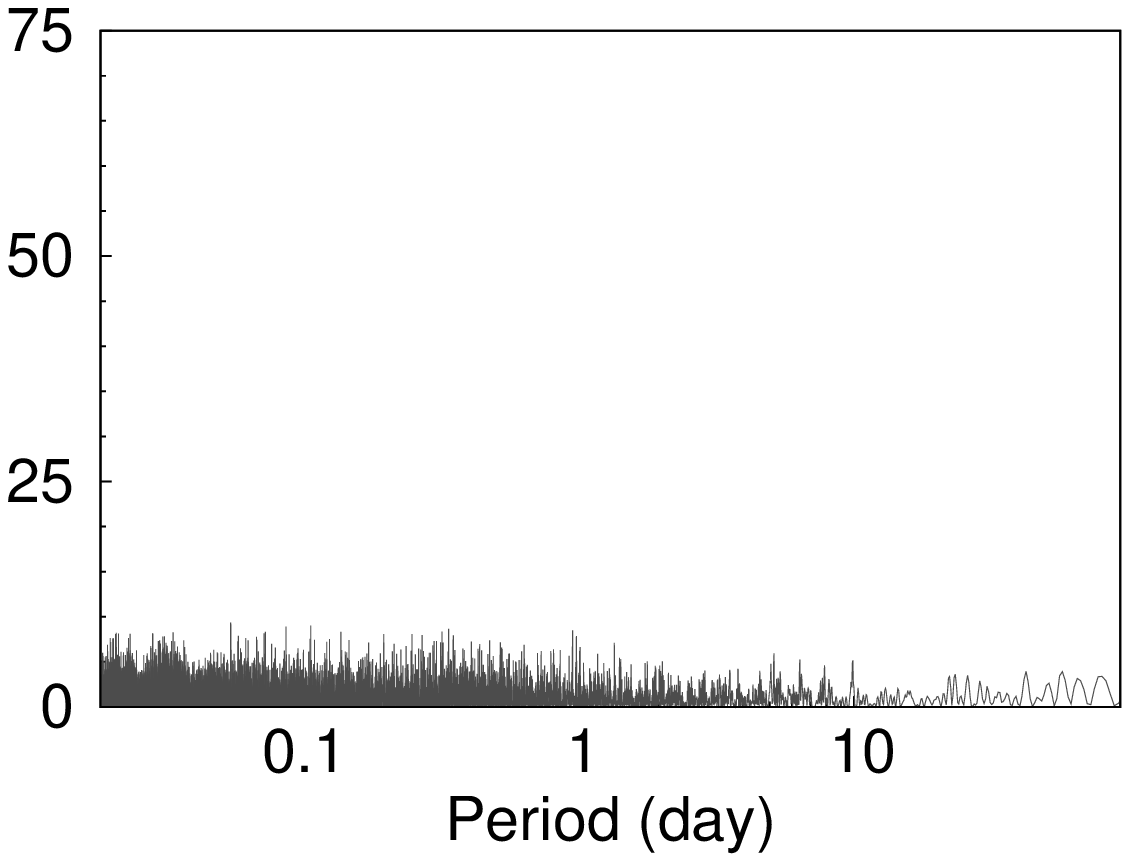} \\
\includegraphics[width= 3.3 cm,height = 4.2 cm, angle=270]{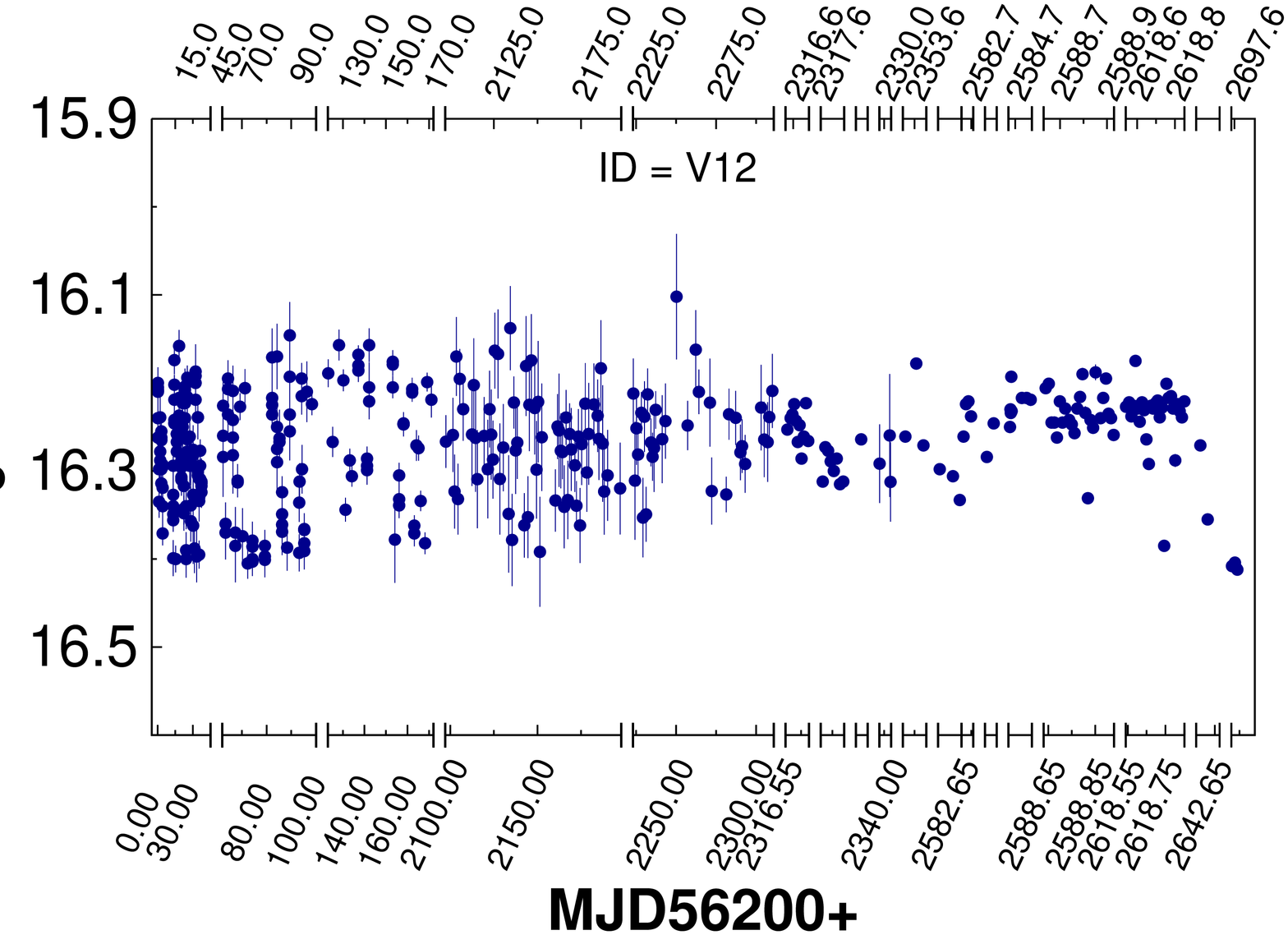}
\includegraphics[width= 3.2 cm,height = 4.2 cm, angle=270]{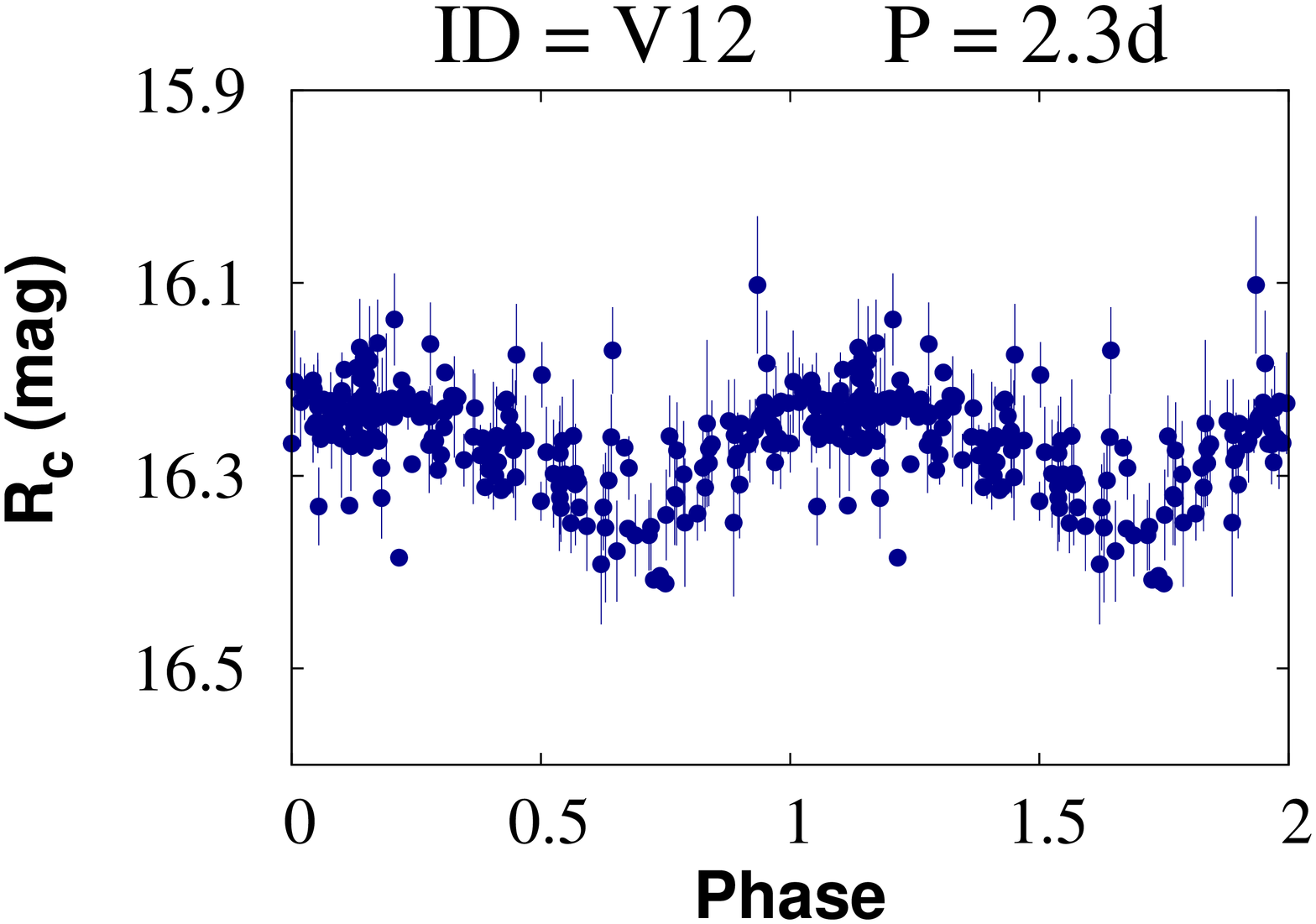}
\caption{Top left panel : Lomb-Scargle power spectrum for the star V12. Top right panel : Power spectra of its shuffled LC. The highest peak at 2.3 days is taken as the estimated period. Bottom left panel : LC of the star V12. Data gaps are represented with vertical gaps along the axis. Bottom right panel : Phase-folded LC of the star V12 with the estimated period of 2.3 days.}
 \label{fig : PS}
\end{figure}

We performed differential photometry to identify variables in the $18^{\prime}.4 \times 18^{\prime}.4$ FOV of the Sh 2-190 region \citep[for details see ][]{2020MNRAS.493..267S}. 
All the stars detected in our imaging survey (3461) are used in this analysis.
Phase-folded LCs based on the period estimation from the \textit{Period\footnote{http://www.starlink.rl.ac.uk/docs/sun167.htx/sun167.html}} software using 
the Lomb-Scargle (LS) periodogram \citep{1976Ap&SS..39..447L,1982ApJ...263..835S}
were used to identify the periodic variables \citep[see for details, ][]{2020MNRAS.493..267S}. 
The uncertainty in period estimation ($\delta$P) can be determined from the full width at half maximum (FWHM)  of the main peak of the window function ($\nu_{FWHM}$),
i.e, $\delta$P $=$ $\nu_{FWHM}\times$P$^{2}$  \citep[see for details,][]{2004A&A...417..557L}.
For the period range of 0.5d - 5d, the estimated errors are found to be in between $\sim$ 0.001d - 0.02d.
To verify whether the peak corresponding to the estimated period in the LS power spectrum has significant signal, 
it has been related to a false alarm probability (FAP). FAP is a probability 
of finding a peak  similar to the estimated period in a random data set \citep{2009A&A...502..883R}. 
To estimate this, we have reshuffled the original LC of the periodic variable and generated its LS power spectra.
The maximum peak in this power spectra is then compared with the original one.
This was repeated 1000 times for each of the peoriodic variables to extimate their FAP, i.e,
for FAP = 1\%, the power of 10 randomized LCs has exceeded the power of the original LC.
All the periodic variables identified in this study have FAP $<$ 0.1\%. 
As an example, we show the power spectra of the periodic variable V12 before and after shuffling its
LC in the top-left and top-right panels of Figure \ref {fig : PS}, respectively. 
Clearly, there is no significant peak in the power spectra of the shuffled LC.  
The peak with maximum power is found at 2.3 days for this variable. 
The time-series and the phase-folded LCs of the star V12 are also shown in the bottom-left and 
bottom-right panels of Figure \ref {fig : PS}, respectively.
We also used the
\textit {NASA Exoplanet Archive Periodogram service\footnote{https://exoplanetarchive.ipac.caltech.edu/cgi-bin/Pgram/nph-pgram}}
and PERIOD04\footnote{http://www.univie.ac.at/tops/Period04} \citep{2005CoAst.146...53L} to further verify the periods 
of these stars. Once the periodic variables were identified, the rest of the LCs were  visually inspected to identify the 
non-periodic variables 
on the basis of their systematic variation larger than the scatter in the LC of the comparison star. 

\begin{figure}
\centering
 \includegraphics[width=0.85\columnwidth, angle=0]{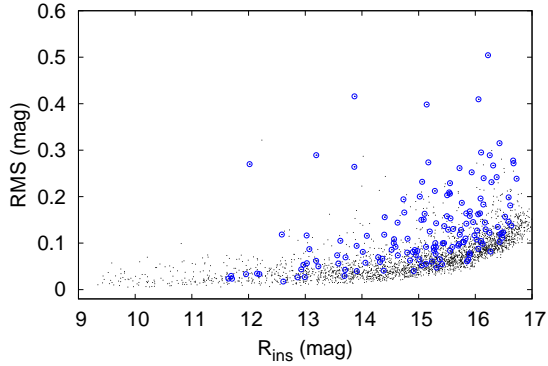}
 \caption{The RMS dispersion of instrumental magnitude as a function of mean instrumental magnitude
of the stars in the Sh 2-190 region (black dots). Open circles are the identified variable stars.}
 \label{fig : RMS}
\end{figure}

In Figure \ref{fig : RMS} we show the RMS dispersion of all the target stars as a function of mean instrumental magnitude. 
The dispersion increases towards the fainter end due to the increase in photometric uncertainty. Identified variables are shown with blue open circles.
Despite having a very high RMS value, some of the stars in Figure \ref{fig : RMS} are not designated as variables due to unusually high photometric errors (bad pixels, the bright background of the nearby star and residual from the cosmic corrections could be possible reasons) compared to the stars of the same magnitude bin. Once again these stars were visually checked to ascertain the variability.

\begin{figure}[h]
\centering
\includegraphics[width=0.99\columnwidth, angle= 0]{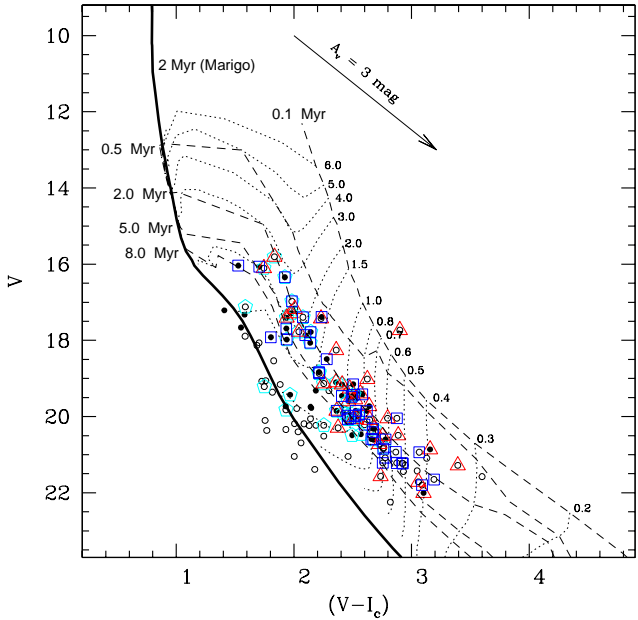}
\caption{$V/(V-I_{c})$ CMD for the variable stars in the Sh 2-190 region. 
Filled and open circles represent periodic and non-periodic variables, respectively. 
Cyan pentagons represent variable stars with membership probability higher than 80\% as derived from the PM data. 
Class\,{\sc ii} and Class\,{\sc iii} YSOs are plotted with red triangle and blue square symbols, respectively. 
The black thick curve shows the 2 Myr post-MS isochrone for the solar metallicity obtained from \citet{2008A&A...482..883M}, while the black 
dashed and dotted curves represent the PMS isochrones for various ages and evolutionary tracks for various masses, respectively \citep{2000A&A...358..593S}.  
All the isochrones and evolutionary tracks are corrected for the distance (2.1 kpc) and reddening (A$_{V} =$ 2.46 mag).
The arrow indicates the reddening vector.}
 \label{fig : V_CMD}
\end{figure}

By adopting the aforementioned procedure, we identified 134  variables 
(57 as  periodic and 77 as non-periodic) from a sample of 3461 stars. 
Remaining stars  are considered as non-variables in the present analysis. 
The period and amplitude of the variables range between 12 hrs-80 days and 0.1-2.2 mag, respectively.
The catalogue by \citet{2017MNRAS.468.2684P} has been used to cross-identify 85 variables as
probable PMS stars. A 1-arcsec search radius was adopted for the cross identification. 
Thirty seven and forty eight PMS variables as per classification by \citet{2017MNRAS.468.2684P} are found to 
be Class\,{\sc ii}  and  Class\,{\sc iii} sources, respectively. 
Almost half of the PMS variables (i.e., 45) are found to be periodic in nature and the rest have irregular brightness 
variation. The identification number, coordinates, period and other parameters of these PMS variables are 
listed in Table \ref{Variables}.
The remaining 49 variables could  be MS/field stars and their LC will be discussed in a separate study.

\subsection{Determination of Physical Parameters of the variables}
\subsubsection{Through HR diagram}
\label{HR_diagram}

Figure \ref{fig : V_CMD} shows  the $V/(V-I_{c})$ color-magnitude diagram (CMD) for all the identified variables along with the post-MS and PMS isochrones. 
We have detected variability down to a very deep magnitude limit of $V\sim22$ mag.
The Class \,{\sc ii} and Class \,{\sc iii} PMS variables are indicated by red triangle and blue square symbols, respectively. 
The location of the 36 variables identified as members of the Sh 2-190 region (cf. Section \ref{membership}) are also shown with cyan pentagons in Figure \ref{fig : V_CMD}.
Clearly, almost all of the variables having counterparts in the published catalog of YSOs  \citep{2017MNRAS.468.2684P}
are located above the MS, as expected for PMS objects, on the CMD. With a few exceptions at fainter magnitudes where the parallax errors are high, a majority of the members also fall in the PMS phase. These confirm that our variables are in the PMS phase.

\begin{figure}[h]
\centering
\includegraphics[width=0.45\columnwidth, angle= 0]{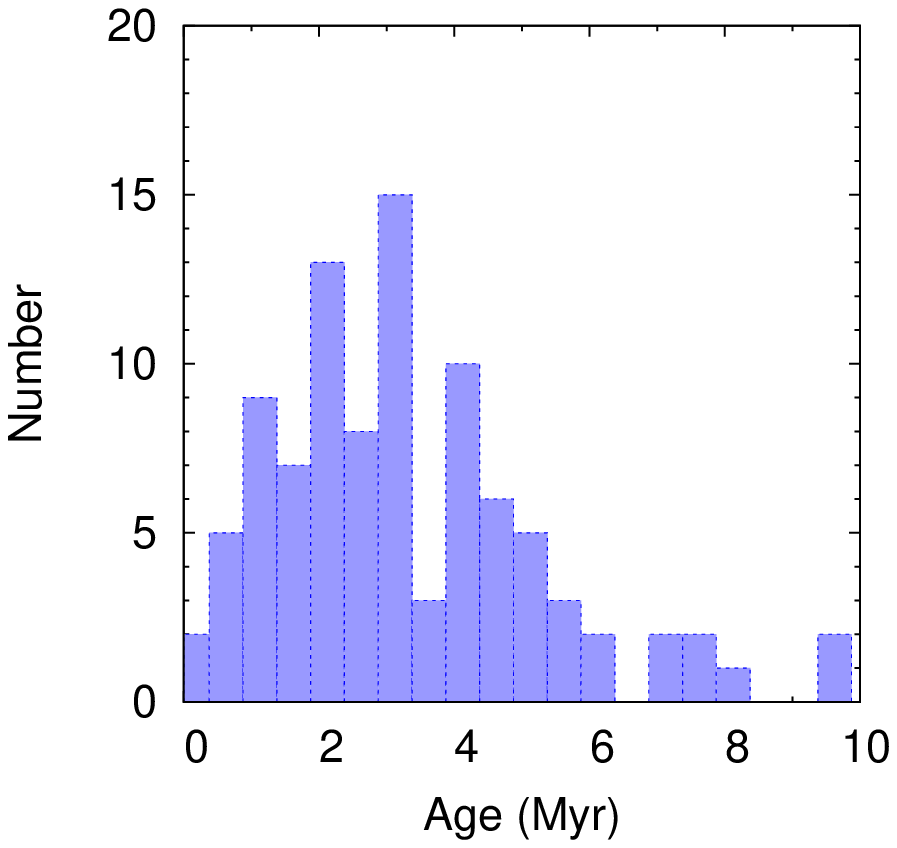}
\hspace{0.02 cm}
\includegraphics[width=0.415\columnwidth, angle= 0]{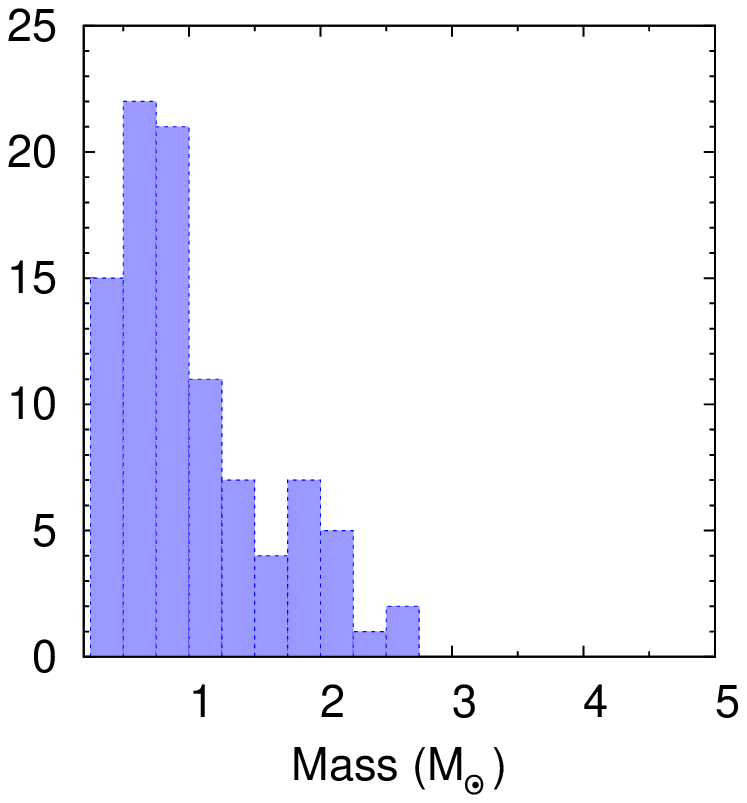}
\caption{Age and mass frequency distribution for PMS variables.}
 \label{fig: Hist_age_mass}
\end{figure}

We have estimated the age and mass of each PMS variable from
their position in the CMD \citep[for details, cf. ][]{2009MNRAS.396..964C, 2017MNRAS.467.2943S} and results are listed
in the Table \ref{Variables}.  Figure \ref{fig: Hist_age_mass} shows the age and mass distribution of these PMS variables. 
The age distribution shows a peak around 3 Myr and an age spread of up to 6 Myr with some objects as old as 10 Myr. The
average age and mass of the 85 PMS variables associated with Sh 2-190 are estimated to be $2.6\pm1.5$ Myrs and $0.9\pm0.6$  M$_\odot$, respectively.

\subsubsection{Through Spectral energy distribution}
\label{SED}

To characterize and understand the physical nature of the PMS variables, we investigated their SEDs using the grid models and fitting tools of
\citet{2003ApJ...598.1079W, 2003ApJ...591.1049W, 2004ApJ...617.1177W} and \citet{2006ApJS..167..256R, 2007ApJS..169..328R}, respectively.
This method has been widely used in our previous studies \citep[e.g.,][and references therein]{2020MNRAS.493..267S,2020MNRAS.498.2309S, 2020ApJ...896...29K, 2020ApJ...891...81P}. 
We fitted the SEDs of 84 PMS variables for which at least five band photometric data are available among the
optical, NIR (2MASS) and MIR (WISE, $Spitzer$) data.
The SED fitting tool fits each of the models to the data allowing the distance and extinction as free parameters.
The input  distance range  of the IC 1805 region is taken as 1.9 - 2.3 kpc keeping in mind the error 
associated with distance estimation (see Section \ref{distance and age of cluster}).
As this region is highly nebulous, we varied the $A_V$ by putting foreground reddening of the
IC 1805 region as a lower limit (2.5 mag). For the upper limit, we put a very large $A_V$ value, i.e., 30 mag, to accommodate embedded sources
\citep[See also,][]{2017MNRAS.467.2943S,2012ApJ...755...20S,2013MNRAS.432.3445J,2014MNRAS.443.1614P}. 
Instead of the formal errors in the catalog we set photometric uncertainties of 10\% for the optical and 20\% for both the NIR and MIR data to 
avoid any possible biases caused by underestimated flux uncertainties during fitting. 
We have used the relative probability distribution for the stages of all the `well-fit' models to obtain the 
physical parameters of the PMS variables. The well-fit models for each source
are defined by $\chi^{2}- \chi^{2}_{min}$ $\leq$ 2$N_{data}$, where $\chi^{2}_{min}$ is the goodness-of-fit parameter 
for the best-fit model and $N_{data}$ is the number of input data points.

\begin{figure}
\centering
\includegraphics[width=0.7\columnwidth, angle= 0]{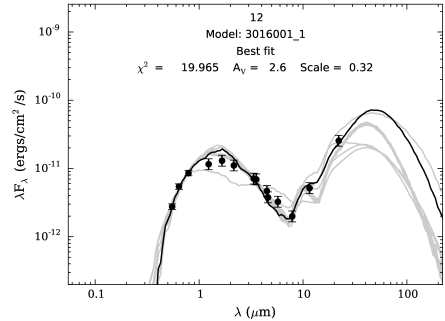}
\caption{Sample SED of PMS variable  created by the SED fitting tools of
\citet{2007ApJS..169..328R}. The black curve shows the best fit and the grey curves show the subsequent
well fits. The filled circles with error bars denote the input flux values.}
 \label{fig:SED}
\end{figure}

In Figure \ref{fig:SED}, we show the sample SED for a PMS variable 
where the solid black curves represent the best fit and the grey
curves are the subsequent good fits. From the well-fit models for each source derived from the
SED fitting tool, we calculated the $\chi^2$ weighted model parameters
such as mass, age, accretion rate, disc mass etc. of the PMS variables
\citep[see for details,][]{2020MNRAS.493..267S} and are listed in Table \ref{Variables}. 
Here we would like to mention that the disc mass obtained from SED fitting may not account for all the mass of the circumstellar 
disc as we do not have photometric data that well covers all the disc extension emissions. Thus, this estimation could be a lower limit of the disc mass. 
The average age and mass of the 84 PMS variables are found to 
be $2.7\pm1.6$ Myrs and $2.1\pm1.2$ M$_\odot$,  respectively. These parameters from the SEDs are comparable within errors to those 
obtained from the CMD ($2.6\pm1.5$ Myrs and $0.9\pm0.6$  M$_\odot$; cf. Section \ref{HR_diagram}).
These estimated physical parameters also indicate that the PMS variable are low mass T Tauri stars.

\subsection{Disc indicators}
\label{disc_indicators}
  
We have used $\Delta (I_{c}-K_{s})$ and $([3.6]-[4.5])$  indices as disc indicators for the present study \citep[for details, cf.][]{2020MNRAS.493..267S}. 
These color indices are sensitive to the inner and outer part of the stellar disc, respectively  \citep{2000AJ....120.3162L,2010A&A...515A..13R}. 
The index $\Delta (I_{c}-K_{s})$ is expressed as \citep[cf.][]{1998AJ....116.1816H}:

\begin{equation}
\Delta(I_{c}-K_{s}) = (I_{c}-K_{s})_{obs} - (A_{I_{c}} - A_{K_{s}}) - (I_{c}-K_{s})_{0} 
\end{equation}

where $(I_{c}-K_{s})_{obs}$ is the observed color of the star, $(I_{c}-K_{s})_{0}$ is its intrinsic color  and $A_{I_{c}}$ and
$A_{K_{s}}$ are interstellar extinctions in the $I_{c}$ and $K_{s}$ bands, respectively. The extinction is normal towards this direction with A$_{V} =$ 2.46 mag \citep{1990ApJ...353..174S,1993AJ....106.1947H,1983JRASC..77...40J,2013A&A...554A...3S}.
The intrinsic color, $(I-K)_{0}$, were taken from the PMS isochrones of \citet{2000A&A...358..593S}.

\subsection{X-ray luminosity}

We converted the X-ray count rates provided in \citet{2014yCat..22130001T} of the identified PMS variables
to the X-ray fluxes using the {\scriptsize MEKAL} model of PIMMS software\footnote{Distributed by NASA’s High Energy Astrophysics Science Research Center; http://heasarc.gsfc.nasa.gov/docs/software/tools/pimms.html.}. The required input parameters in this model are the temperature of the emitting coronal gas ({\textit k}T), 
hydrogen column density ($N_{H}$) along the direction of the source and abundance of the PMS stars. 
We adopted a thermal plasma model with {\textit {k}}T = 2.7{\textit {k}}eV and 
the abundance 0.4$\ast$Z$_\odot$, that are typical for PMS stars \citep{2005ApJS..160..401P}. The extinction value of A$_{V} =$ 2.46 mag
towards Sh 2-190 (cf. Section \ref{distance and age of cluster}) yields a hydrogen column density 
$N_{H}$ = 5.48$\times$10$^{21}$ cm$^{-2}$ \citep{2009MNRAS.400.2050G}.
Finally, the PIMMS X-ray fluxes were corrected for a distance of 2.1 Kpc to get the $L_X$ of individual stars. 
The $L_{bol}$ of the PMS variables were taken from the PMS isochrones of \citet{2000A&A...358..593S} according to their age and mass as derived by their CMD. 
The uncertainty in this approach could be due to
potential differences in {\textit {k}}T  or A$_{V}$ in individual stars.

\begin{figure}[h]
\centering 
\includegraphics[width= 4.4 cm,height = 7.7 cm, angle= 270]{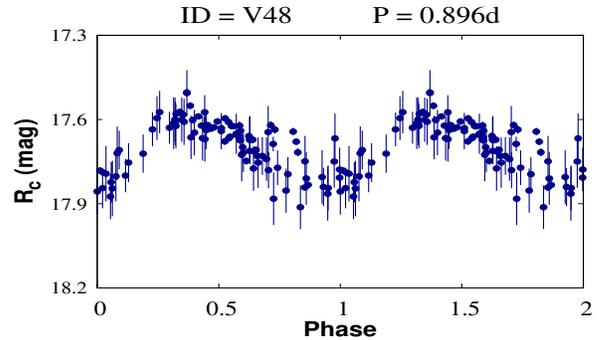}
\caption{Phase folded LCs of 17 Class\,{\sc ii} periodic variables.
The identification numbers and periods (days) of the corresponding stars are given on the top of
each panel. Complete set of LCs are provided in the electronic form only.}
 \label{Class_II_P}
\end{figure}

\begin{figure}
\centering
\includegraphics[width= 5.0 cm,height = 8.0 cm, angle=270]{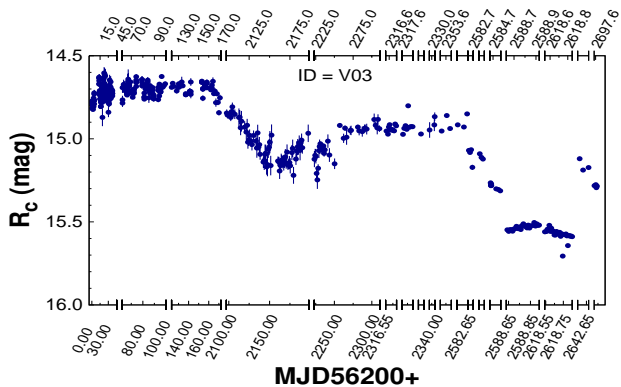}
\caption{LCs of 20 Class\,{\sc ii} non-periodic variables (CTTSs). When there are data gaps they are represented with vertical gaps along the axis. Corresponding identification numbers are given in each panel. A complete set of LCs are provided in the electronic form only.} 
 \label{class_II_NP}
\end{figure}

\section{Discussion}
\label{PMS}

In this study, we will be focusing  on the 85 identified PMS variables. We will discuss their LCs and the physical processes responsible for
their variability in the ensuing sections.

\subsection{Light curves of PMS variables}
\label{LC PMS}
We will discuss and compare the LCs of 37 Class \,{\sc ii} and 48 Class \,{\sc iii} variables in the following sub-sections.

\subsubsection{Class \,{\sc ii} variables}
\label{CTT}

\begin{figure}[h]
\centering
\includegraphics[width= 5 cm,height = 8 cm, angle=270]{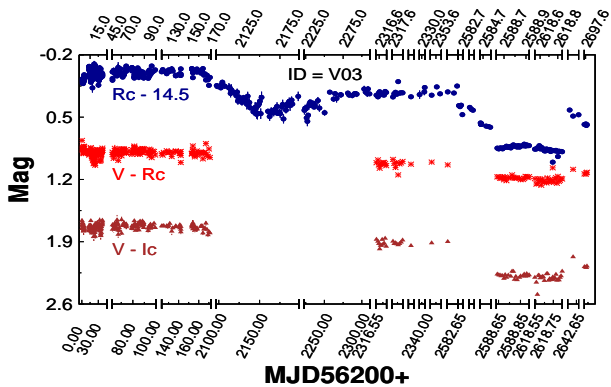}
\caption{Color evolution of a few selected CTTSs variables. The R$_{c}$ band LC, V-R$_{c}$ and V-I$_{c}$ color curves are shown with blue filled circle, red stars and brown triangles, respectively. When there are data gaps they are represented with vertical gaps along the axis. A complete set of LCs is provided in the electronic form only.}
\label{Fig: color_1}
\end{figure}

The LCs of 17 Class\,{\sc ii} periodic variables are shown in Figure  \ref{Class_II_P}. 
The period of these variables ranges from 0.9 days to 24 days with a median value of
3.1 days. Almost half of these variables have a period longer than 5 days. 
Longer periods are common in the stars with circumstellar disc \citep[e.g.,][]{1993AJ....106..372E,2013MNRAS.430.1433A}.
The LCs of 20 non-periodic Class\,{\sc ii} variables are shown in Figure \ref{class_II_NP}.
The amplitude of these variables ranges from 0.25 to 2.2 mag with a median value of 0.81 mag.
Almost half of these variables have amplitude larger than 1 mag.
In the LCs of V03 and V04, there are huge ($\sim$ 0.5-1.0 mag) dips indicative of occultation events that span for months. 
These occultation events are more sensitive at shorter wavelengths as can be seen in the  Figure \ref{Fig: color_1}. 
Although there are large time gaps in some photometric bands, around MJD 2150+56200 and MJD 2588+56200 for V03 and around MJD 75+56200 and MJD 130+56200 days for V04, 
we note that as the star gets fainter their colors also become redder, which can be seen from both $V$-$I_{c}$ and $V$-$R_{c}$ colors. 
These events could be due to dust occultations. Apart from the occultation events, these LCs are mostly stable with some minute variations.
In some other LCs (e.g., V08, V10, V62, V74 and V80) also, we see significant dips but due to lack of continuous data it is difficult to comment on their nature. 
In the cases of V62, V74 and V80, there are strong inter-day variations along with systematic increasing and decreasing trends at different epochs. 
The peak-to-trough amplitude of variations in these stars are $>$ 1.2 mag with V62 having 2.2 mag variation. 
V40, V51, V91 and V111 also show a daily variation of the ``stochastic" nature, which may be caused by
superposition of variable extinction and stochastic accretion \citep{2014AJ....147...82C}. 
In contrast, V30, V31, V93, V101 and V107  show an increase or decrease in the brightness throughout or in some part of their LCs. The above mentioned characteristics are typical of CTTSs \citep[e.g.,][]{2014ApJS..211....3P}.
These 20 Class\,{\sc ii} non-periodic variables are thus further classified as CTTSs \citep[see also,][]{2019A&A...627A.135B}. These classifications are given in Table \ref{Variables}.

\vspace{0.5 cm}

\begin{figure}
\vspace{0.15 cm}
\centering         
\includegraphics[width= 4.4 cm,height = 7.7 cm, angle= 270]{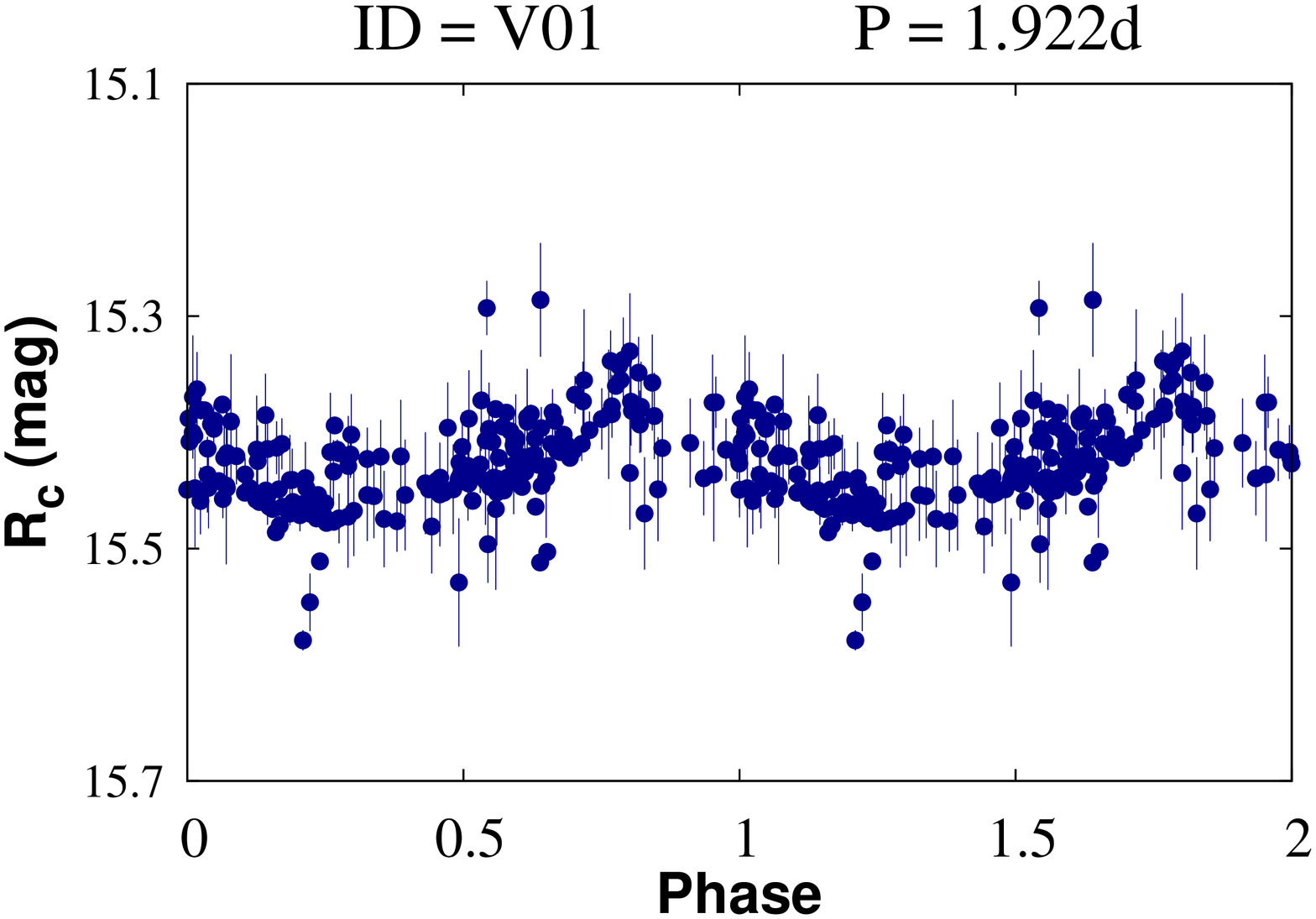}
\caption{Phase folded LCs of 28 Class\,{\sc iii} periodic  variables (WTTSs).
The identification numbers and periods (days) of the corresponding stars are given on the top of
each panel. A complete set of LCs is provided in the electronic form only.}
 \label{Class_III_P}
\end{figure}

\begin{figure}
\centering
\includegraphics[width= 5.0 cm,height = 8.0 cm, angle=270]{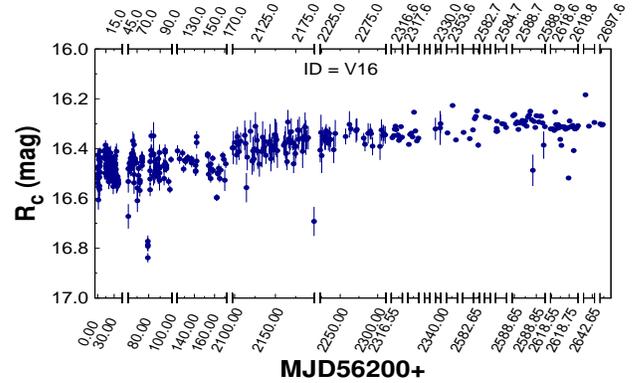}
\caption{ LCs of 20 Class\,{\sc iii} non-periodic variables. When there are data gaps they are represented with vertical gaps along the axis.
Corresponding identification numbers are given in each panel. A complete set of LCs are provided in the electronic form only.}
\label{Class_III_NP}
\end{figure}

\subsubsection{Class \,{\sc iii} variables}

The LCs of the 28 Class\,{\sc iii} periodic variables are shown in the Figure  \ref{Class_III_P}.
The period of these sources ranges between 0.5 days to 24 days with a median value of 1.6 days.
Twenty one (75\%) out of the twenty eight stars have period less than 5 days.
The Class\,{\sc iii} sources are showing amplitude variations in the range $\sim$0.1-0.7 mag and a majority of them (21 i.e., 75\%) have amplitudes $<$ 0.4 mag.
These characteristics are typical of WTTSs where variability is mostly caused by the irregular distribution of cool spots on the stellar surface. Hence,
these 28 variables are further classified as WTTSs \citep[see also,][]{2019A&A...627A.135B}. These classifications are given in Table \ref{Variables}.
The LCs of the 20 non-periodic Class\,{\sc iii}  variables are shown in the Figure \ref{Class_III_NP}. 
Their amplitudes range from 0.26 to 1 mag with median value of 0.55 mag. More than 63\% (i.e. 12) of these variables have amplitudes
$\leq$ 0.7 mag.

\subsubsection{Comparison between the light curves of Class \,{\sc ii} and Class \,{\sc iii} variables}

From the previous sub-sections, it is evident that 
Class\,{\sc ii} sources have in general a longer period as compared to the Class\,{\sc iii} sources.
Class\,{\sc iii} sources are more evolved and have less disc material around these stars. 
According to the disc-locking model, this allows Class\,{\sc iii} sources to spin-up freely without any regulation imposed by the disc \citep[][]{1991ApJ...370L..39K,1994ApJ...429..797S}. 

\begin{figure}[h]
\centering
\includegraphics[width=0.85\columnwidth, angle= 0]{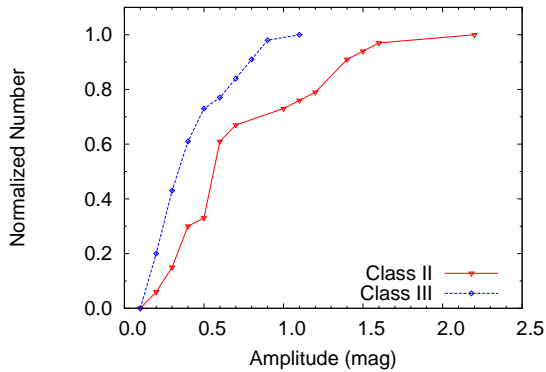}
\caption{Normalized cumulative amplitude distribution of Class\,{\sc ii} and Class\,{\sc iii} variables. }
 \label{fig: CAD}
\end{figure}

The strength of the flux variation depends on the related physical mechanism.
The flux variations in Class\,{\sc ii} and Class\,{\sc iii}  objects are mostly governed by the disc 
phenomenon (accretion, extinction etc.) and spot modulation, respectively. In general, these different mechanisms  cause the Class\,{\sc ii} objects to vary with larger amplitude relative to the Class\,{\sc iii} objects. In our sample, the amplitude in Class\,{\sc ii} sources ranges from 0.18 to 2.2 mag with a  median 0.58 mag while Class\,{\sc iii} sources have amplitudes in the range 0.11 to 1.0 mag with a  median amplitude 0.36 mag. 
To statistically verify whether the Class\,{\sc ii} and Class\,{\sc iii} objects differ in terms of their variability amplitude,
we plotted the normalized cumulative amplitude distribution of these sources in Figure \ref{fig: CAD}.
The figure manifests that Class\,{\sc ii}  variables 
have larger amplitudes of variation as compared to the Class\,{\sc iii} sources with 99\% confidence level as calculated by the Kolmogorov-Smirnov test. 
Similar trends have been found previously for the variables identified in the IC 5070 \citep{2019A&A...627A.135B} and Sh 2-170 \citep{2020MNRAS.493..267S} star-forming regions.
Also, the LCs of Class\,{\sc ii}  variables show more dynamic and different temporal features which include either a single or a combination of the phenomenon 
like short duration fadings, long term dippers induced by circumstellar extinctions, sharp increases in luminosity (bursters) and 
stochastic changes in brightness caused by variable accretion \citep{2014AJ....147...82C}. 

\subsection {Role of disc in stellar variability}
\label{disclocking}

\citet{1993AJ....106..372E} have found that stars with longer periods are surrounded by an accretion
disc while the stars lacking accretion disc are predominantly fast rotators. This distribution can be 
explained if the stellar angular momentum is regulated during the disc accretion phase by a mechanism that 
balances the spin-up torque applied by the accretion of high
angular momentum material from the disc. The magnetic star-disc interaction between the stellar magnetosphere and the circumstellar 
disc has been proposed to explain this effective removal of angular momentum from PMS stars during the first $\sim$ 10 Myr of their evolution \citep[i.e., disc-locking,][]{1991ApJ...370L..39K,1994ApJ...429..797S,1995RMxAC...1..293N,1995ApJ...447..813O}.

\begin{figure}[h]
\centering
\includegraphics[width=0.24\textwidth]{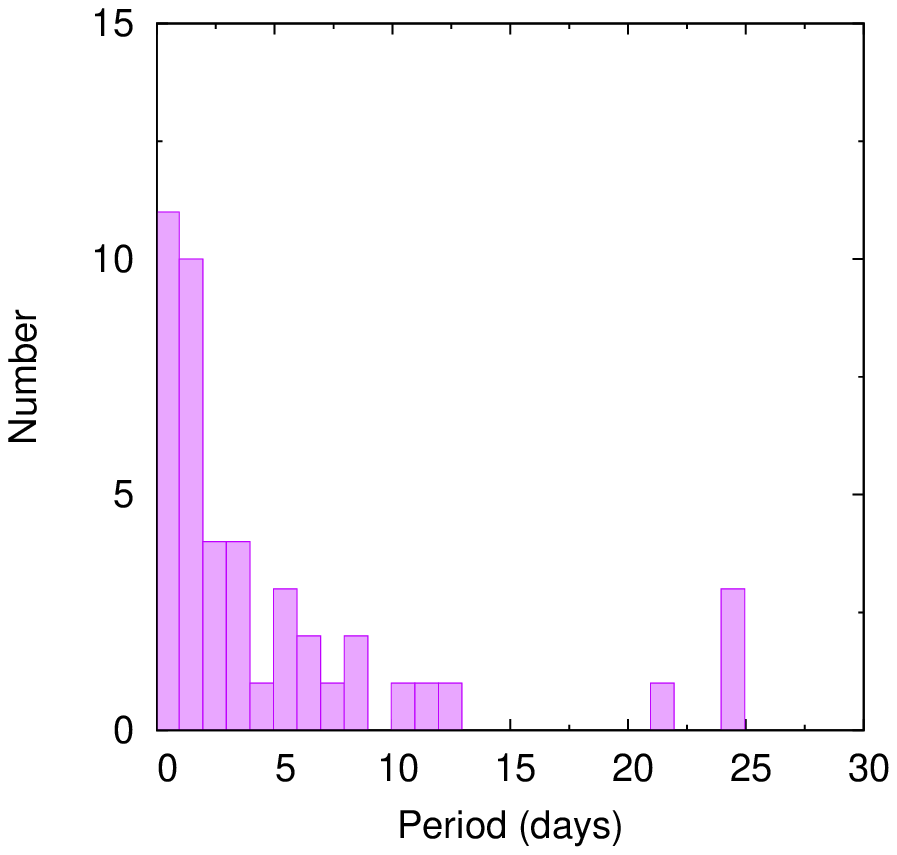}
\includegraphics[width=0.223\textwidth]{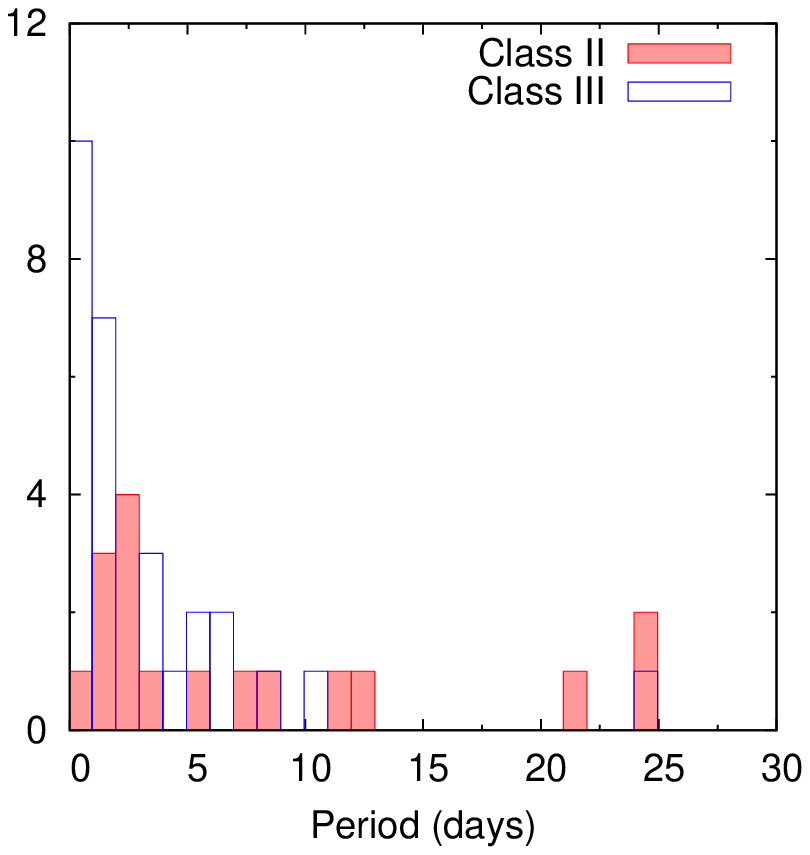}
 \caption{Left panel: Period distribution of all periodic PMS variables. Right panel:
 Period distribution of Class\,{\sc ii} and Class\,{\sc iii} variables.}
 \label{fig : period distribution}
\end{figure}

The disc-locking mechanism can be verified from the bimodal
distribution of variability period in young stars as the disc-locked
slow rotators can explain the separate period distribution from the
usual ones \citep[][]{1991ApJ...370L..39K,1994ApJ...429..797S}.
However, there have been several conflicting pieces of evidence regarding this, for example, 
some authors \citep[e.g.,][]{2002A&A...396..513H,2005A&A...430.1005L,2020MNRAS.493..267S} have found the bimodal period distribution which  favors the disc-locking mechanism, while others  \citep{2004AJ....127.2228M} have found the unimodal period distribution. 
To further test this mechanism, in Figure \ref{fig : period distribution}, we show the period distribution for all 45 periodic PMS variables 
(with 17 Class\,{\sc iii} and 28 Class\,{\sc iii} sources presented separately on the right)
identified in the Sh 2-190 region. Although the distribution looks unimodal for all the periodic PMS variables, we can see different period distributions for the Class\,{\sc ii} and Class\,{\sc iii} sources. While the Class\,{\sc iii} sources show a peak period of around 1 day, the Class\,{\sc ii} sources have a peak of around 3 days. 
Besides the small number of stars in the sample, we have performed Kolmogorov-Smirnov (KS) test to the period distribution of the Class\,{\sc ii} and Class\,{\sc iii} objects. 
It returned a $p$-value of 0.16 indicating that the probability of these objects representing different population is 84\%. 
Longer periods for the younger objects hint towards the disc-locking mechanism. 
A statistically significant sample of objects is required to draw a definitive conclusion.

\begin{figure*}
\centering
 
 \vspace{0.1 cm}
  \includegraphics[width=6.2cm, height=4cm, angle=0]{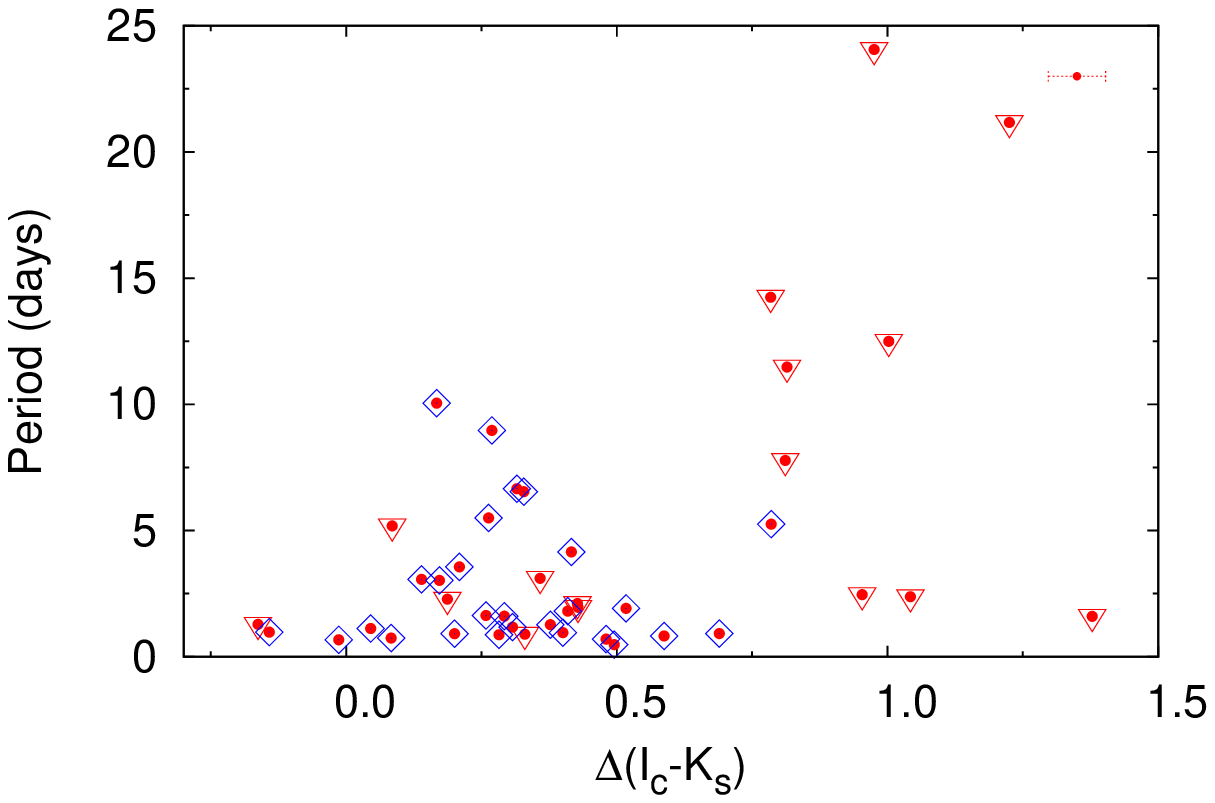}
 \hspace{0.6 cm}
 \includegraphics[width=6cm, height=4cm, angle=0]{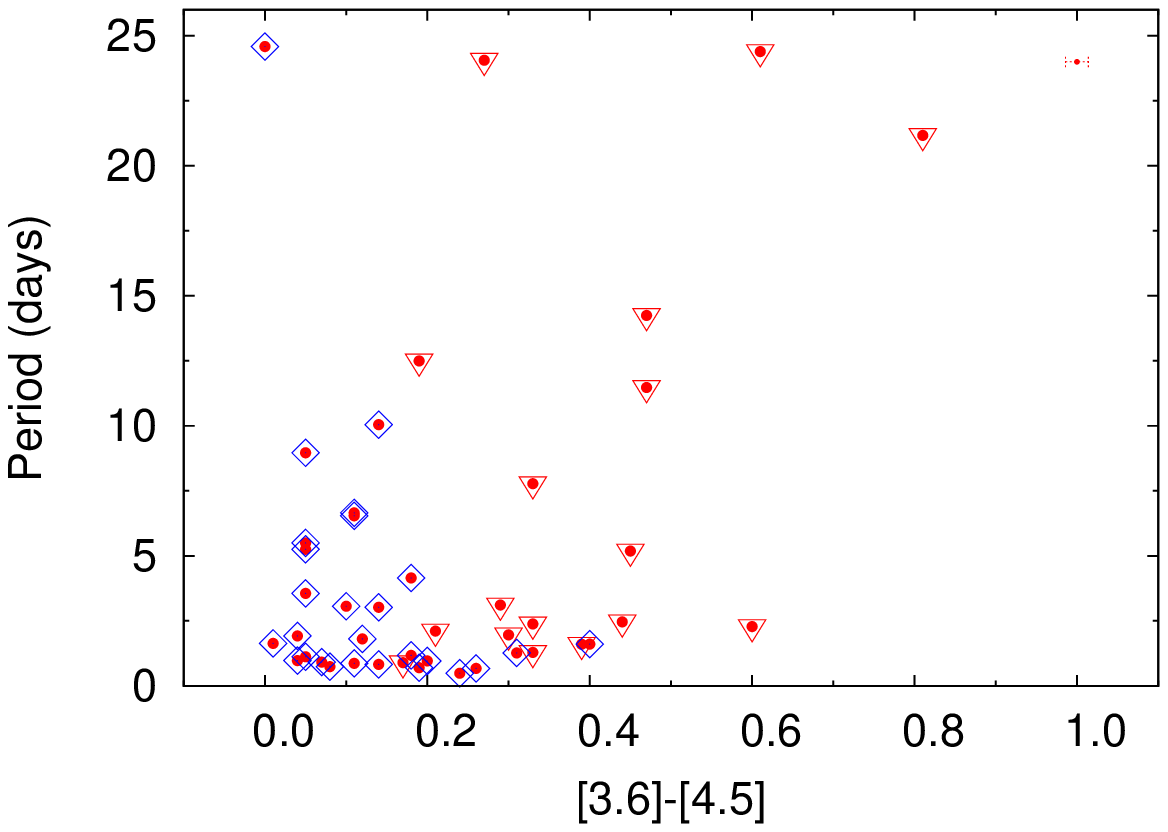} \\
 \vspace{0.2cm}
 \hspace{-0.2 cm}
 \includegraphics[width=6.0cm, height=4.0cm, angle=0]{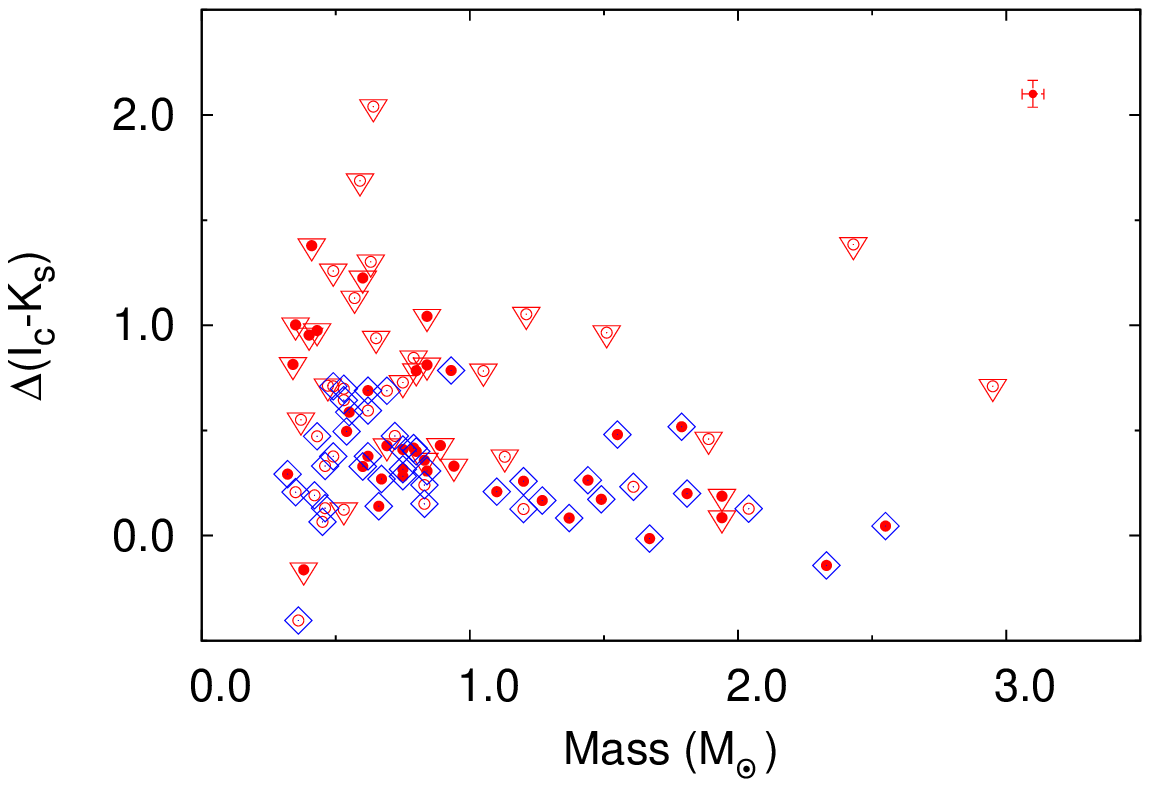}
 \hspace{0.9 cm}
 \vspace{-0.2cm}
 \includegraphics[width=6cm, height=3.9cm, angle=0]{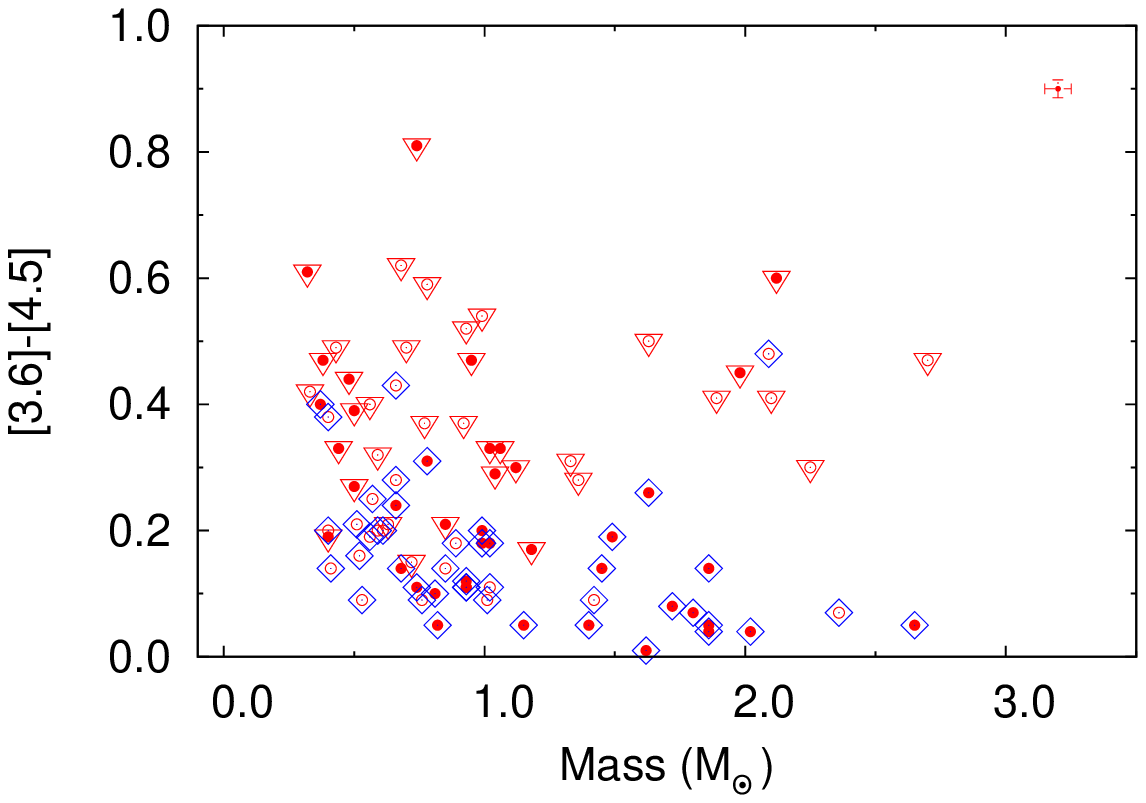}
 \caption{Upper panels : Period as a function of NIR excess $\Delta (I_{c}-K_{s})$ \& MIR
excess $[[3.6]-[4.5]]$ Lower panels : NIR excess $\Delta (I_{c}-K_{s})$ and MIR excess $[[3.6]-[4.5]]$ as a function of stellar mass. The Class\,{\sc ii} and Class\,{\sc iii} sources are represented with inverted triangles and diamonds, respectively.
Filled and open circles represent periodic and non-periodic variables, respectively. Corresponding mean errors are shown with error bars.}
 \label{fig : SED_correlation}
\end{figure*}

The disc-locking mechanism can be further verified from the correlation between rotation period and size of the disc of PMS variables \citep{2002A&A...396..513H}. 
In Figure \ref{fig : SED_correlation} (upper panels), we plot rotation periods as a
function of the disc indicators $\Delta (I_{c}-K_{s})$ and $[[3.6]-[4.5]]$ (see Section \ref{disc_indicators}). 
Although we do not see any clear trends in these plots but
they roughly suggest that the mean rotational speed becomes less when there is a increase in  NIR/MIR excess.
 The period versus $\Delta (I_{c}-K_{s})$ plot suggest that most of the 
slow rotators (having periods $>$ 10 days) are Class\,{\sc ii} sources  having
higher value of $\Delta (I_{c}-K_{s})$ index ($>$ 0.75 mag),
whereas Class\,{\sc iii} sources with $\Delta (I_{c}-K_{s})$  $\sim$ 0.3 mag have periods $\leqslant$ 10 days. 
Similar results have also been found previously for different regions \citep[e.g.,][]{2000AJ....120..349H,2002A&A...396..513H, 2005A&A...430.1005L,2006ApJ...646..297R,2020MNRAS.493..267S}.
Besides the small number of the periodic stars, these results suggest that the presence of a disc regulates stellar rotation
in a way that the younger stars having a thicker disc are slow rotators.
To check the mass dependence of $\Delta (I_{c}-K_{s})$ and $[[3.6]-[4.5]]$ indices, we plot these parameters
in the lower panels of Figure \ref{fig : SED_correlation}. 
Although we need more data points to conclude, but it appears that the higher mass stars have relatively lesser 
IR excess than the lower mass stars.

\begin{figure}
\vspace{0.3 cm}
\centering
\includegraphics[width=0.35\textwidth]{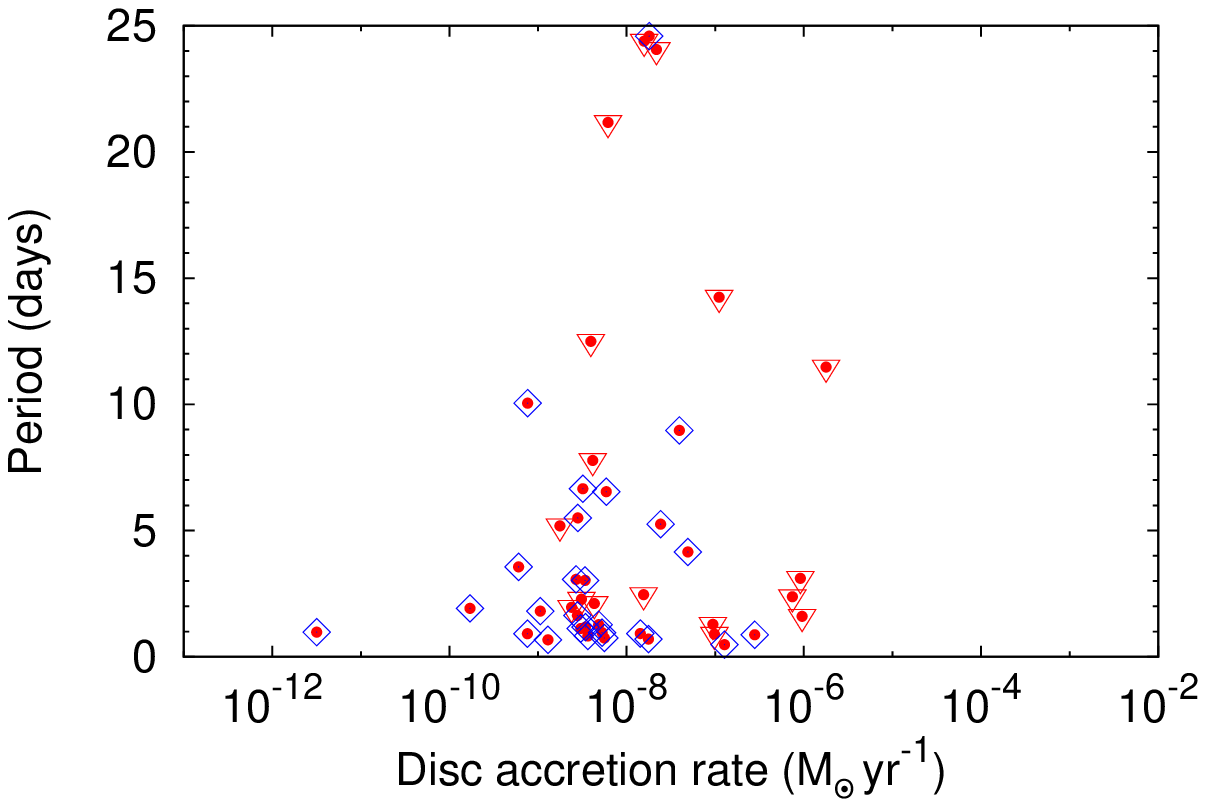} \\
\hspace{-0.2 cm}
\includegraphics[width=0.35\textwidth]{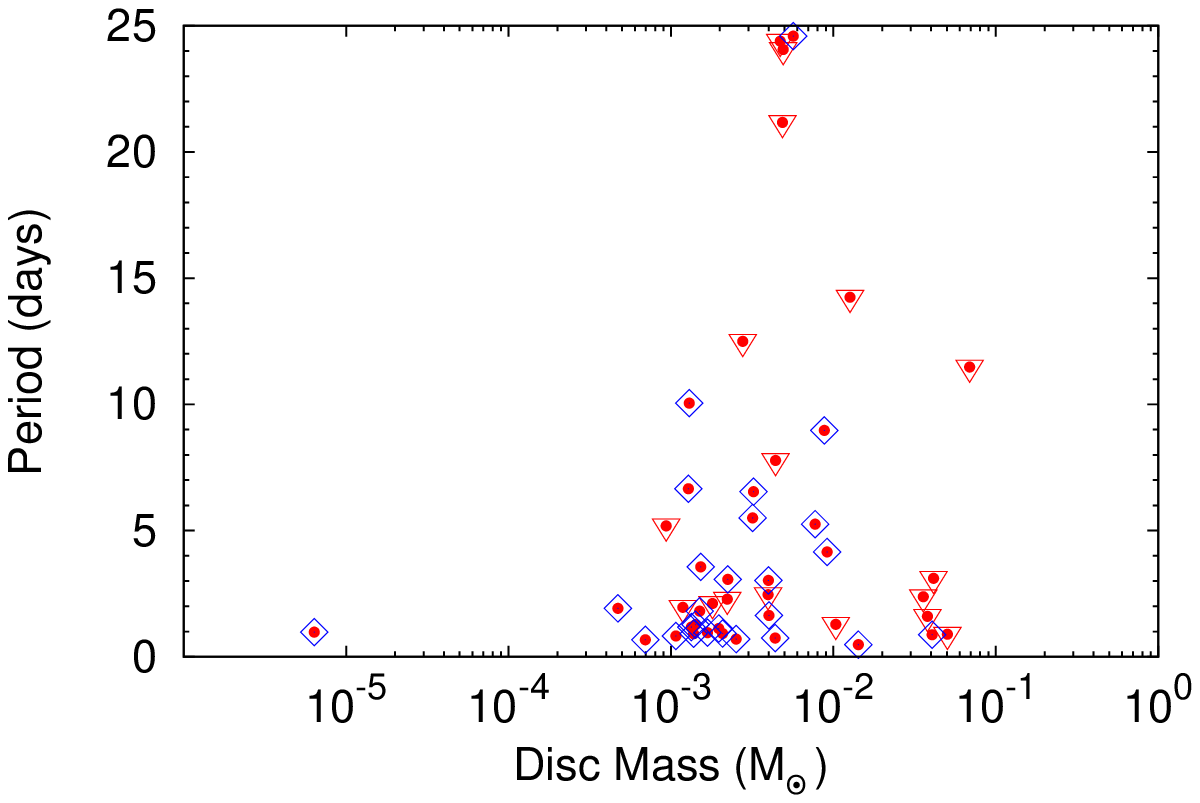}
\caption{ Period as a function of disc accretion rate and disc mass. The symbols are the same as in Figure \ref{fig : SED_correlation}.}
\label{fig : Period_NIR-MIR_&_NIR-MIR_Mass}
\end{figure}

Since, in PMS stellar evolution, mass accretion takes place from the circumstellar disc on to the star \citep{1994ApJ...429..797S, 2005A&A...430.1005L}, the rotation of stars with active accretion is likely to be more regulated by the disc and they have longer period as compared to disc-less stars where accretion is inactive.
We plot period as a function of disc accretion rate and disc mass of the PMS variables in Figure \ref{fig : Period_NIR-MIR_&_NIR-MIR_Mass}. 
Though we do not see any dependence of rotation period on the disc mass or disc accretion rate 
a lager sample of PMS variables will be helpful to conclude on this.
For a given rotation period, the value of disc mass and accretion rate spans over few orders of magnitude. 
Previous authors \citep[for example,][]{2006ApJ...647L.155F,2017A&A...599A..23V} 
have investigated any possible correlation between rotation period and accretion rate using the UV excess as an indicator of accretion and found that slowly rotating stars are more likely to have lager UV excess. \citet{2017A&A...599A..23V} found that CTTSs with large UV excess are mostly slow rotators
while WTTSs are distributed over the whole period range and have smaller UV excess. When comparing the accretion rate of the CTTSs with their rotation period they found a diverse range of accretion values corresponding to a rotation period 
and speculated that different sets of mechanisms are responsible for regulating
stellar rotation and accretion rates. While the accretion rate is regulated by the small scale magnetic field structure 
near the stellar surface, star-disc coupling and
angular momentum regulation is dominated by the large-scale magnetic field structure \citep{2012ApJ...755...97G}.

\begin{figure*}
\centering
\includegraphics[width=6.2cm, height=4cm, angle=0]{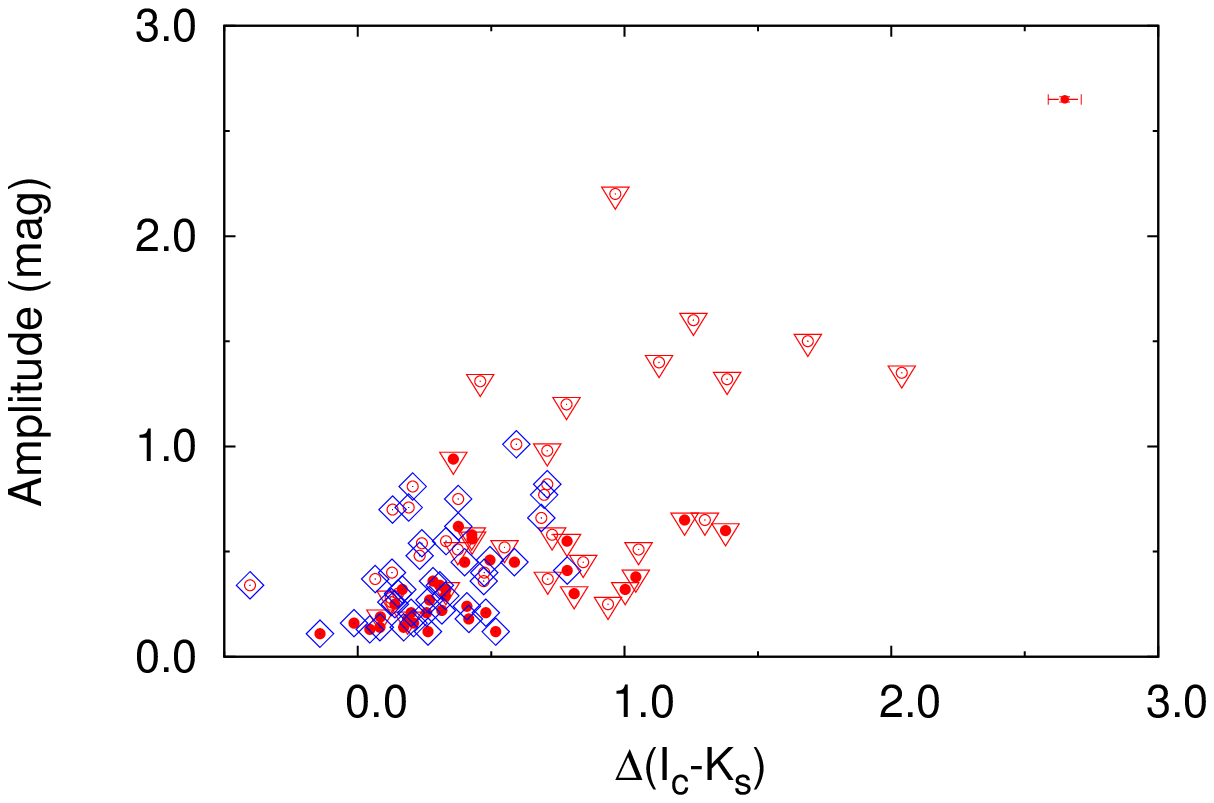}
\hspace{0.5cm }
\includegraphics[width=6cm, height=4cm, angle=0]{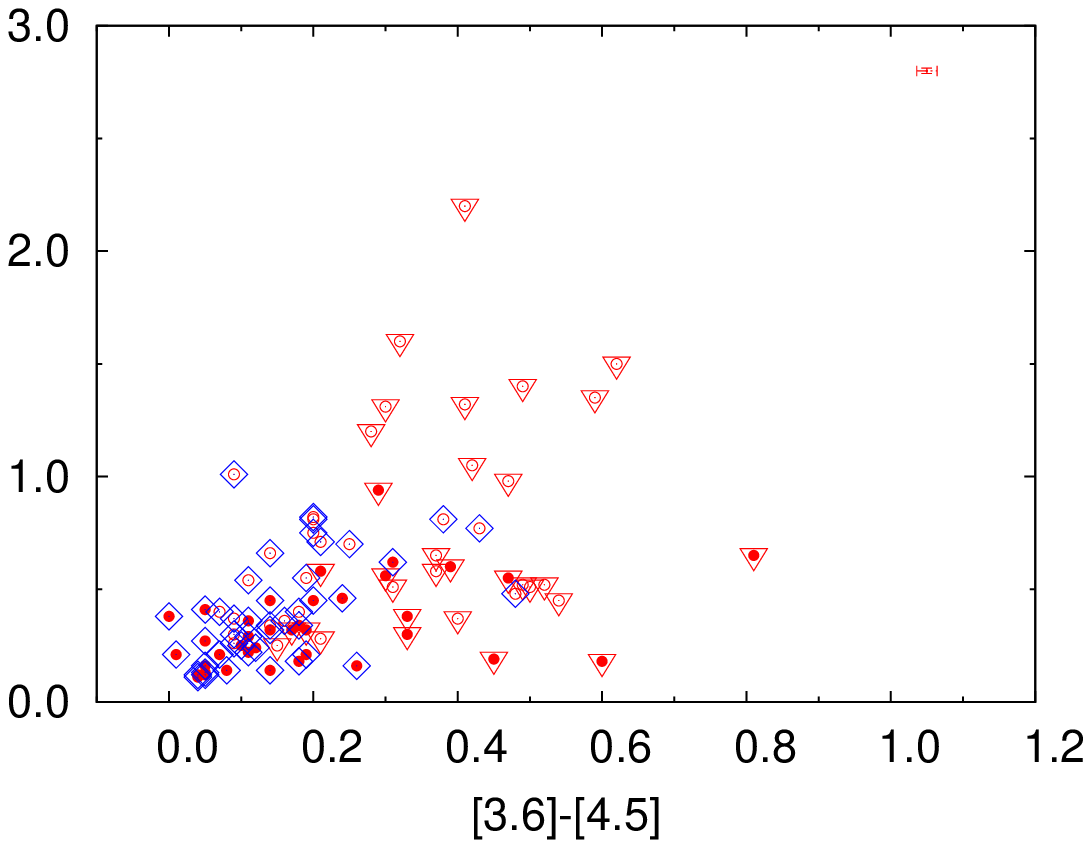}  \\
 \vspace{0.05 cm}
\hspace{-0.2cm }
\includegraphics[width=6.2cm, height=3.9cm, angle=0]{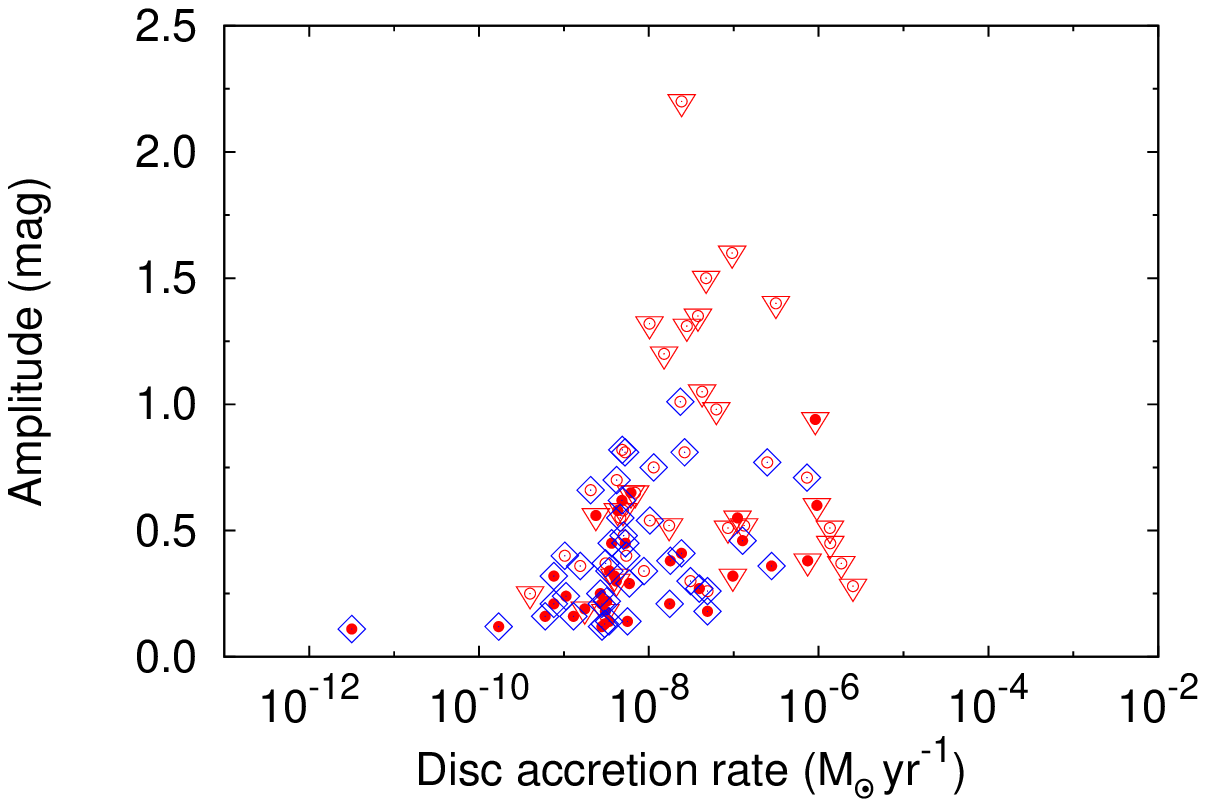}
 \hspace{0.5 cm}
 \includegraphics[width=6cm, height=3.9cm, angle=0]{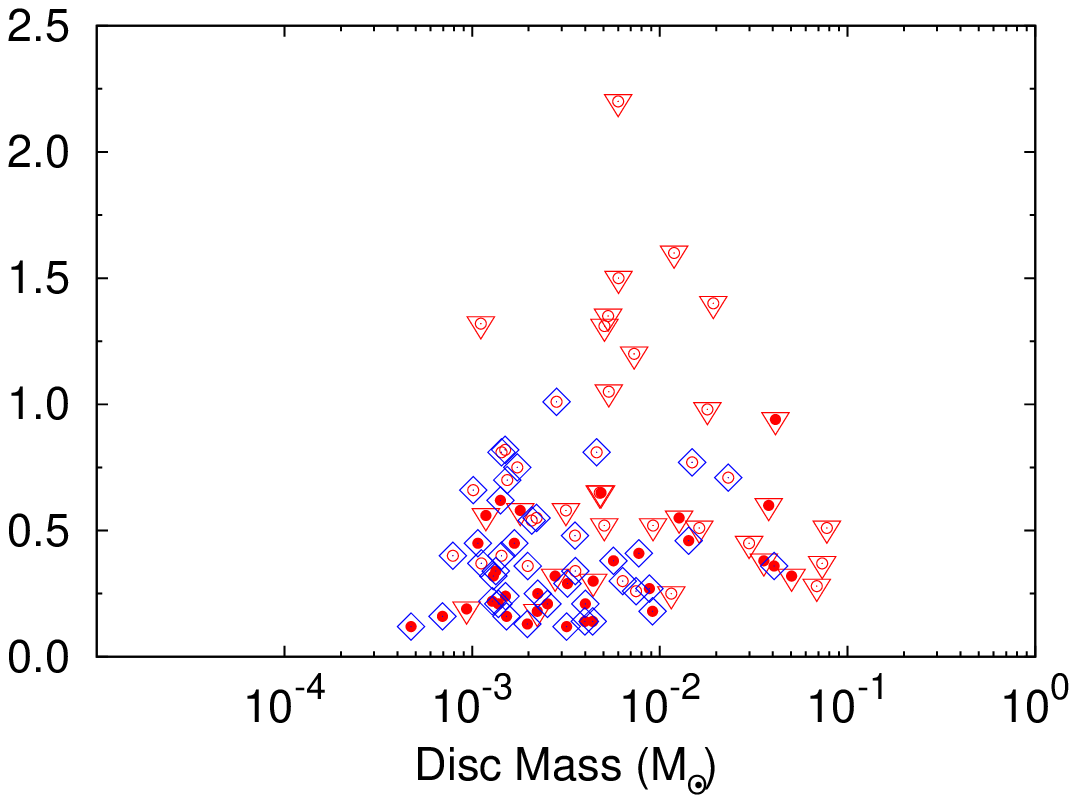} \\
 \caption{Amplitude of variation as a function of NIR excess $\Delta (I_{c}-K_{s})$ and MIR
excess $[[3.6]-[4.5]]$ (upper panels) and as a function of disc accretion rate and disc mass  (lower panels).  
The symbols are the same as in Figure \ref{fig : SED_correlation}.}
\label{fig : Amplitude_NIR-MIR-Accretion-Discmass}
\end{figure*}

As the variable accretion and extinction by circumstellar disc significantly enhance the 
brightness variation in PMS stars, the presence of thicker disc and active accretion are also expected to influence the
amplitude of brightness variation.
The amplitude of variation as a function  of $\Delta (I_{c}-K_{s})$ and $[[3.6]-[4.5]]$ are plotted in the upper panels of Figure \ref{fig : Amplitude_NIR-MIR-Accretion-Discmass}.  
Clearly, the amplitude of variation increases with disc indicators, which means the presence of thicker disc 
and envelop induces larger amplitude variations. 
The plots also show that the Class\,{\sc ii} sources have active circumstellar discs as compared to Class\,{\sc iii} sources.
Lower panels of the Figure \ref{fig : Amplitude_NIR-MIR-Accretion-Discmass} suggest  that
the stars with higher disc accretion activity ($>10^{-8}$ M$_\odot~ yr^{-1}$) induces higher
amplitude of variation. The amplitude of variation also seems to depends on the mass of the circumstellar
disc in the sense that stars having discs with mass $\geqslant$ 2$\times10^{-3}$ M$_{\odot}$ show larger amplitude of variation as compared to those with lower disc mass.
Hence, it appears that the Class\,{\sc ii} sources exhibit more active and dynamic variability because of their accretion activities. 
\citet{2002A&A...396..513H} have also reported a good correlation of RMS amplitude with $\Delta(I_{c}-K_{s})$  in Orion Nebula Cluster.
Similar results were also found in our previous studies on Sh 2-170 \citep{2020MNRAS.493..267S}.

\begin{figure}[h]
\centering
\includegraphics[width=0.35\textwidth]{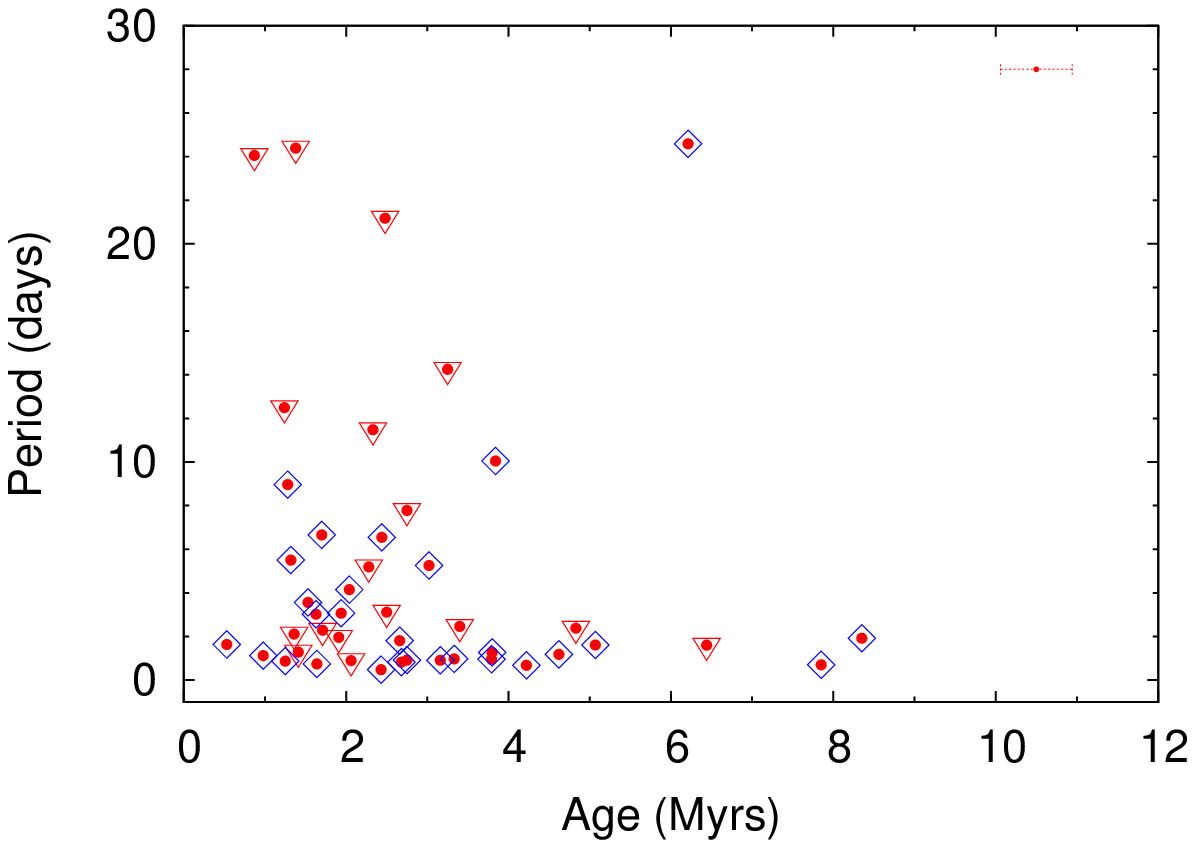}
\hspace{0.5 cm}
\includegraphics[width=0.35\textwidth]{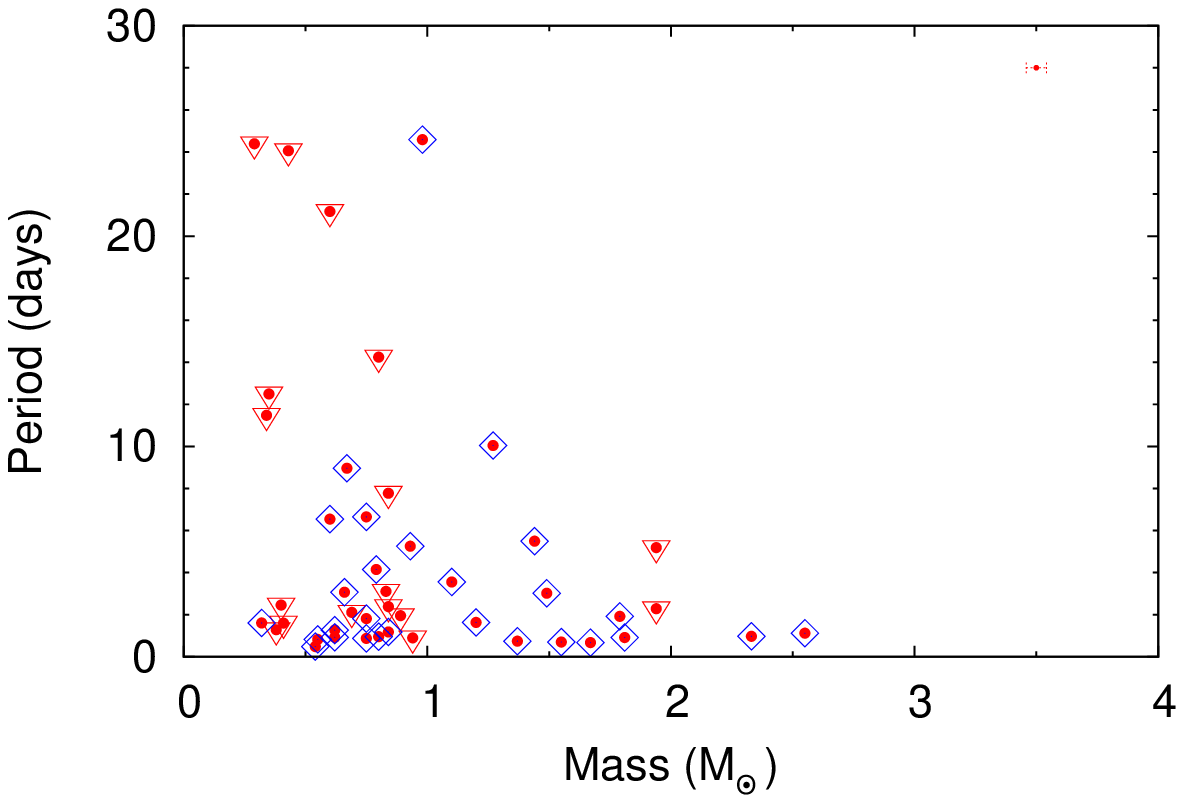}
 \caption{Period as a function of age and mass of PMS variables.  Corresponding mean errors are shown with error bars in the
top right corner of each panel. The symbols are the same as in Figure \ref{fig : SED_correlation}.}
 \label{fig : Period_Age-Mass}
\end{figure}

\subsection {Dependence of stellar variability on the age and mass of stars}
\label{age_mass_correlation}

\begin{figure}
\centering
\includegraphics[width=0.35\textwidth]{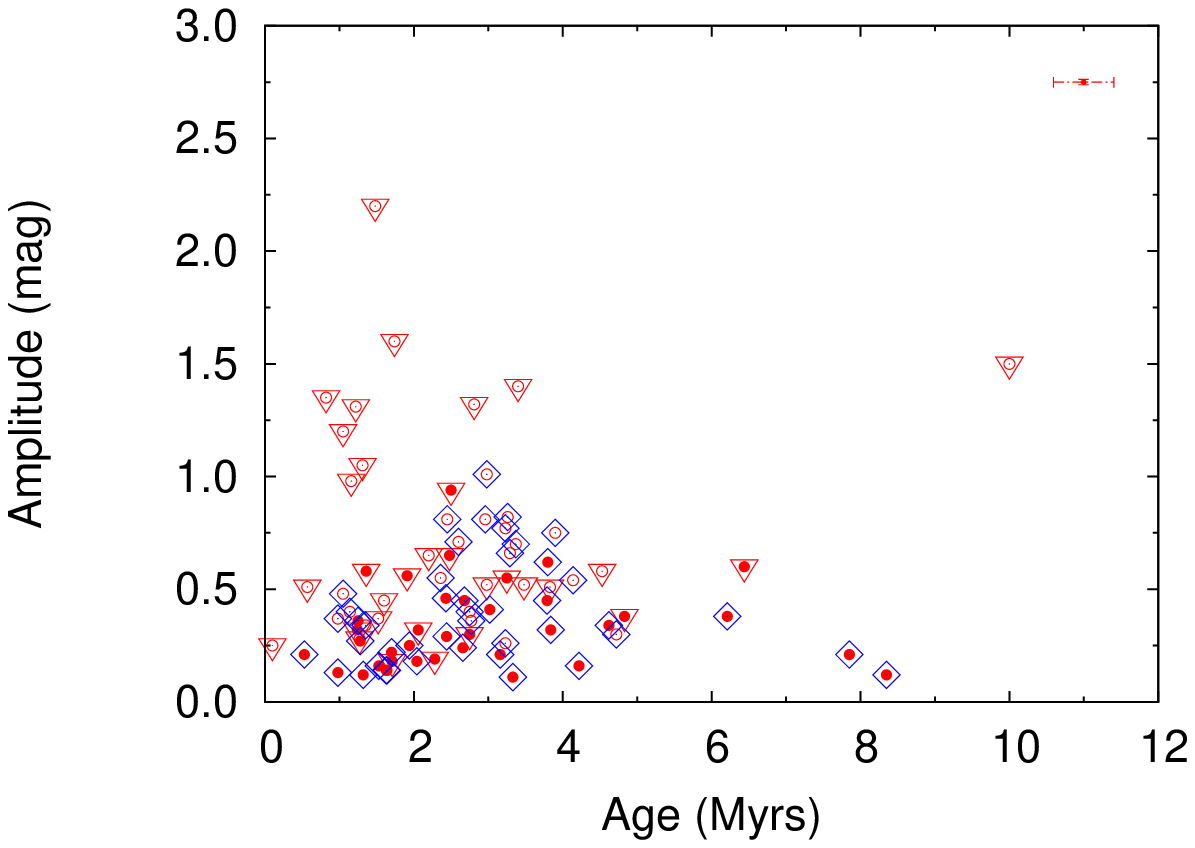} \\
\includegraphics[width=0.35\textwidth]{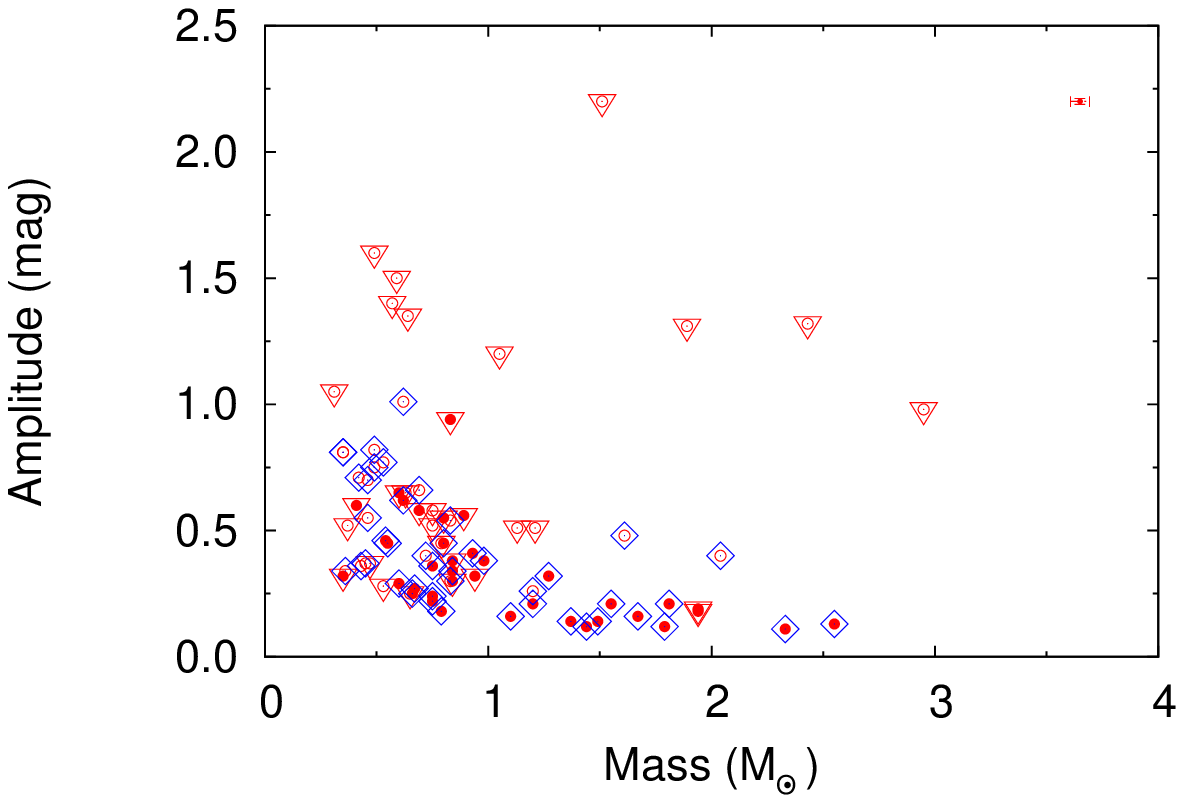}
 \caption{Amplitude of variations as a function of age and mass of PMS variables. The symbols are the same as in Figure \ref{fig : SED_correlation}.  Corresponding mean errors are shown with error bars.}
 \label{fig : Amplitude_Age-Mass}
\end{figure}

The age and mass of PMS stars are very important parameters that constrain the physical processes of the evolution 
of the central star and the circumstellar disc. 
The rate of the mass-accretion usually depends on the initial mass of the PMS stars. During the evolution, 
PMS stars also increase their mass through the accretion of material from the circumstellar disc.  
The disc subsequently dissipates resulting fewer activities related to accretion processes. 
To understand the role of age and mass on stellar variability, we plotted the period and amplitude of the PMS variables as a function of their age and mass in Figure \ref{fig : Period_Age-Mass}
and Figure \ref{fig : Amplitude_Age-Mass}, respectively.
The upper panel of Figure \ref{fig : Period_Age-Mass} indicates that stars with periods up to $\sim$4 days
are uniformly distributed over the entire range of ages,
whereas most of the PMS variables having periods $>$ 4 days are younger than $\sim$4 Myr.
\citet{2016MNRAS.456.2505L} have also reported a similar result that PMS stars
with age $\geqslant$ 3 Myr are relatively fast rotators.

The lower panel of Figure \ref{fig : Period_Age-Mass} displays the dependence of period on the stellar mass.
Here also we can see a decreasing trend in the distribution, i.e,
out of 12 rotators having period $>$ 6 days, 11 have mass $<$ 1.2 M$_\odot$, whereas all the stars more massive than 
$>$ 1.2 M$_\odot$ are fast rotators (period $<$ 6 days). 
Previously, \citet{2017A&A...603A.106R} and \citet{2012ApJ...747...51H} have found that the PMS stars with mass 
roughly below  0.5 M$_\odot$ rotate slower than their massive counterparts in their study related with CygOB2 and NGC 6530 regions, respectively. 
On the contrary, \citet{2010MNRAS.403..545L} have found that the 
lower mass stars ($<$ 0.4 M$_\odot$) rotate relatively faster than the higher mass stars ($>$ 0.4 M$_\odot$) in the CepOB3 region. 
\citet{2017A&A...599A..23V} have also found similar result in the case of NGC 2264. They divided the stars in three mass bins i.e., 
(i) M $<$ 0.4 M$_\odot$; (ii) 0.4 M$_\odot$ $\leq$ M $\leq$ 1 M$_\odot$; (iii) M $>$ 1 M$_\odot$, and found that lowest mass group 
consists a peak of fast rotators (P $=$ 1-2 days) while for higher mass groups an emerging peak around P $=$ 3-4 days are seen. 
Stars more massive than 1.4 M$_\odot$ rotate faster as they have largely radiative interiors unlike their less massive counterparts 
which spend long time along the convective track during their PMS evolution. Convection brakes the star by powering stellar 
winds that carry angular momentum. Massive stars lack this mechanism and experience different rotational evolution from less massive stars. 
To check whether this effect is present in NGC 2264, \citet{2017A&A...599A..23V} further divided their 3rd mass group into below and above 1.4 M$_\odot$ and 
found that the latter rotate faster with a median period of 3 days while the stars in the former group are slow rotators with median period of 5 days. 
In the present study, we also observe that the higher mass stars (M $>$ 1.2 M$_\odot$) are fast rotators. Recently, \citet{2020MNRAS.493..267S} and \citet{2012MNRAS.427.1449L,2016MNRAS.456.2505L} have  also found similar results. 
\citet{2000AJ....120..349H} and \citet{2005MNRAS.358..341L} have also found  strong correlation between stellar mass and rotation rate in the case of ONC and IC 348, respectively.  \citet{2005MNRAS.358..341L} concluded that the strong mass dependence of rotation period seen in ONC \citep{2002A&A...396..513H} 
may well be a common feature of PMS populations. 

Since younger stars should have higher disc fraction, we may see a larger amplitude variation in the younger population.
In the top panel of Figure \ref{fig : Amplitude_Age-Mass},
we can observe a decrease in the amplitude of the PMS variables with the increase in their age. 
The decreasing trend with age supports previous findings \citep{2011MNRAS.418.1346L,2012MNRAS.427.1449L,2016MNRAS.456.2505L,2020MNRAS.493..267S} 
that a significant amount of disc is dissipated by $\sim$ 5 Myr. This is also in accordance with the result obtained by \citet{2001ApJ...553L.153H}.
The bottom panel shows amplitude as a function of stellar mass. Although the trend is not clear in the case of Class\,{\sc ii} objects 
but it appears that the amplitude of variation decreases with increase in the mass of Class\,{\sc iii} objects (blue diamonds). 
This result seems to indicate that relatively massive PMS stars have smaller spots on their surface. 
As the variability in Class\,{\sc iii} sources (or WTTSs) sources are mostly regulated by cool spots on their photosphere 
the size of the spot is proportional to amplitude of variation. With increasing mass a star develops a larger radiative core and the 
convective envelope becomes thinner, causing less efficient dynamo mechanism, and consequently the spot size reduces  \citep{2008sust.book.....T,2009A&ARv..17..251S}.

\subsection{X-ray activities in PMS stars}
\label{correlation_with_X-ray}

\begin{figure}
\centering
\includegraphics[width=0.35\textwidth]{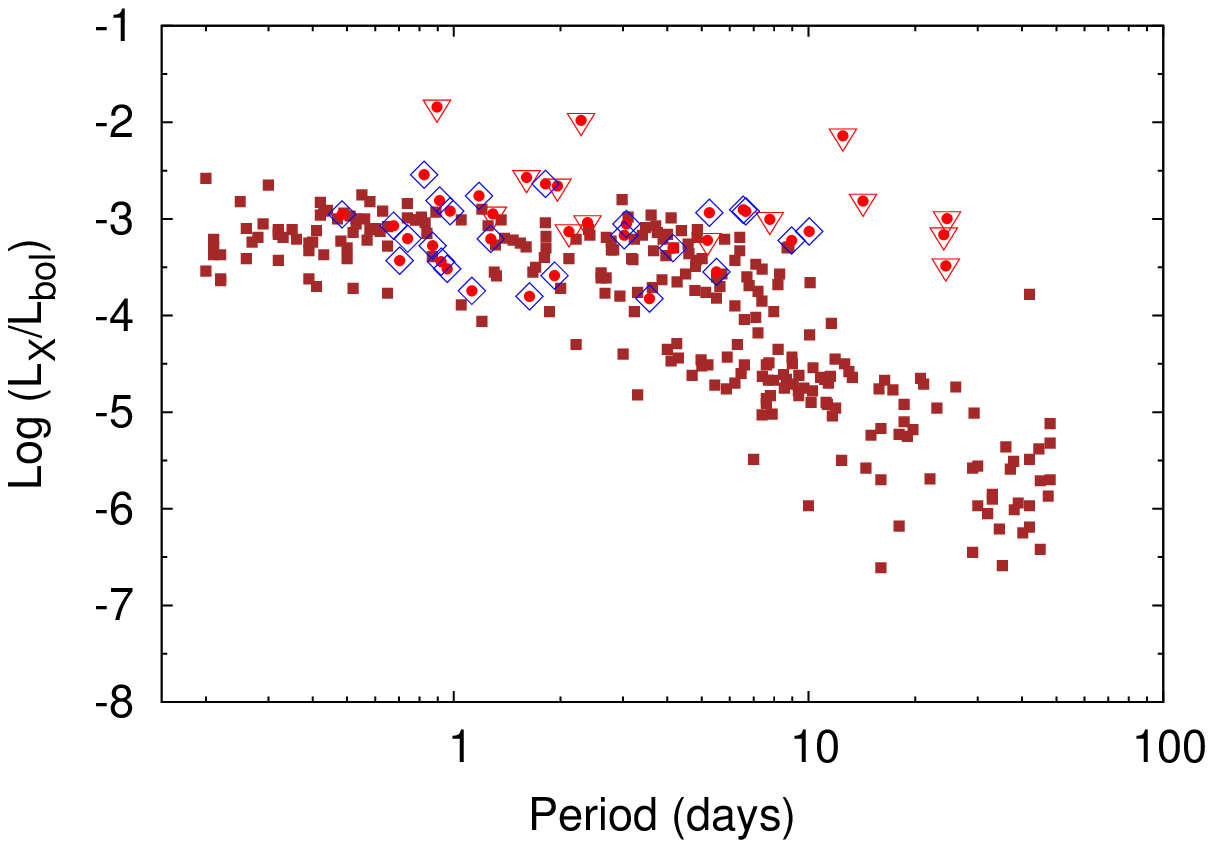}
 \hspace{0.5 cm}
\includegraphics[width=0.35\textwidth]{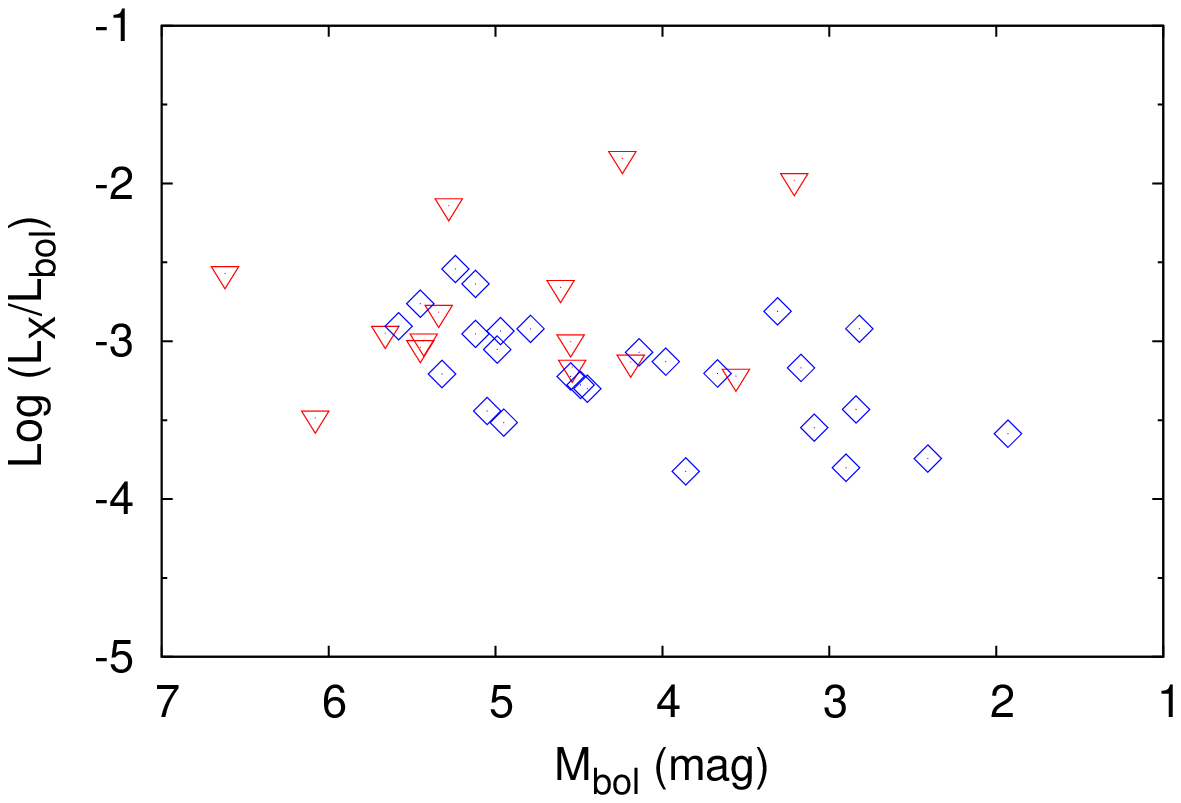}

\caption{Fractional X-ray luminosity of the periodic PMS sources as a function of period (top panel) and bolometric magnitude  (bottom panel). 
The brown square symbols in the upper panel represent the MS samples taken from \citet{2003A&A...397..147P}. Other symbols are the same as in Figure \ref{fig : SED_correlation}.}.
\label{fig : X-ray_correlation}
\end{figure}

Both accreting and non-accreting PMS stars exhibit strong X-ray emission produced mainly by magnetic activity 
in hot corona while an additional soft component is produced by accretion shock in the case of accreting stars.
The magnetic activity is believed to be generated by a dynamo mechanism at the interface of the radiative and convective zones
that is driven by rotational and convective motions \citep[$\alpha~\Omega$ dynamo:][]{1955ApJ...122..293P}. 
For the MS stars, it is found that X-ray luminosity increases with increasing rotation speed 
until a star reaches a saturation level \citep[log ($L_X/L_{bol}$) $\approx$ -3,][]{2003A&A...397..147P}. 
However, the period-magnetic activity relation in PMS stars is not well established.  
To explore this,  we have plotted  fractional X-ray luminosity, i.e., ($L_X/L_{bol}$), 
as a function of rotation period of PMS stars  in the upper panel of Figure \ref{fig : X-ray_correlation}. 
 The distribution is roughly flat at the saturation level, log ($L_X/L_{bol}$) = -3.05.
 This is comparable with the saturation value of the MS sample of 
\citet{2003A&A...397..147P} shown with brown square symbols in the plot.
\citet{2012A&A...539A..64A} have found similar results in IC348 and concluded that more or less chaotic nature of convective dynamo is responsible for large scatter in magnetic activity among TTSs. The scatter of $L_X/L_{bol}$ in Figure \ref{fig : X-ray_correlation} is similar to that of IC348 
which shares a similar age with IC 1805 ($\approx$ 3 Myrs). On the other hand, Orion which is younger ($\approx$ 
1 Myrs) shows a similar distribution with a larger scatter of $L_X/L_{bol}$ \citep [see also,][] {2012A&A...539A..64A}.
In case of NGC 6530, \citet{2012ApJ...747...51H} have found 
fractional X-ray luminosities of the stars are more or less flat with rotation period, approximately at the saturation level. 
However, the fastest rotators show  lower  X-ray luminosities, suggestive of the so-called ``super-saturation" where X-ray luminosity decreases with increasing rotation speed.

\begin{figure}
\centering
 \includegraphics[width=0.35\textwidth]{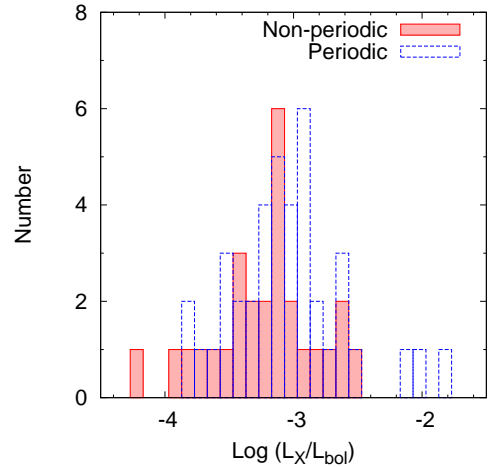}
\caption{ Distribution of log ($L_X/L_{bol}$) of periodic (blue bordered histogram) and non-periodic (red shaded histogram) PMS stars.}
\label{fig : X-ray_hist}
\end{figure}

 To check the dependence of X-ray activity on stellar mass we have plotted fractional X-ray luminosity as a function of 
bolometric magnitude ($M_{bol}$) in the lower panel of Figure \ref{fig : X-ray_correlation}. 
Although it is difficult to conclude from this small sample size, it appears that there is a weak anti-correlation for  the  Class\,{\sc iii} objects in the sense that $L_X/L_{bol}$ 
decreases as the stars get brighter or massive.
This suggests that lower mass stars, which are mostly convective, produce a larger fraction of X-ray 
in comparison to their massive counterparts. More studies on larger sample size are required to confirm this result.

We have also explored the possibility if there is any difference in the X-ray activity of PMS stars with (periodic) and without (non-periodic) rotation period. 
The distribution of log ($L_X/L_{bol}$) of the periodic and non-periodic PMS stars are shown with blue-bordered and red shaded histograms, respectively in Figure \ref{fig : X-ray_hist}. 
The distribution of periodic PMS stars peaks at a larger value of $L_X/L_{bol}$ relative to non-periodic PMS stars. 
The average values of the fractional X-ray luminosity ($L_X/L_{bol}$) in periodic stars (0.17$\pm$0.28\%) is almost double as compared to non-periodic (0.09$\pm$0.07\%) stars.
In our sample of WTTSs and CTTSs (cf. Section \ref{LC PMS}),  the former (mean $L_X/L_{bol}$ = 0.09$\pm$0.07\%) are also found to be more X-ray active than the latter (mean $L_X/L_{bol}$ = 0.04$\pm$0.03\%).
This may be due to the fact that the rapid
stellar rotation in the later stage of stellar evolution produces a stronger magnetic field through an
$\alpha~\Omega$ type dynamo (for stars with radiative core and convective envelope) or through a distributed turbulent dynamo
(for fully convective stars) \citep{2004AJ....127.3537S}.

\section{Summary and conclusion}
\label{conclusion}

In this paper we have presented the results of the multi-epoch and deep imaging survey in  $V$, $R_{c}$, $I_{c}$ bands
to understand the characteristics of PMS variables in the Sh2-190 region. Following are the main results.

\begin{itemize}
\item
We have identified 134 variables in this region.
Eighty-five of them are found to be probable PMS stars, whereas the remaining stars could be MS/field population.
Out of the 85 PMS variables, 37 and 48 are classified as Class\,{\sc ii} and Class\,{\sc iii} sources, respectively. 
LCs of 17 Class\,{\sc ii}  and 28 Class\,{\sc iii} variables show periodicity. 

\item
The majority of the PMS variables have mass and age in the range of 0.1 $\leq$ M/M$_\odot$ $\leq$ 4.0 and  0.3 - 2.0 Myrs,
respectively, and hence should be TTSs. The rotation period of the  PMS variables ranges from 12 hrs to 24 days, whereas
the amplitude of variation spans from 0.1 mag to 2.2 mag. The amplitude is larger in  Class\,{\sc ii} sources (up to 2.2 mag) as compared to Class\,{\sc iii} sources (upto $\sim$ 1.0 mag).

\item 
In general,it appears that Class\,{\sc ii} sources show longer period of variability as compared to Class\,{\sc iii} sources. 
Also, the period distribution of Class\,{\sc ii} sources peaks at $\sim$ 3 days, while Class\,{\sc iii} sources peak at 1 day.
Stars with smaller IR excess seem to rotate faster in comparison to stars with larger IR excess.

\item
The amplitude of variation in the PMS stars shows a increasing trend with increase in the disc indicators. This suggests that 
Class\,{\sc ii} sources with circumstellar disc exhibit more active and dynamic activities as compared to Class\,{\sc iii} sources. 
The amplitude of variability seems to be influenced by disc mass and disc accretion rate in the sense that 
the presence of massive discs ($\gtrsim$ 2 $\times 10^{-3}$ M$_\odot$) and higher disc accretion rate ($\gtrsim 10^{-8}$ M$_\odot yr{^{-1}}$) induces higher amplitude of variation.

\item
The different period distribution of Class\,{\sc ii} and Class\,{\sc iii} sources and the dependence of rotation period on IR excess are compatible with the `disc-locking' model. This model suggests that the rotation of PMS stars is regulated by the presence of circumstellar discs, that when a star is disc-locked, its rotation speed doesn't change and that when the star is released from the locked-up disc, it can spin up with its contraction.

\item
With the increase in stellar mass in the range of $\sim$0.5-2.5 M$_\odot$ the period of PMS variables decreases. The amplitude of variation of the Class\,{\sc iii} objects show a decreasing trend with increasing mass. 
This  suggests that with increasing mass a star develops the radiative core quickly and convective envelope becomes thinner, 
which results in the reduction in spot size and causes lesser amplitude of variation in stars whose variability is regulated by spot modulation.

\item
These results favor the preposition that cool spots on WTTSs are mostly responsible for their variations, while
hot spots on CTTSs caused by variable mass accretion from the inner disc and/or variable extinction events 
contribute to their larger amplitudes and more irregular behaviors. 

\item
Fractional X-ray luminosity ($L_X/L_{bol}$)  as a function of rotation period shows a flat distribution at log ($L_X/L_{bol}$) $\approx$ -3.0, 
which indicates that PMS stars in the Sh 2-190 region are at the saturation level reported for the MS stars. 
The log ($L_X/L_{bol}$) of  younger Class \,{\sc ii} sources is generally less as compared to that of comparatively older 
Class \,{\sc iii} sources hinting towards a less significant role of the stellar disc-related mechanisms in X-ray generation.

\end{itemize}

\section*{Acknowledgments}

We are very thankful to the anonymous referee for constructive suggestions$/$comments. The observations reported in this paper were obtained by using the
1.3 m Devasthal Fast Optical Telescope (DFOT, India), the 0.81m Tenagara automated telescope (South Arizona) (Observations were remotely done in the National Central University, Taiwan)
and Zwicky Transient Facility. Based on observations obtained with the Samuel Oschin 48-inch Telescope at 
the Palomar Observatory as part of the Zwicky Transient Facility project. ZTF is supported by the National 
Science Foundation under Grant No. AST-1440341 and a collaboration including Caltech, IPAC, the Weizmann 
Institute for Science, the Oskar Klein Center at Stockholm University, the University of Maryland, the University 
of Washington, Deutsches Elektronen-Synchrotron and Humboldt University, Los Alamos National Laboratories, 
the TANGO Consortium of Taiwan, the University of Wisconsin at Milwaukee, and Lawrence Berkeley National 
Laboratories. The operations are conducted by COO, IPAC, and UW. This work made use of data from the Two 
Micron All Sky Survey (a joint project of the University of Massachusetts and the Infrared Processing and Analysis 
Center/California Institute of Technology, funded by the National Aeronautics and Space Administration and the 
National Science Foundation), and archival data obtained with the $Spitzer$ Space Telescope and Wide Infrared 
Survey Explorer (operated by the Jet Propulsion Laboratory, California Institute of Technology, under contract with 
the NASA. This publication also made use of data from the European Space Agency (ESA) mission $Gaia$ (https://www.cosmos.esa.int/gaia), processed by the $Gaia$ Data Processing and Analysis Consortium (DPAC, https://www.cosmos.esa.int/web/gaia/dpac/consortium). \\

{\textit Software :} ESO-MIDAS \citep{1992ASPC...25..120B}, IRAF \citep{1986SPIE..627..733T,1993ASPC...52..173T}, DAOPHOT-II software \citep{1987PASP...99..191S}, period  \citep{1976Ap&SS..39..447L,1982ApJ...263..835S}.

\bibliography{sh2-190v2}{}
\bibliographystyle{aasjournal}

\setcounter{section}{0} 
\section {Figures to be published in electronic form only}

\setcounter{figure}{7} 
\begin{figure}[h]
\centering 
\includegraphics[width = 1.9 cm, height = 2.6 cm, angle= 270]{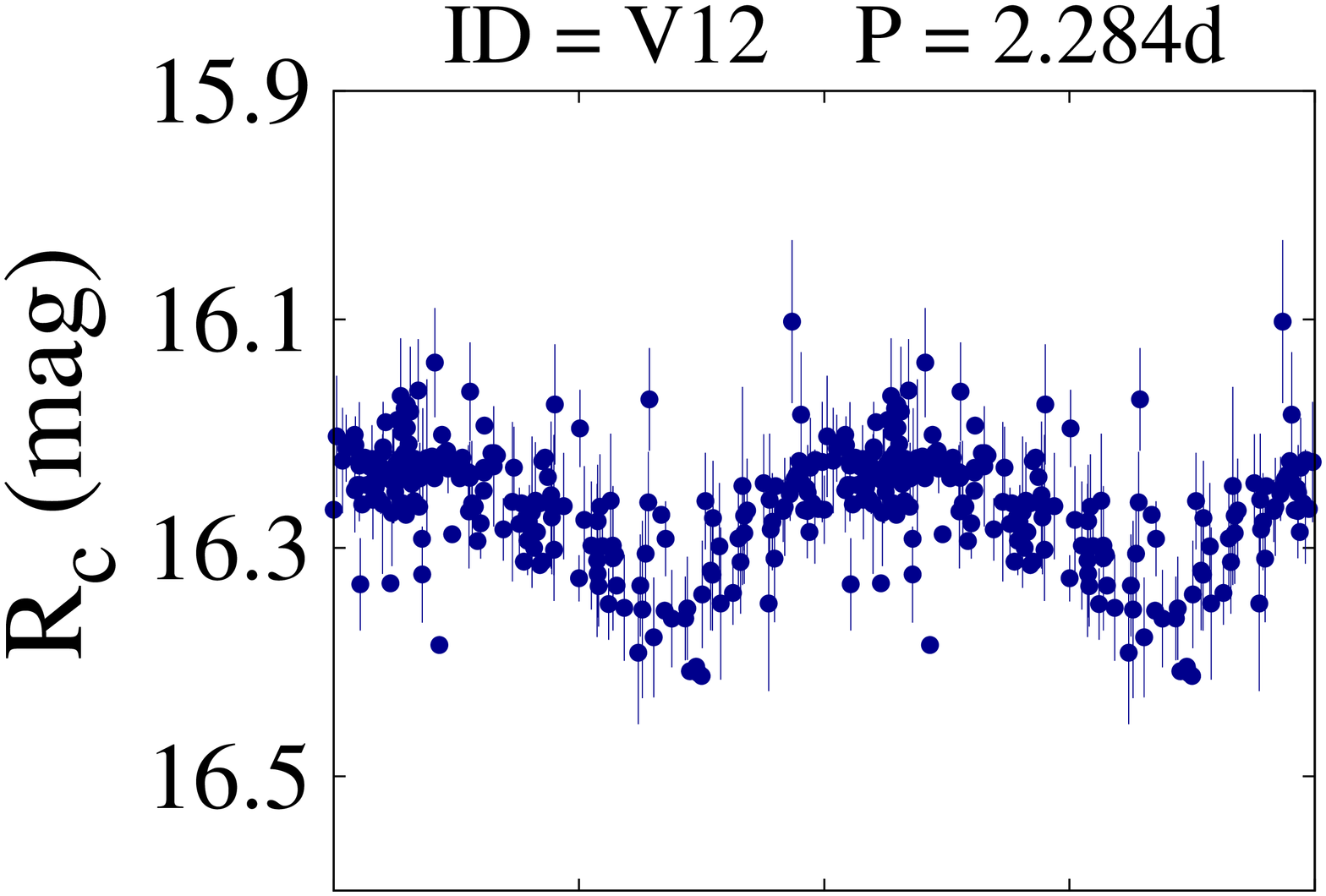}
\hspace{0.03 cm}
\includegraphics[width = 1.9 cm, height = 2.6 cm, angle= 270]{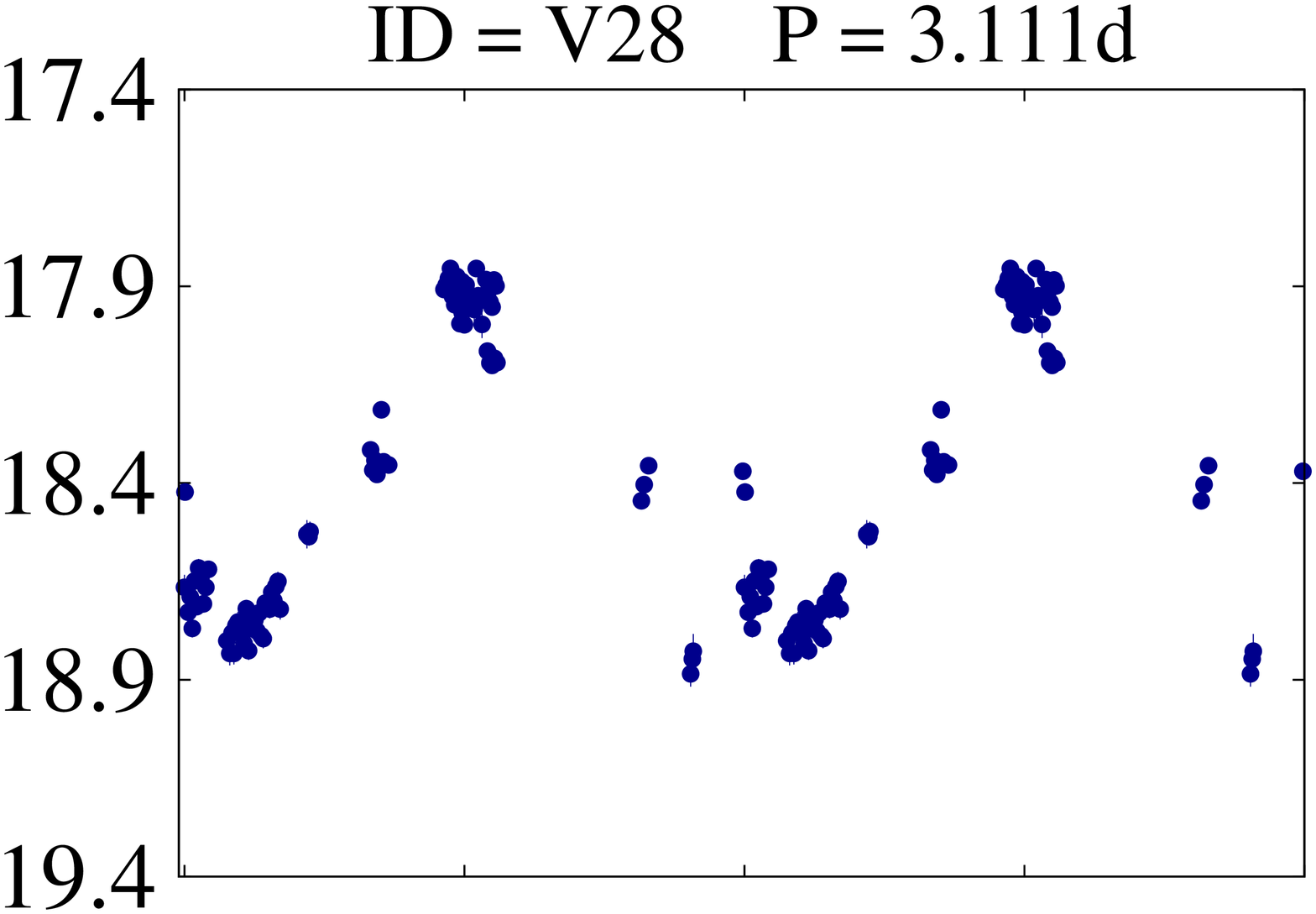}
\hspace{0.03 cm}
\includegraphics[width = 1.9 cm, height = 2.6 cm, angle= 270]{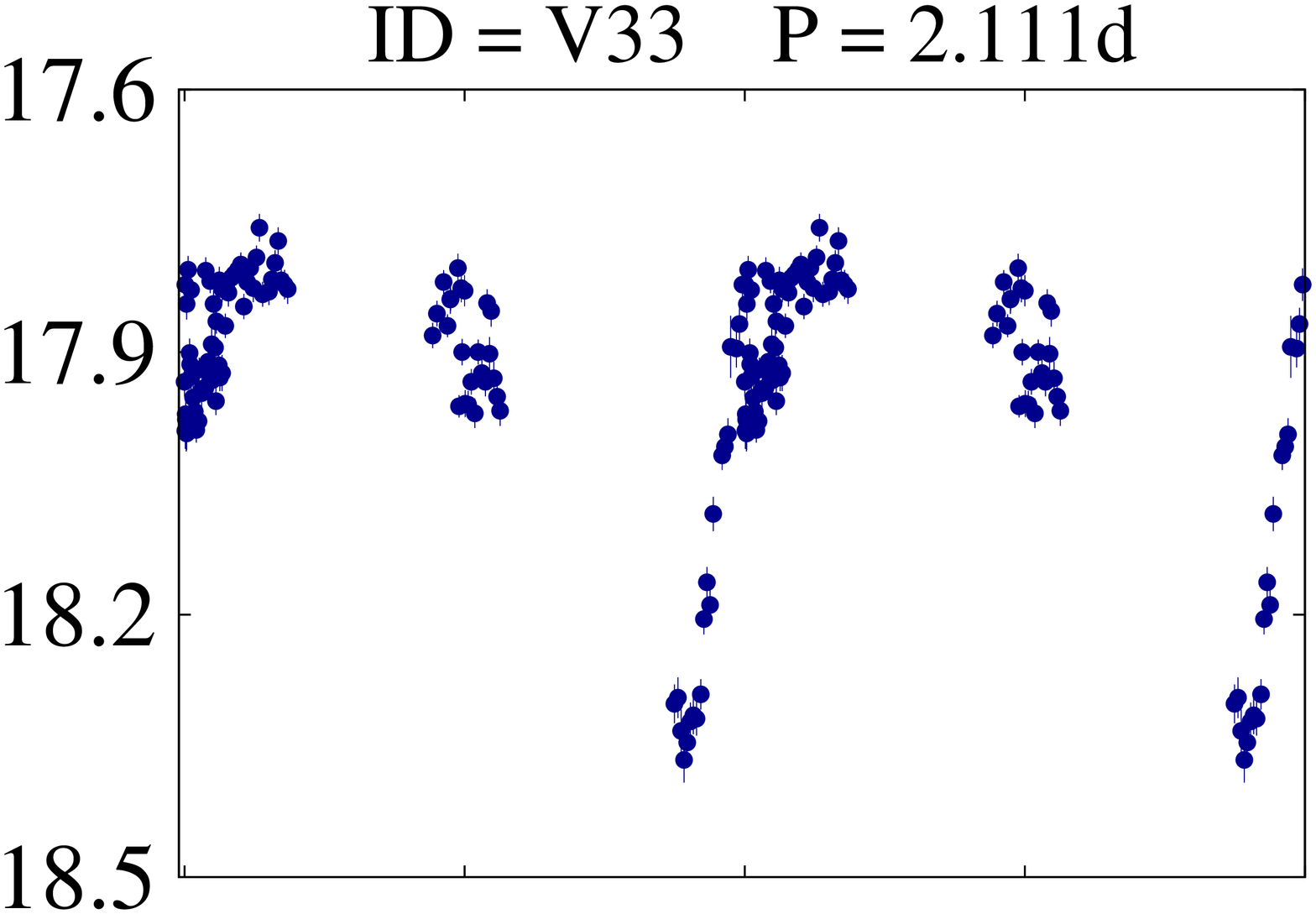}
\hspace{0.03 cm}
\includegraphics[width = 1.9 cm, height = 2.6 cm, angle= 270]{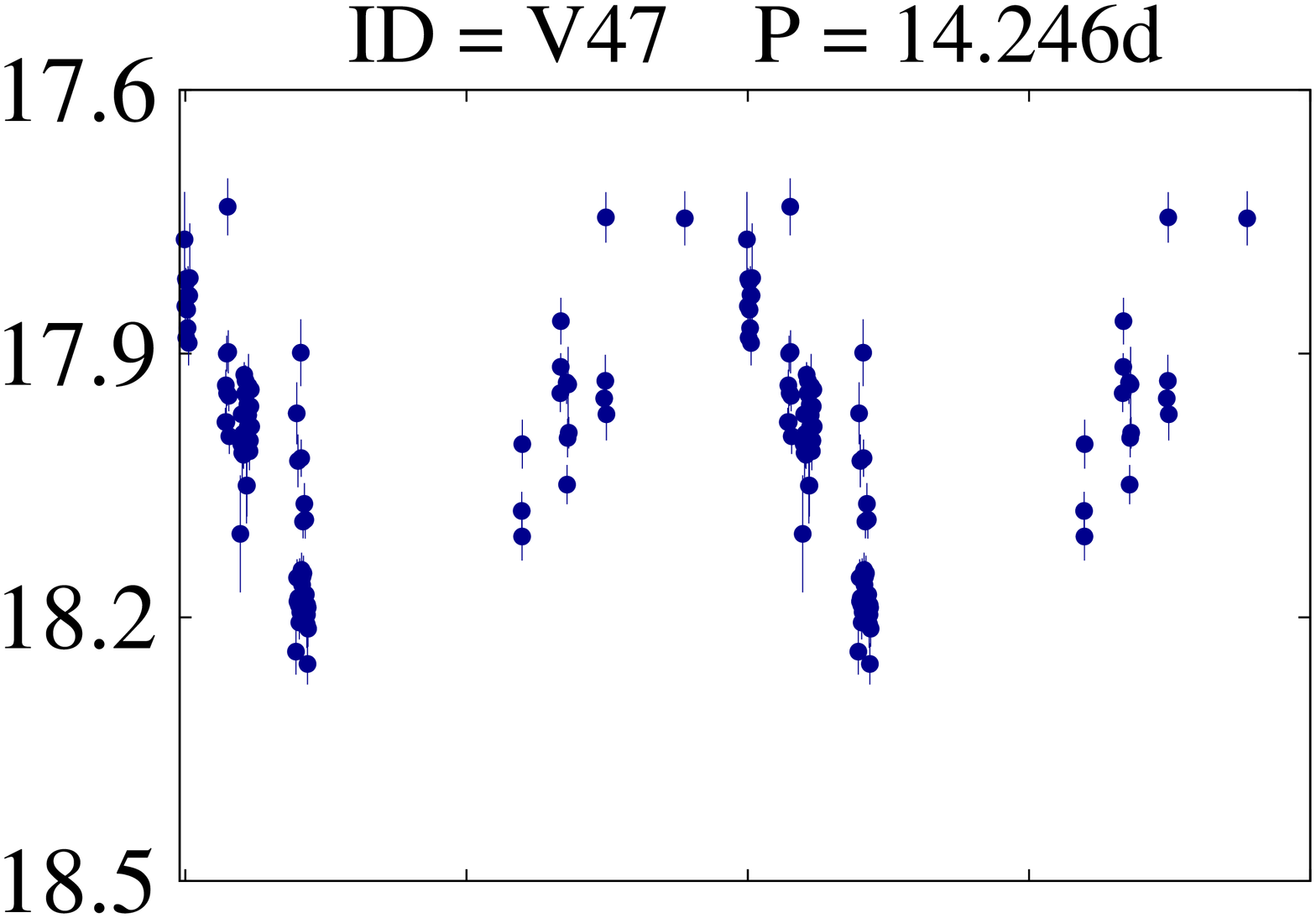} 
\hspace{0.02 cm}
\includegraphics[width = 1.9 cm, height = 2.6 cm, angle= 270]{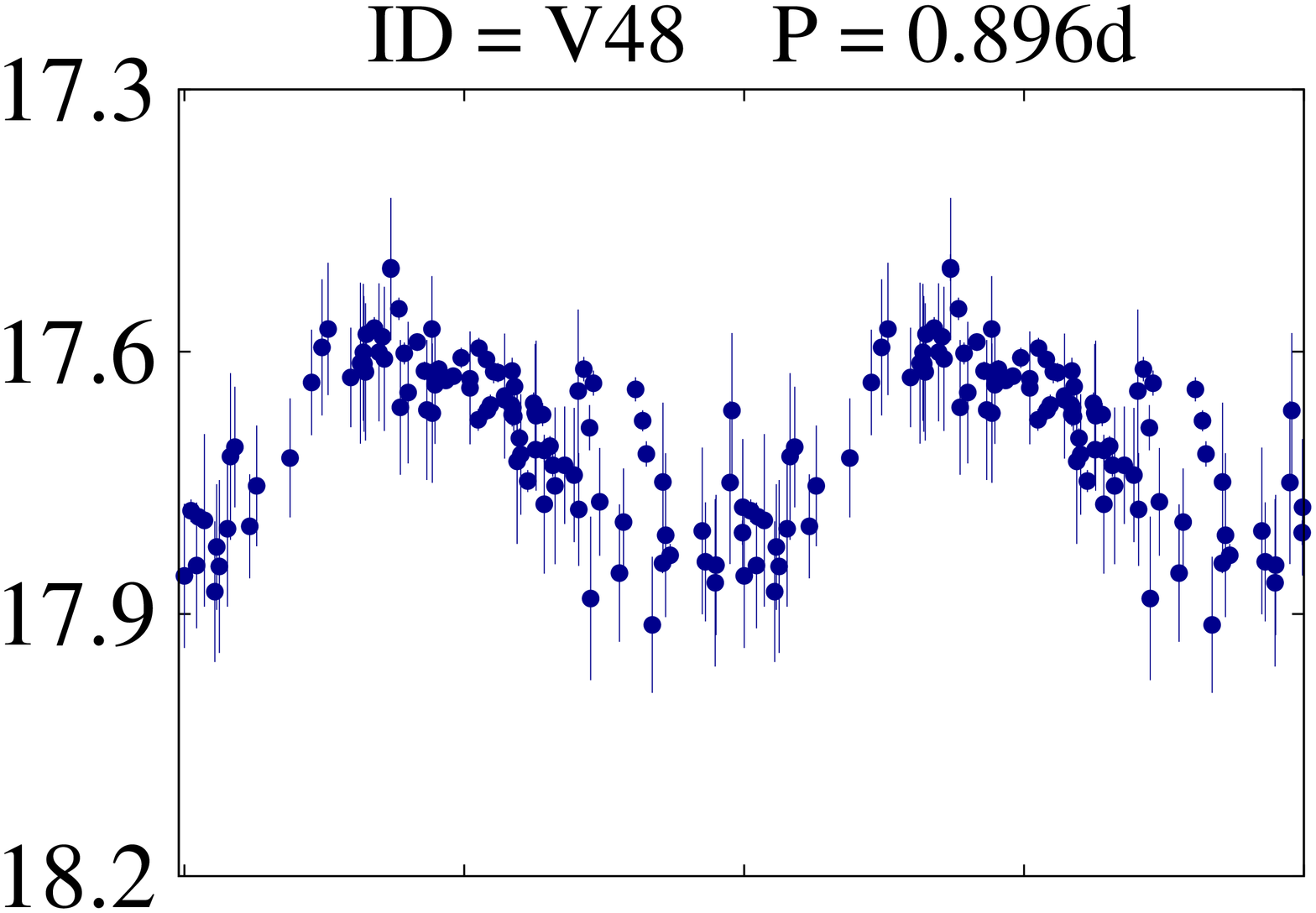} 
\hspace{0.03 cm}
\includegraphics[width = 1.9 cm, height = 2.6 cm, angle= 270]{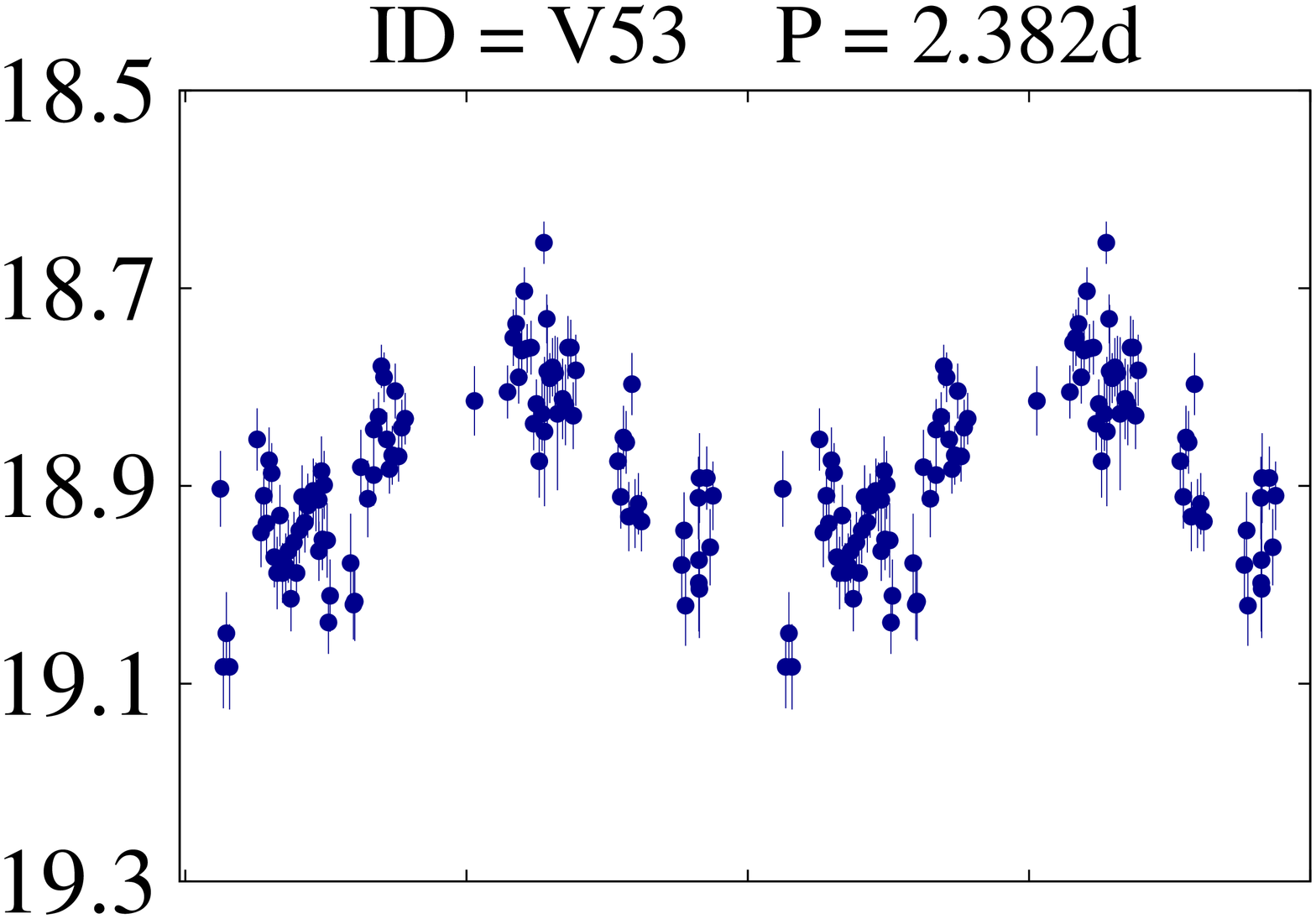} \\ 
\vspace{0.05 cm}
\includegraphics[width = 1.9 cm, height = 2.6 cm, angle= 270]{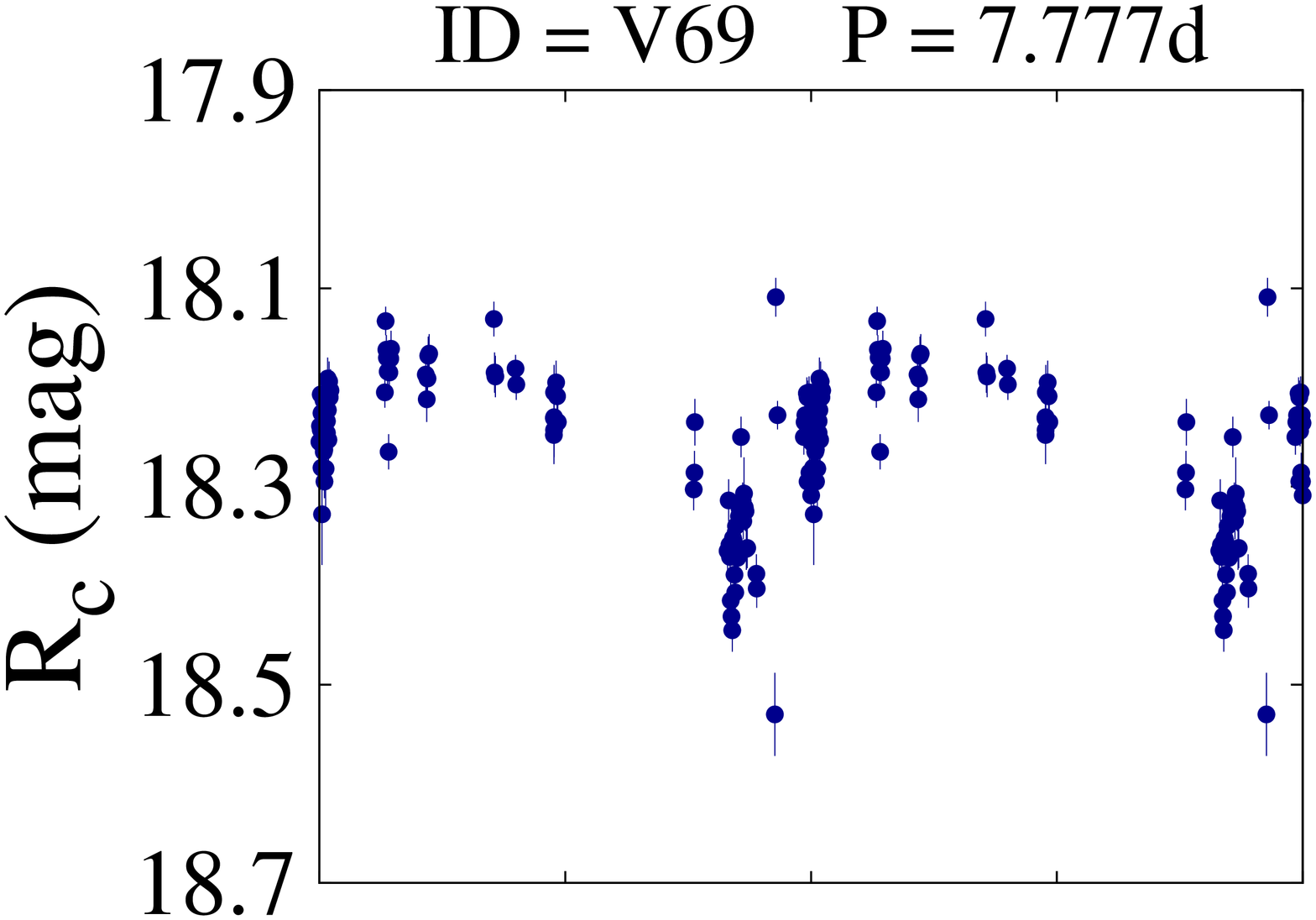}
\hspace{0.03 cm}
\includegraphics[width = 1.9 cm, height = 2.6 cm, angle= 270]{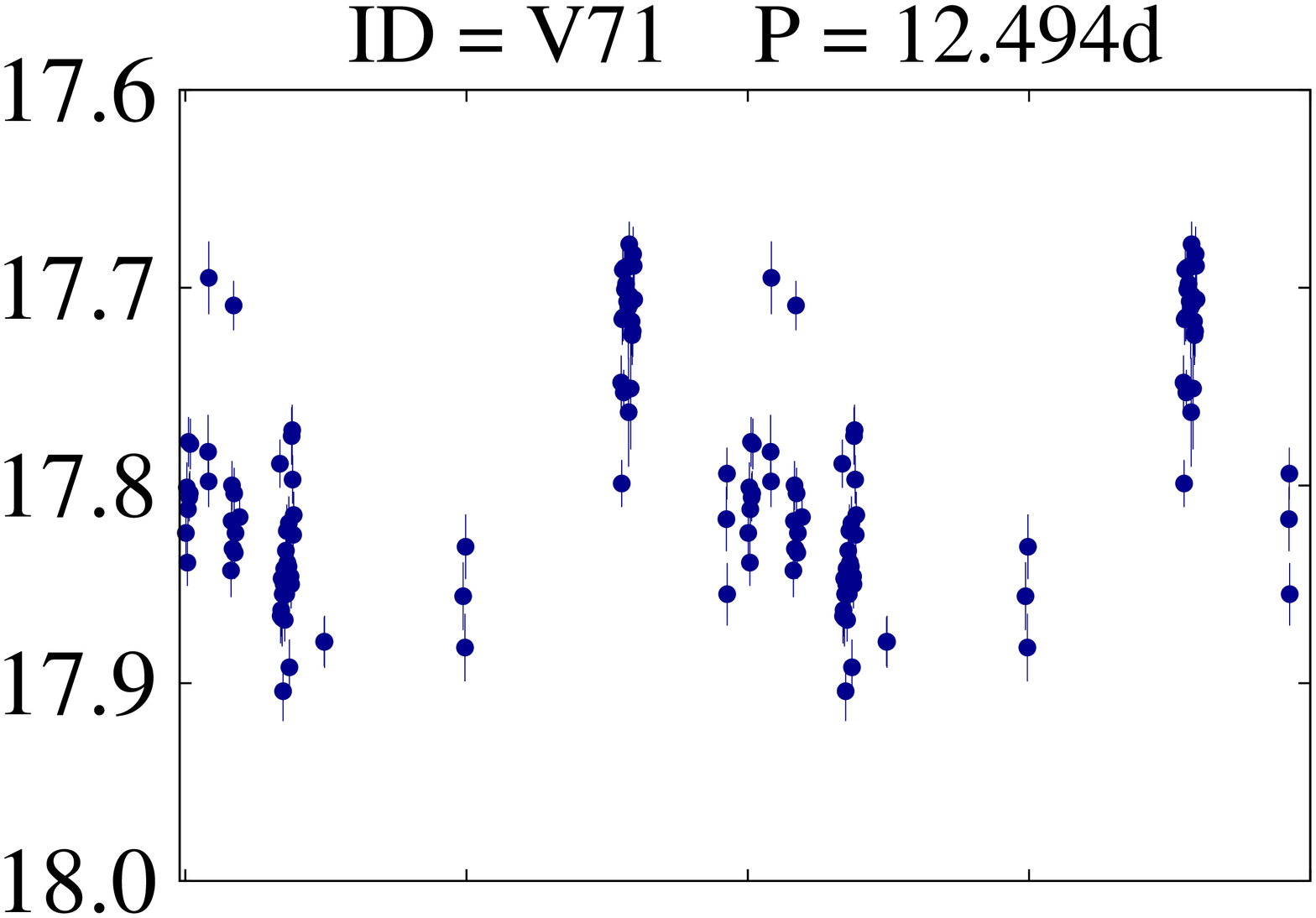}
\hspace{0.03 cm}
\includegraphics[width = 1.9 cm, height = 2.6 cm, angle= 270]{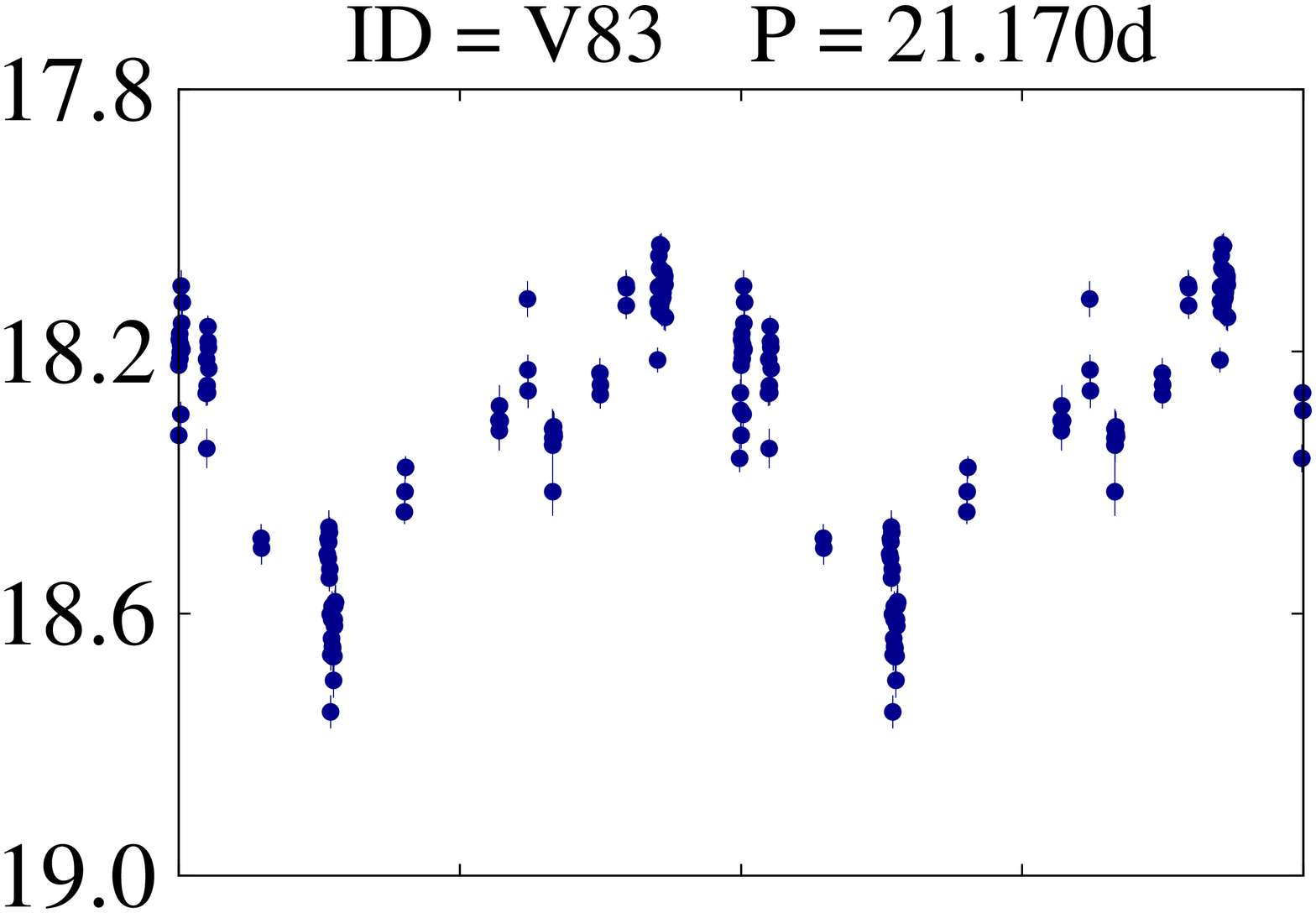}  
\vspace{0.05 cm}
\includegraphics[width = 1.9 cm, height = 2.6 cm, angle= 270]{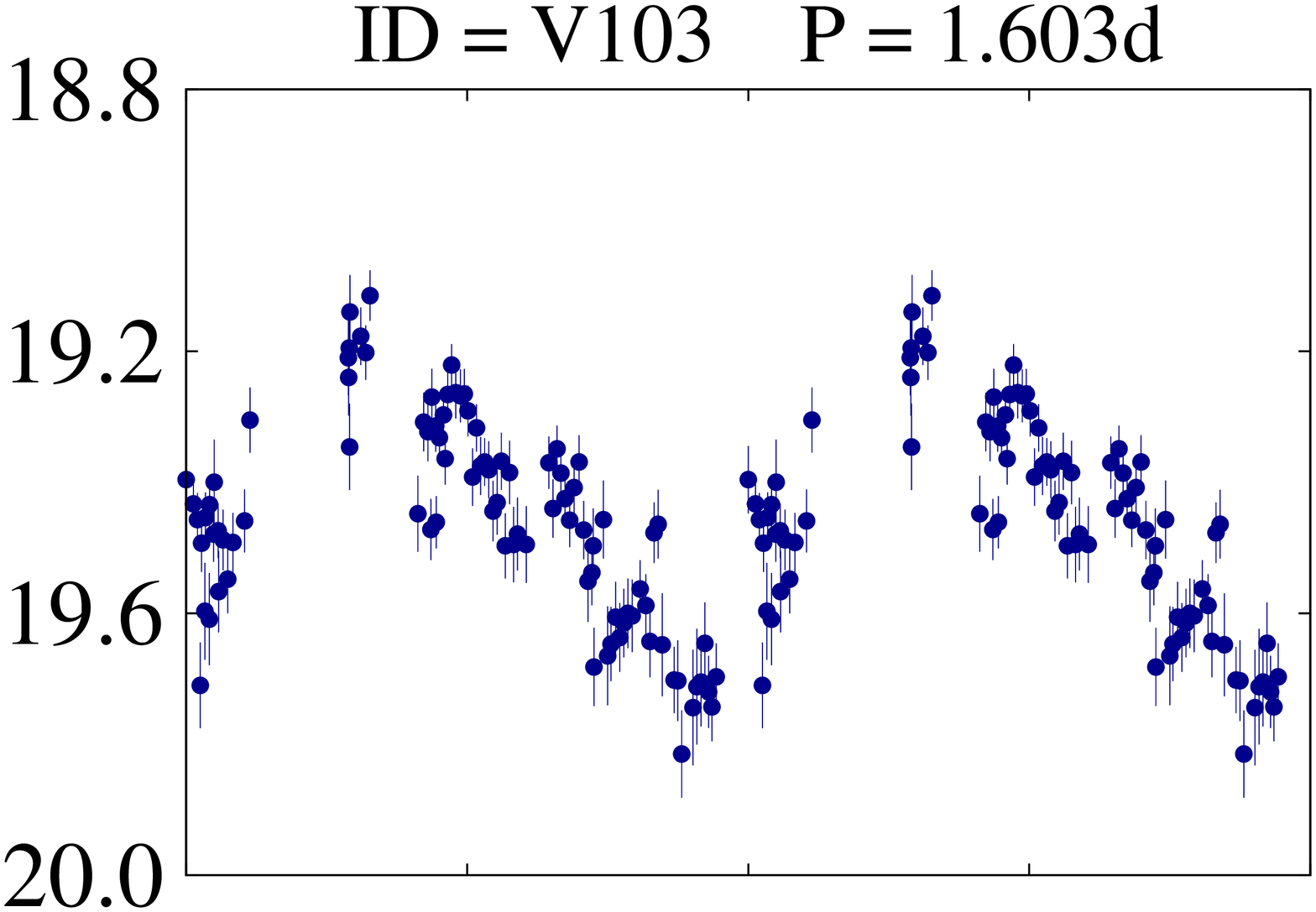}
\hspace{0.03 cm}
\includegraphics[width = 1.9 cm, height = 2.6 cm, angle= 270]{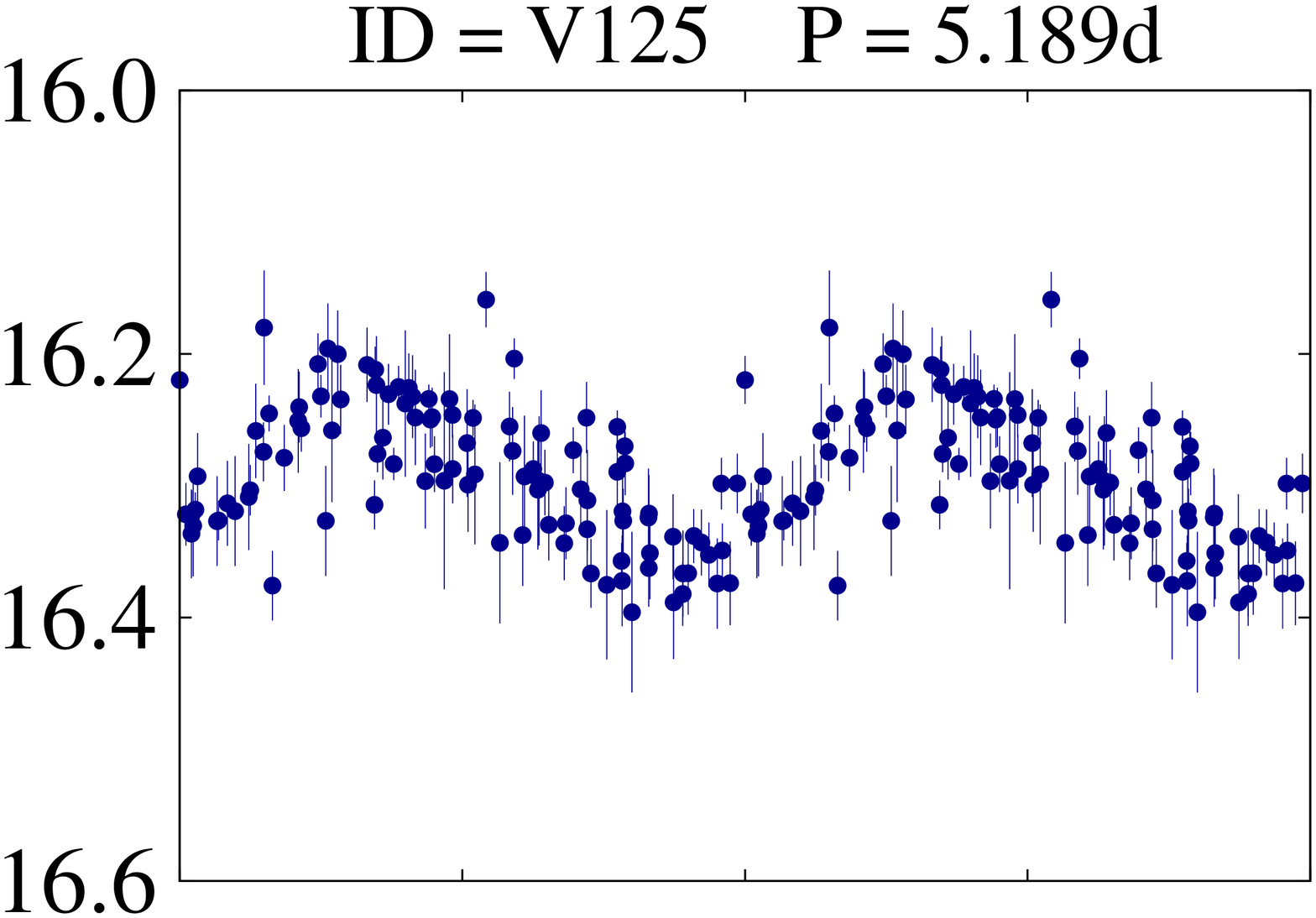}
\hspace{0.03 cm}
\includegraphics[width = 1.9 cm, height = 2.6 cm, angle= 270]{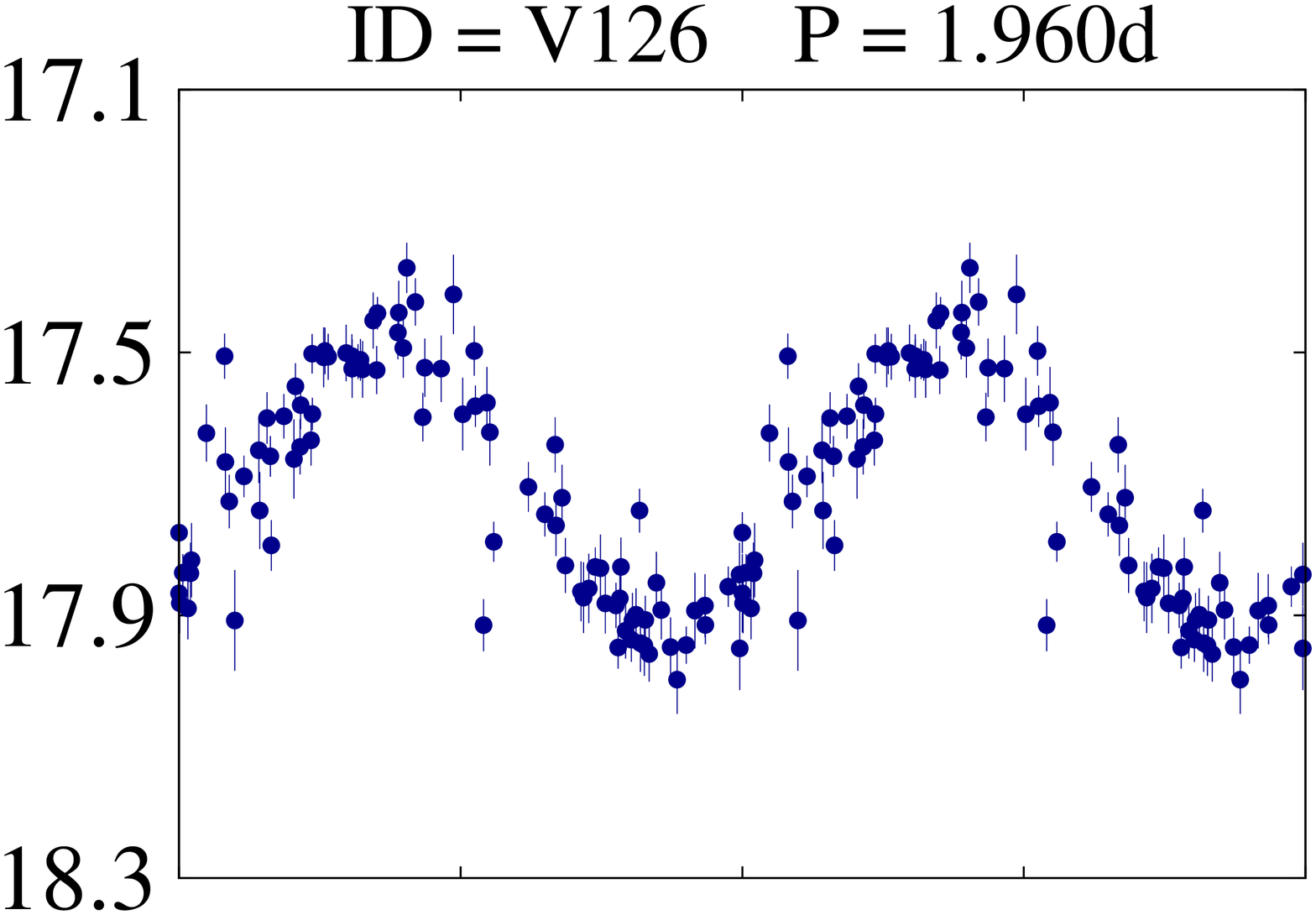}  \\
\vspace{0.05 cm}
\includegraphics[width = 1.9 cm, height = 2.6 cm, angle= 270]{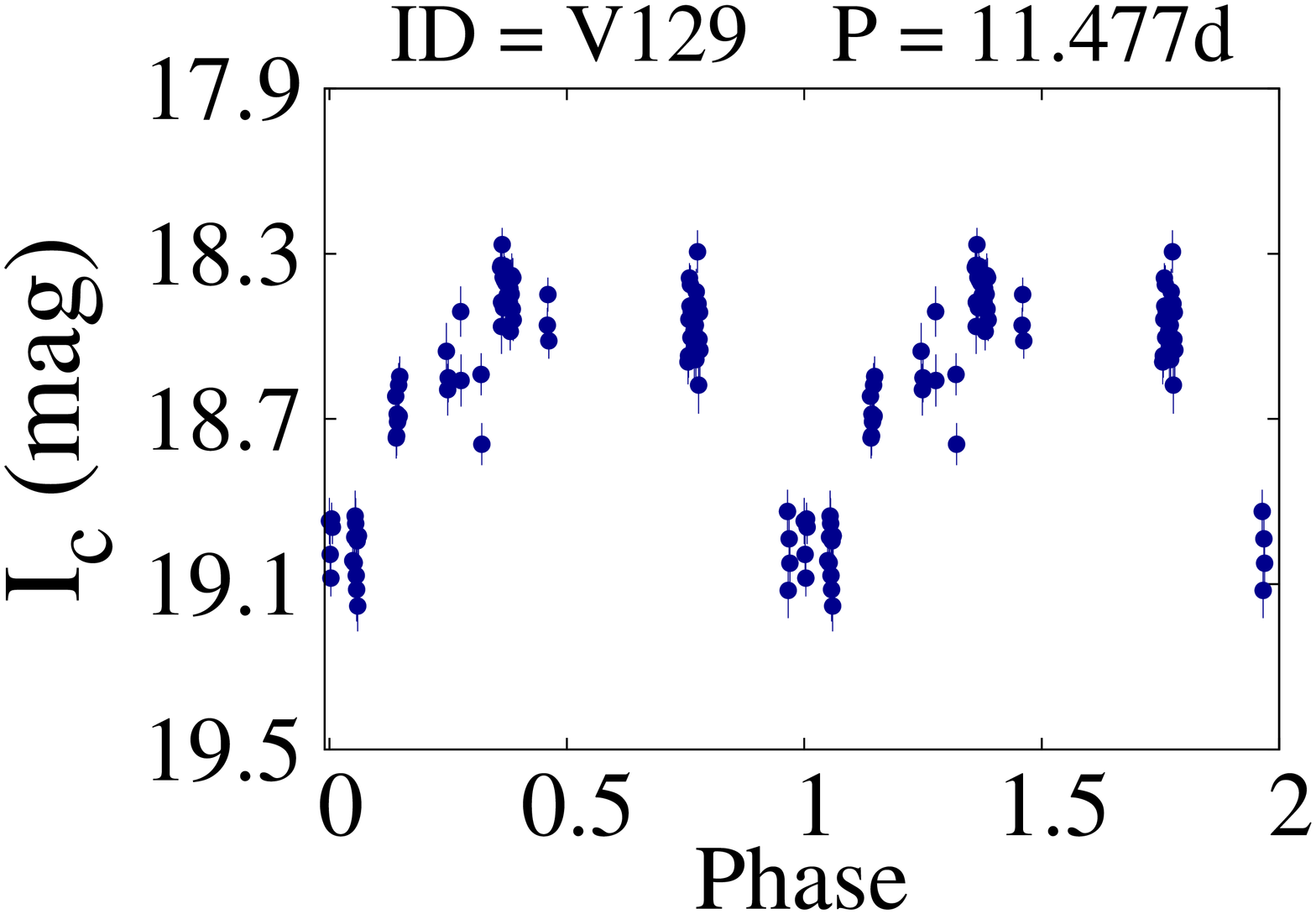}
\hspace{0.03 cm}
\includegraphics[width = 1.9 cm, height = 2.6 cm, angle= 270]{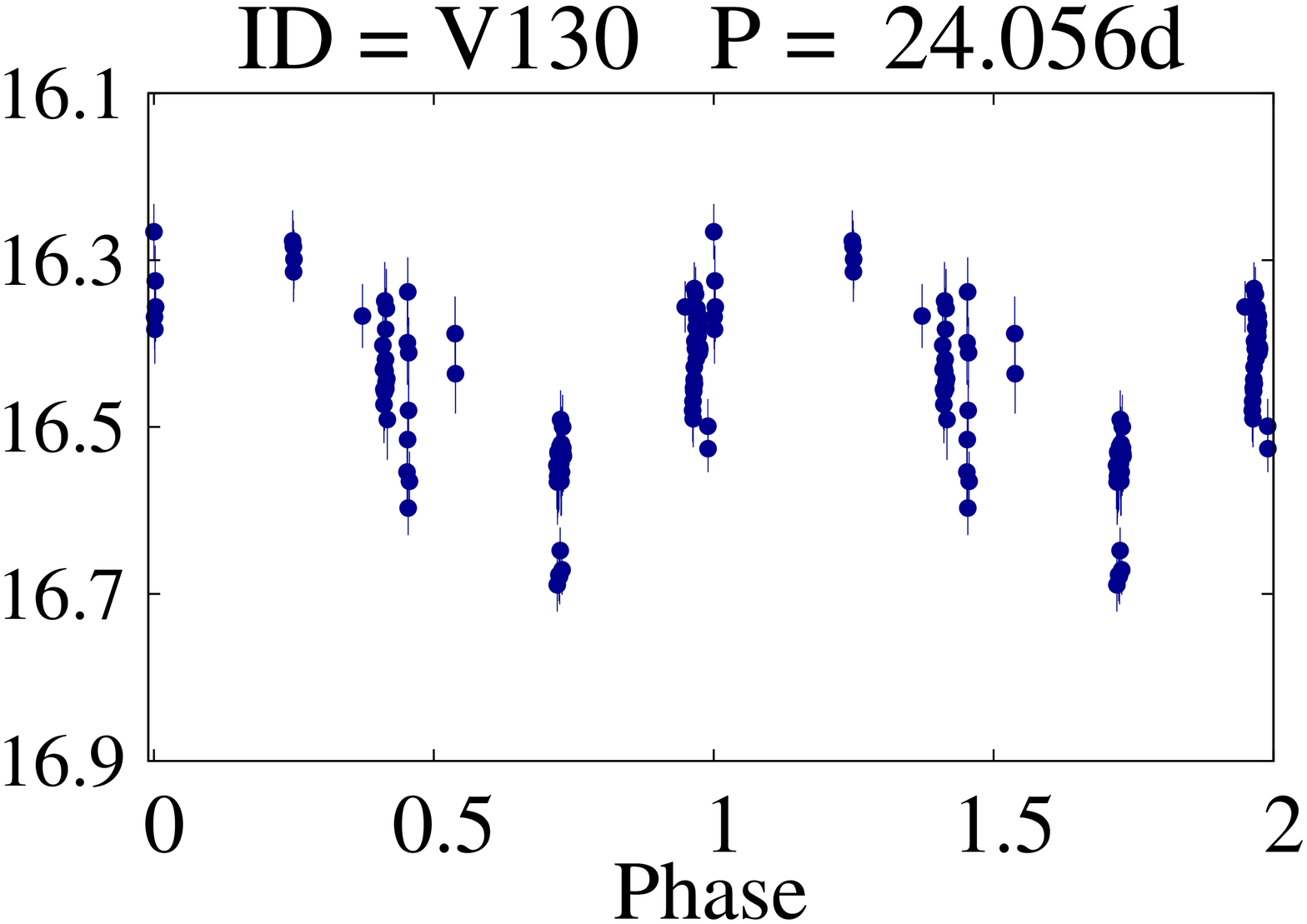}  
\vspace{0.05 cm}
\includegraphics[width = 1.9 cm, height = 2.6 cm, angle= 270]{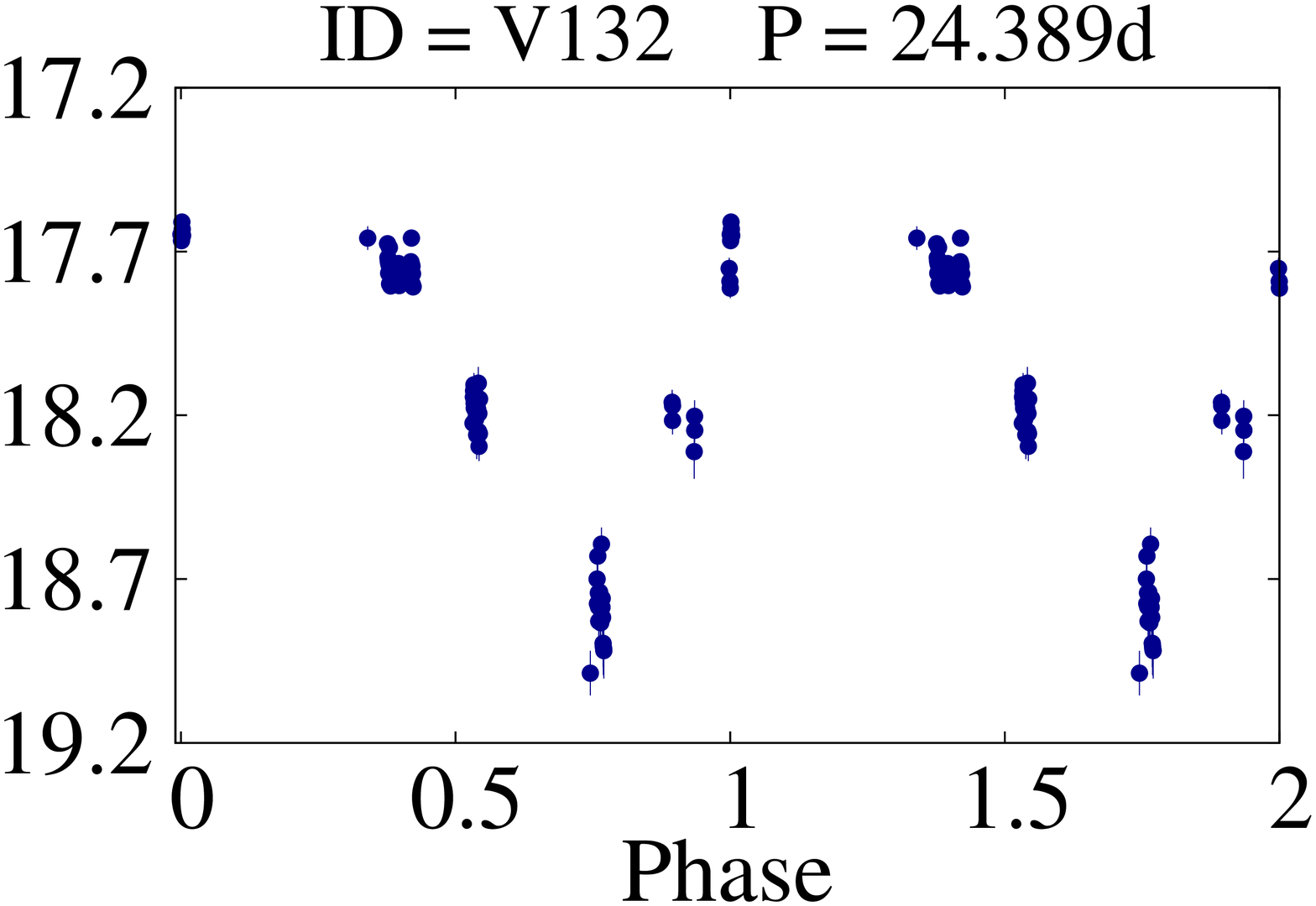}
\hspace{0.03 cm}
\includegraphics[width = 1.9 cm, height = 2.6 cm, angle= 270]{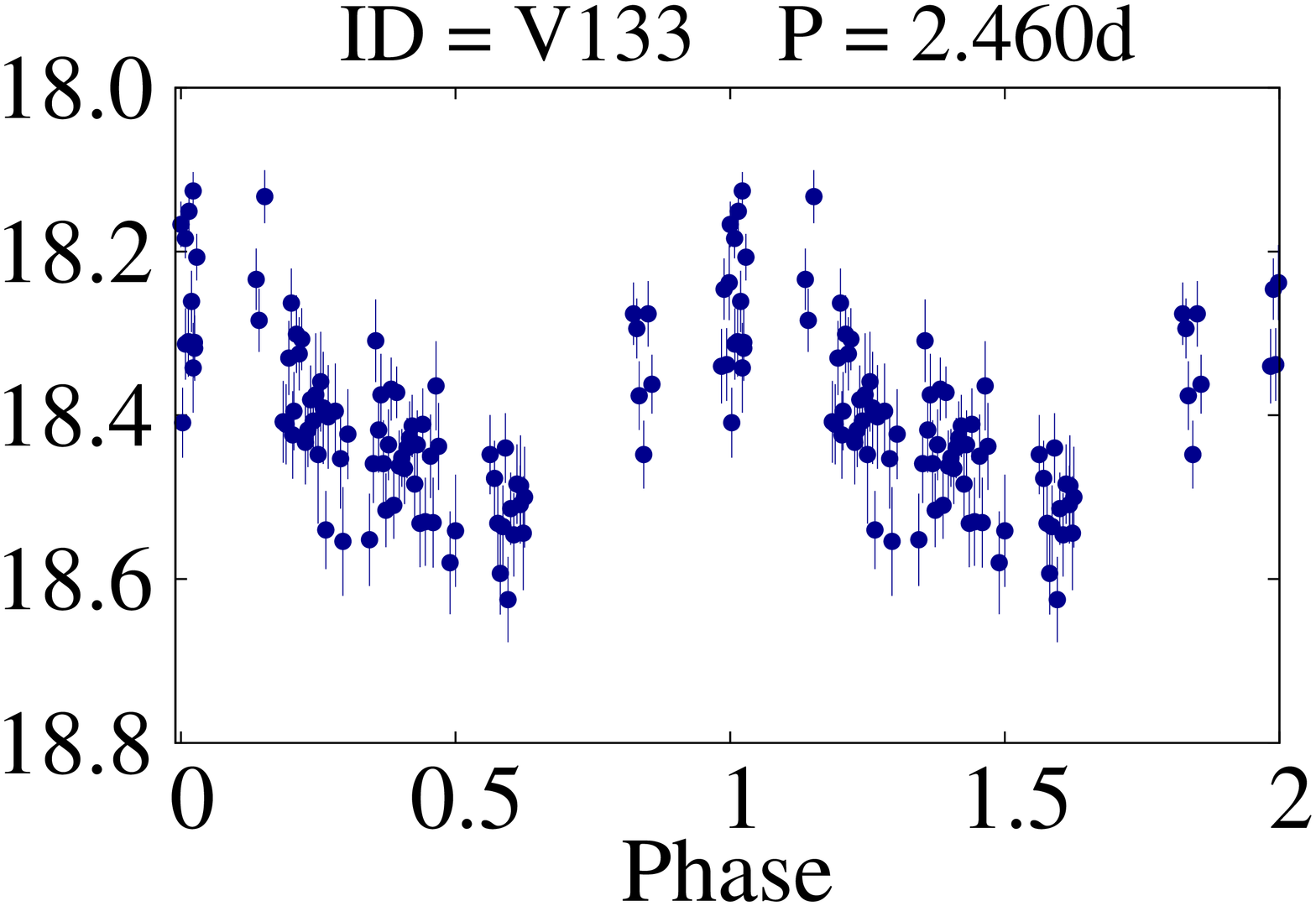}
\hspace{0.03 cm}
\includegraphics[width = 1.9 cm, height = 2.6 cm, angle= 270]{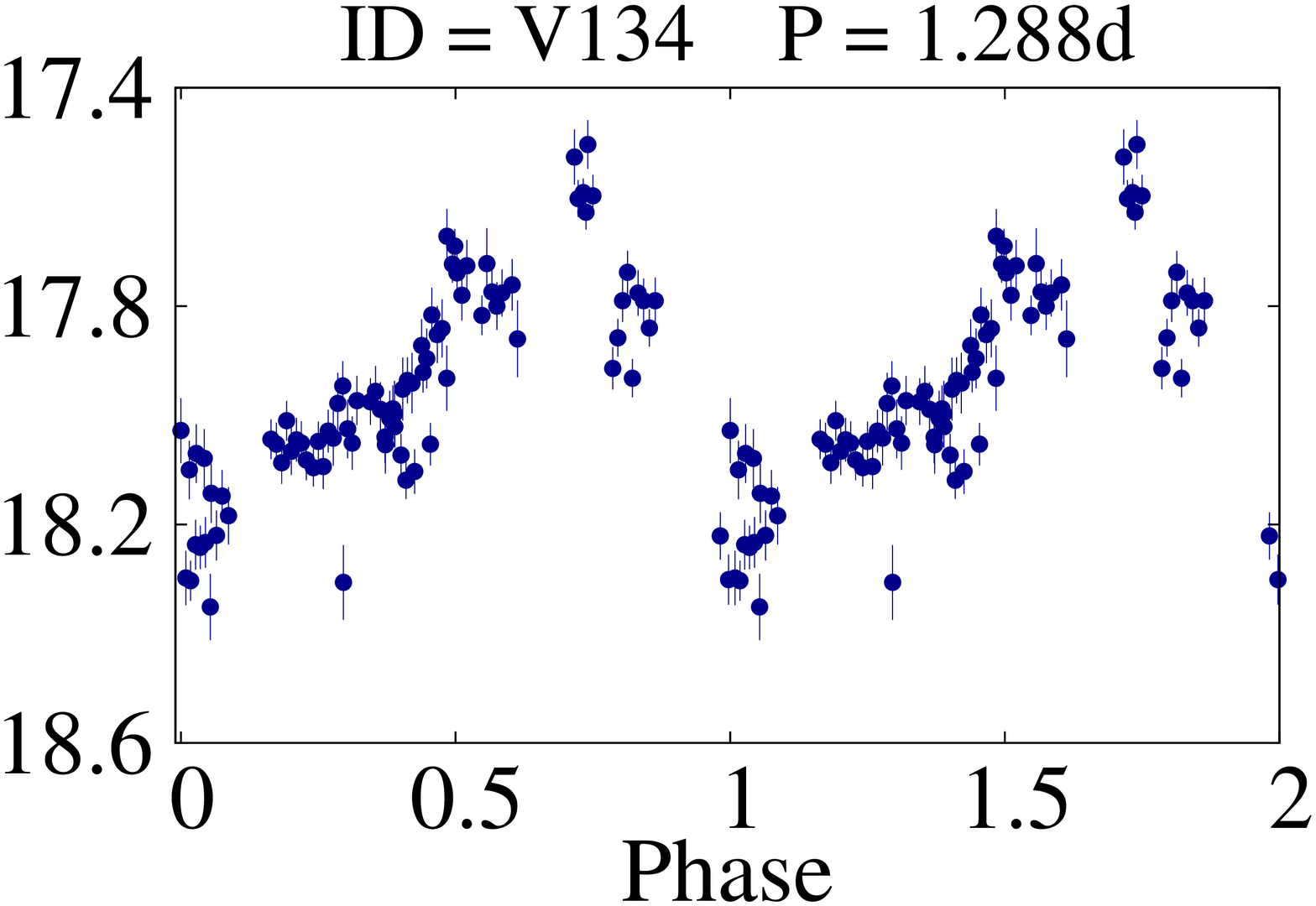}

\caption{Phase folded LCs of 17 Class\,{\sc ii} periodic
 variables.
The identification numbers and periods (days) of the corresponding stars are given on the top of
each panel.}
 \label{fig: LC_Class_II_P}
\end{figure}

\begin{figure*}[h]
\label{fig: LC_CTT_1}
\centering
\includegraphics[width= 4.0 cm,height = 7.0 cm, angle=270]{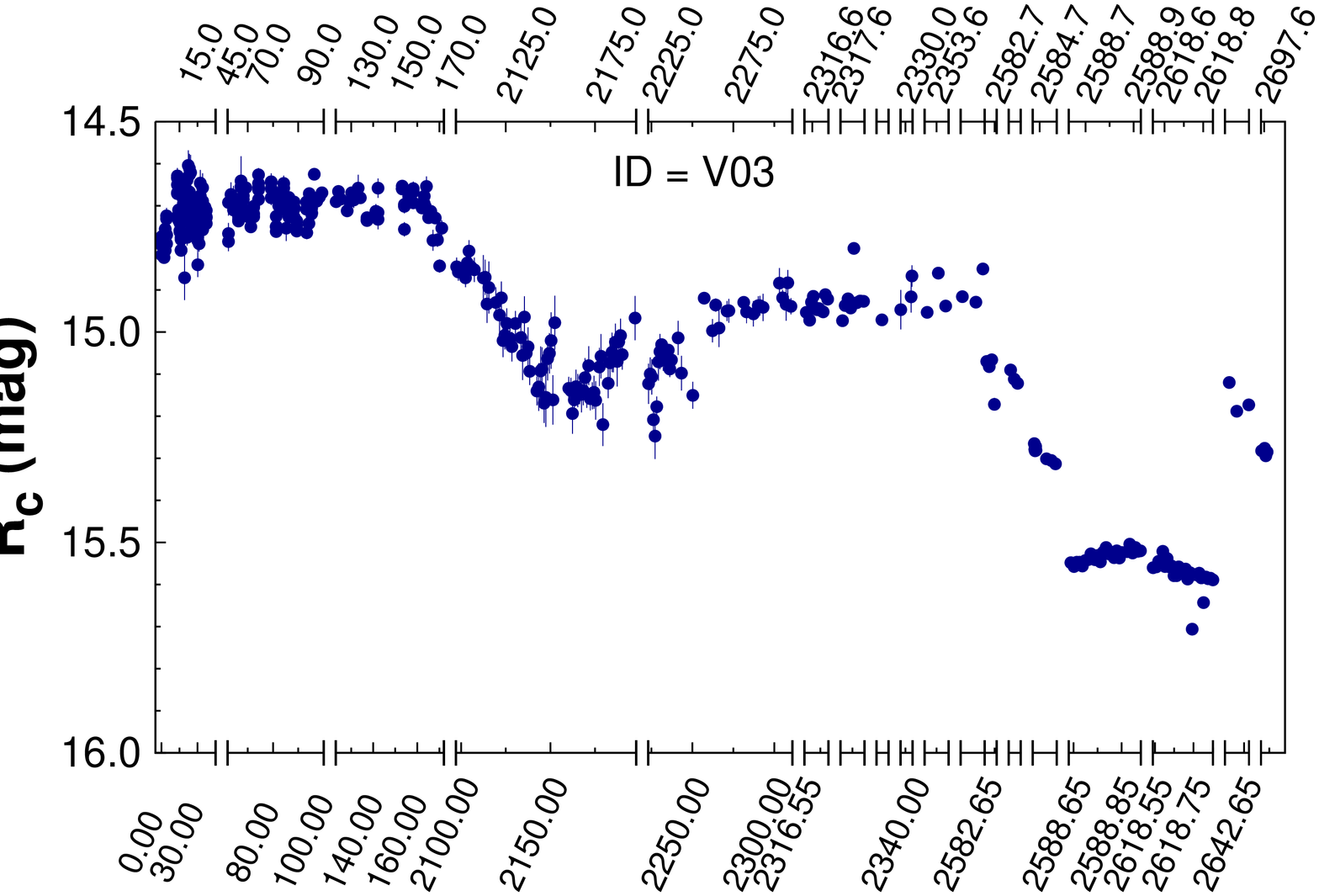}
\hspace{0.1 cm}
\includegraphics[width= 4.0 cm,height = 7.0 cm, angle=270]{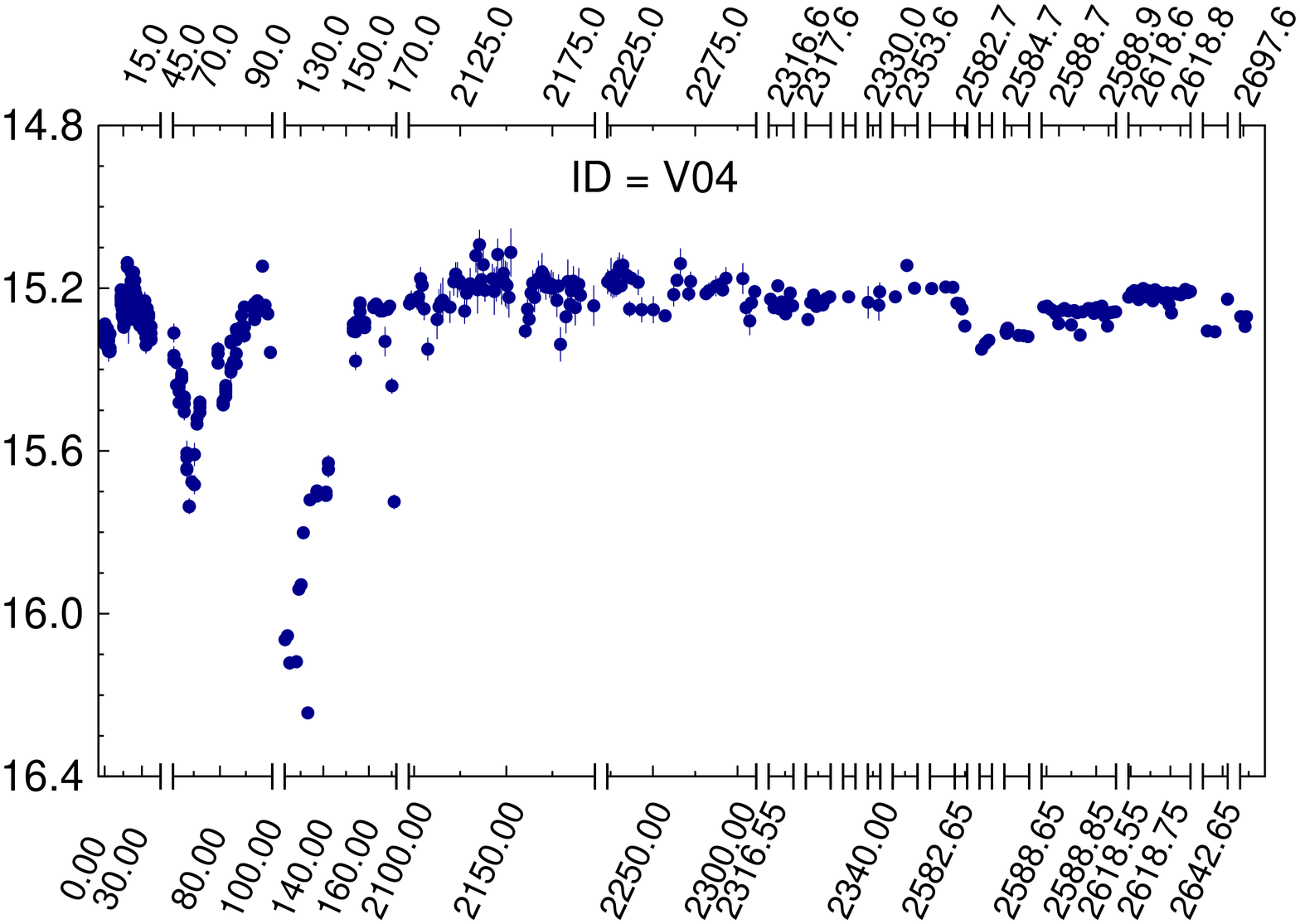} \\
\vspace{0.2 cm}
\includegraphics[width= 4.0 cm,height = 7.0 cm, angle=270]{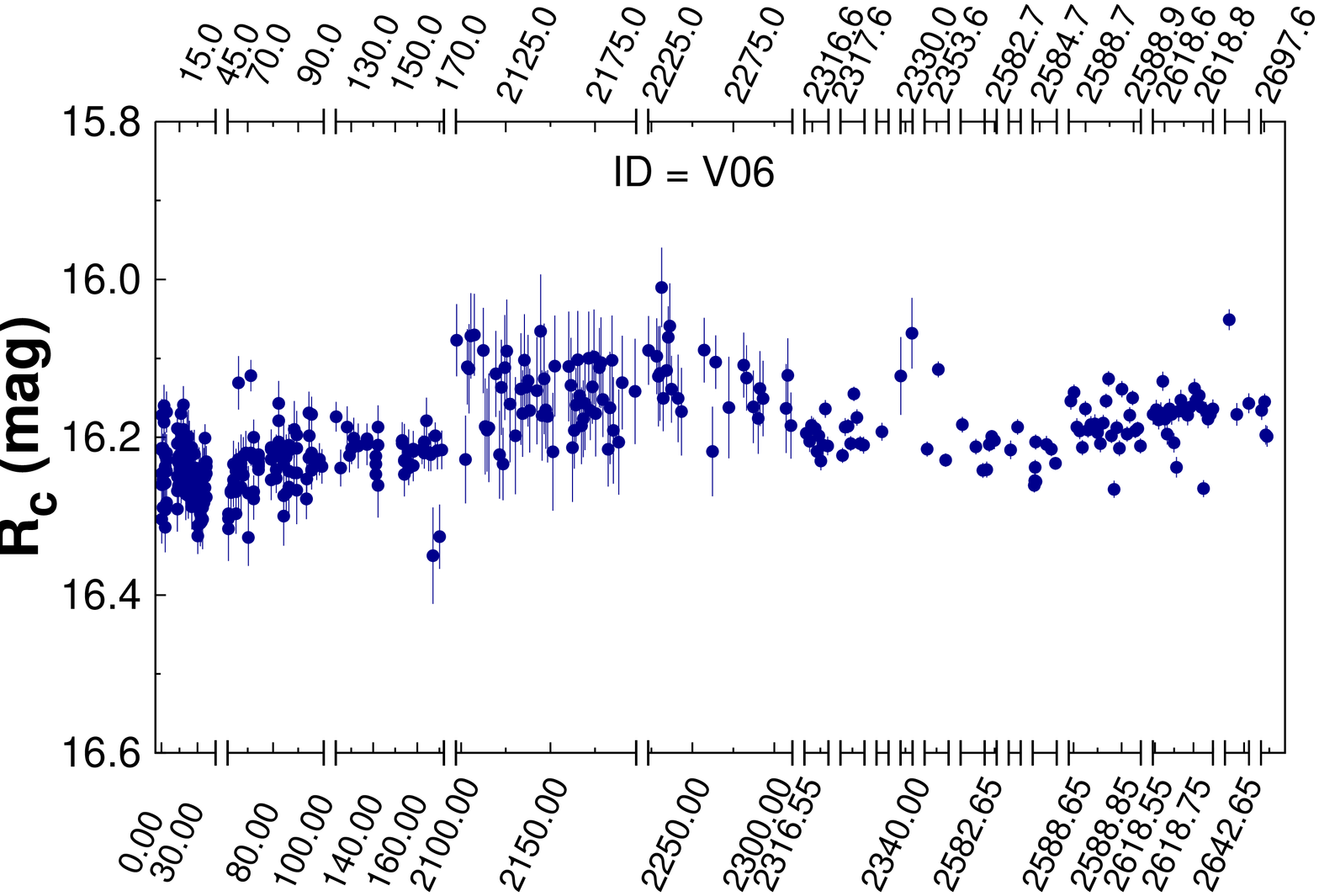}
\hspace{0.1 cm}
\includegraphics[width= 4.0 cm,height = 7.0 cm, angle=270]{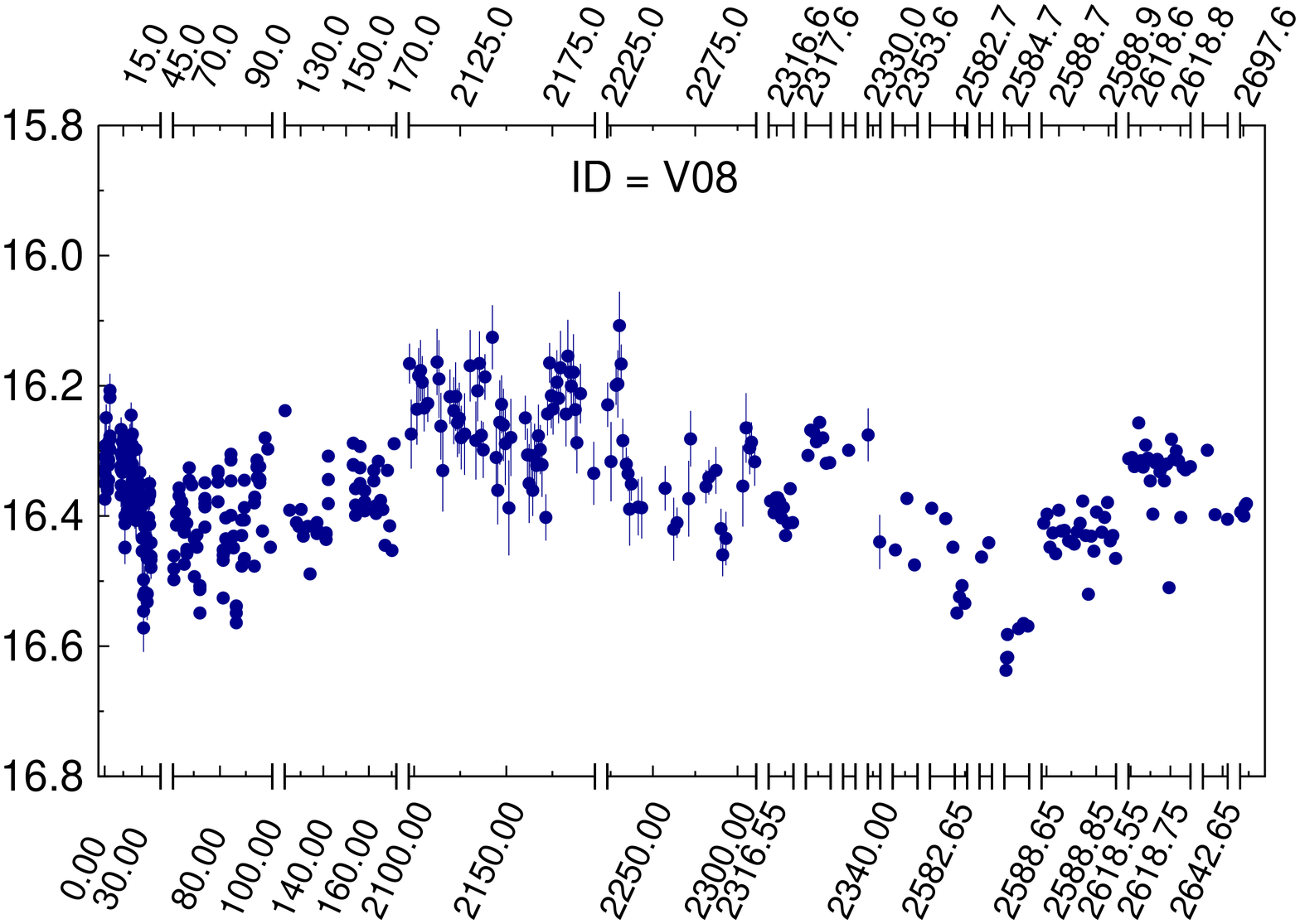} \\
\vspace{0.2 cm}
\includegraphics[width= 4.0 cm,height = 7.0 cm, angle=270]{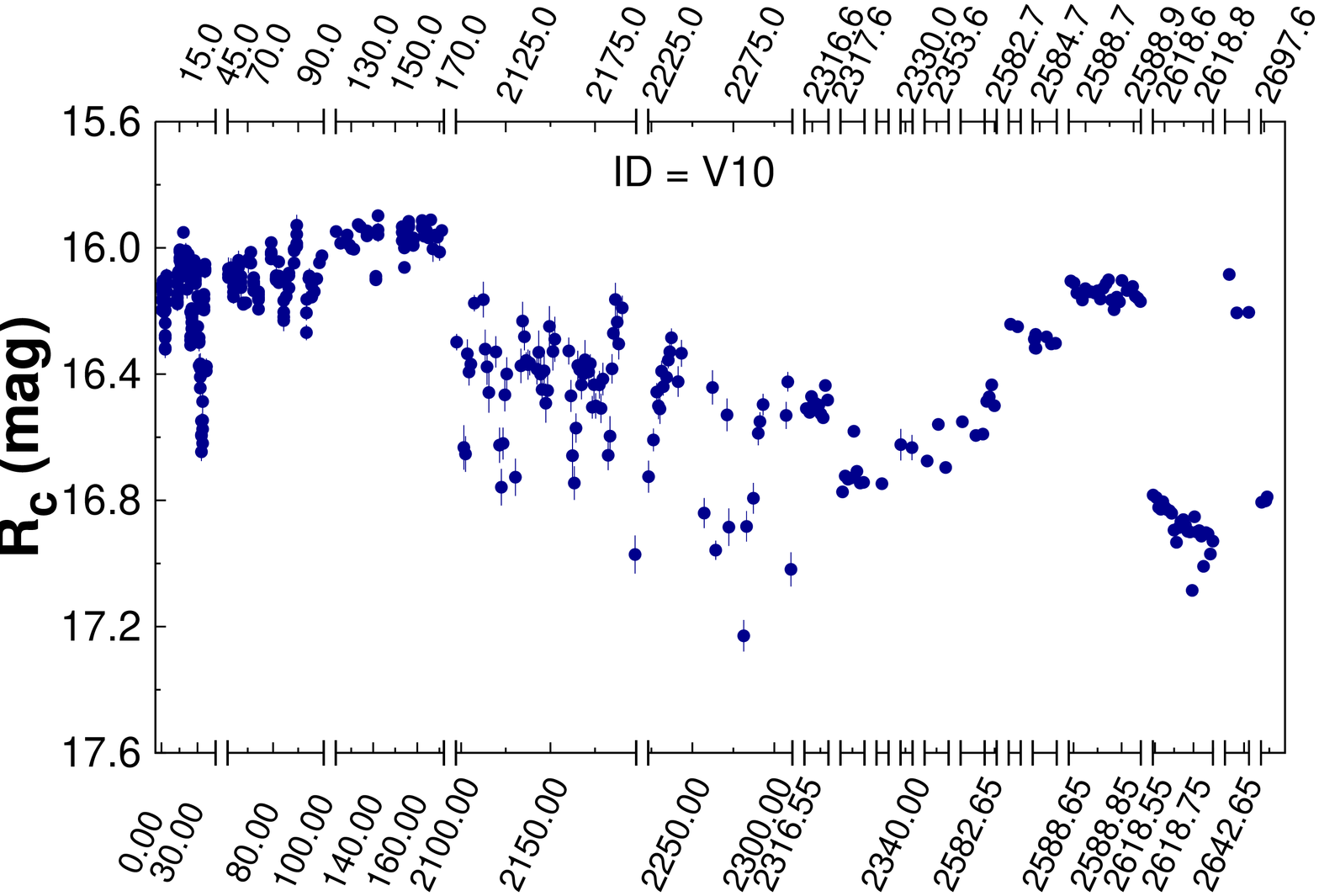}
\hspace{0.1 cm}
\includegraphics[width= 4.0 cm,height = 7.0 cm, angle=270]{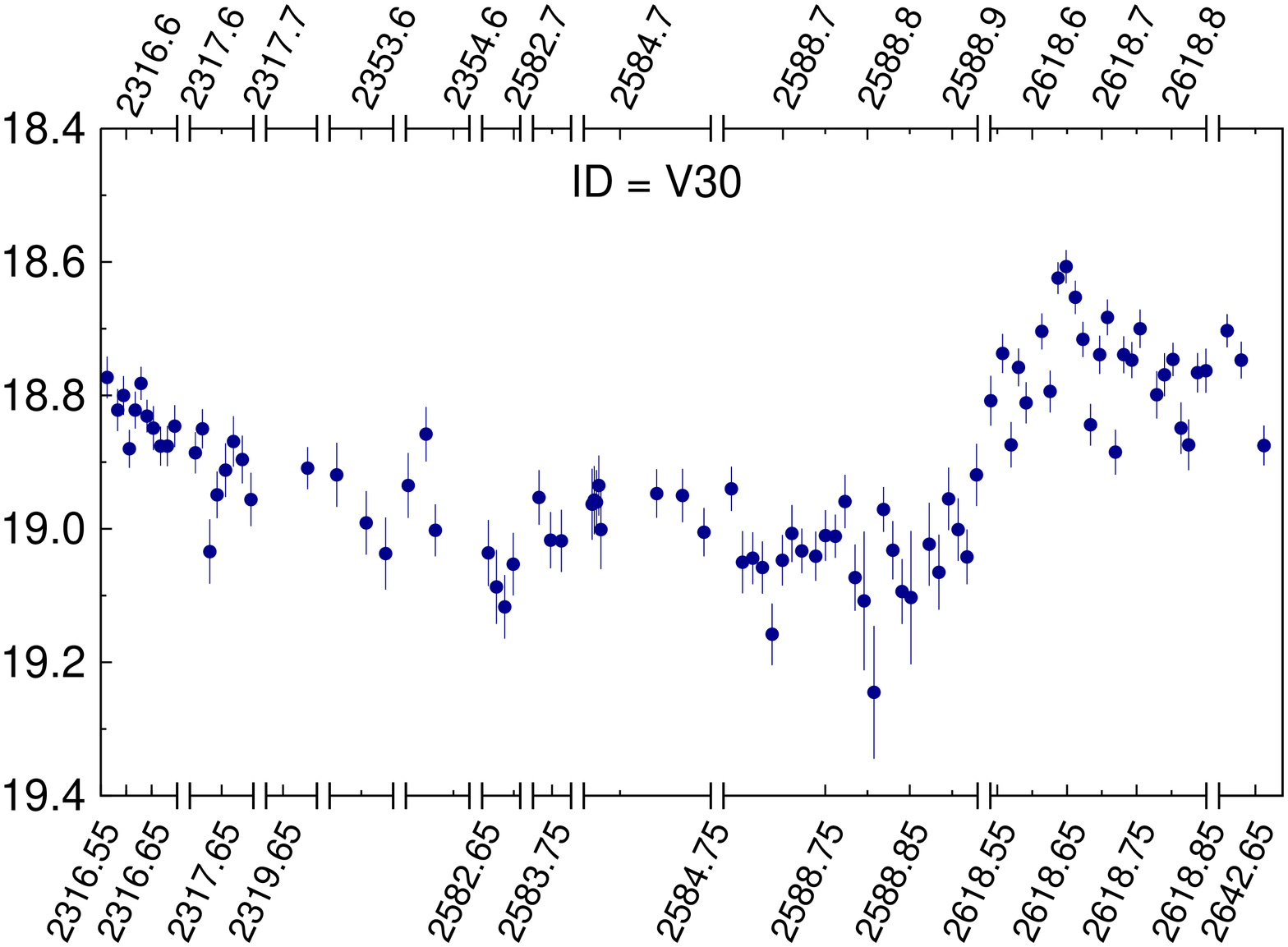} \\
\vspace{0.2 cm}
\includegraphics[width= 4.0 cm,height = 7.0 cm, angle=270]{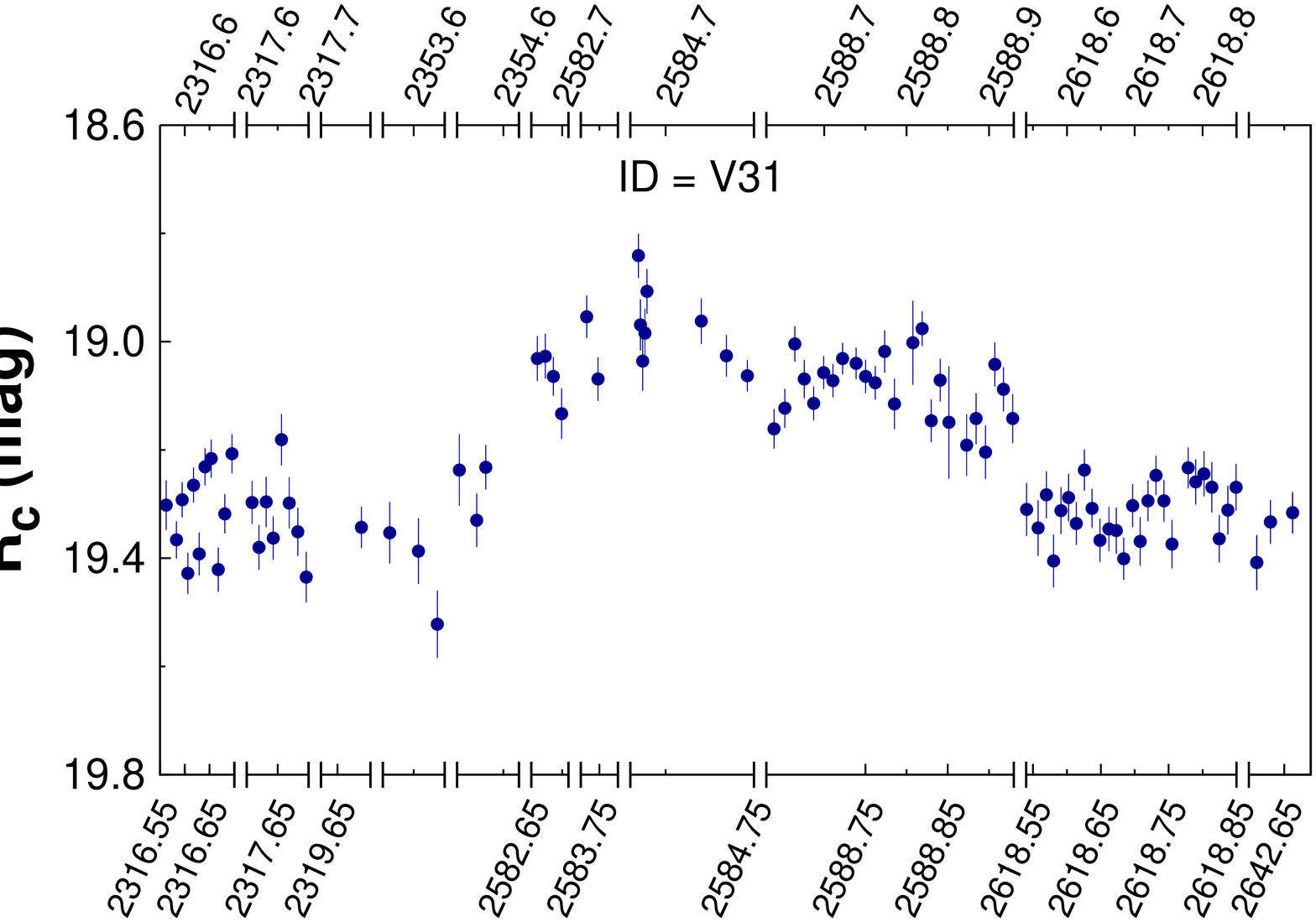}
\hspace{0.1 cm}
\includegraphics[width= 4.0 cm,height = 7.0 cm, angle=270]{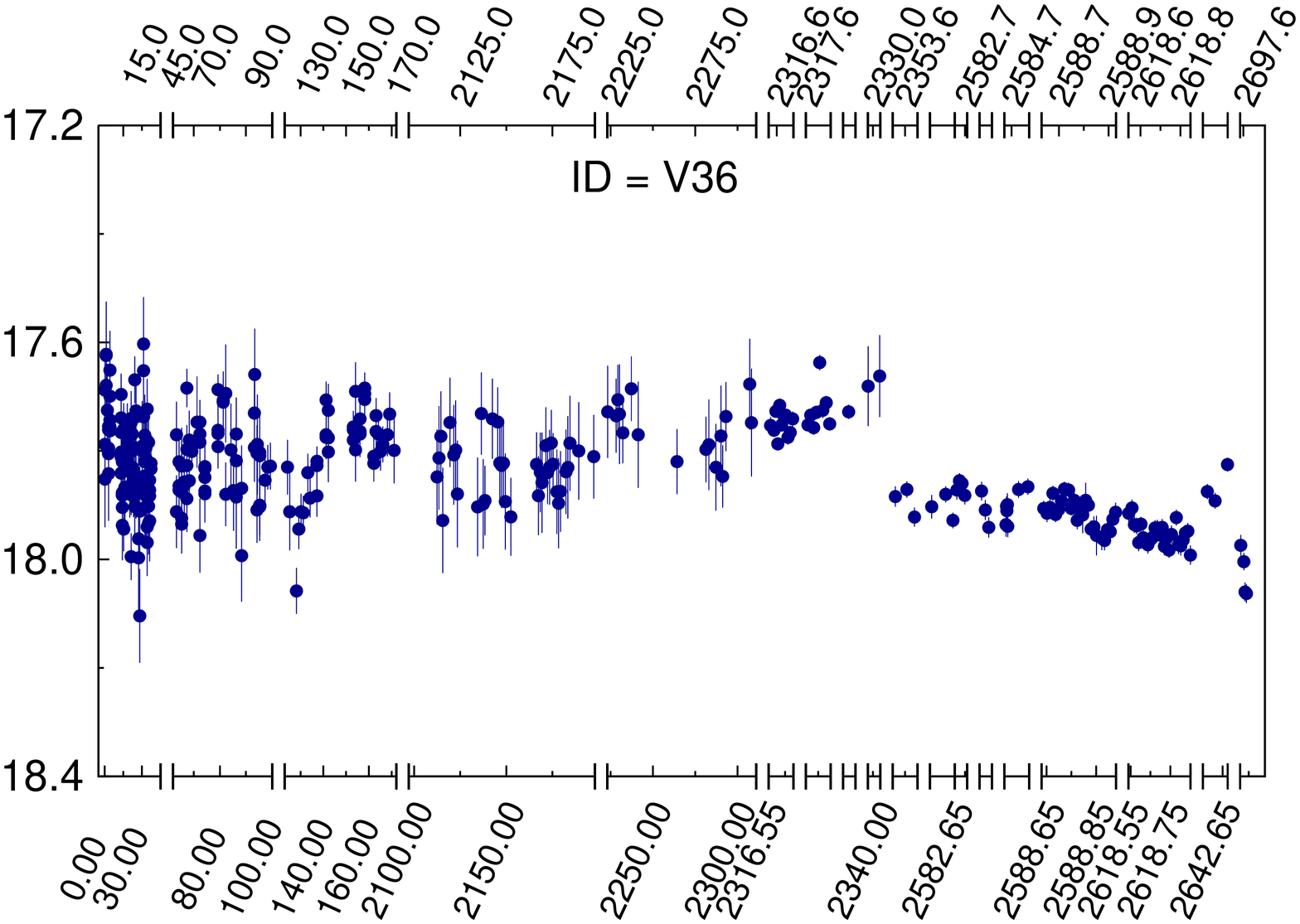} \\
\vspace{0.2 cm}
\includegraphics[width= 4.0 cm,height = 7.0 cm, angle=270]{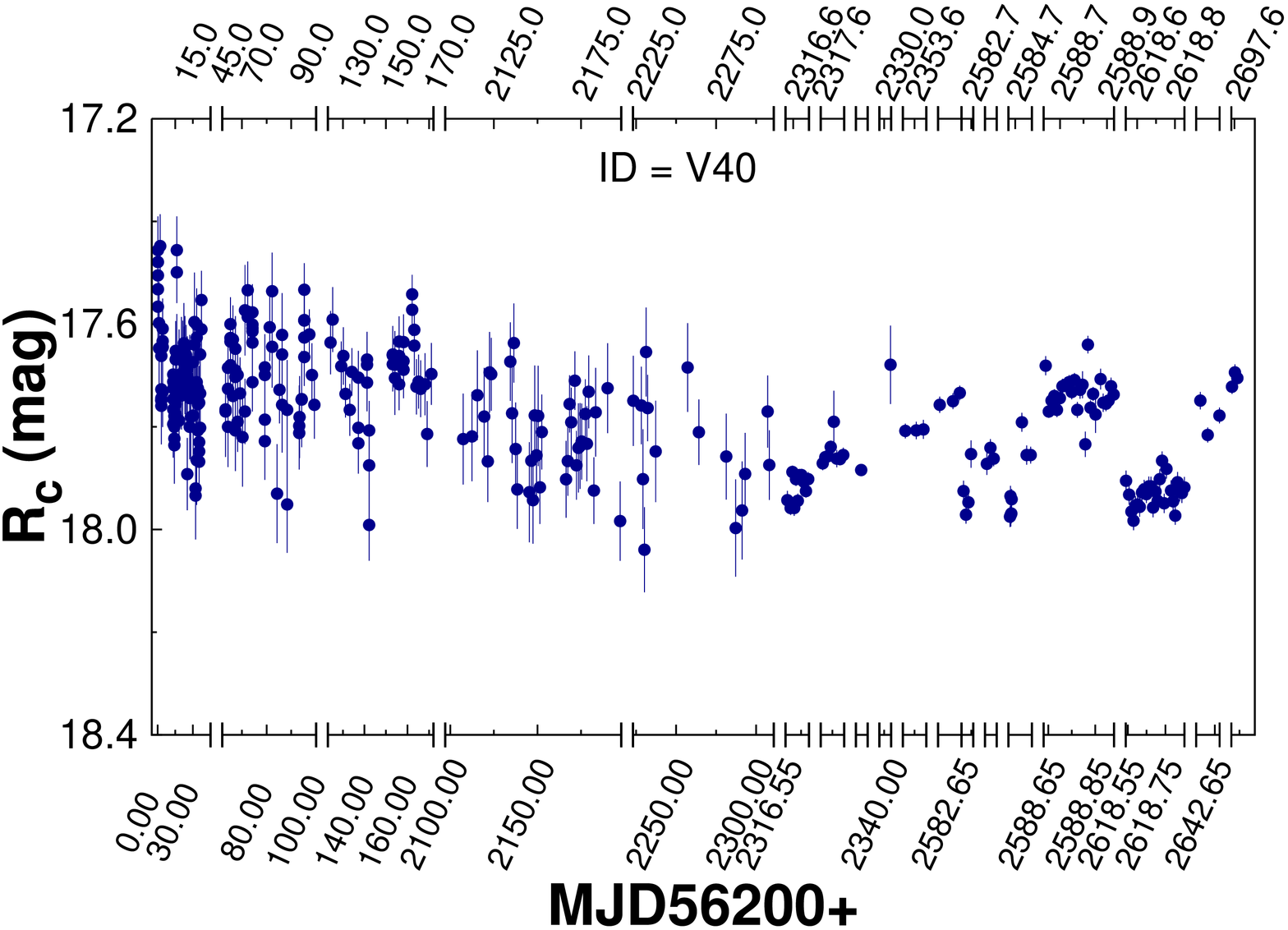}
\hspace{0.1 cm}
\includegraphics[width= 4.0 cm,height = 7.0 cm, angle=270]{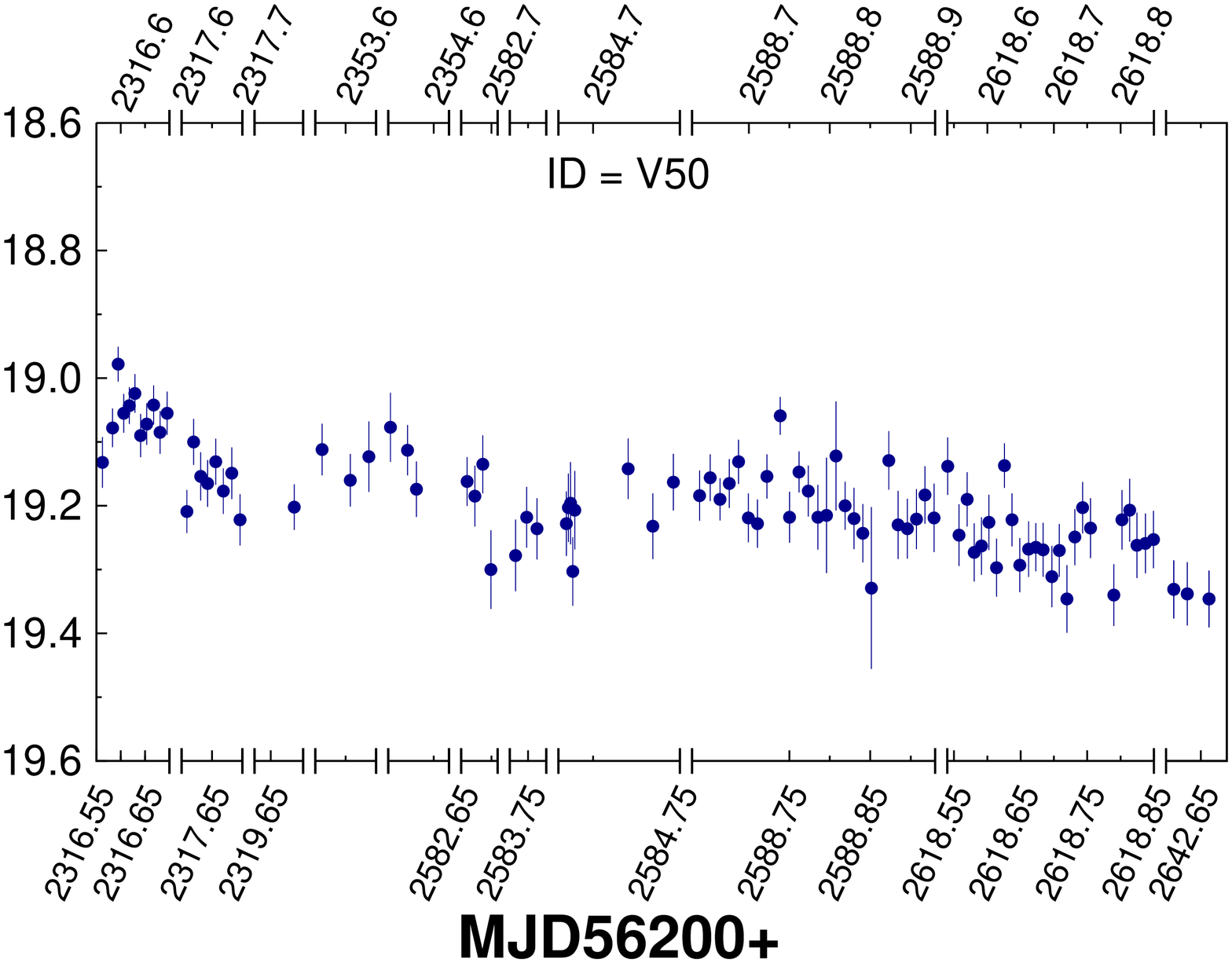} \\

\caption{\scriptsize {LCs of 20 Class\,{\sc ii} non-periodic variables (CTTSs). When there are data gaps they are represented with vertical gaps along the axis. Corresponding identification numbers are given in each panel.}}

\end{figure*}

\setcounter{figure}{8}

\begin{figure*}[h]
\centering
\vspace{0.03 cm}
\includegraphics[width= 4.0 cm,height = 7.0 cm, angle=270]{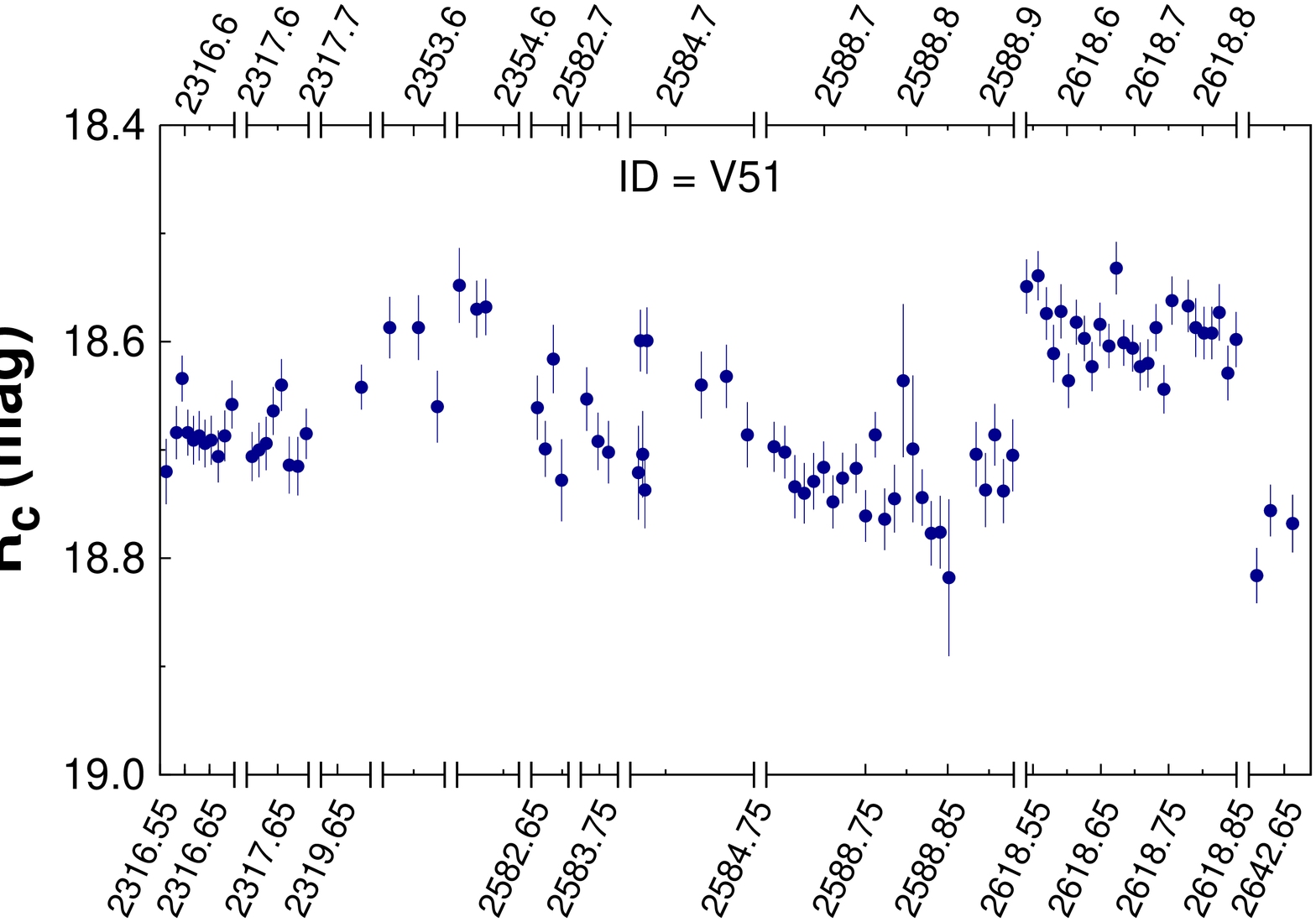}
\hspace{0.1 cm}
\includegraphics[width= 4.0 cm,height = 7.0 cm, angle=270]{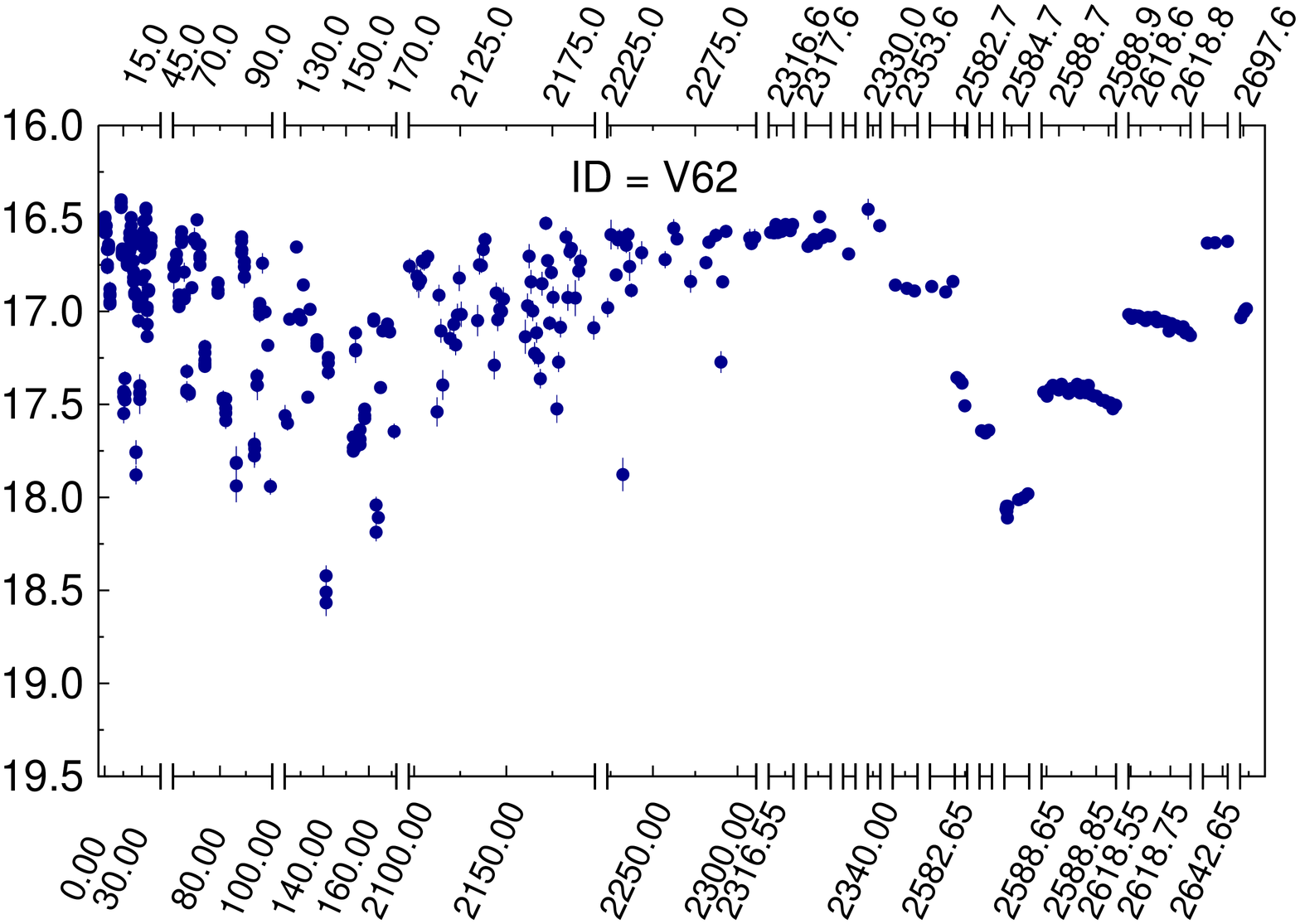} \\
\vspace{0.2 cm}
\includegraphics[width= 4.0 cm,height = 7.0 cm, angle=270]{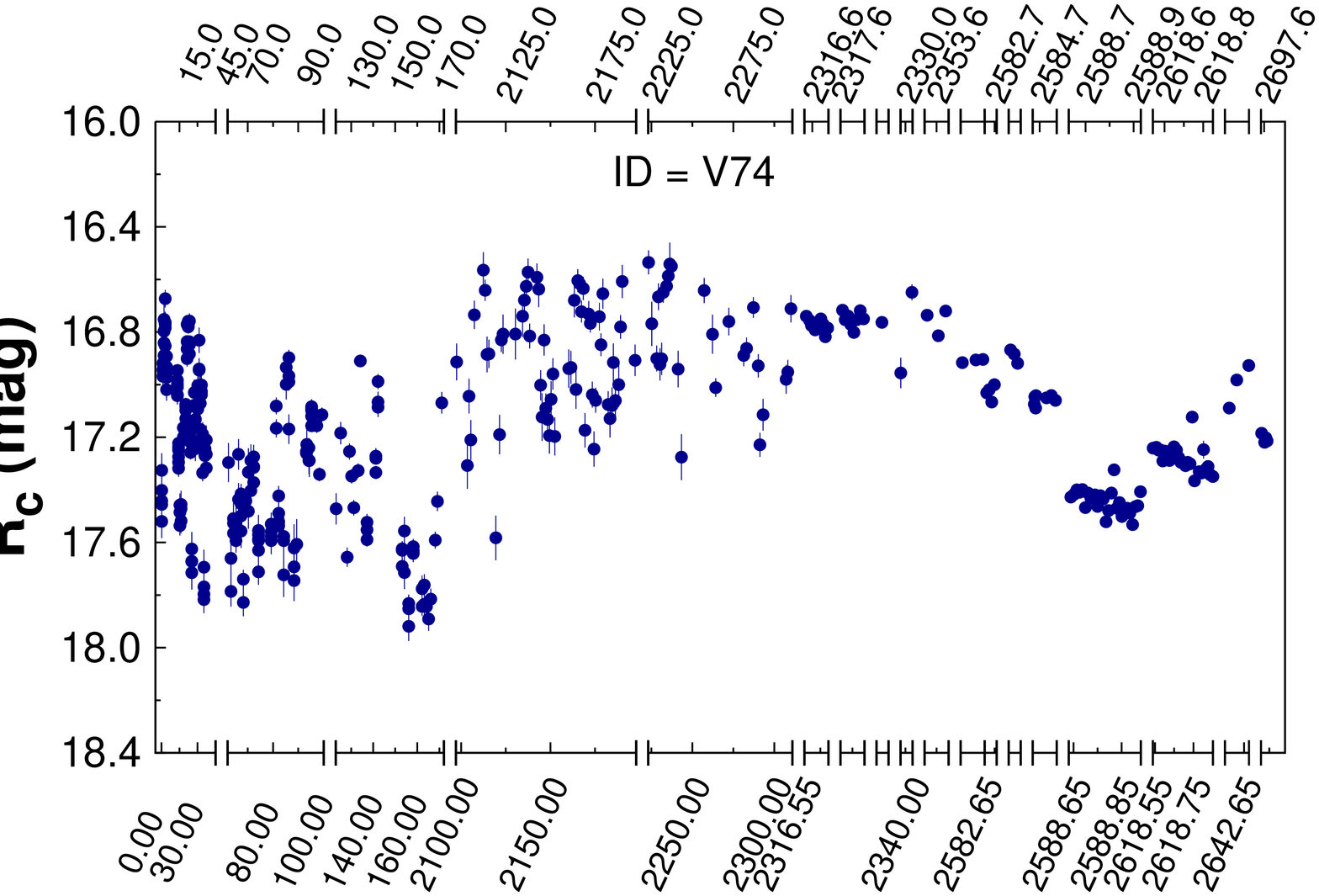}
\hspace{0.1 cm}
\includegraphics[width= 4.0 cm,height = 7.0 cm, angle=270]{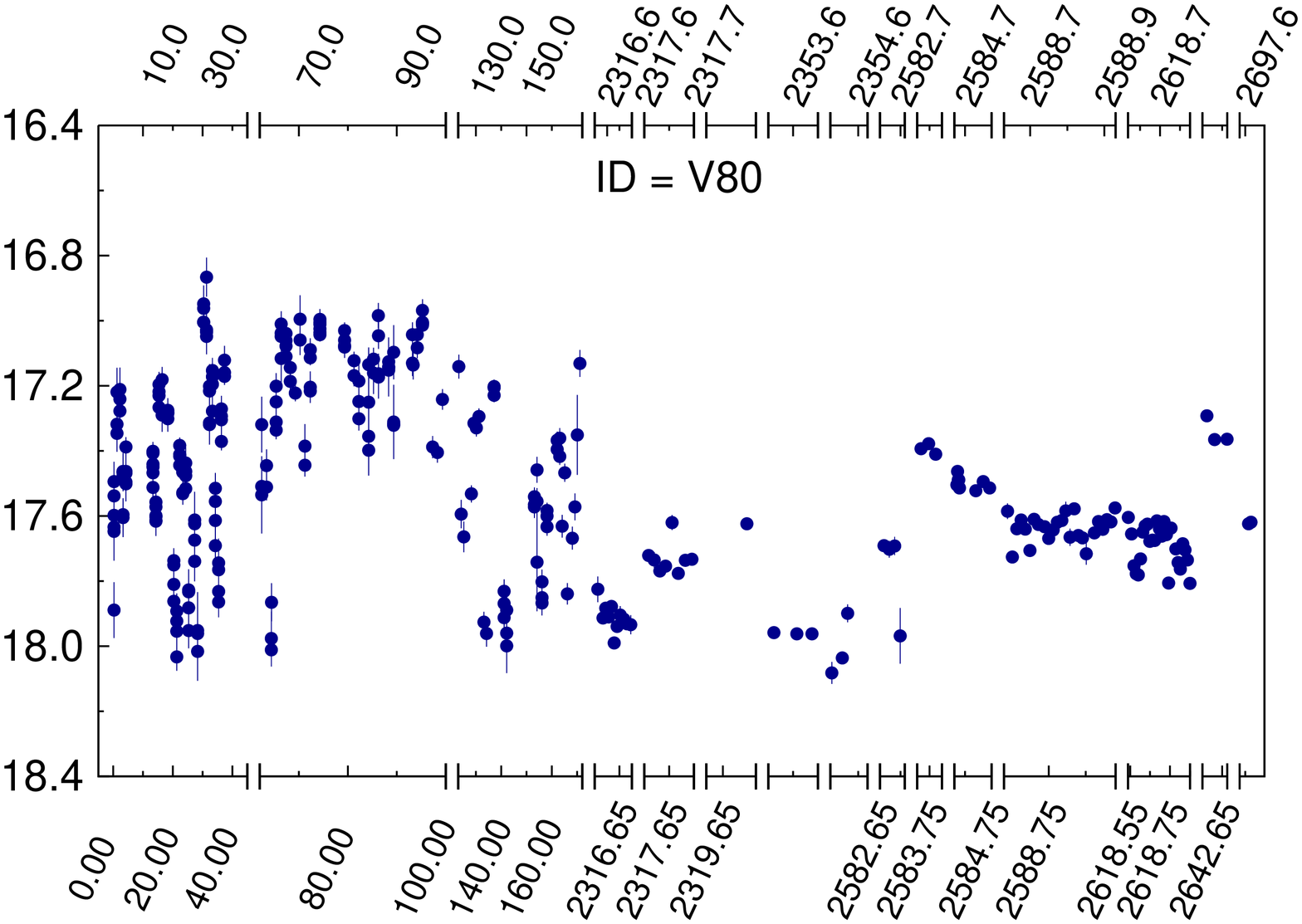} \\
\vspace{0.2 cm}
\includegraphics[width= 4.0 cm,height = 7.0 cm, angle=270]{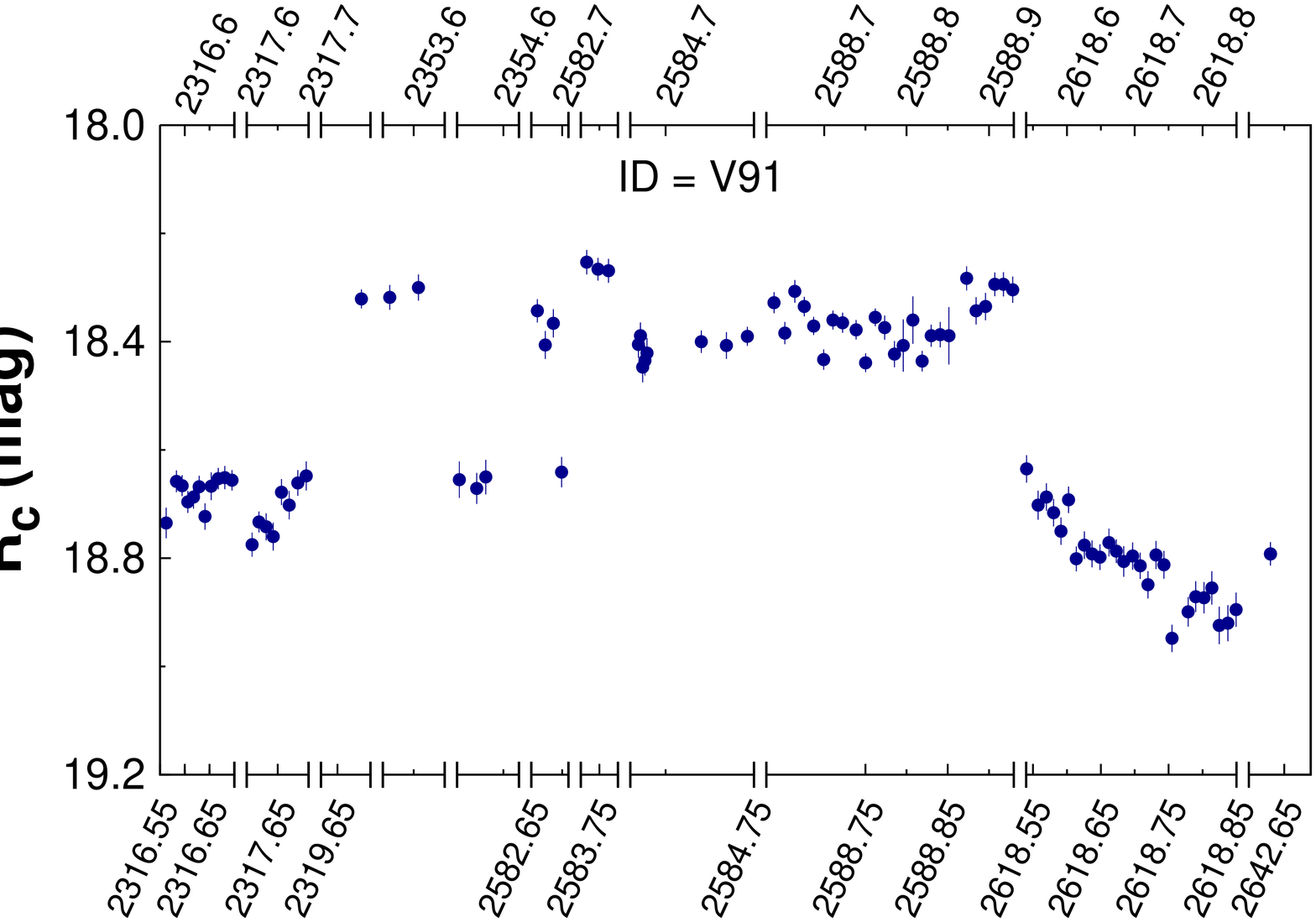}
\hspace{0.1 cm}
\includegraphics[width= 4.0 cm,height = 7.0 cm, angle=270]{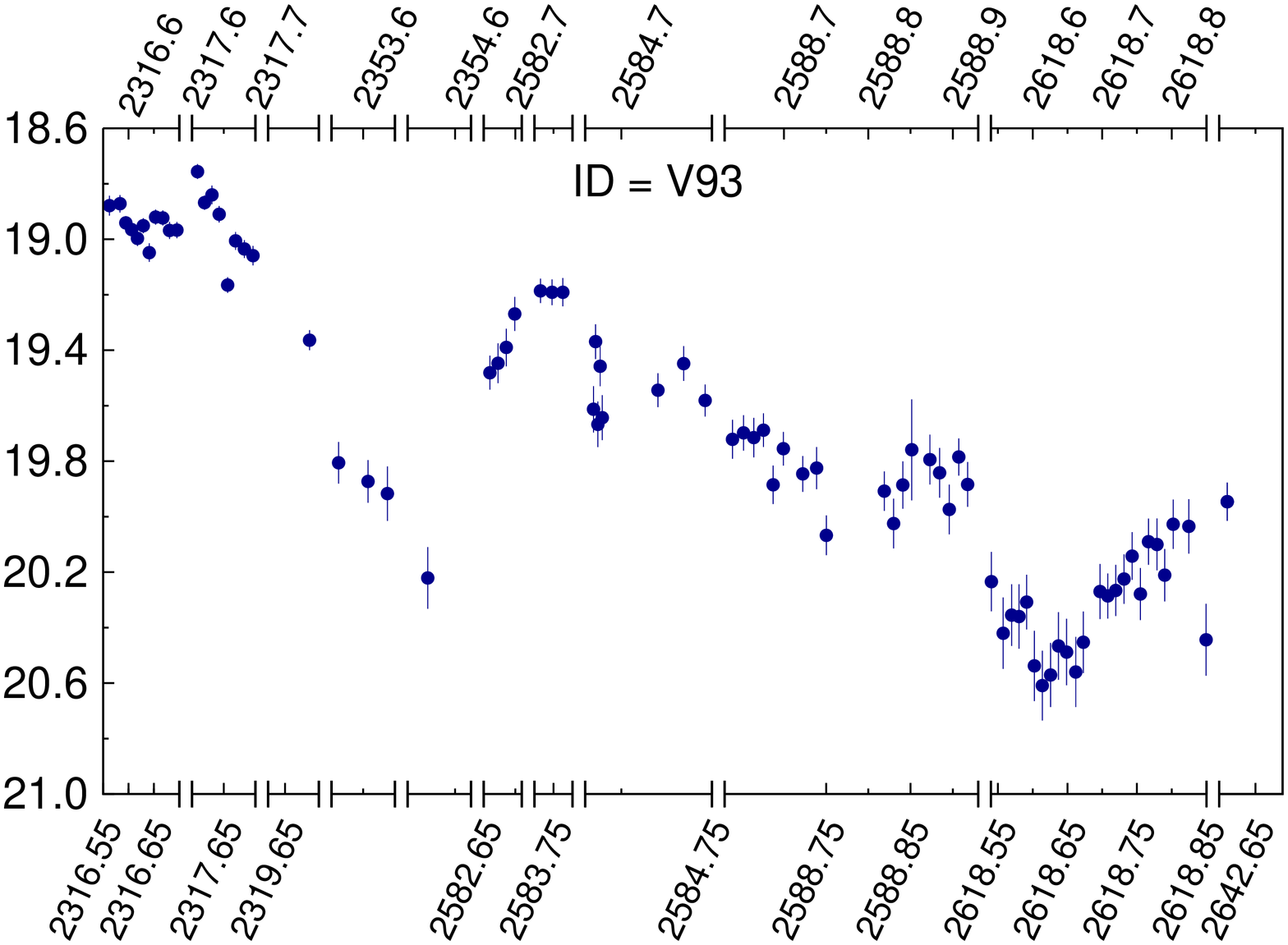} \\
\vspace{0.2 cm}
\includegraphics[width= 4.0 cm,height = 7.0 cm, angle=270]{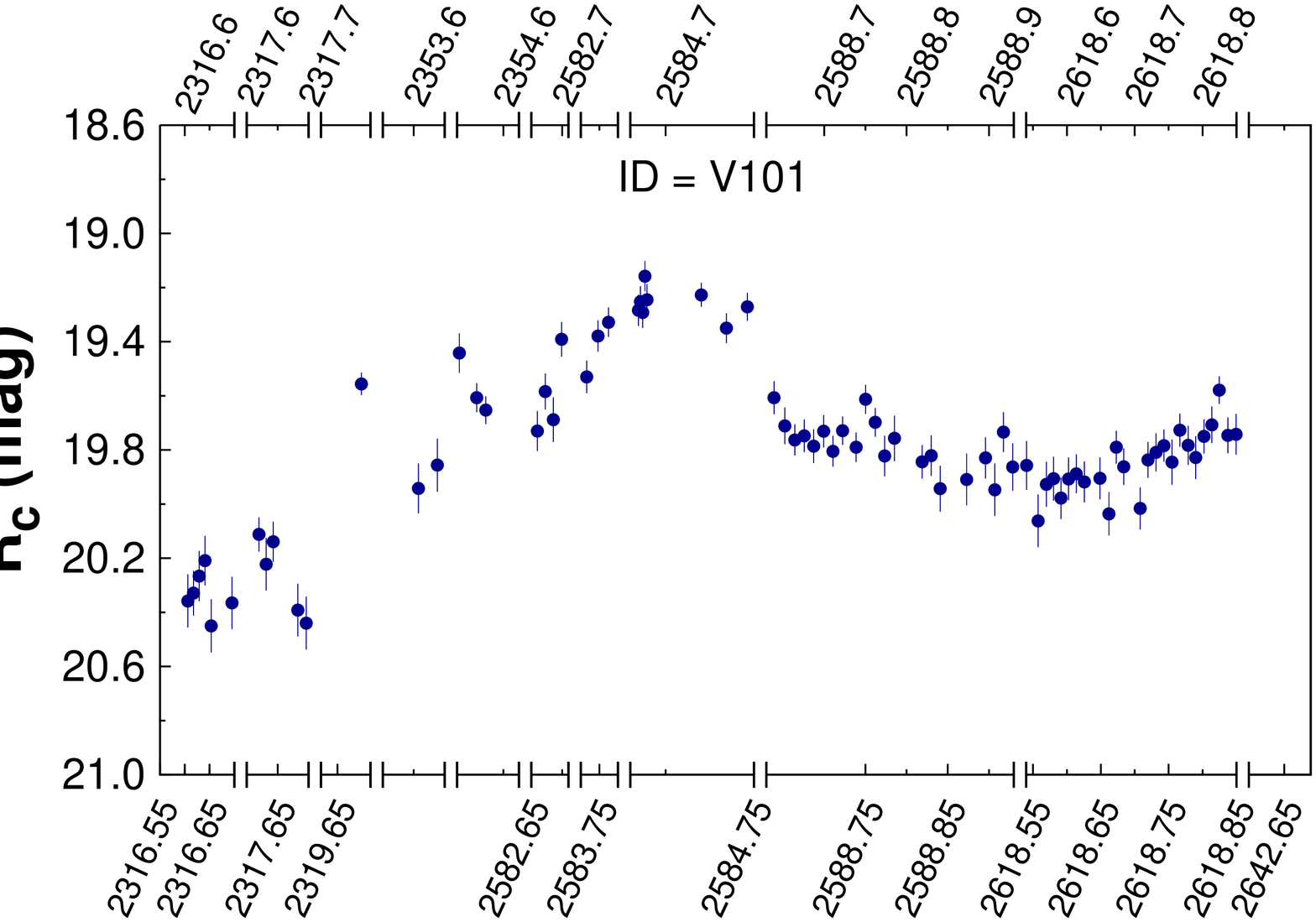}
\hspace{0.1 cm}
\includegraphics[width= 4.0 cm,height = 7.0 cm, angle=270]{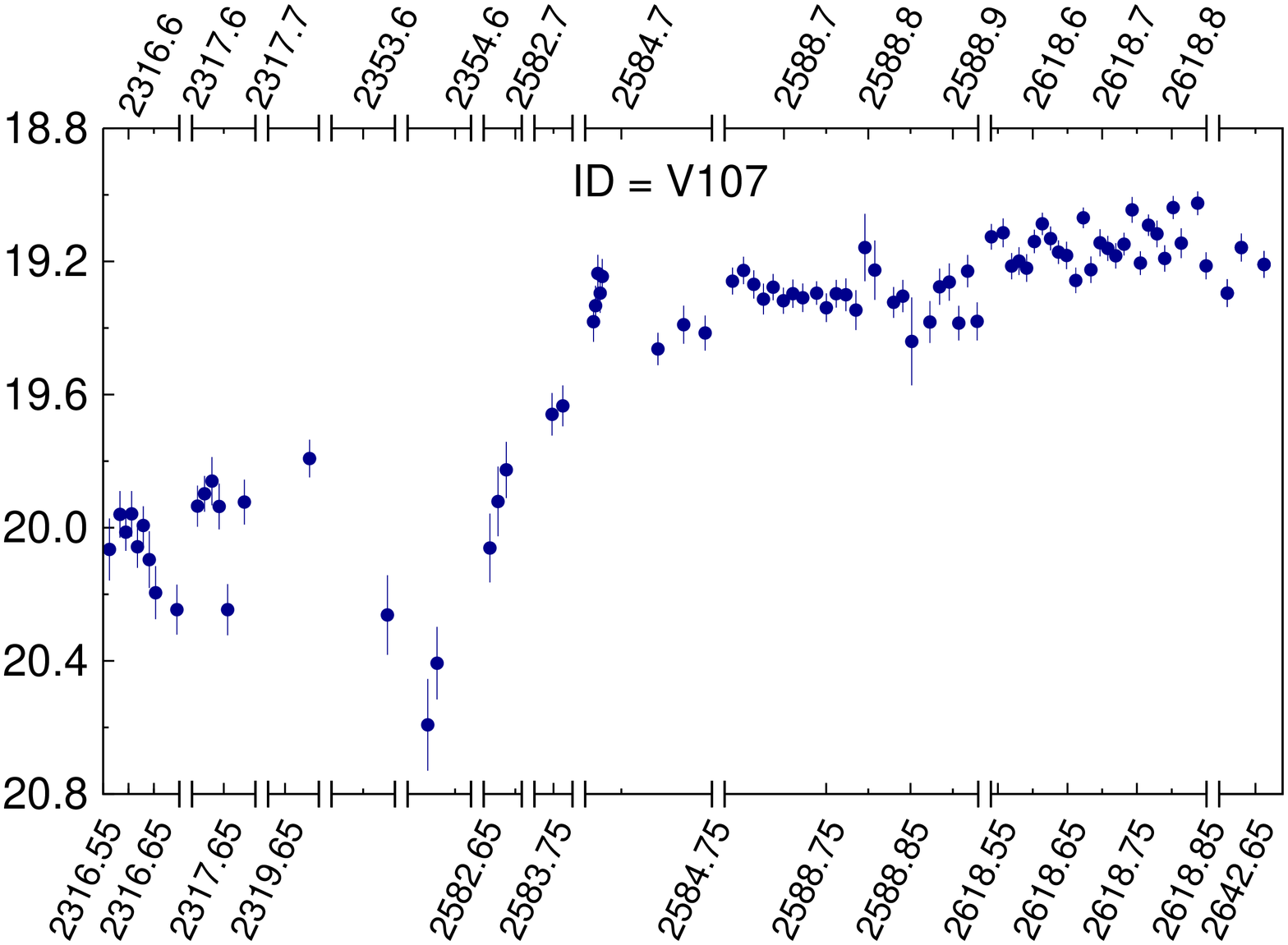} \\
\vspace{0.2 cm}
\includegraphics[width= 4.0 cm,height = 7.0 cm, angle=270]{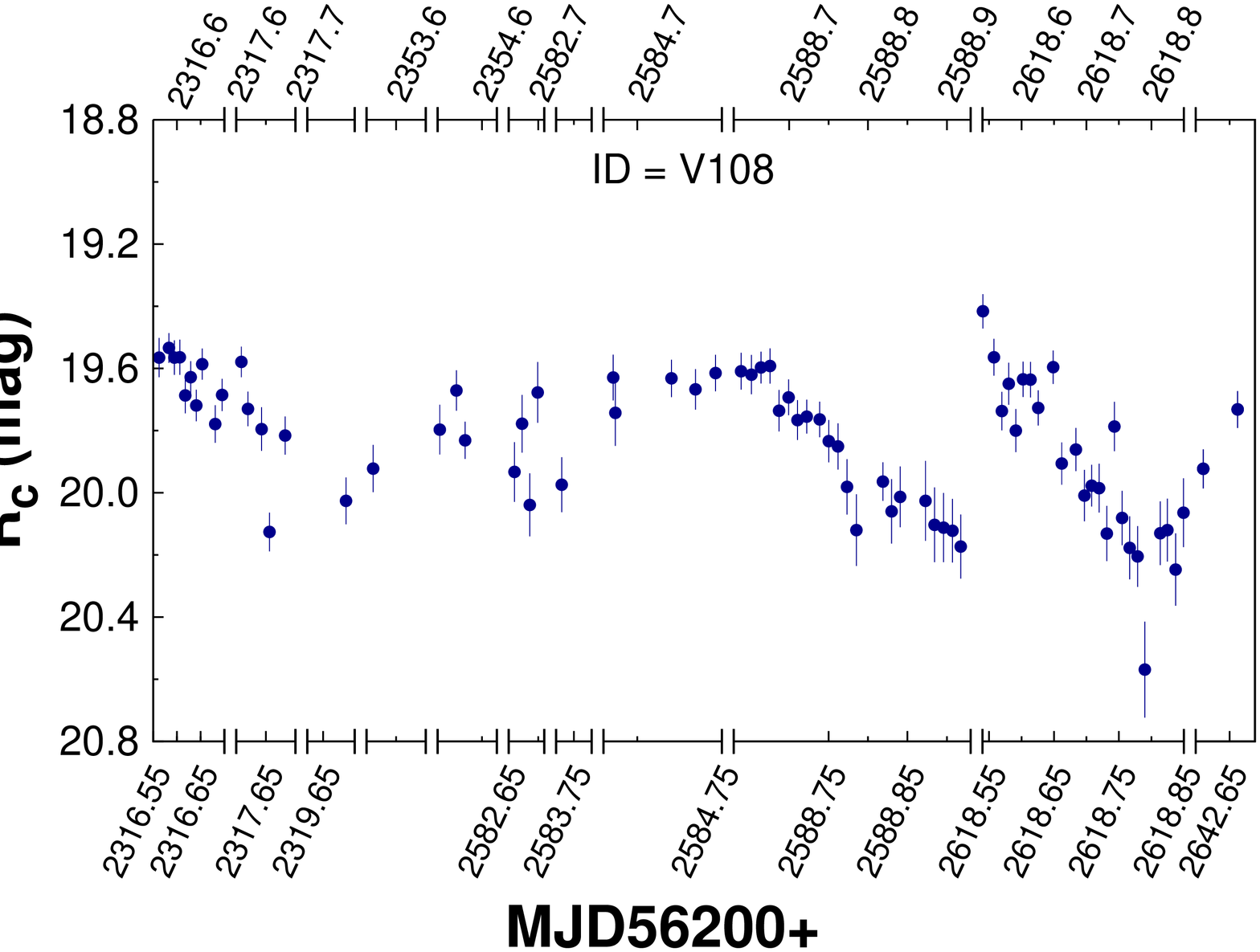}
\hspace{0.1 cm}
\includegraphics[width= 4.0 cm,height = 7.0 cm, angle=270]{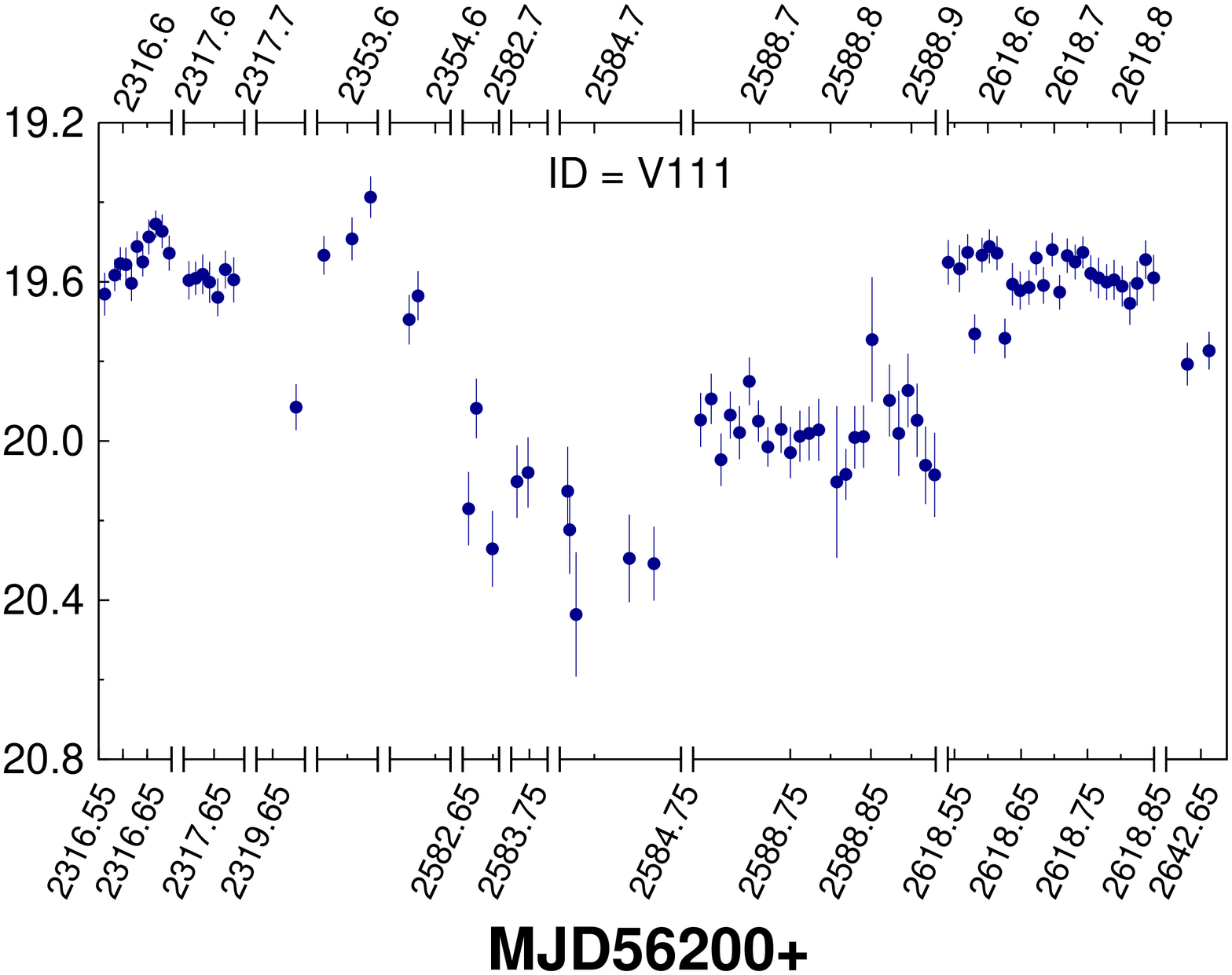} \\

\caption{Contd.}
 \label{fig: LC_CTT}
\end{figure*}

\begin{figure*}[h]
\centering
\includegraphics[width= 4.5 cm,height = 7.5 cm, angle=270]{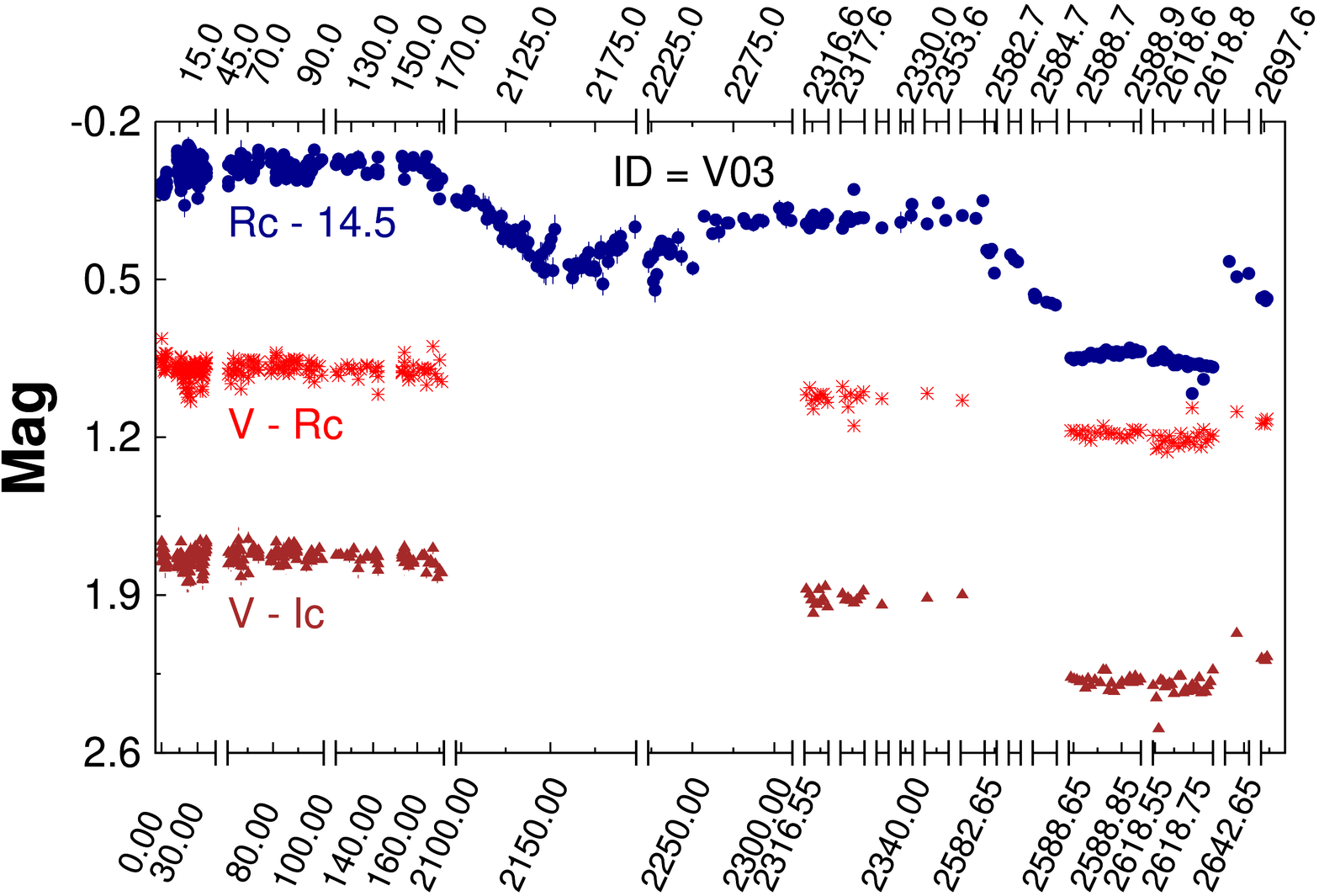}
\hspace{0.1 cm}
\includegraphics[width= 4.5 cm,height = 7.5 cm, angle=270]{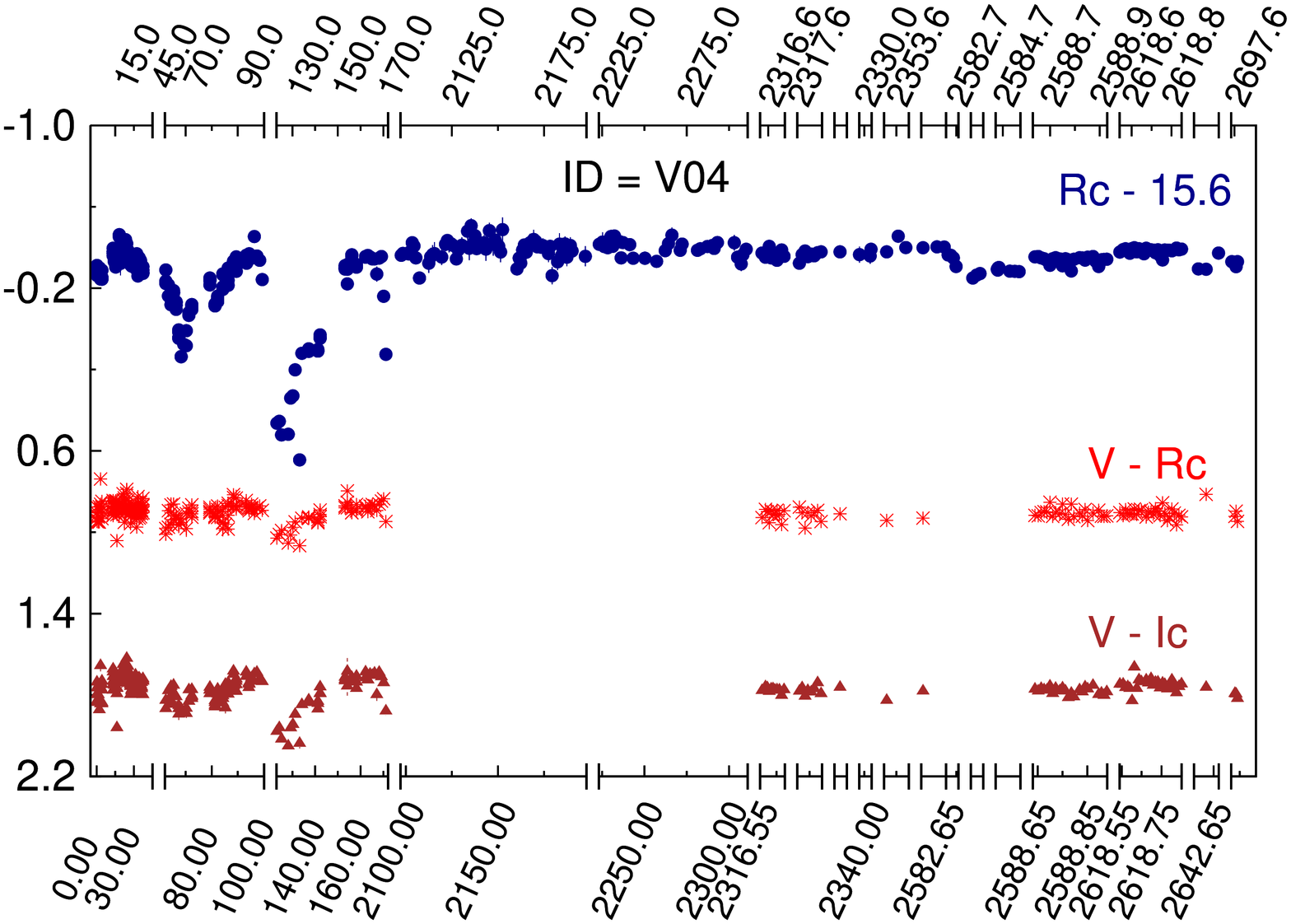} \\
\vspace{0.2 cm}
\includegraphics[width= 4.5 cm,height = 7.5 cm, angle=270]{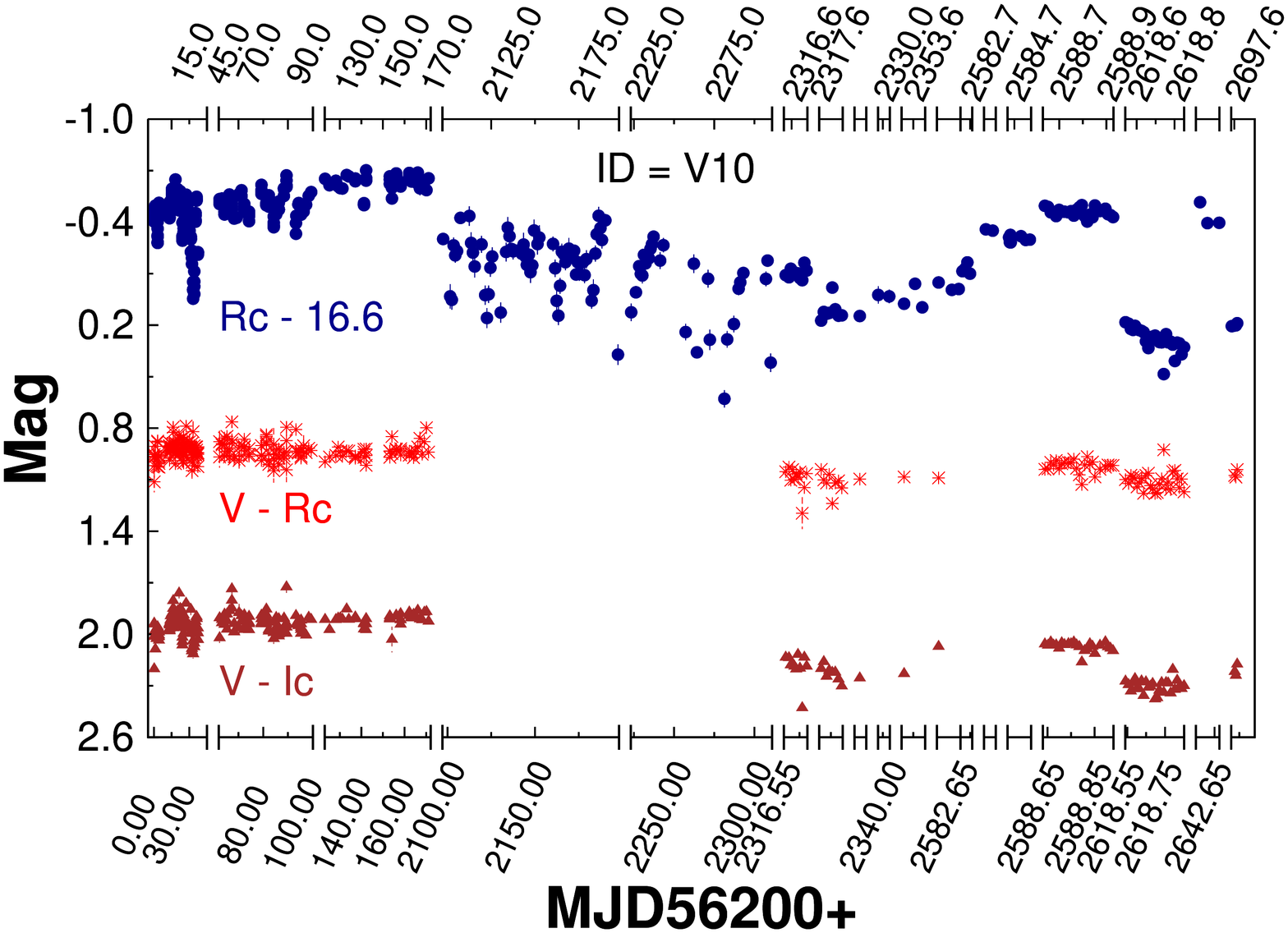}
\hspace{0.1 cm}
\includegraphics[width= 4.5 cm,height = 7.5 cm, angle=270]{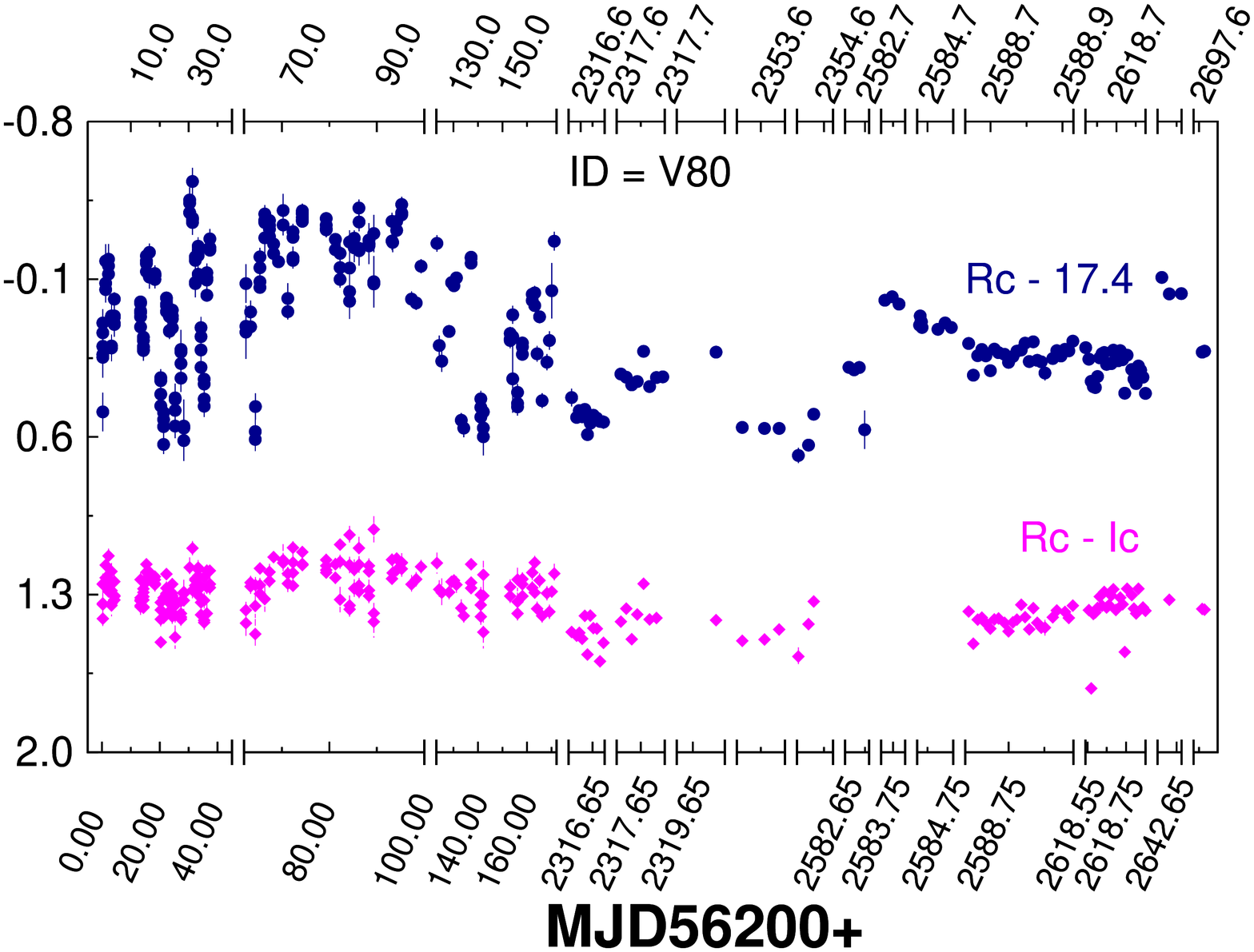} \\
\caption{Color evolution of a few selected CTTSs variables. The R$_{c}$ band LC, V-R$_{c}$, V-I$_{c}$ and R$_{c}$-I$_{c}$ color curves are shown with blue filled circle, red stars, brown triangles and magenta diamonds, respectively. When there are data gaps they are represented with vertical gaps along the axis.}
\label{Fig: color}
\end{figure*}
\vspace{1 cm}

\begin{figure}
\vspace{0.3 cm}
\centering         \vspace{0.05 cm} \hspace{-0.2 cm}
\hspace{-0.1 cm}
\includegraphics[width = 1.9 cm, height = 2.6 cm, angle= 270]{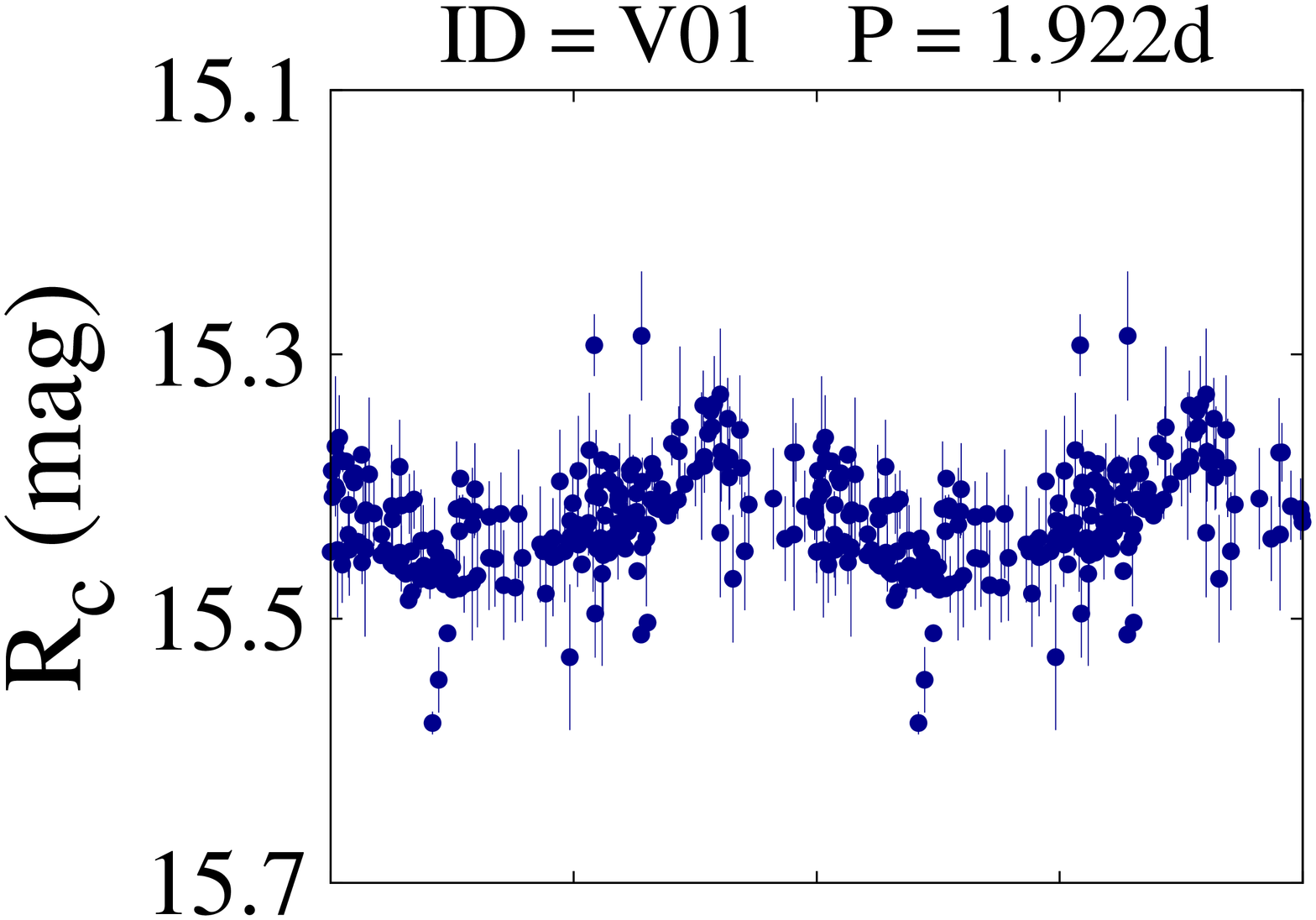}
\hspace{0.03 cm}
\includegraphics[width = 1.9 cm, height = 2.6 cm, angle= 270]{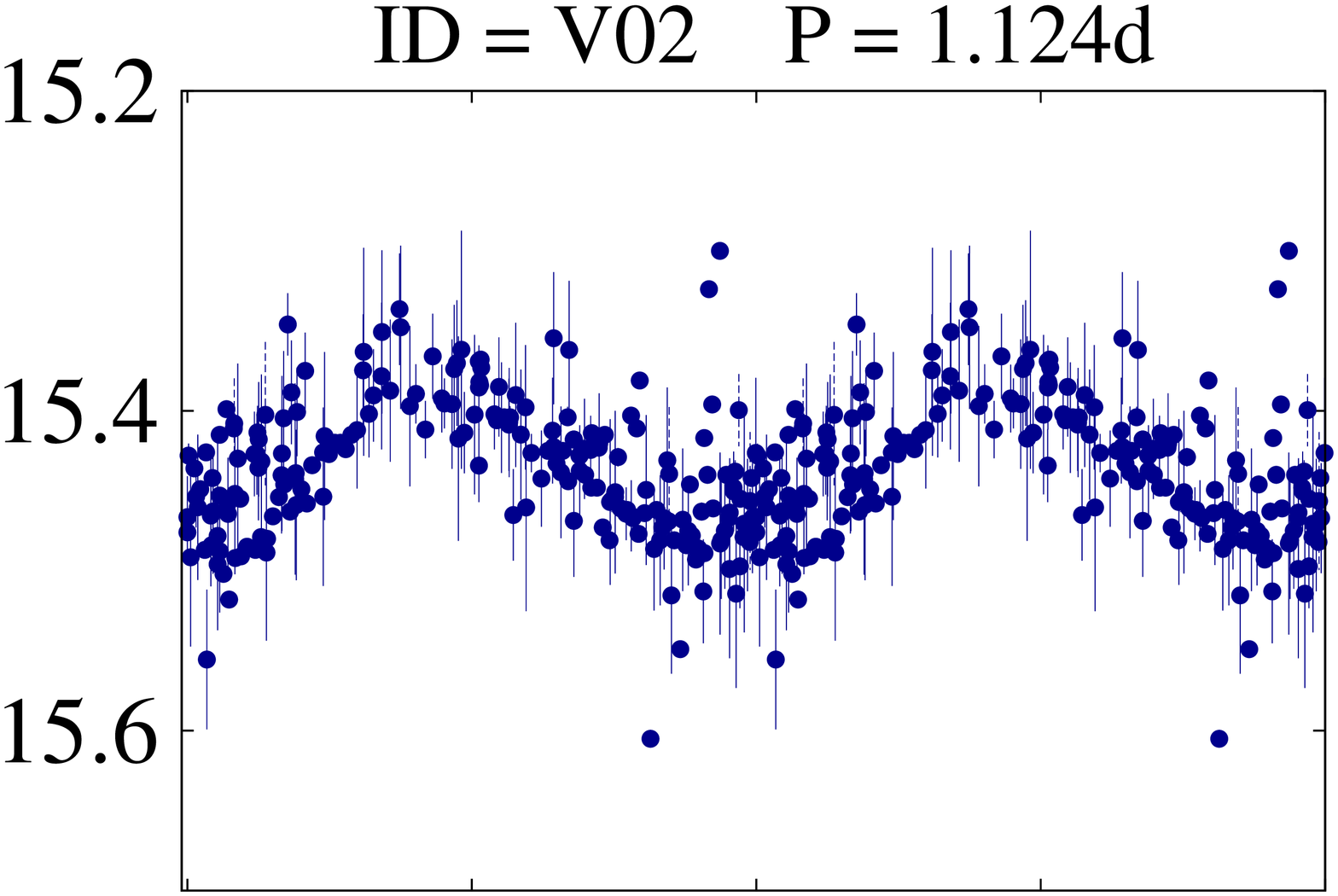}
\hspace{0.03 cm}
\includegraphics[width = 1.9 cm, height = 2.6 cm, angle= 270]{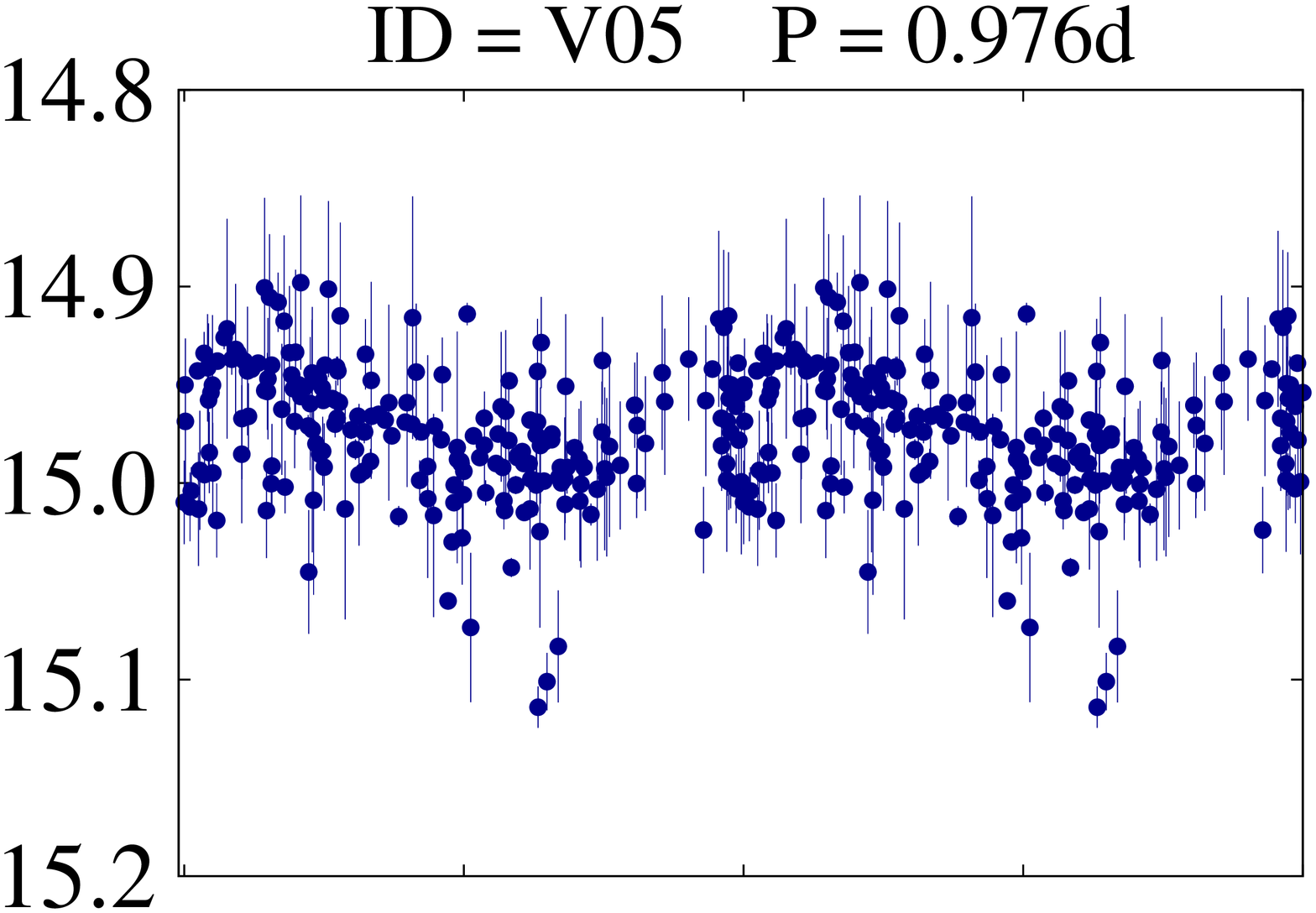}
\hspace{0.03 cm}
\includegraphics[width = 1.9 cm, height = 2.6 cm, angle= 270]{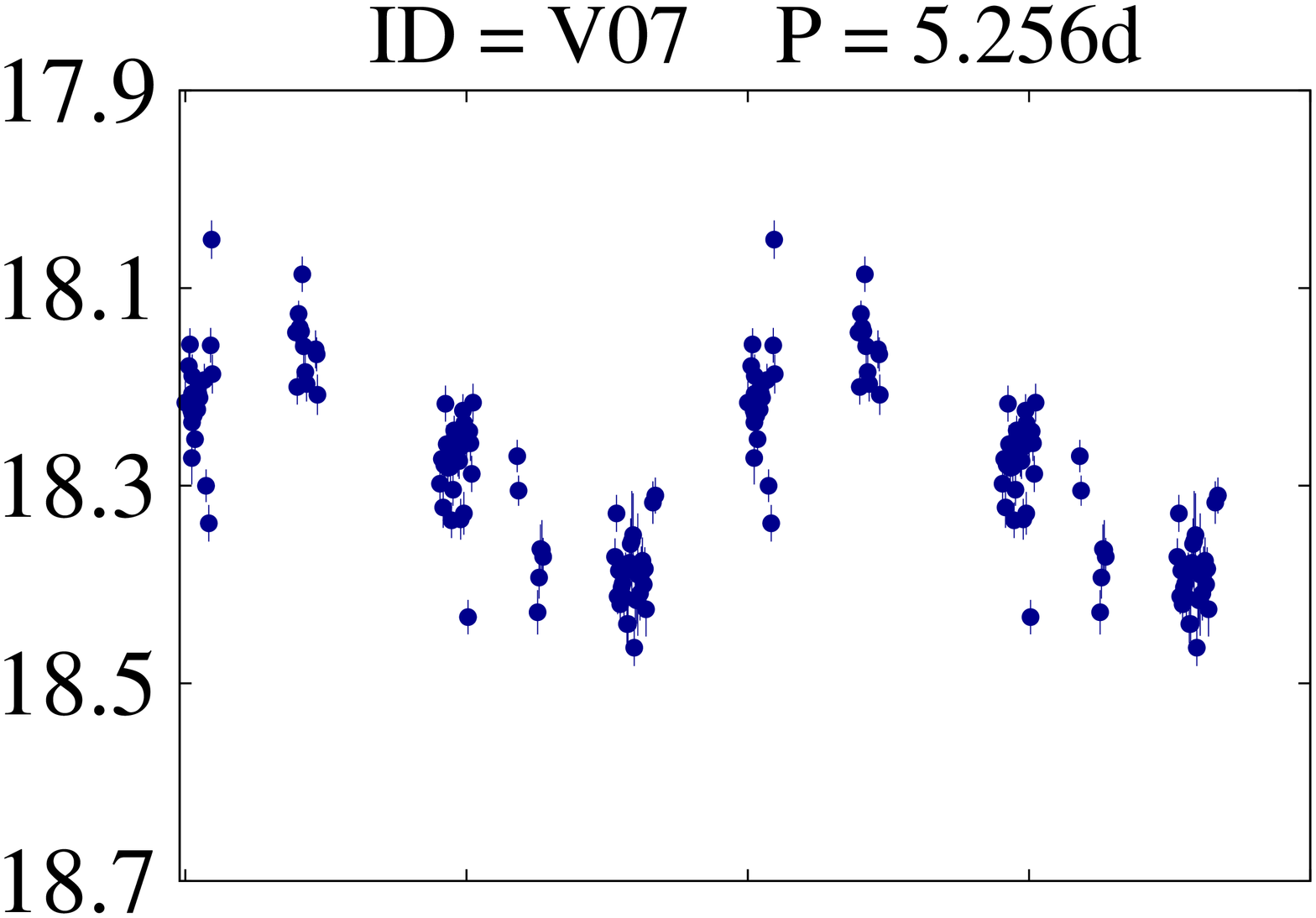}
\hspace{0.03 cm}
\includegraphics[width = 1.9 cm, height = 2.6 cm, angle= 270]{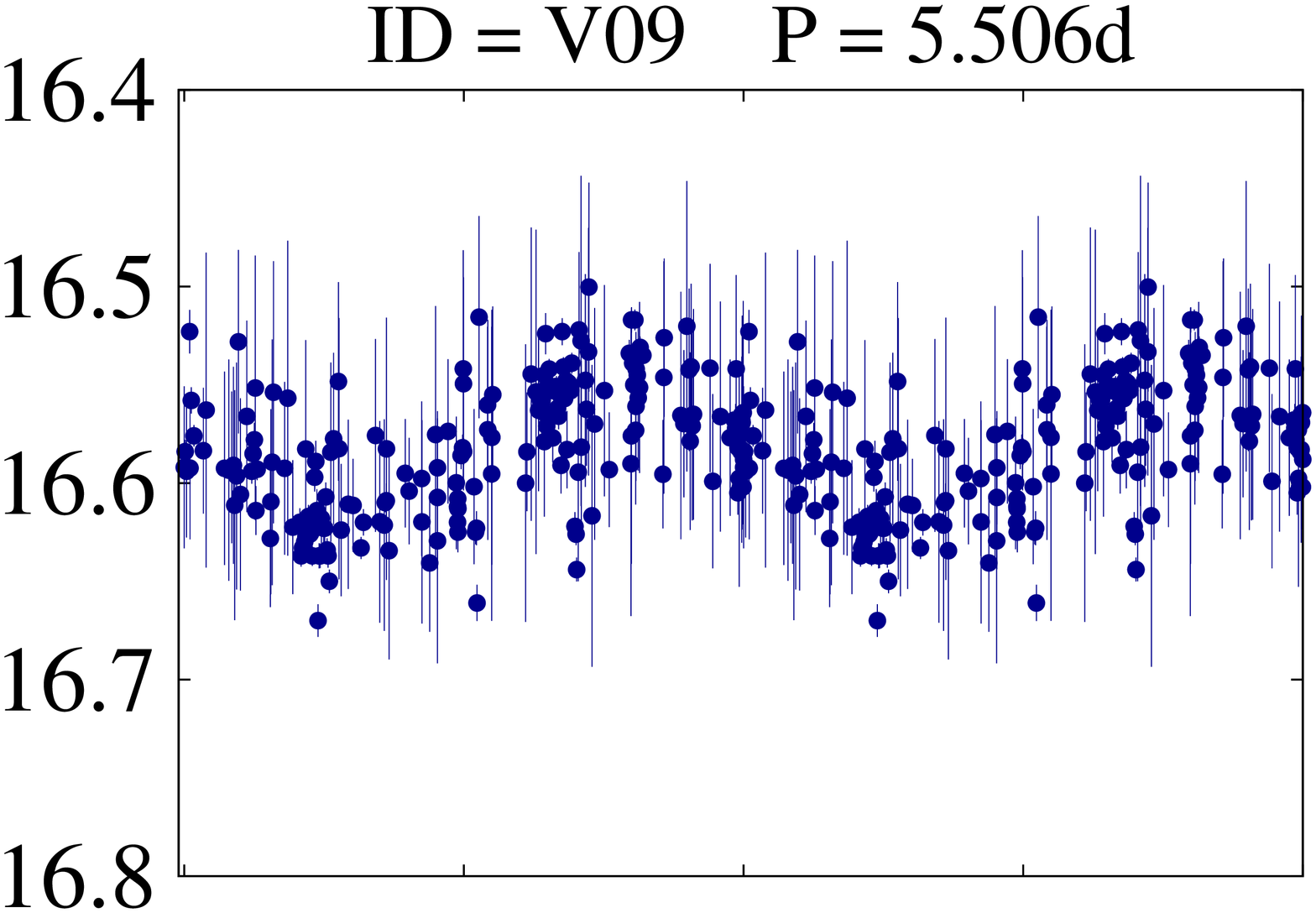} 
\hspace{0.03 cm}
\includegraphics[width = 1.9 cm, height = 2.6 cm, angle= 270]{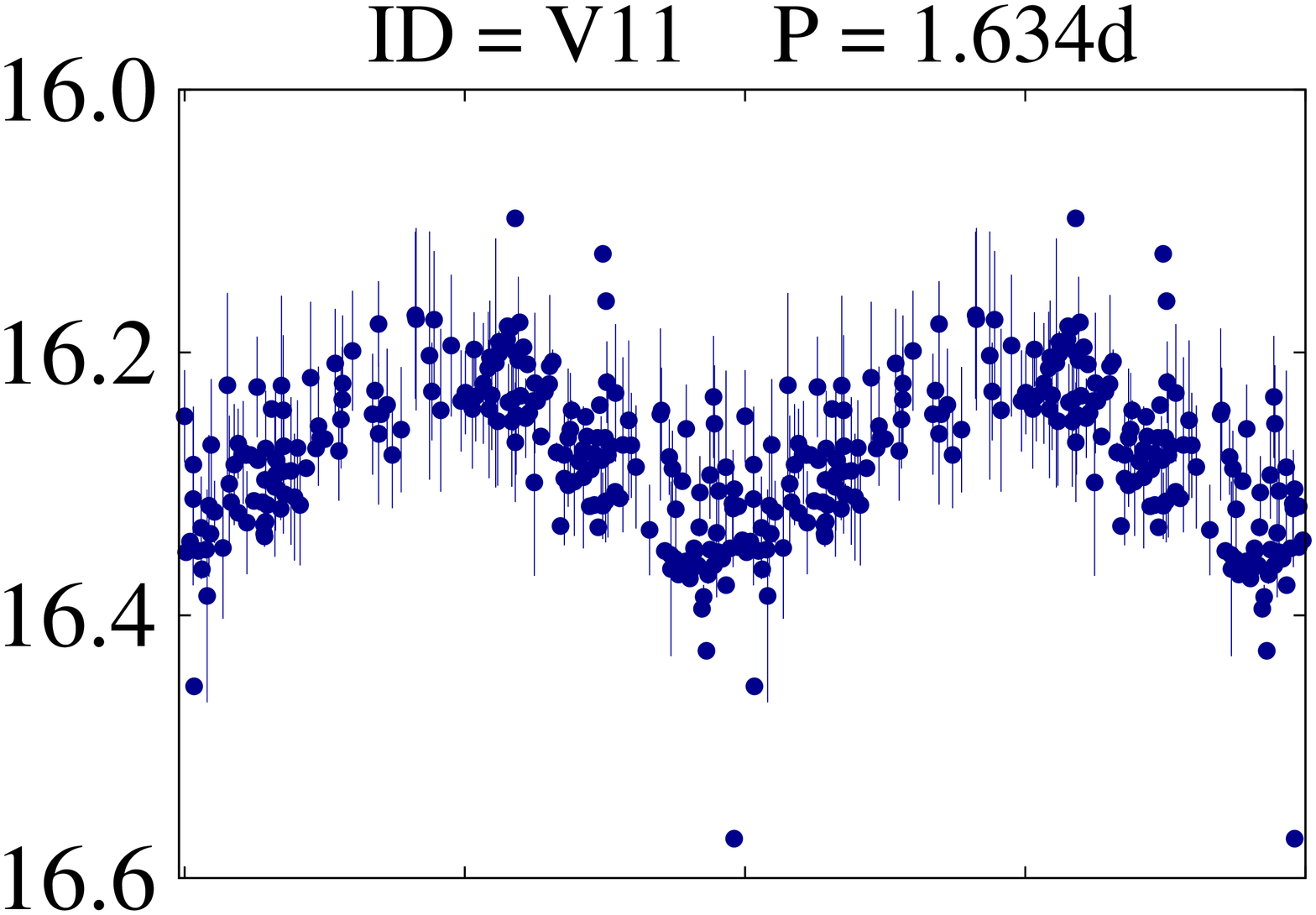} \\
\vspace{0.05 cm}
\includegraphics[width = 1.9 cm, height = 2.6 cm, angle= 270]{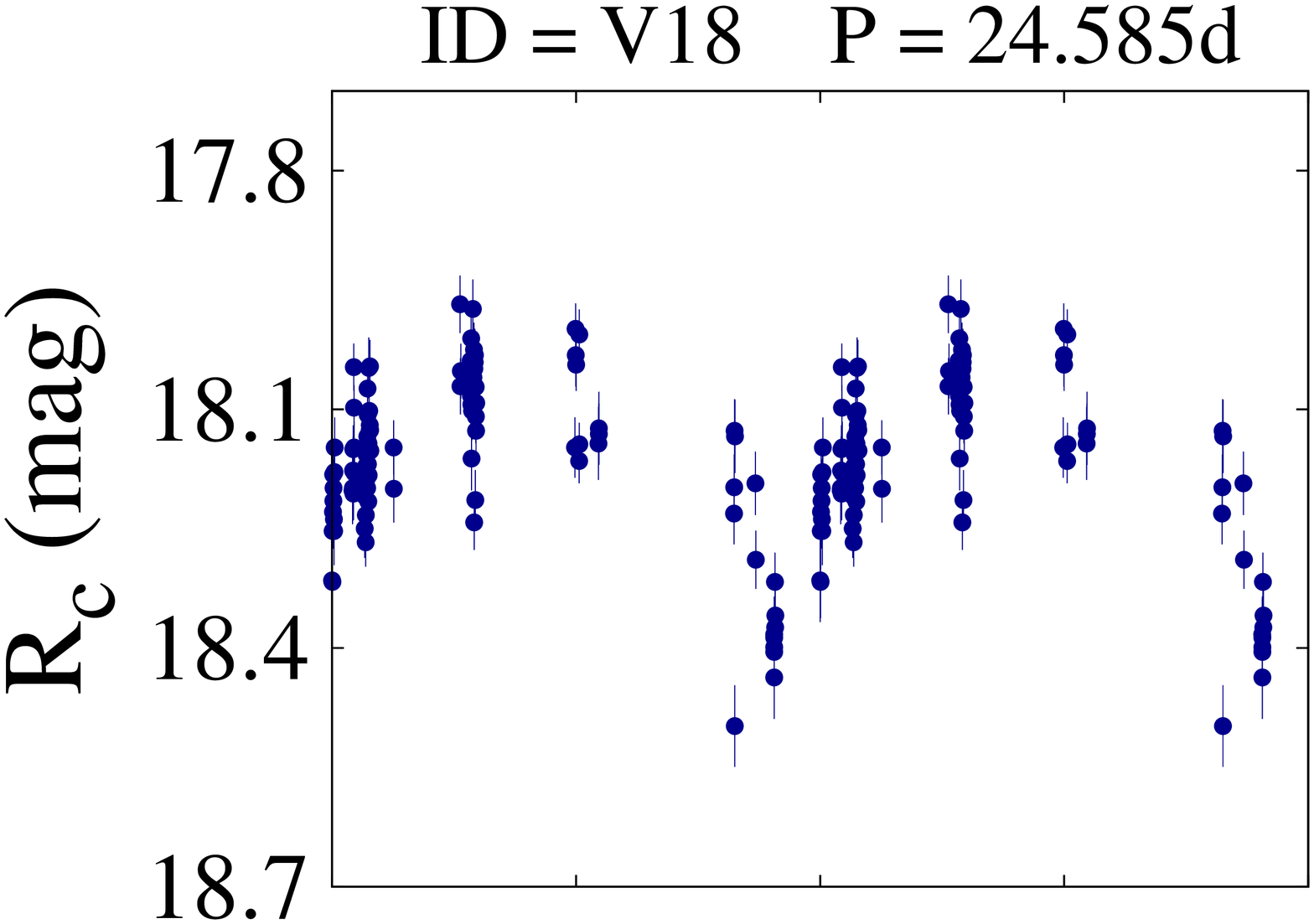}
\hspace{0.03 cm}
\includegraphics[width = 1.9 cm, height = 2.6 cm, angle= 270]{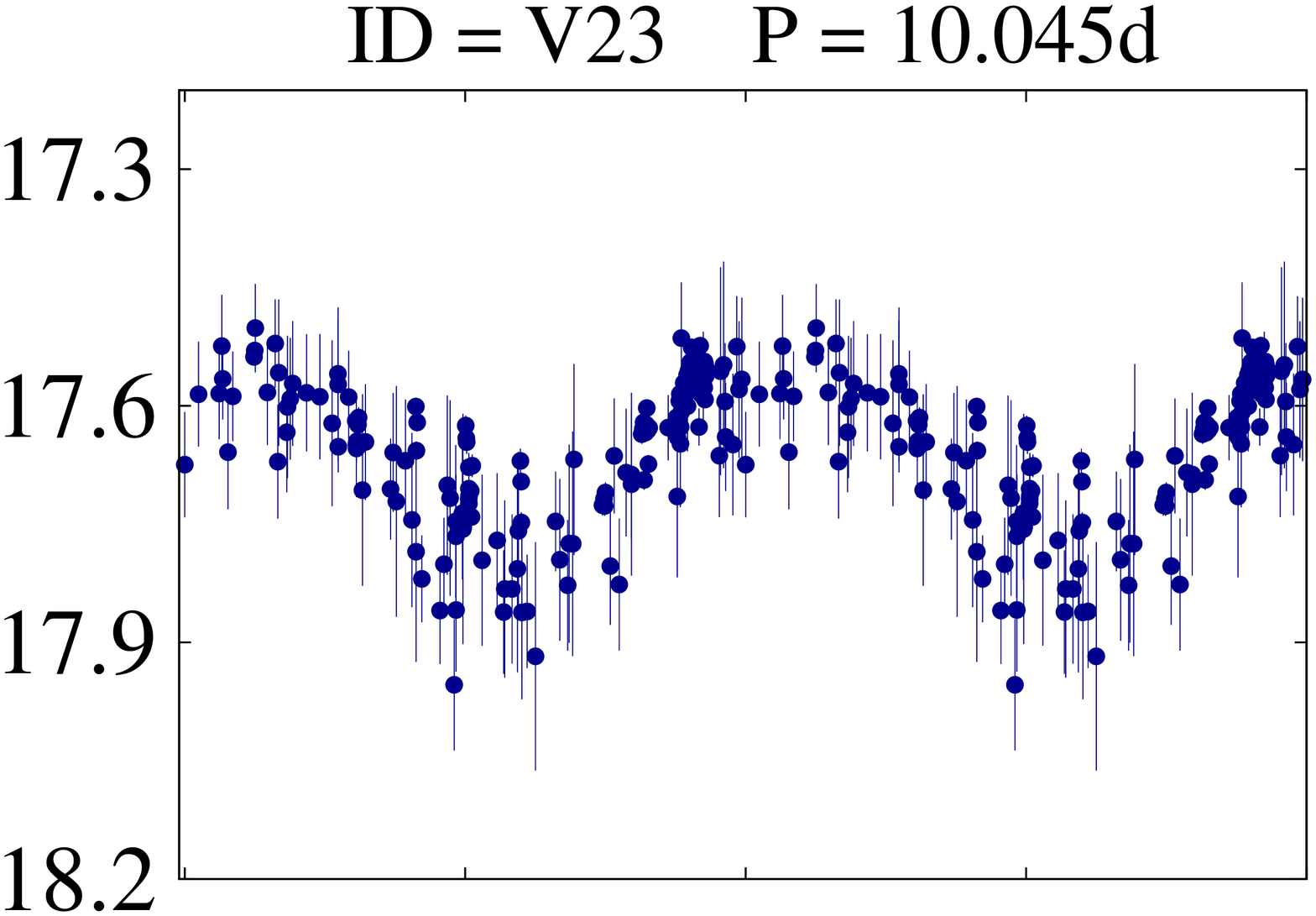}
\hspace{0.03 cm}
\includegraphics[width = 1.9 cm, height = 2.6 cm, angle= 270]{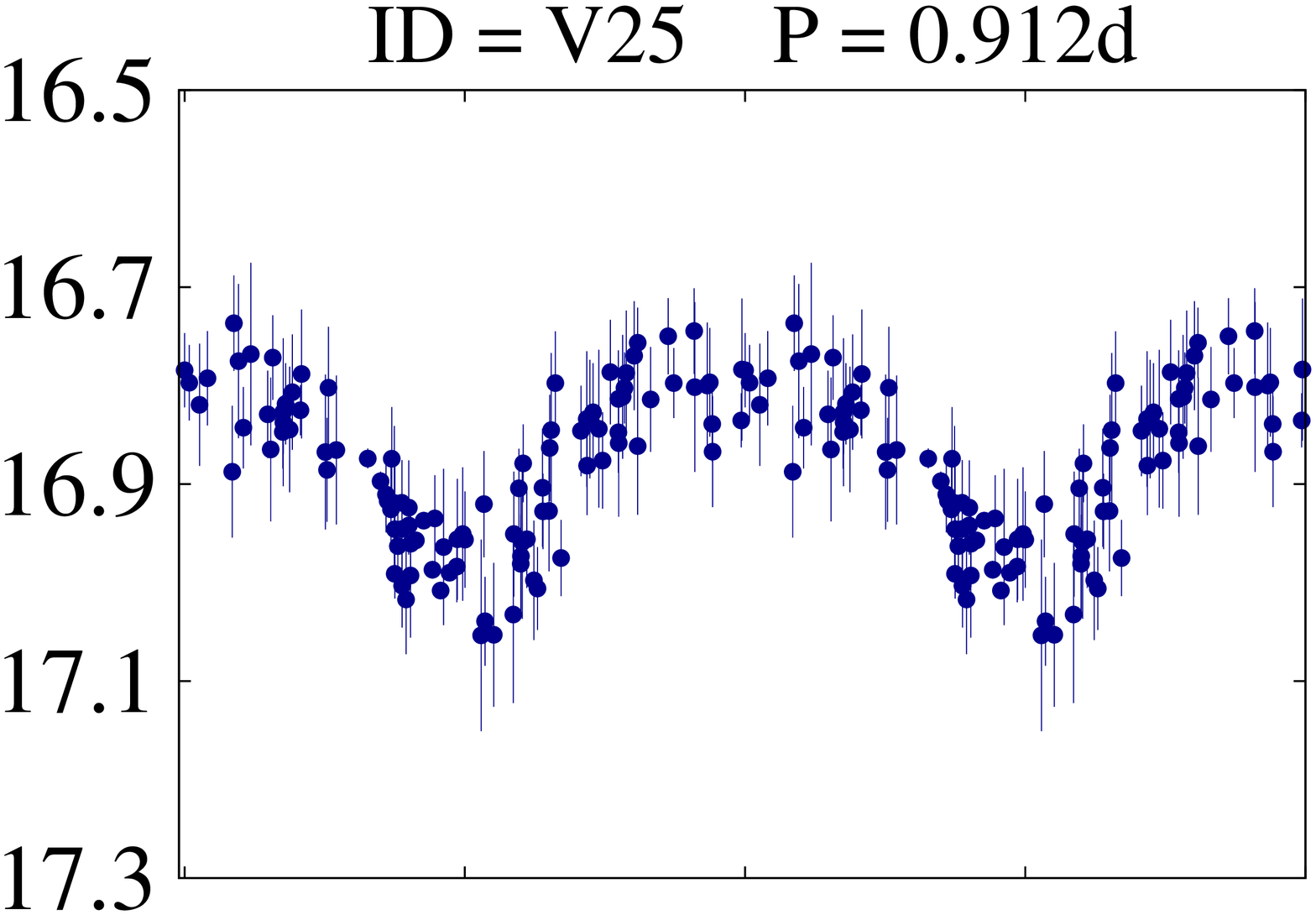}
\hspace{0.03 cm}
\includegraphics[width = 1.9 cm, height = 2.6 cm, angle= 270]{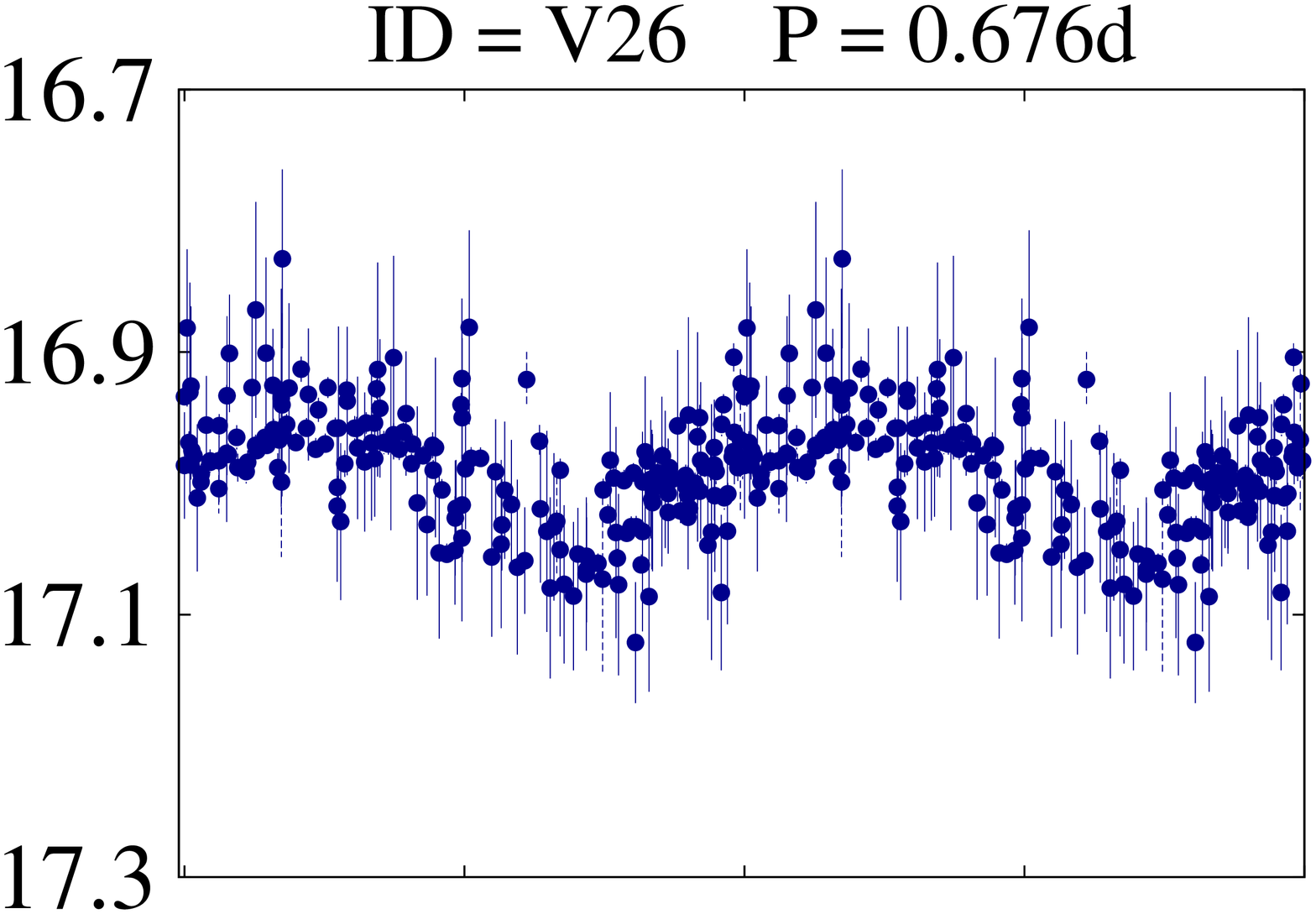} 
\hspace{0.03 cm}
\includegraphics[width = 1.9 cm, height = 2.6 cm, angle= 270]{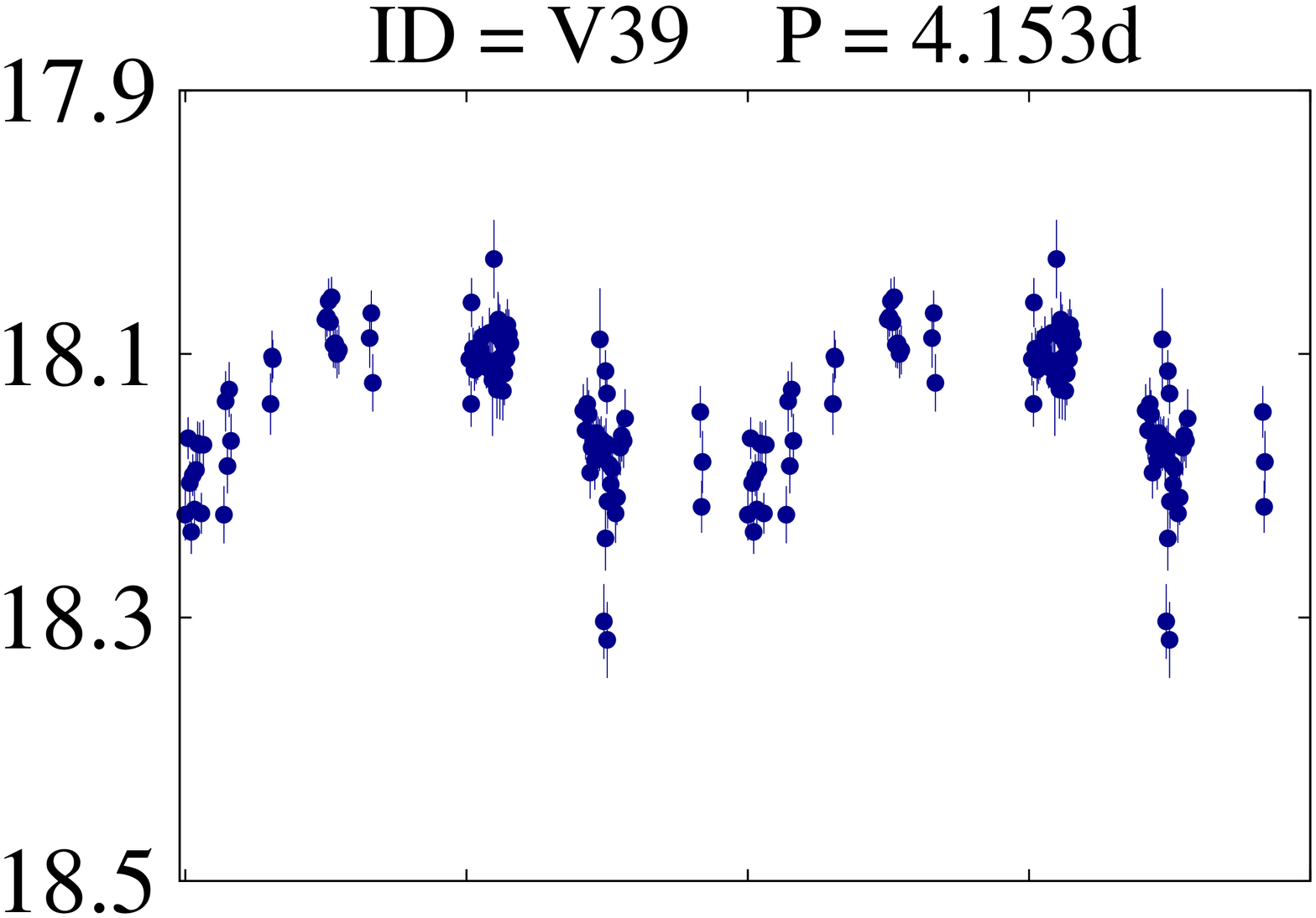} 
\hspace{0.03 cm}
\includegraphics[width = 1.9 cm, height = 2.6 cm, angle= 270]{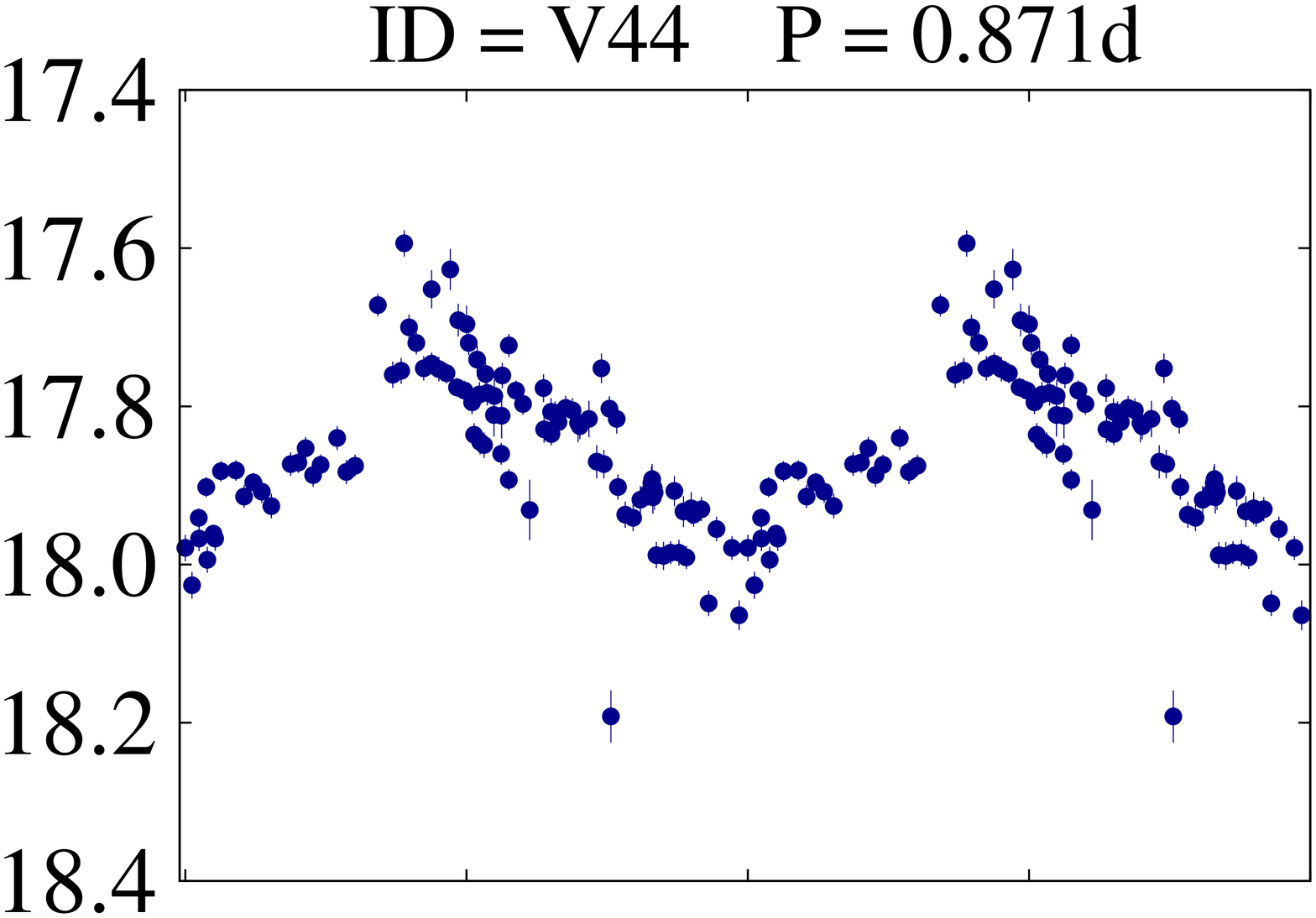} \\
\vspace{0.05 cm}
\includegraphics[width = 1.9 cm, height = 2.6 cm, angle= 270]{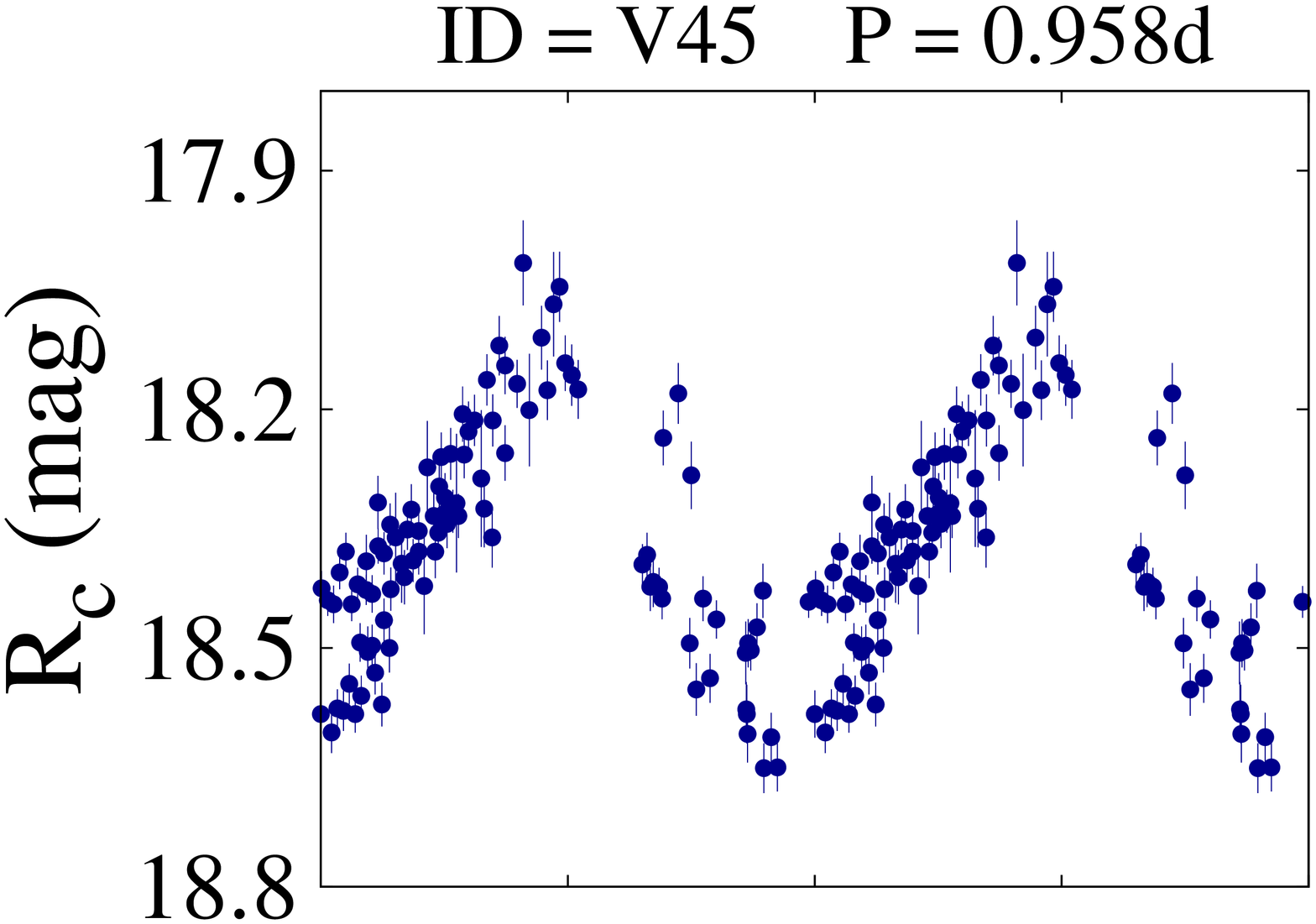}
\hspace{0.03 cm}
\includegraphics[width = 1.9 cm, height = 2.6 cm, angle= 270]{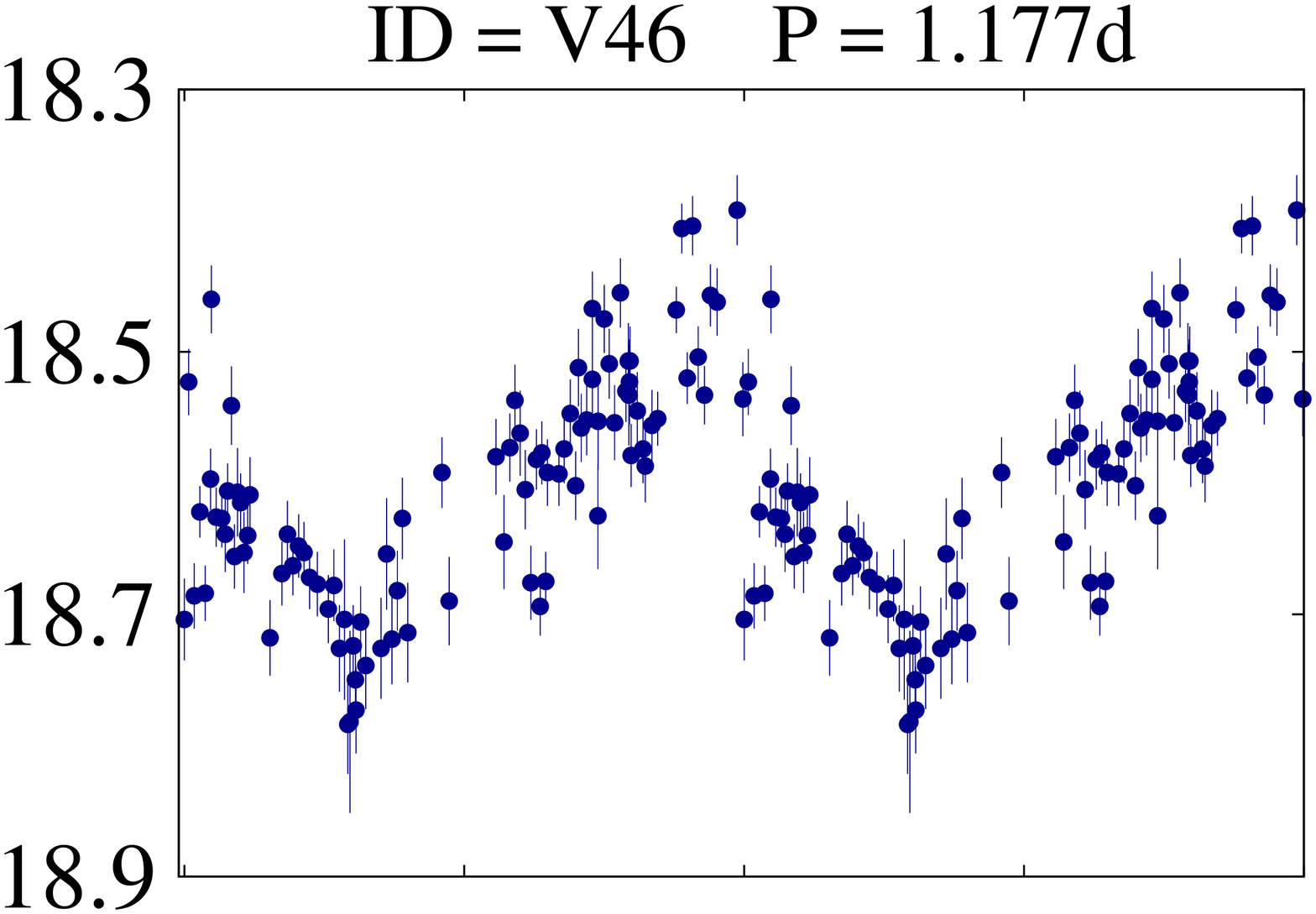}
\hspace{0.03 cm}
\includegraphics[width = 1.9 cm, height = 2.6 cm, angle= 270]{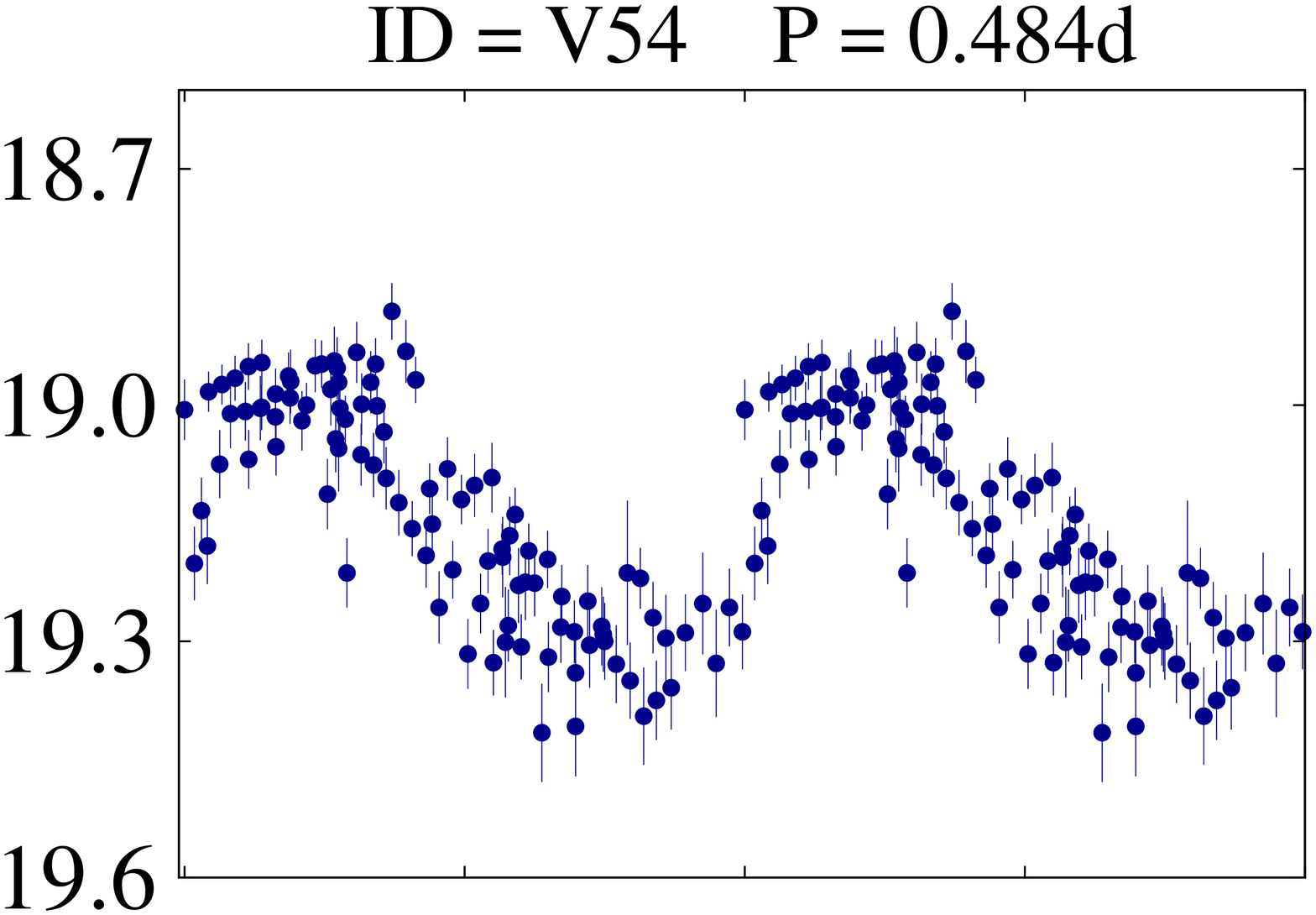}
\hspace{0.03 cm}
\includegraphics[width = 1.9 cm, height = 2.6 cm, angle= 270]{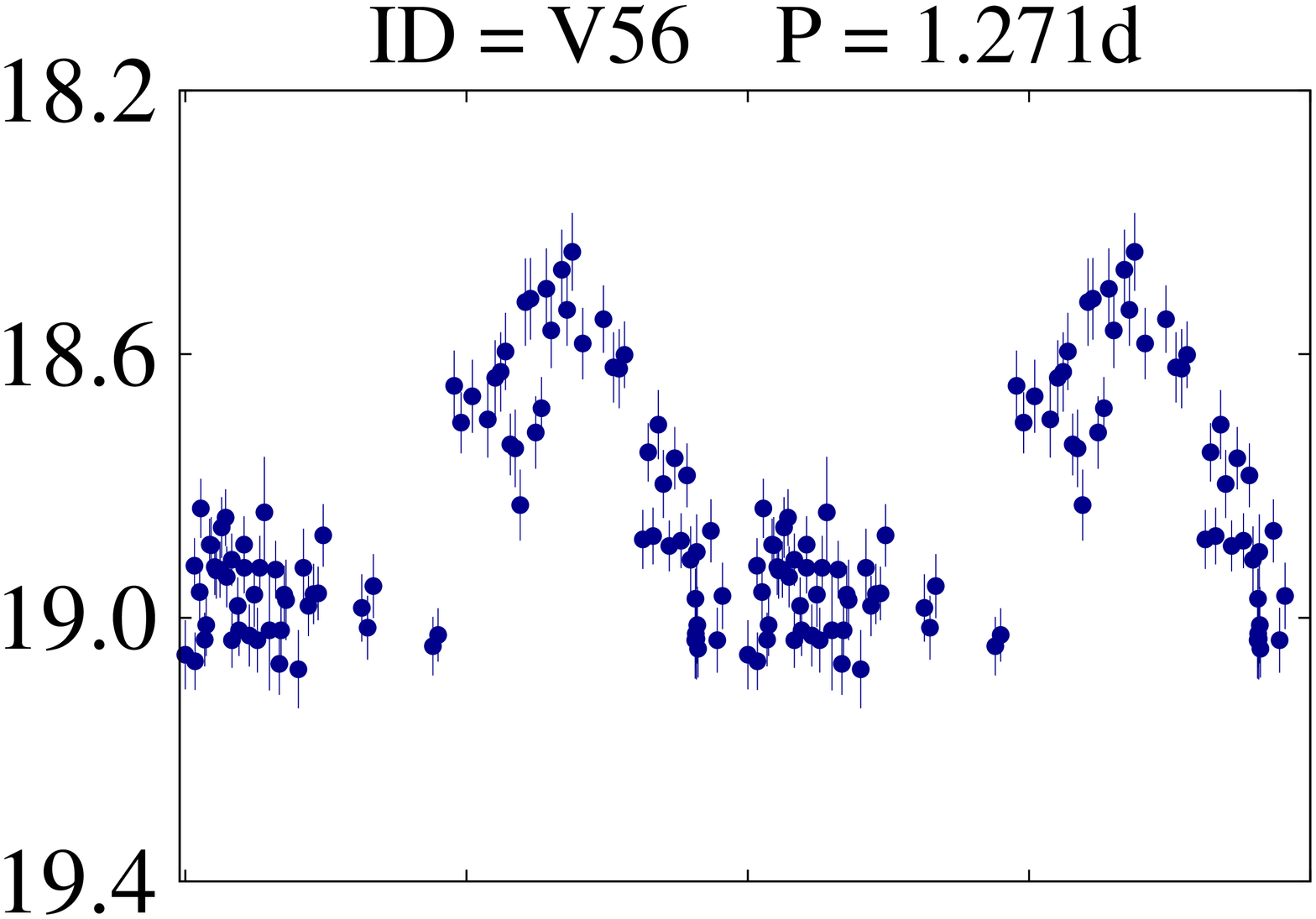} 
\hspace{0.03 cm}
\includegraphics[width = 1.9 cm, height = 2.6 cm, angle= 270]{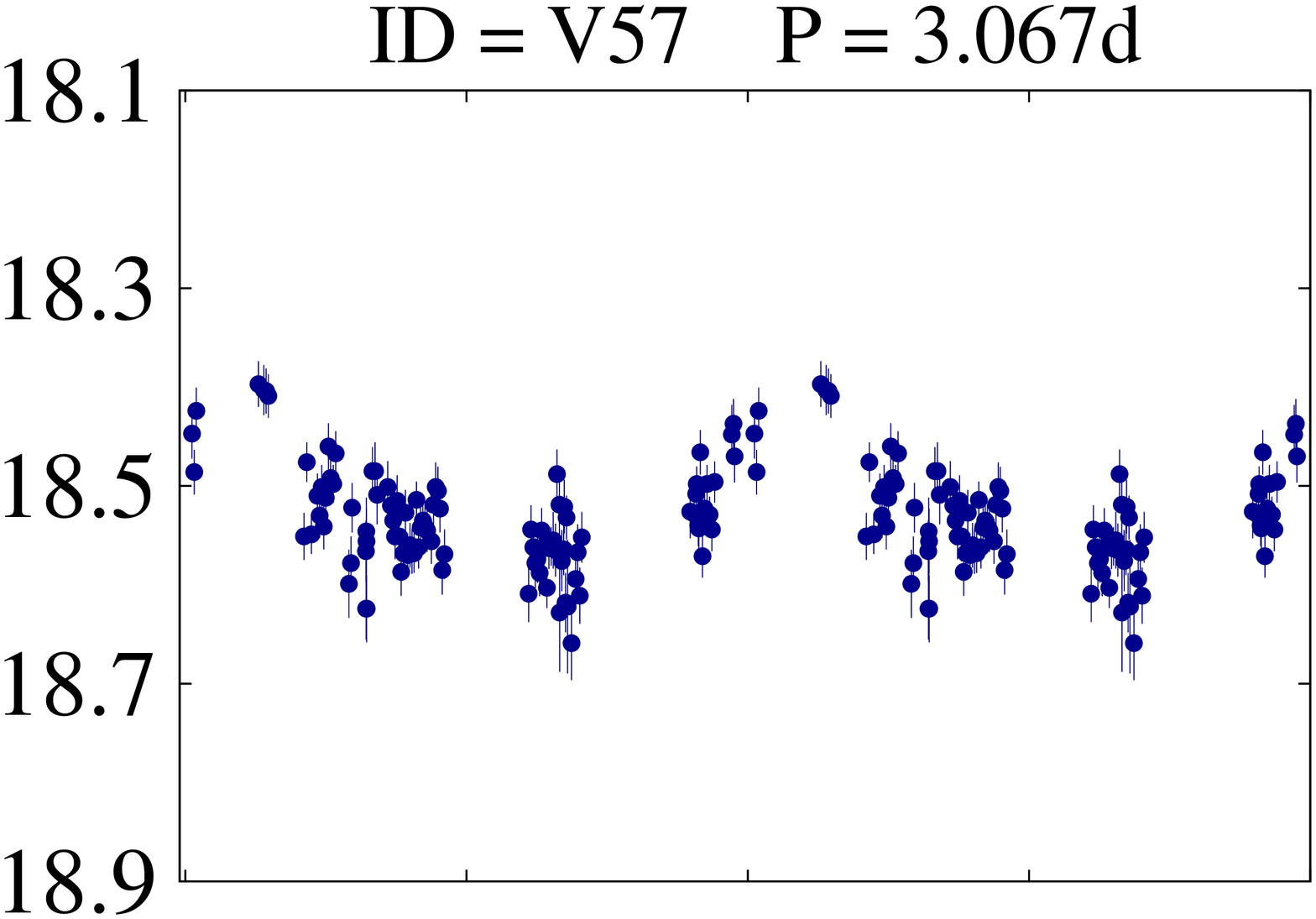}
\hspace{0.03 cm}
\includegraphics[width = 1.9 cm, height = 2.6 cm, angle= 270]{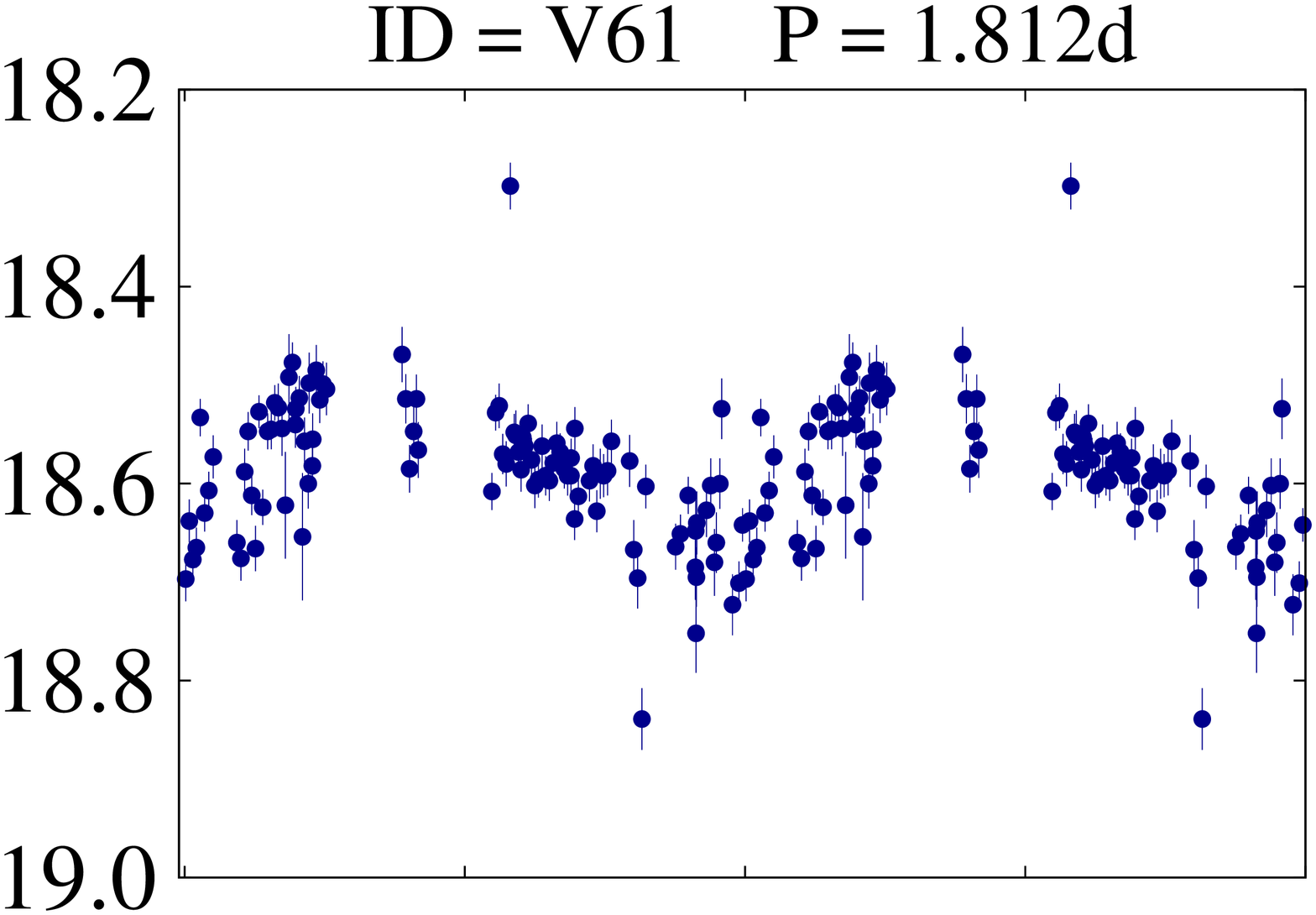} \\
\vspace{0.05 cm}
\includegraphics[width = 1.9 cm, height = 2.6 cm, angle= 270]{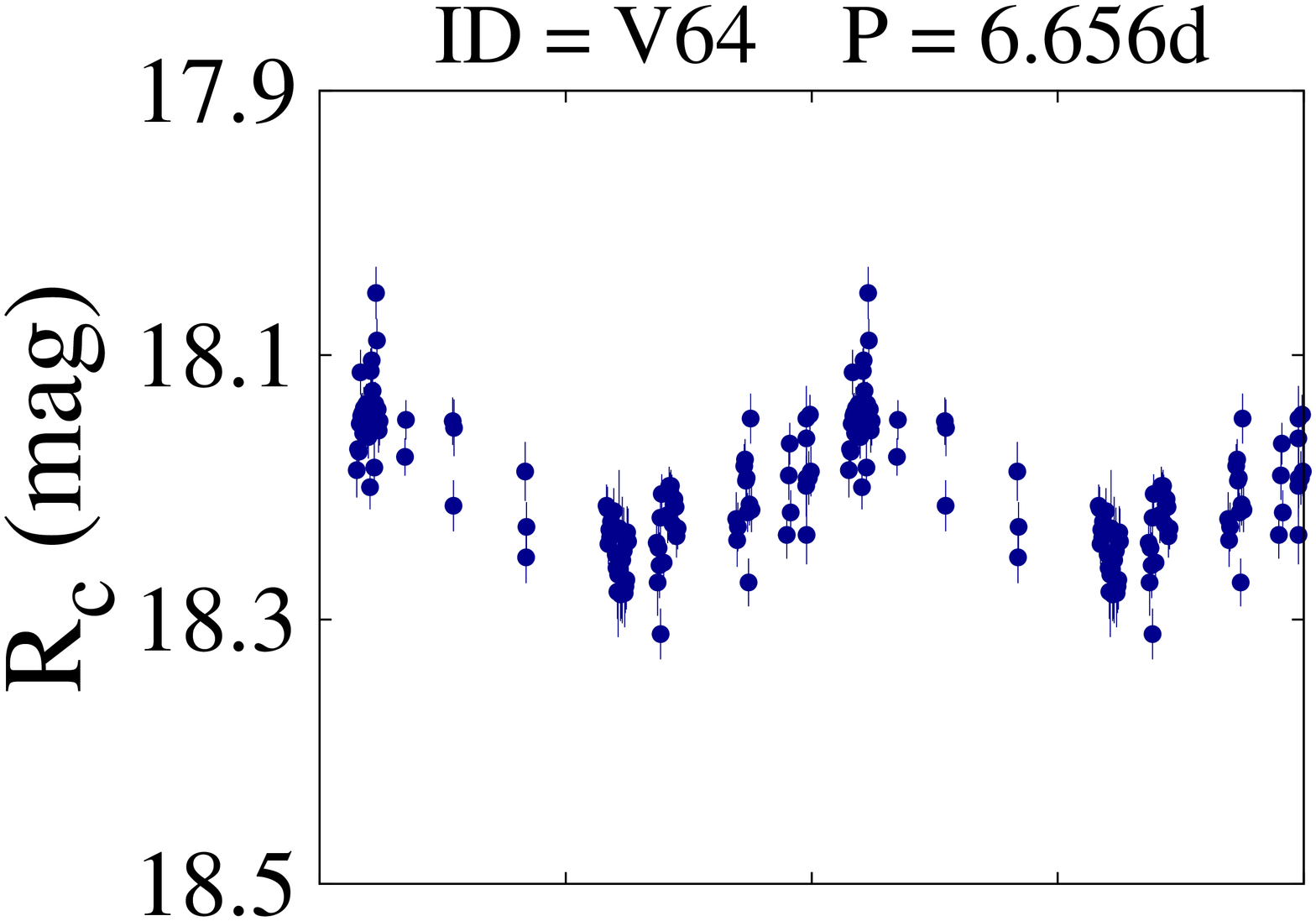}
\hspace{0.03 cm}
\includegraphics[width = 1.9 cm, height = 2.6 cm, angle= 270]{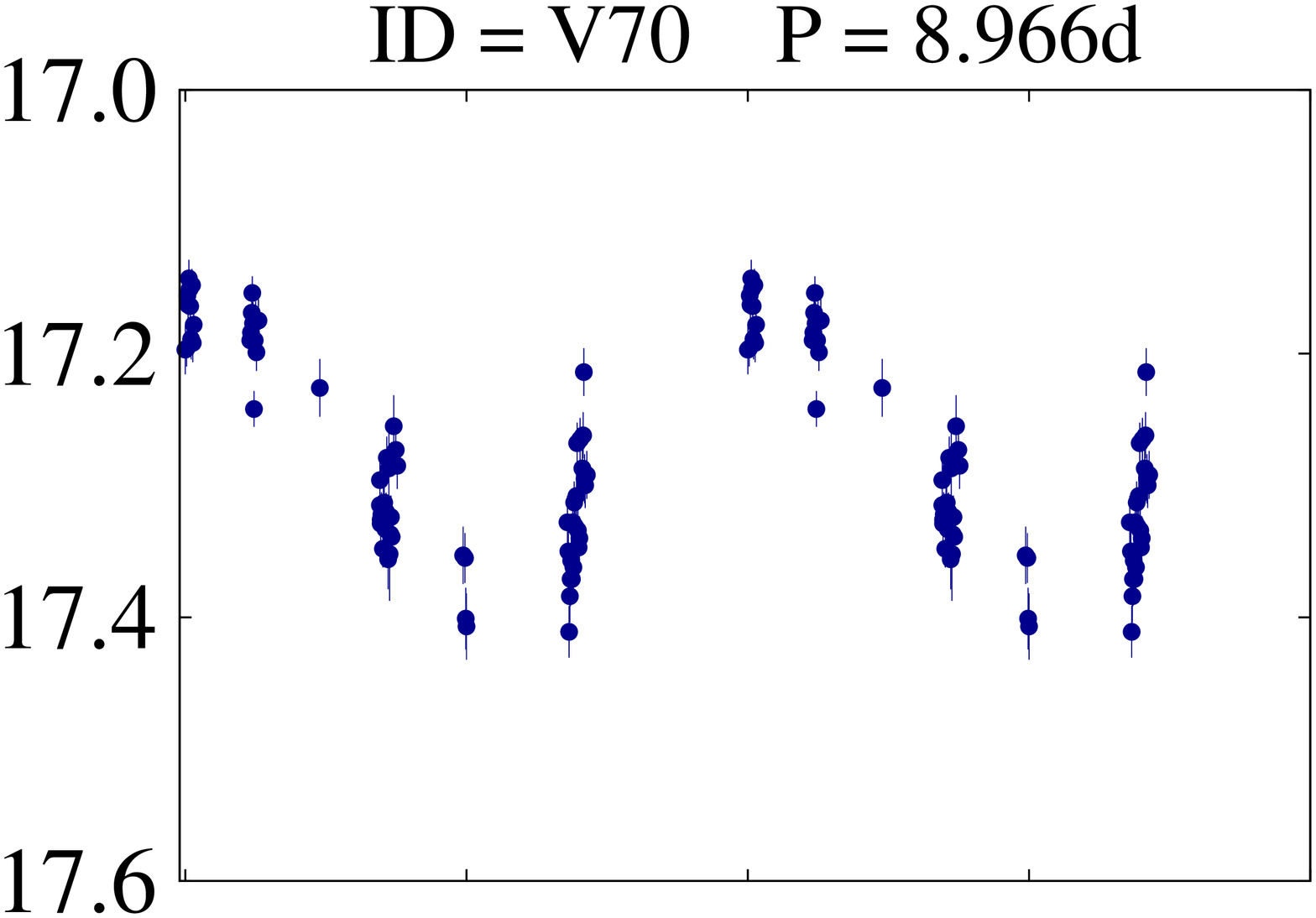}  
\hspace{0.03 cm}
\includegraphics[width = 1.9 cm, height = 2.6 cm, angle= 270]{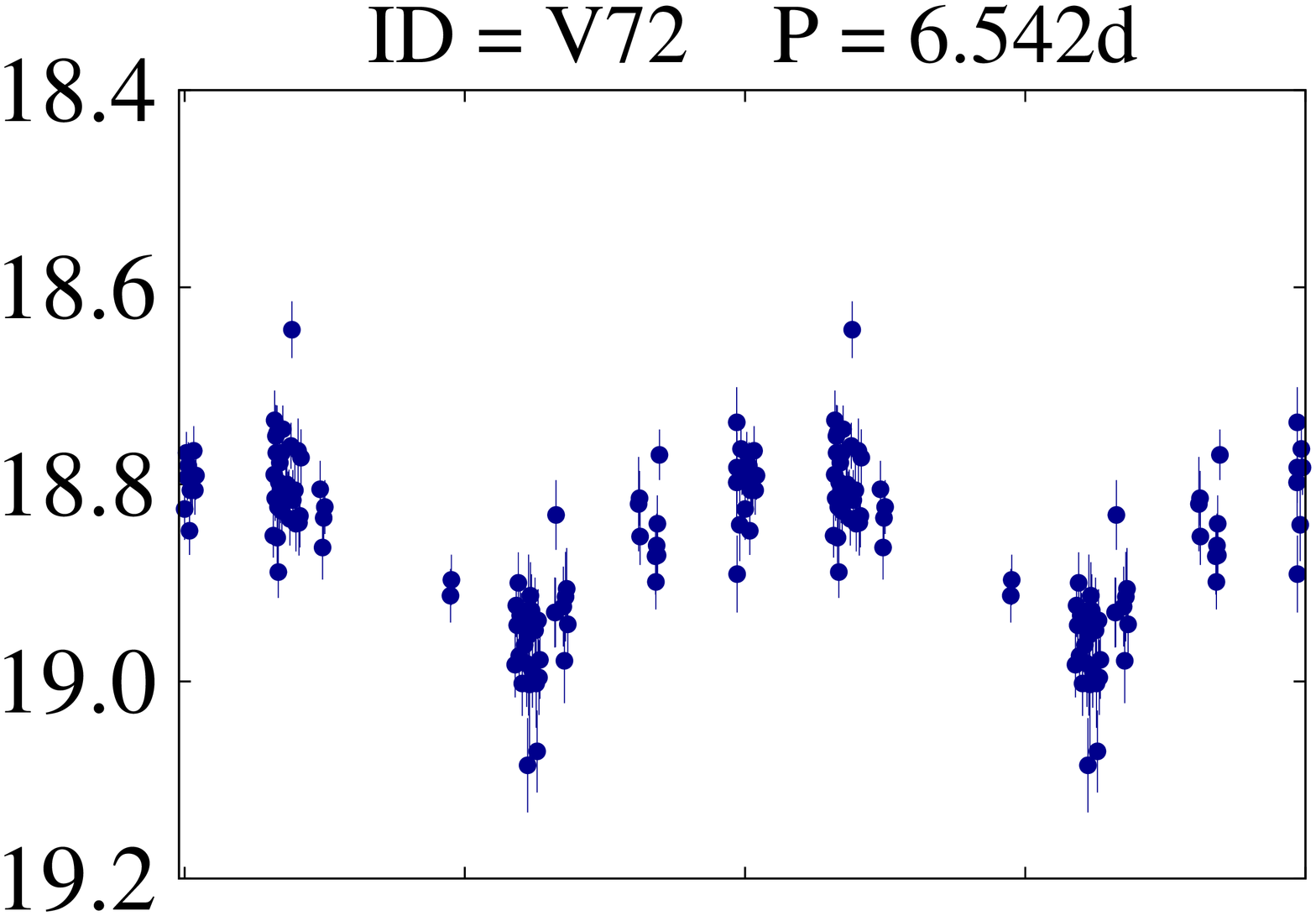}  
\hspace{0.03 cm}
\includegraphics[width = 1.9 cm, height = 2.6 cm, angle= 270]{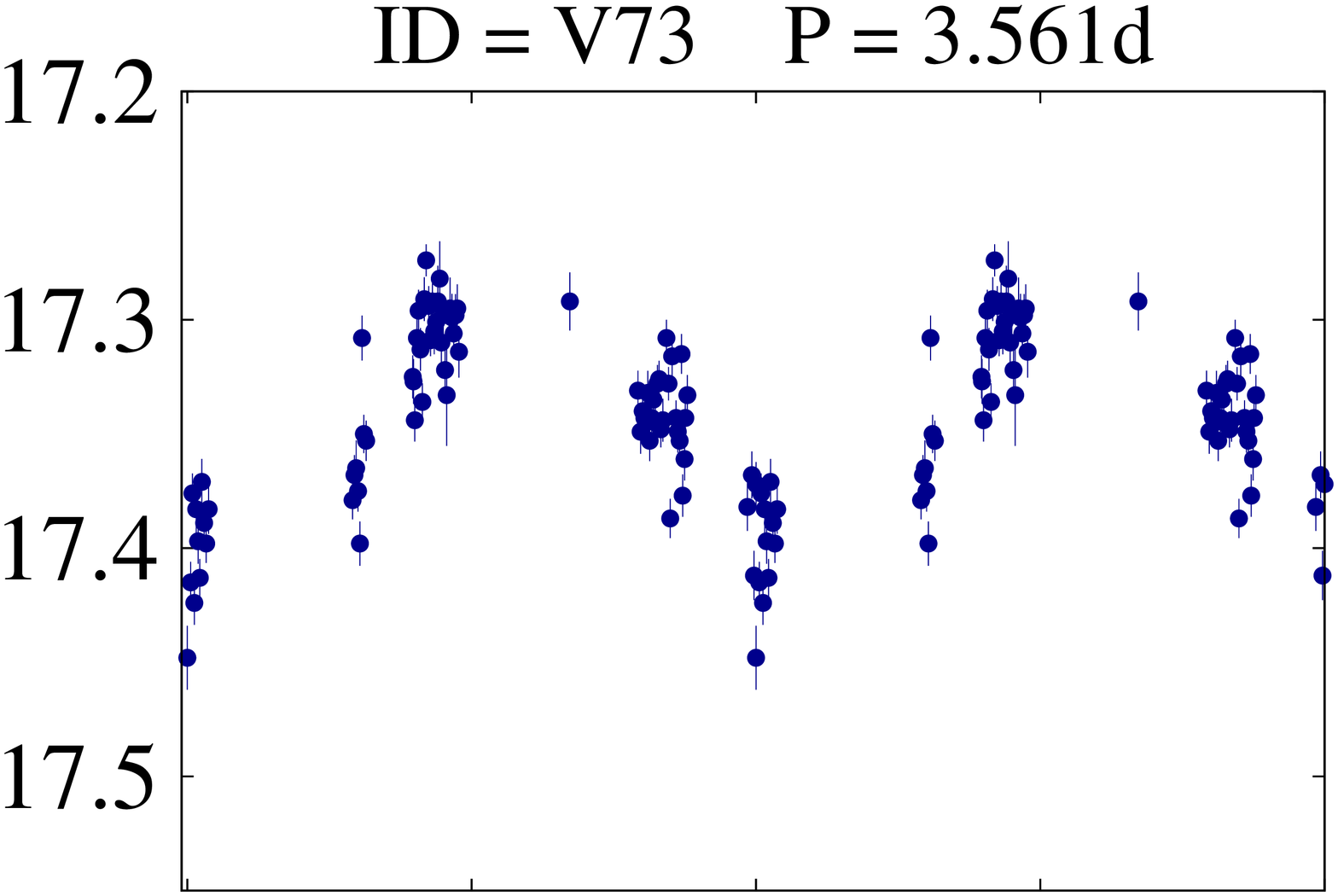}
\hspace{0.03 cm}
\includegraphics[width = 1.9 cm, height = 2.6 cm, angle= 270]{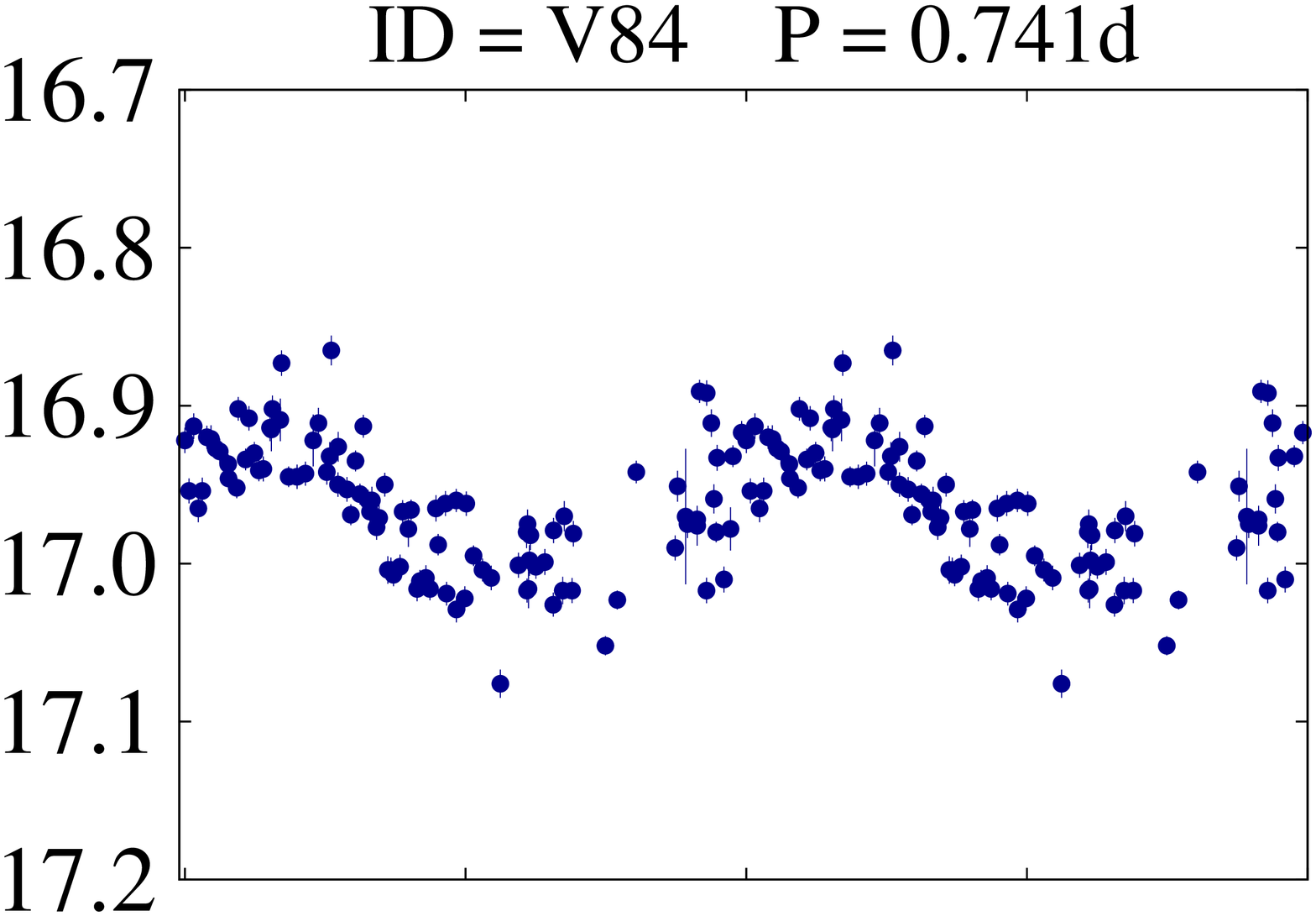}
\hspace{0.03 cm}
\includegraphics[width = 1.9 cm, height = 2.6 cm, angle= 270]{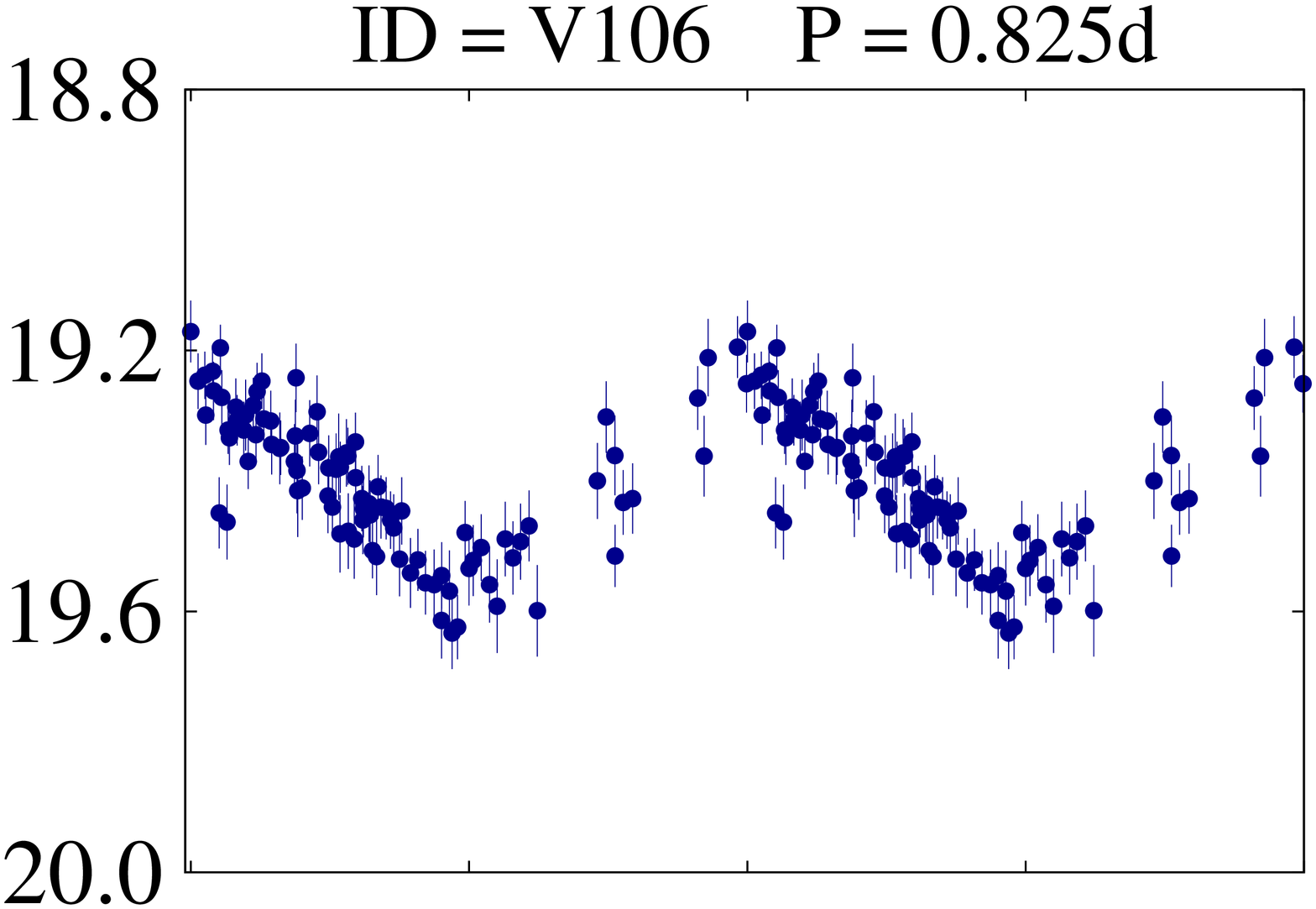} \\  
\vspace{0.05 cm}
\includegraphics[width = 1.9 cm, height = 2.6 cm, angle= 270]{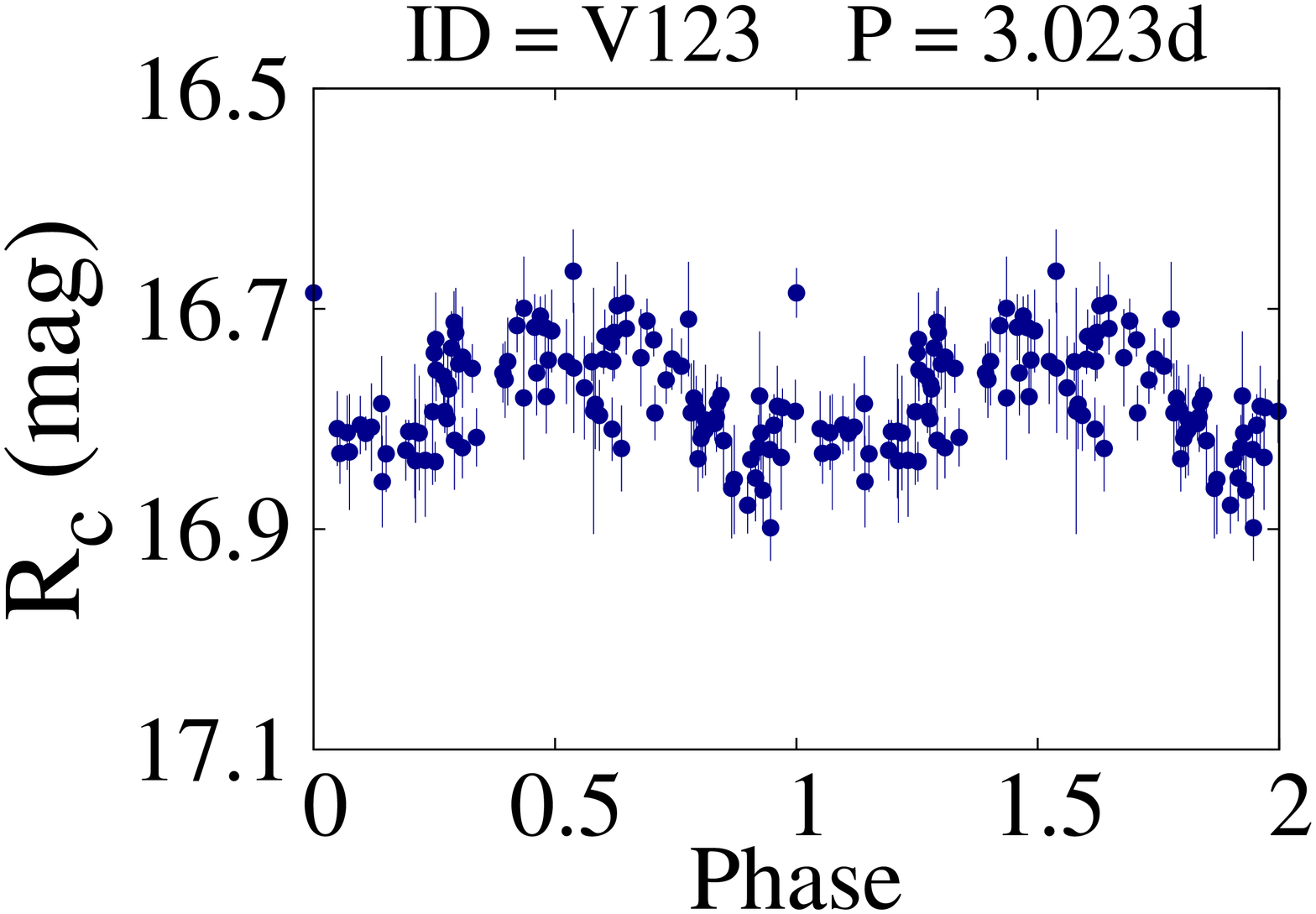}
\hspace{0.03 cm}
\includegraphics[width = 1.9 cm, height = 2.6 cm, angle= 270]{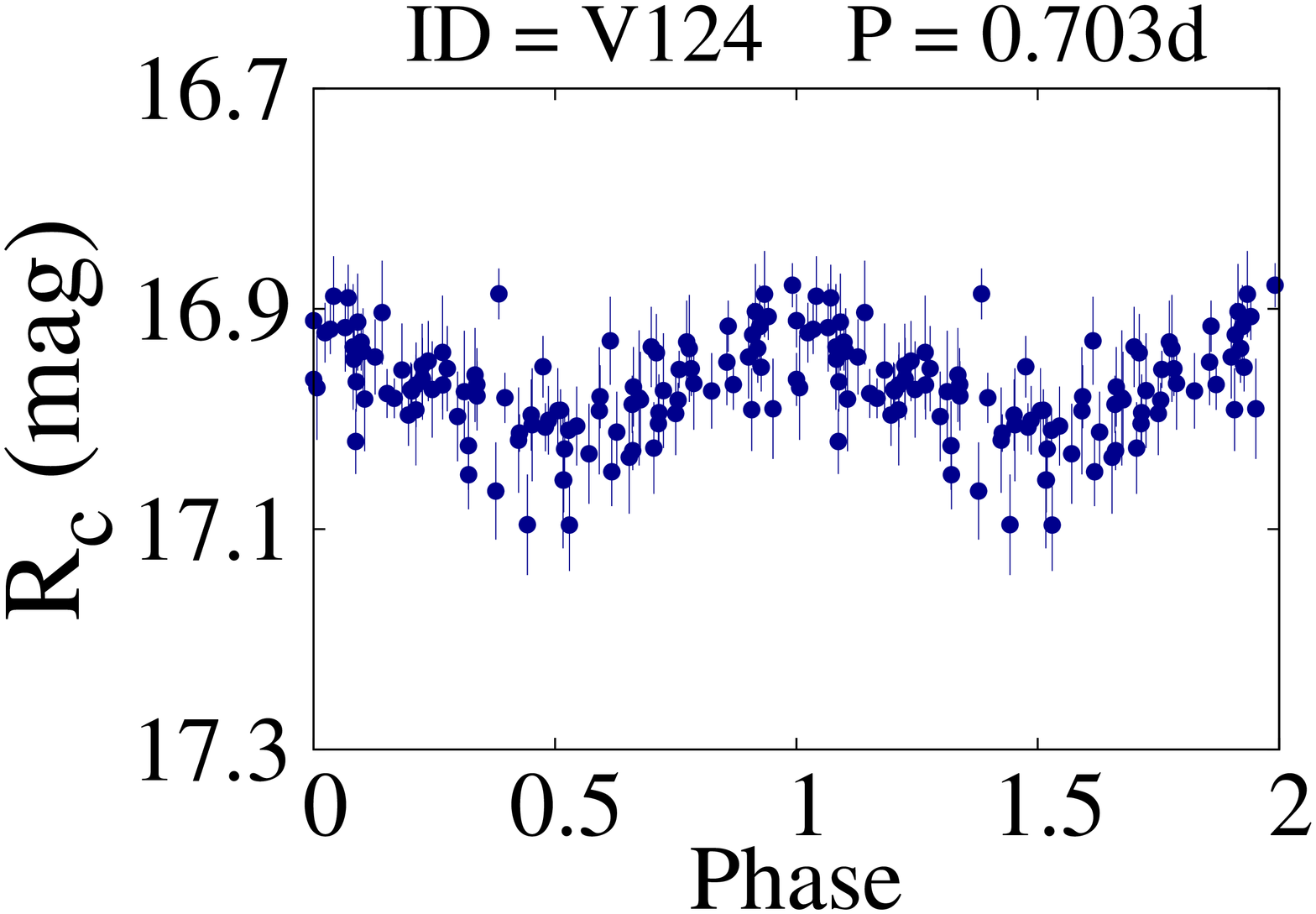}   
\hspace{0.03 cm}
\includegraphics[width = 1.9 cm, height = 2.6 cm, angle= 270]{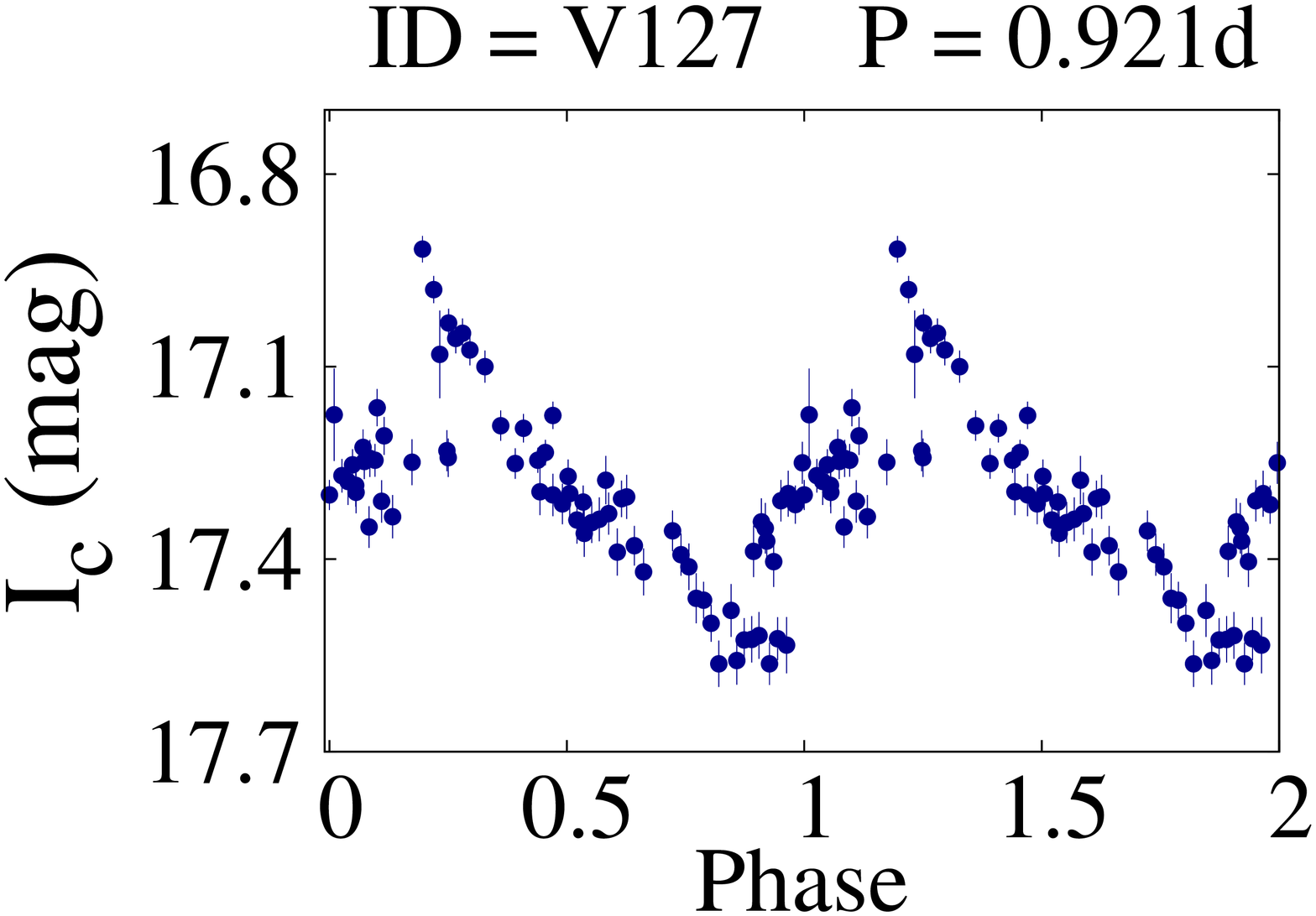}
\hspace{0.03 cm}
\includegraphics[width = 1.9 cm, height = 2.6 cm, angle= 270]{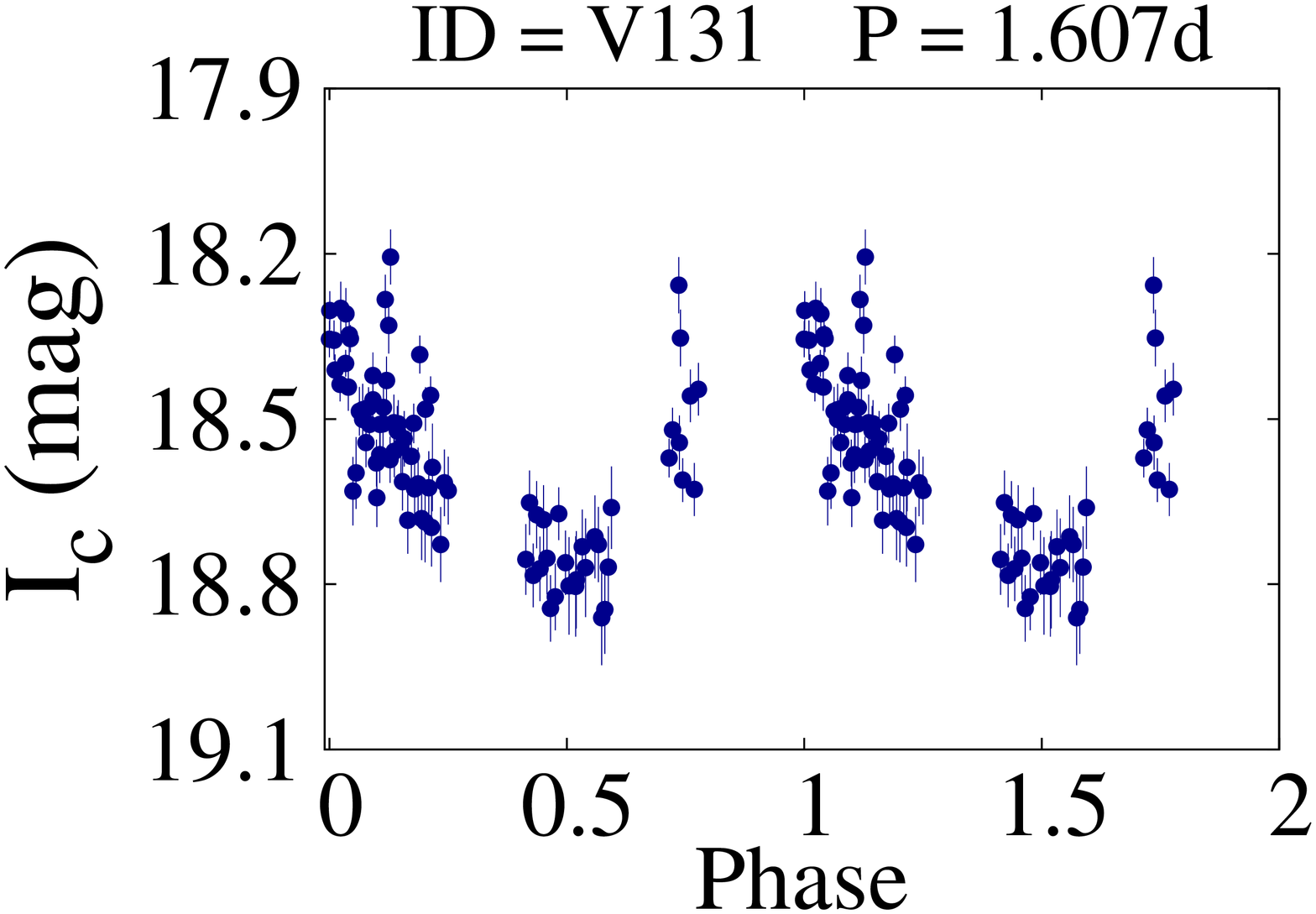}

\caption{Phase folded LCs of 28 Class\,{\sc iii} periodic  variables (WTTSs).
The identification numbers and periods (days) of the corresponding stars are given on the top of
each panel.}
 \label{fig: LC_Class_III_P}
\end{figure}

\begin{figure*}[h]
\centering
\includegraphics[width= 4.0 cm,height = 7.0 cm, angle=270]{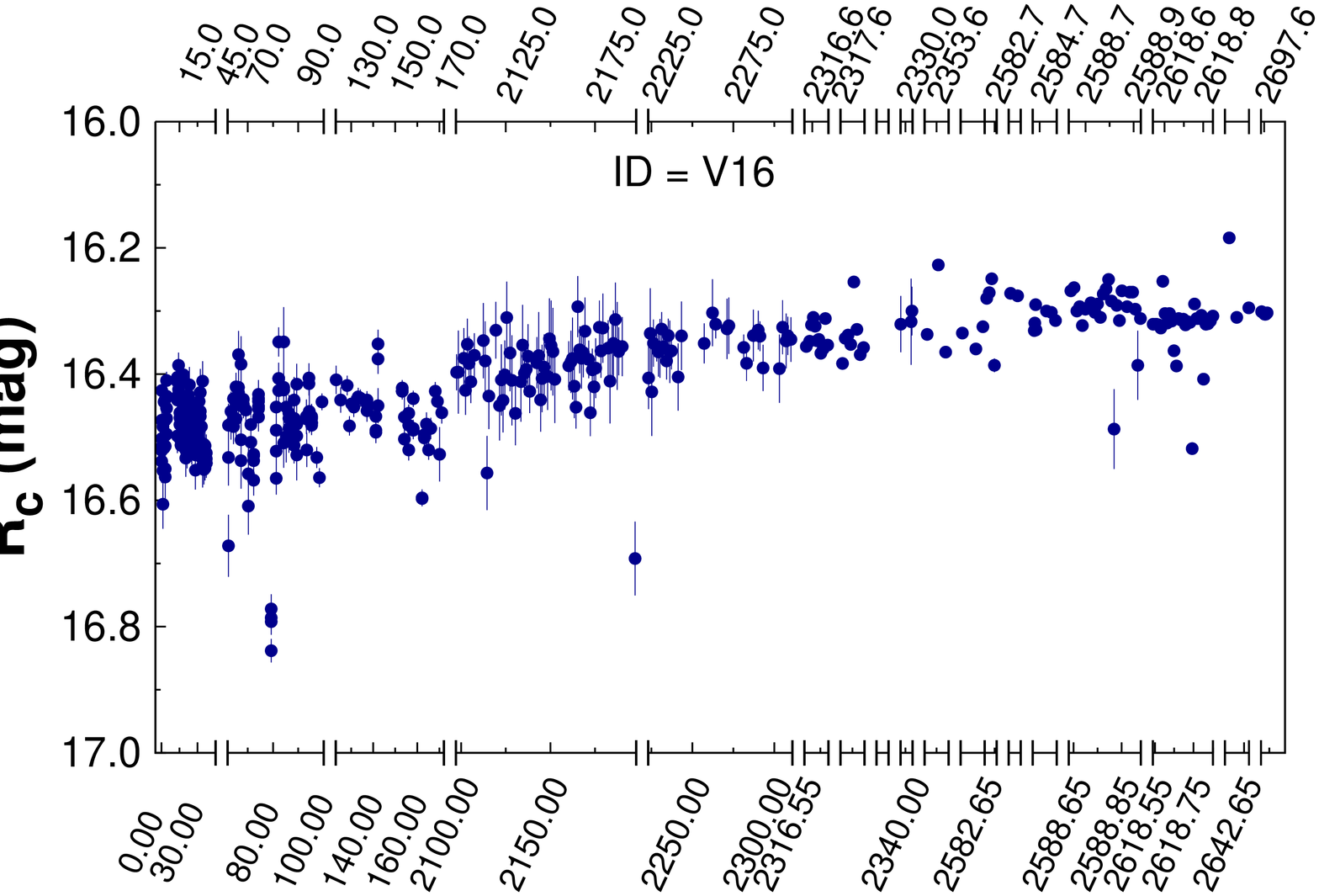}
\hspace{0.1 cm}
\includegraphics[width= 4.0 cm,height = 7.0 cm, angle=270]{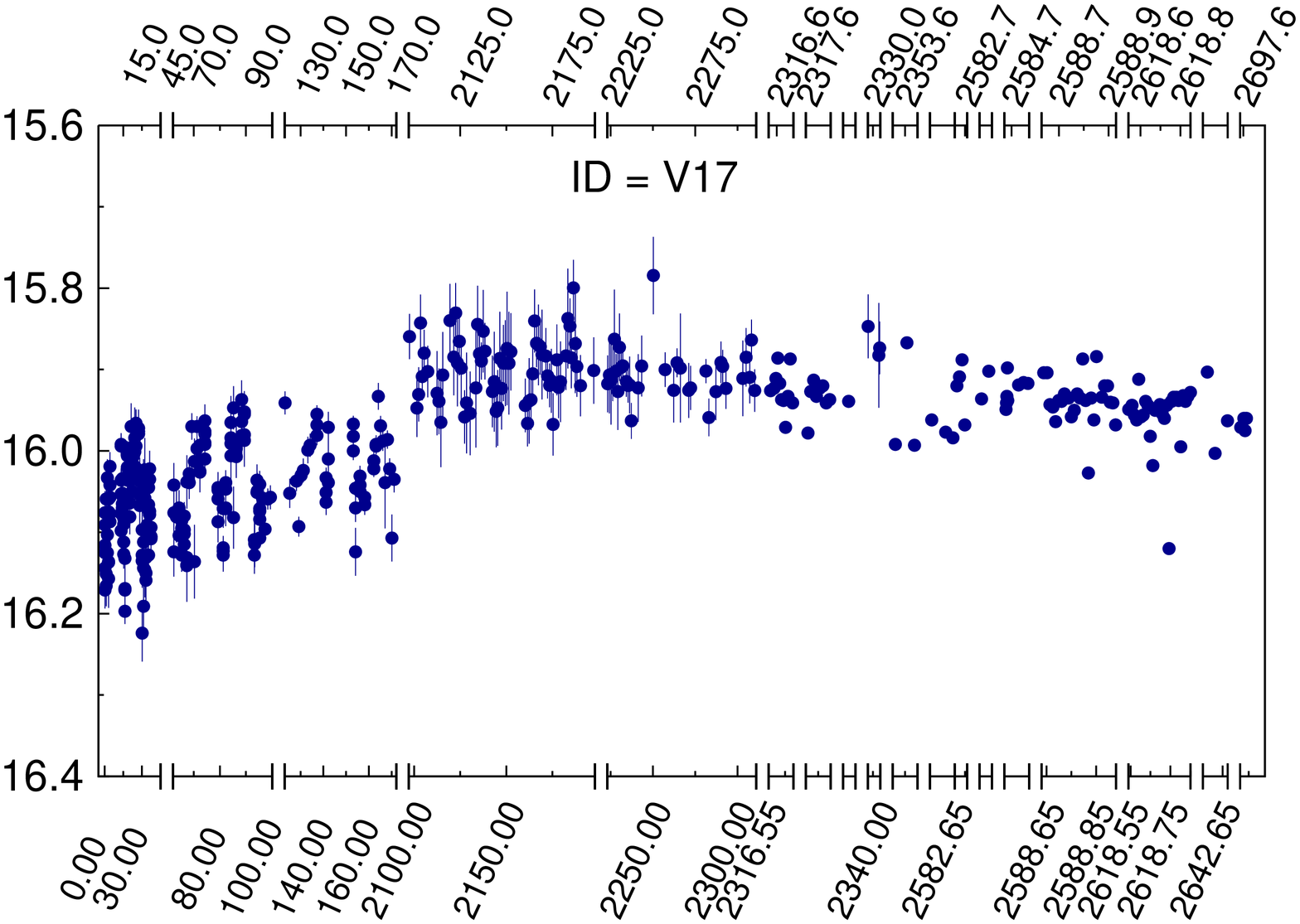} \\
\vspace{0.2 cm}
\includegraphics[width= 4.0 cm,height = 7.0 cm, angle=270]{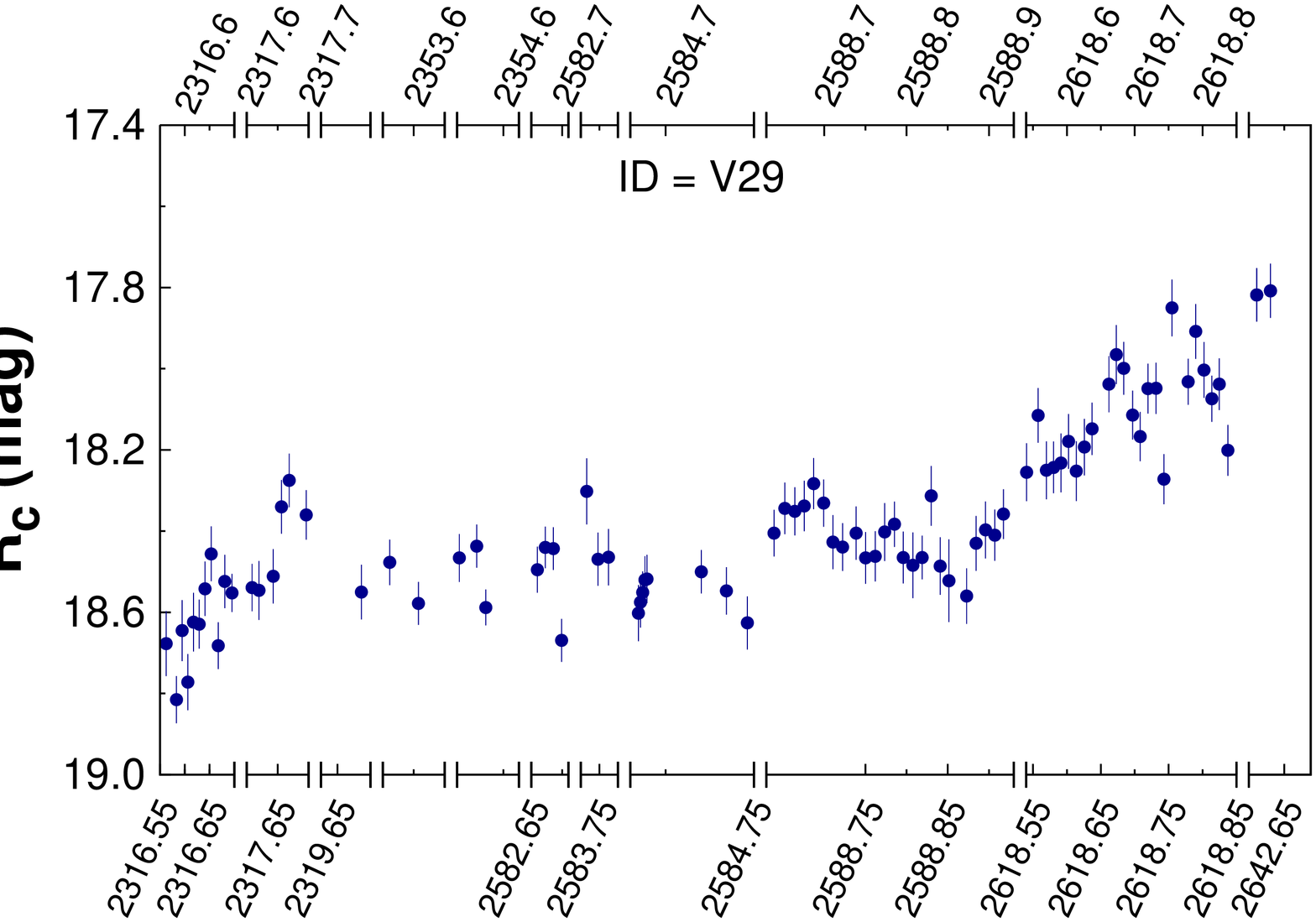}
\hspace{0.1 cm}
\includegraphics[width= 4.0 cm,height = 7.0 cm, angle=270]{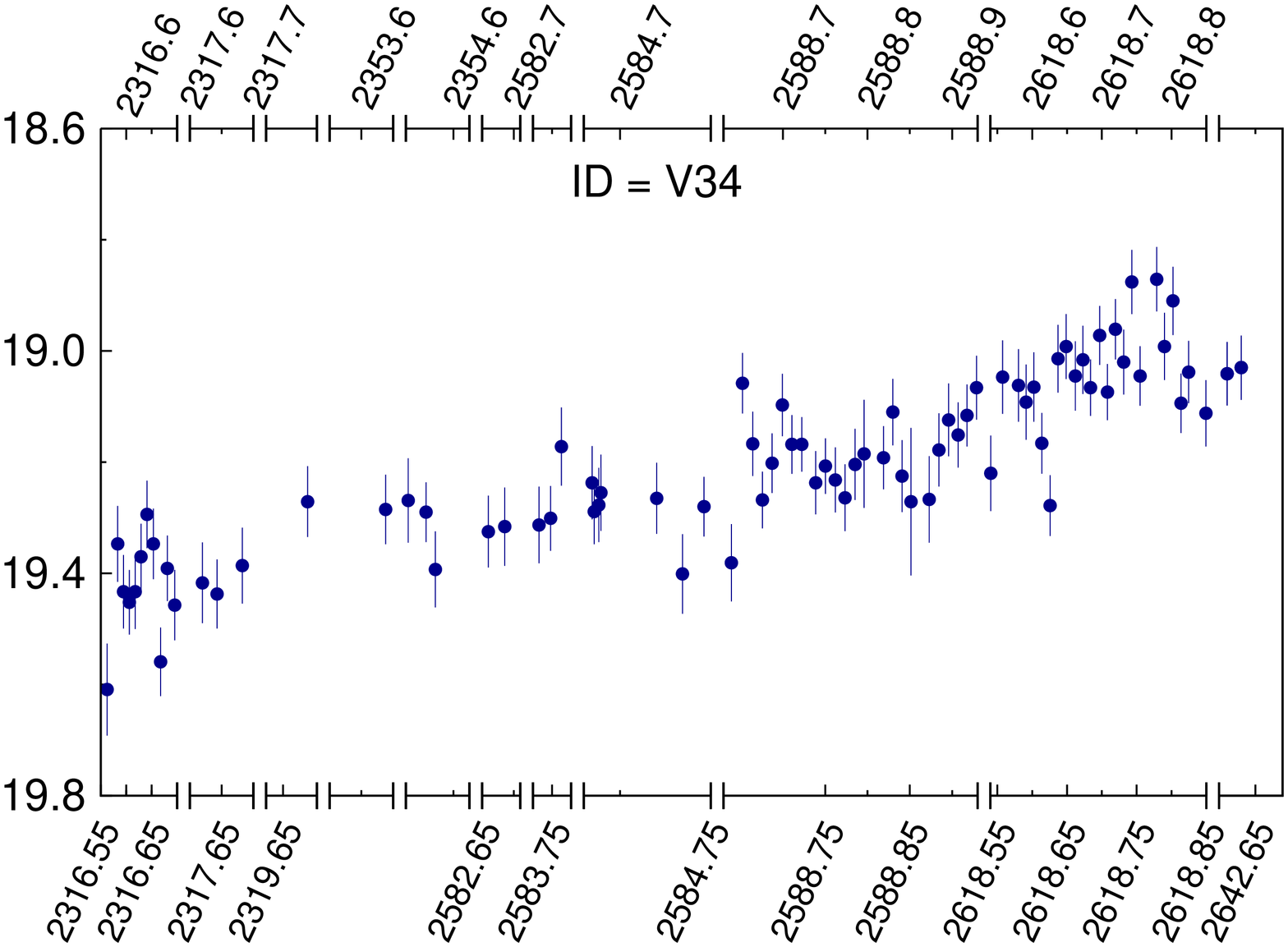} \\
\vspace{0.2 cm}
\includegraphics[width= 4.0 cm,height = 7.0 cm, angle=270]{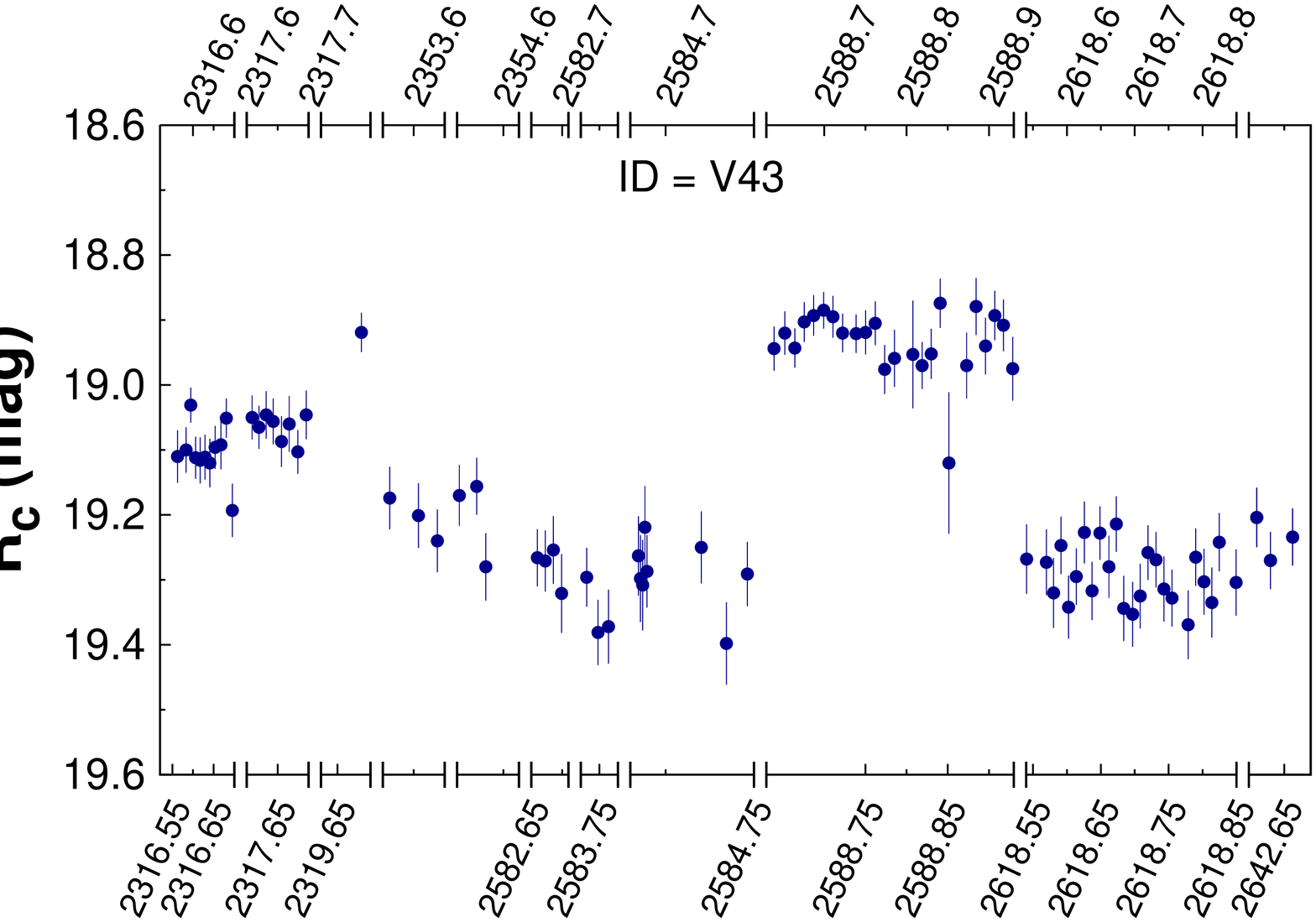}
\hspace{0.1 cm}
\includegraphics[width= 4.0 cm,height = 7.0 cm, angle=270]{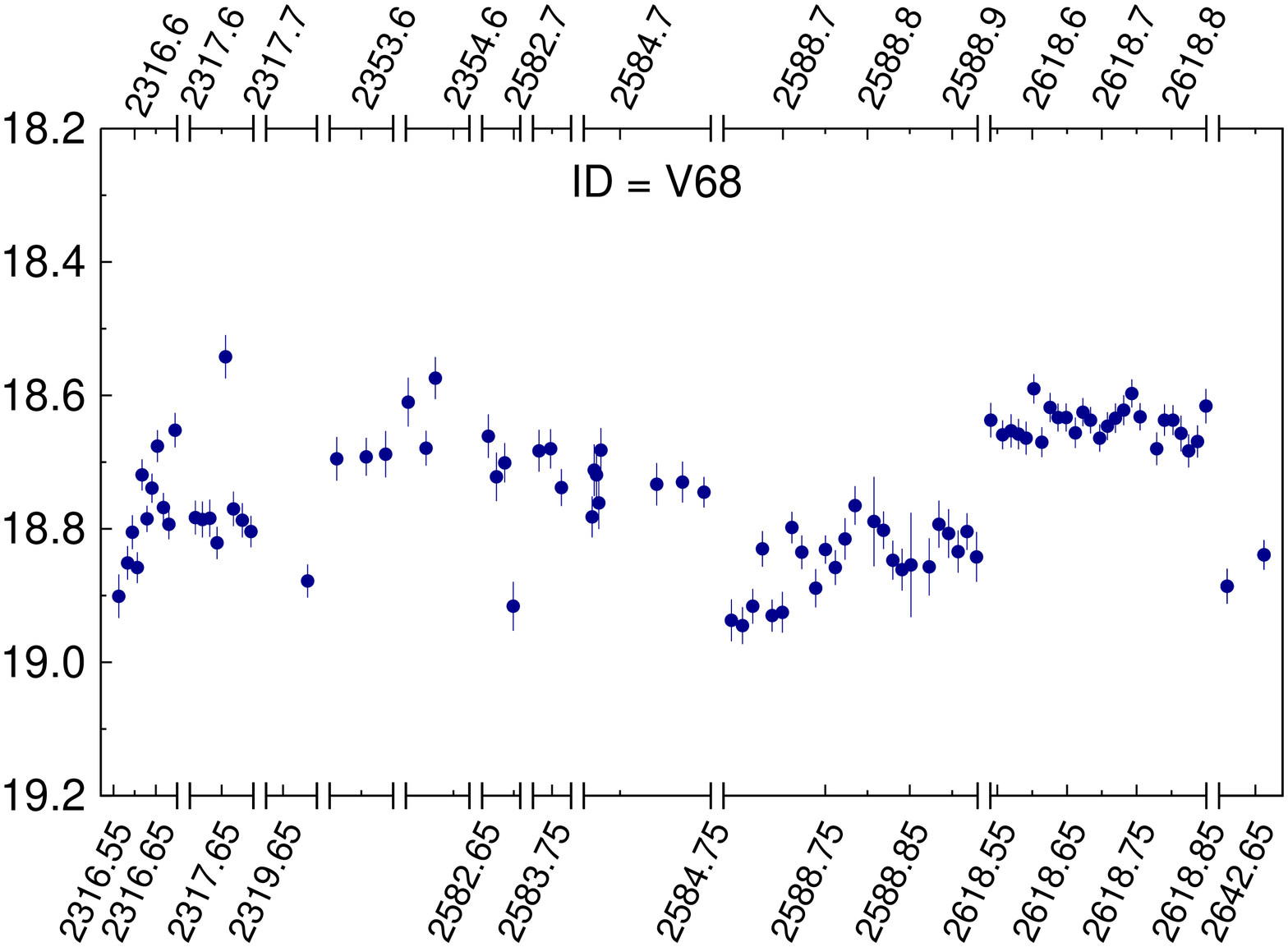} \\
\vspace{0.2 cm}
\includegraphics[width= 4.0 cm,height = 7.0 cm, angle=270]{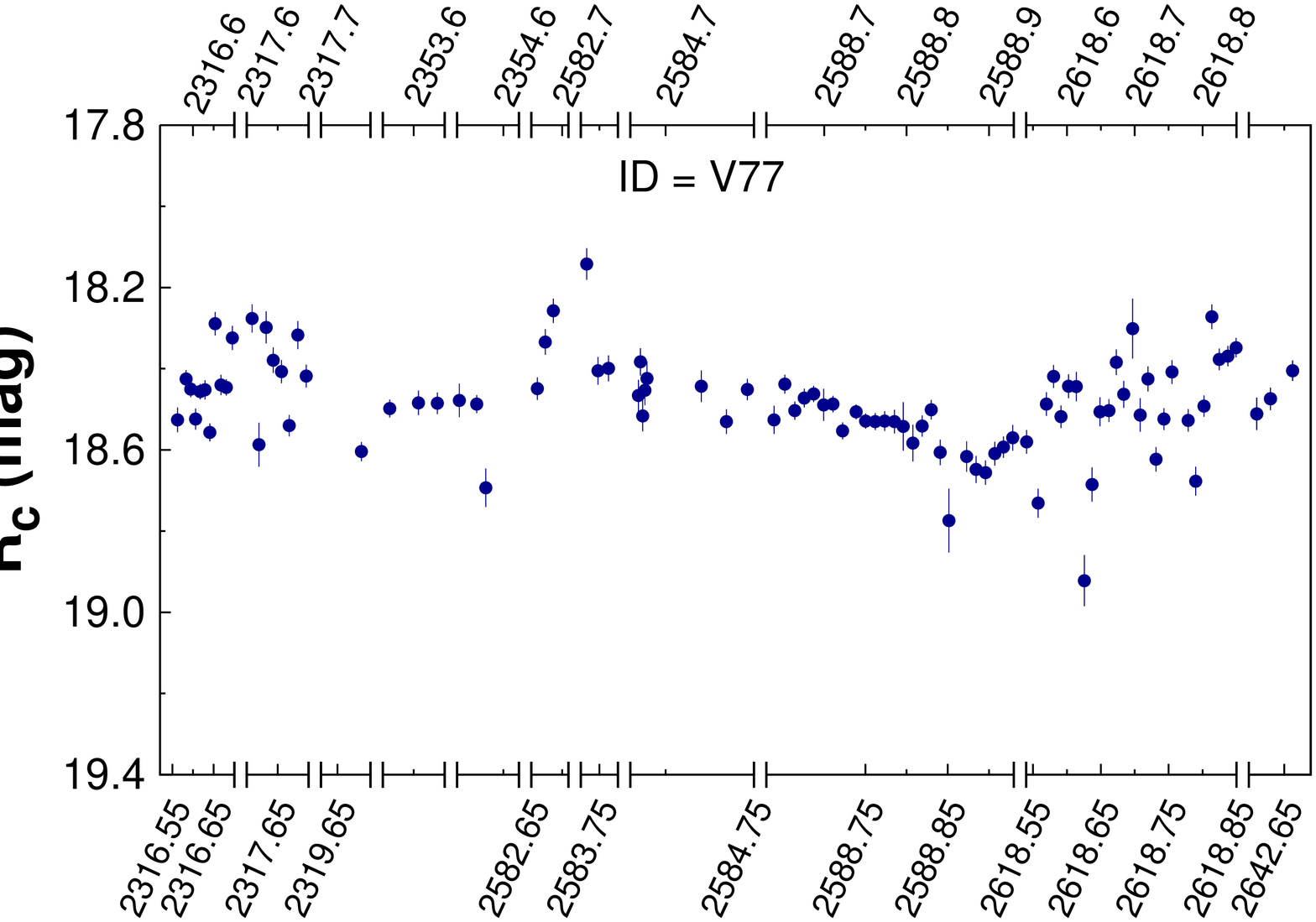}
\hspace{0.1 cm}
\includegraphics[width= 4.0 cm,height = 7.0 cm, angle=270]{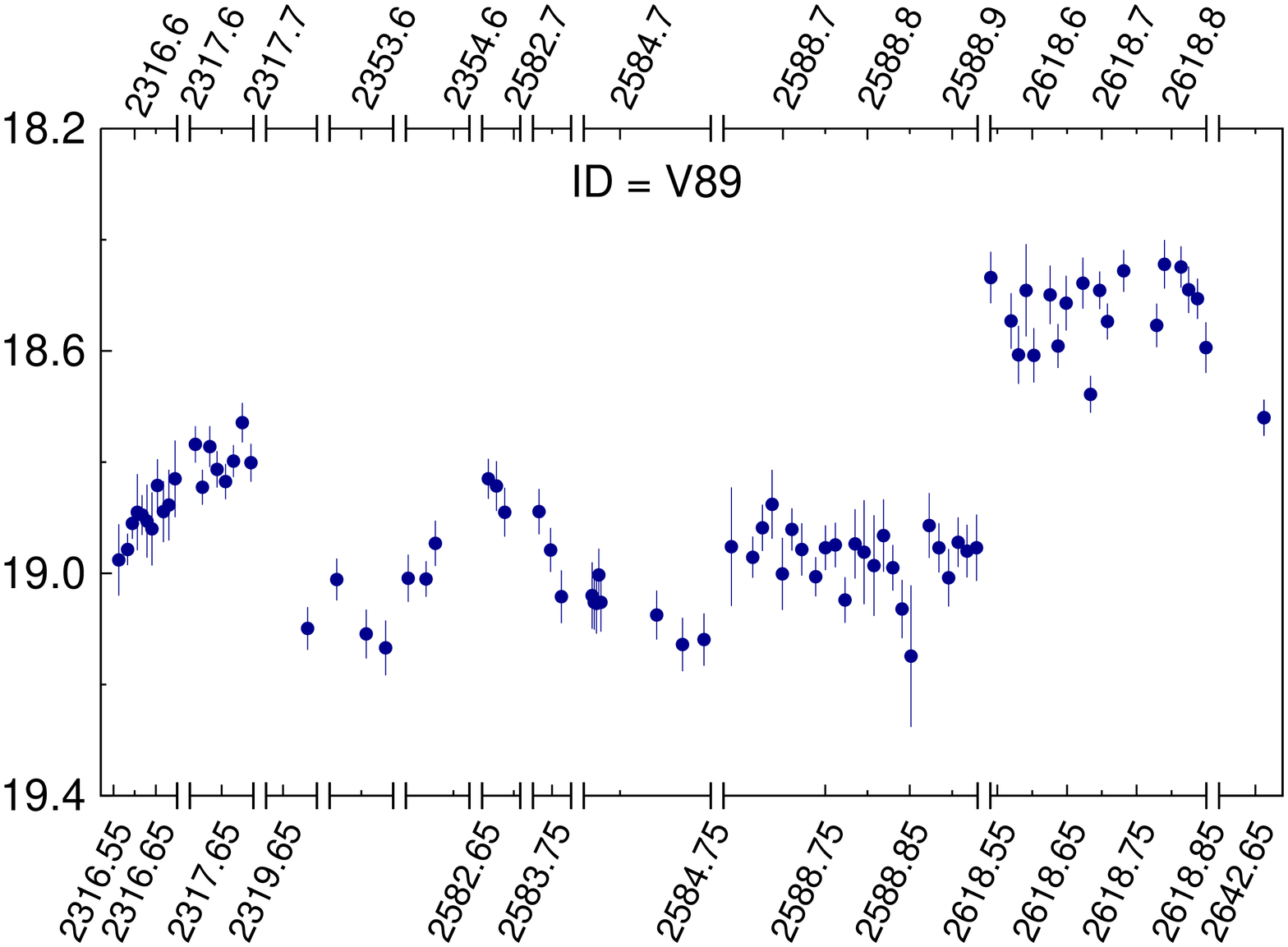} \\
\vspace{0.2 cm}
\includegraphics[width= 4.0 cm,height = 7.0 cm, angle=270]{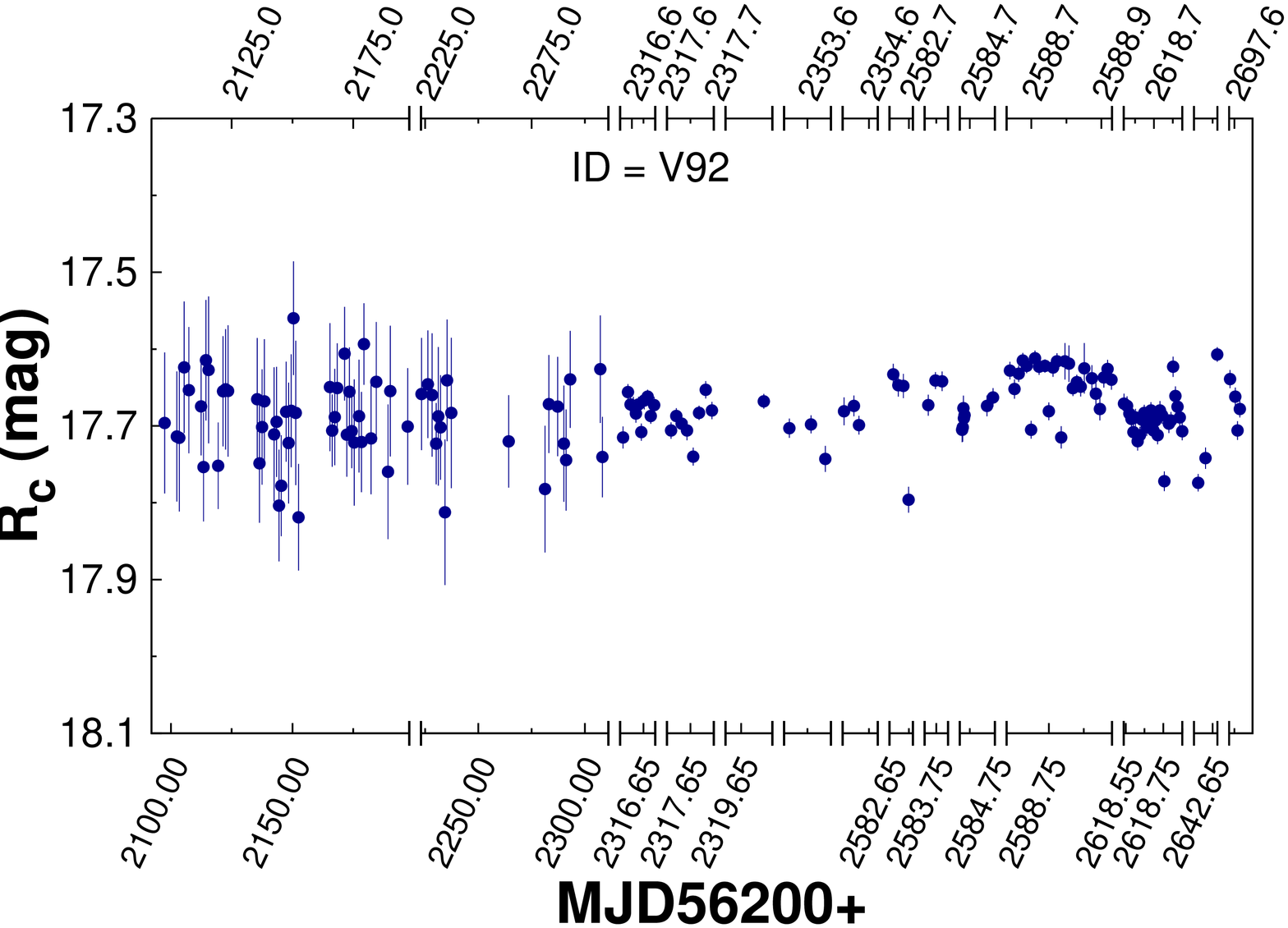}
\hspace{0.1 cm}
\includegraphics[width= 4.0 cm,height = 7.0 cm, angle=270]{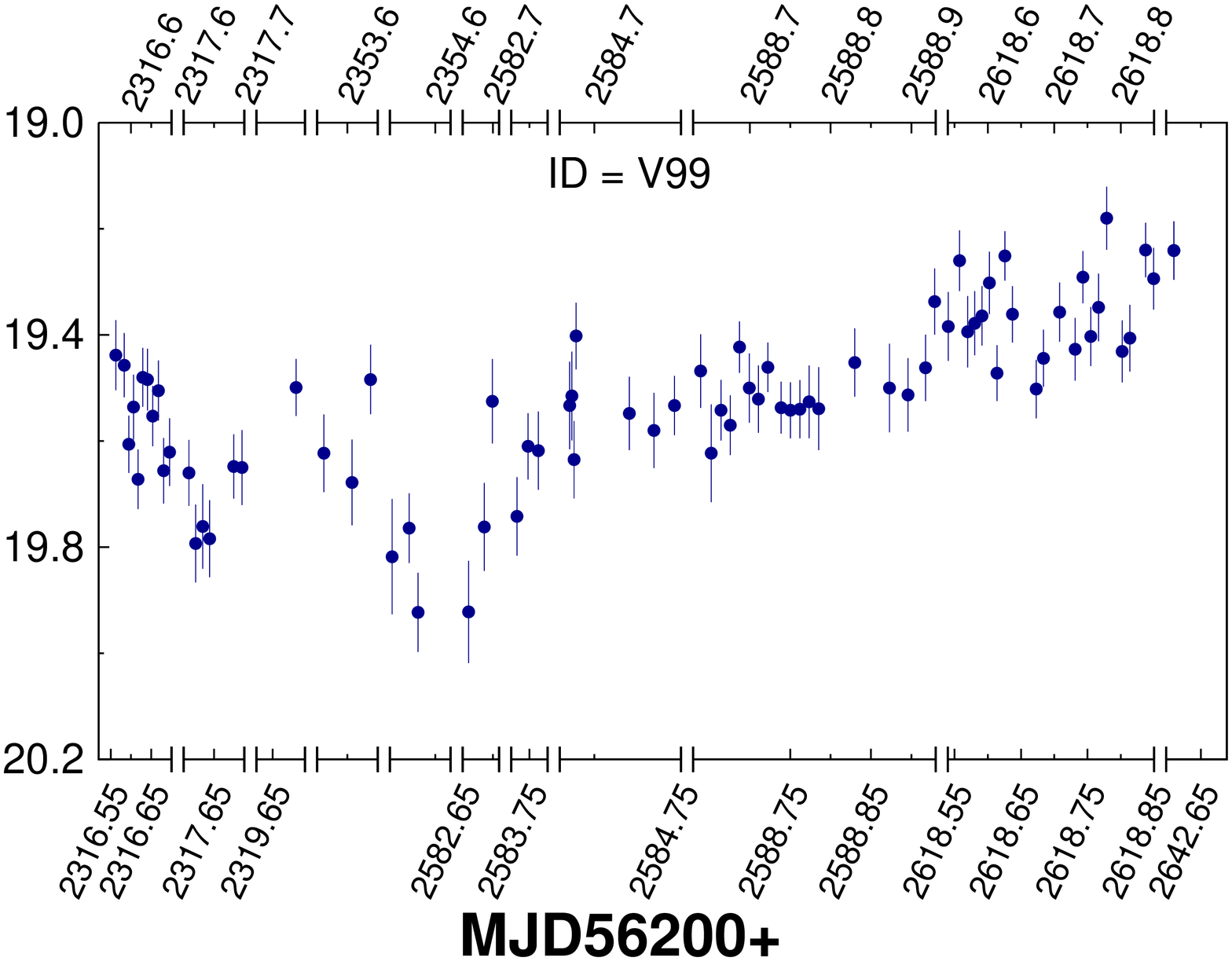} \\

\caption{{\scriptsize LCs of 20 Class\,{\sc iii} non-periodic variables. When there are data gaps they are represented with vertical gaps along the axis.
Corresponding identification numbers are given in each panel. A complete set of LCs are provided in the electronic form only.}}
\end{figure*}

\setcounter{figure}{11}

\begin{figure*}[h]
\centering
\vspace{0.03 cm}
\includegraphics[width= 4.0 cm,height = 7.0 cm, angle=270]{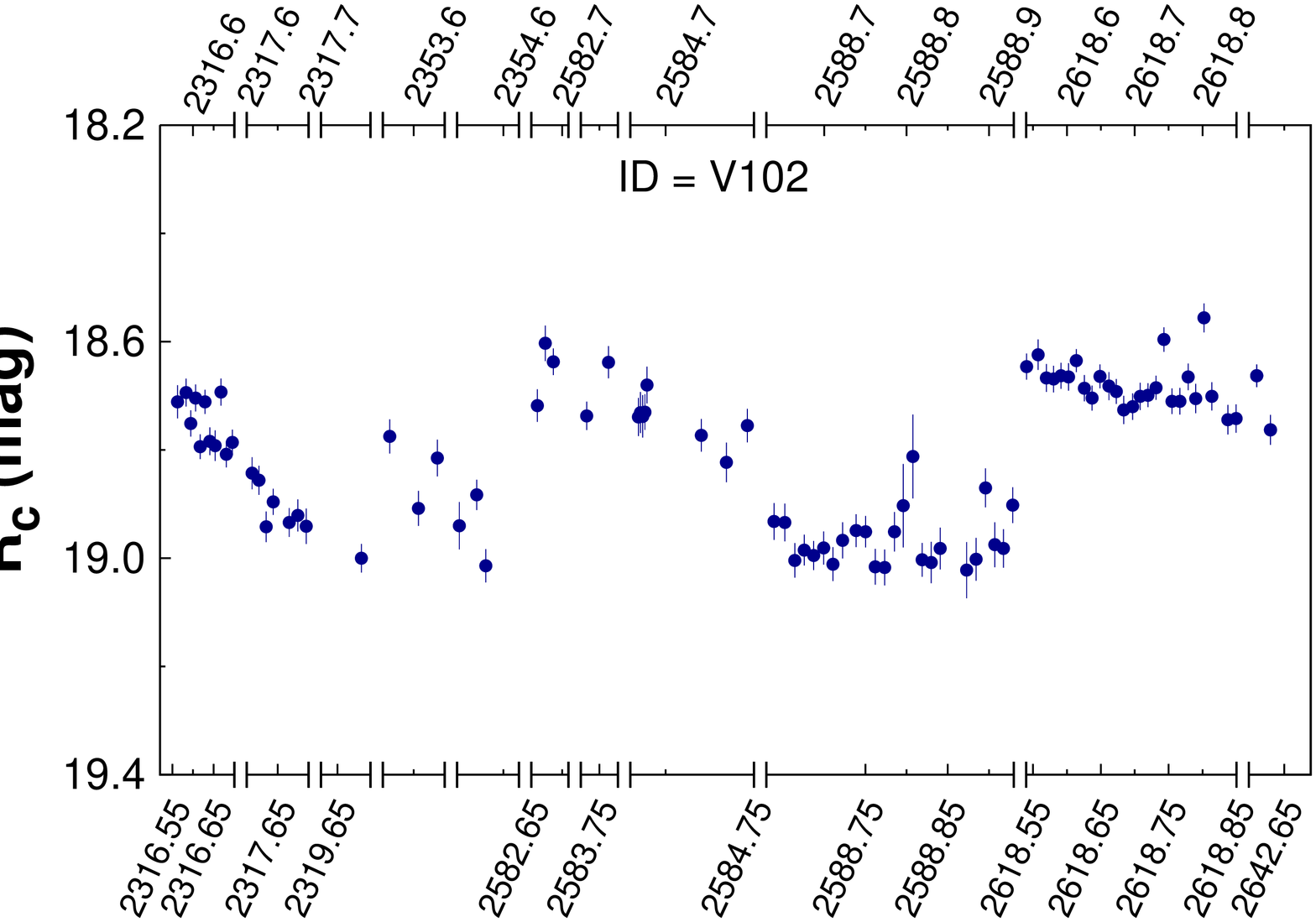}
\hspace{0.1 cm}
\includegraphics[width= 4.0 cm,height = 7.0 cm, angle=270]{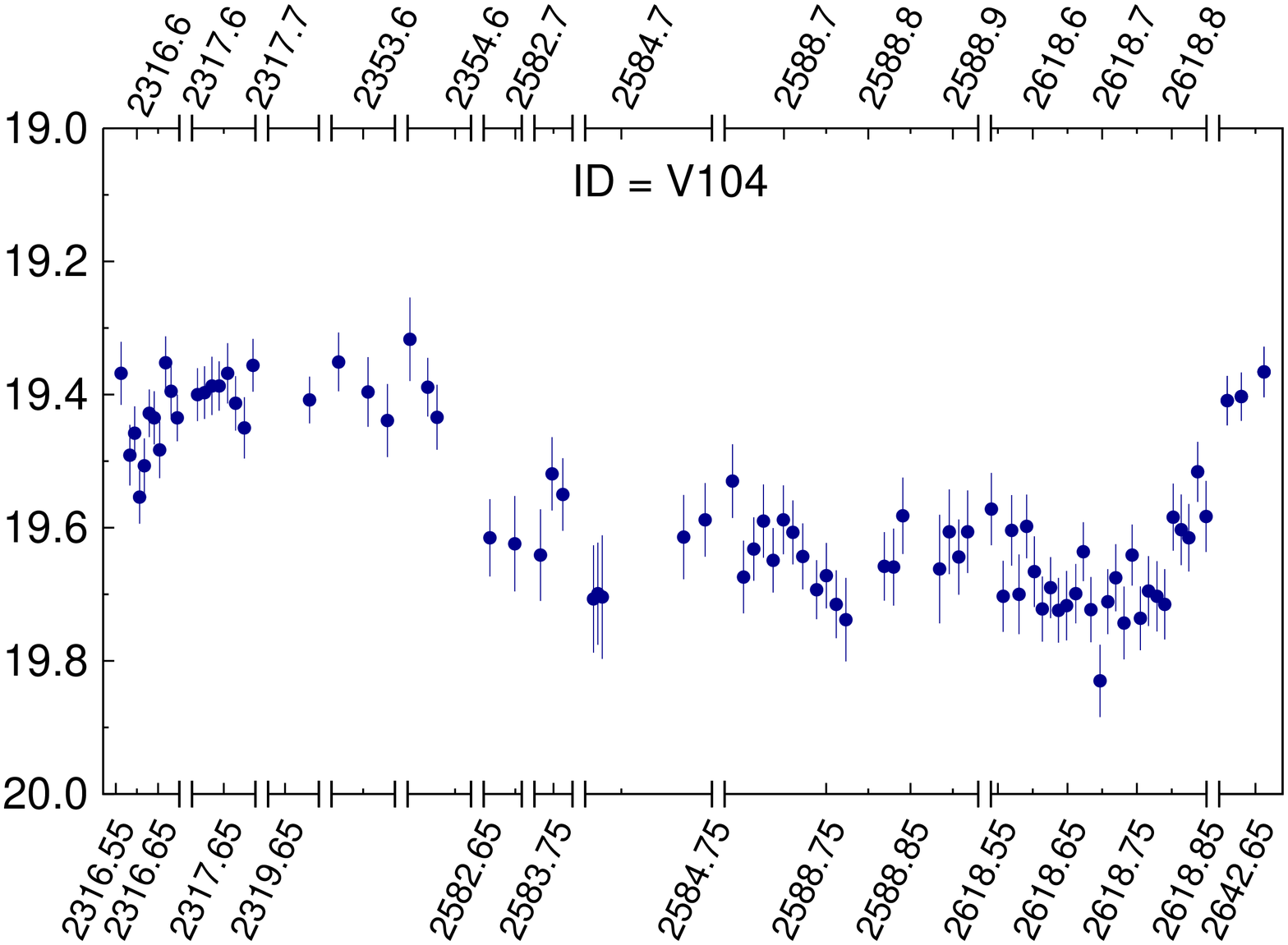} \\
\vspace{0.2 cm}
\includegraphics[width= 4.0 cm,height = 7.0 cm, angle=270]{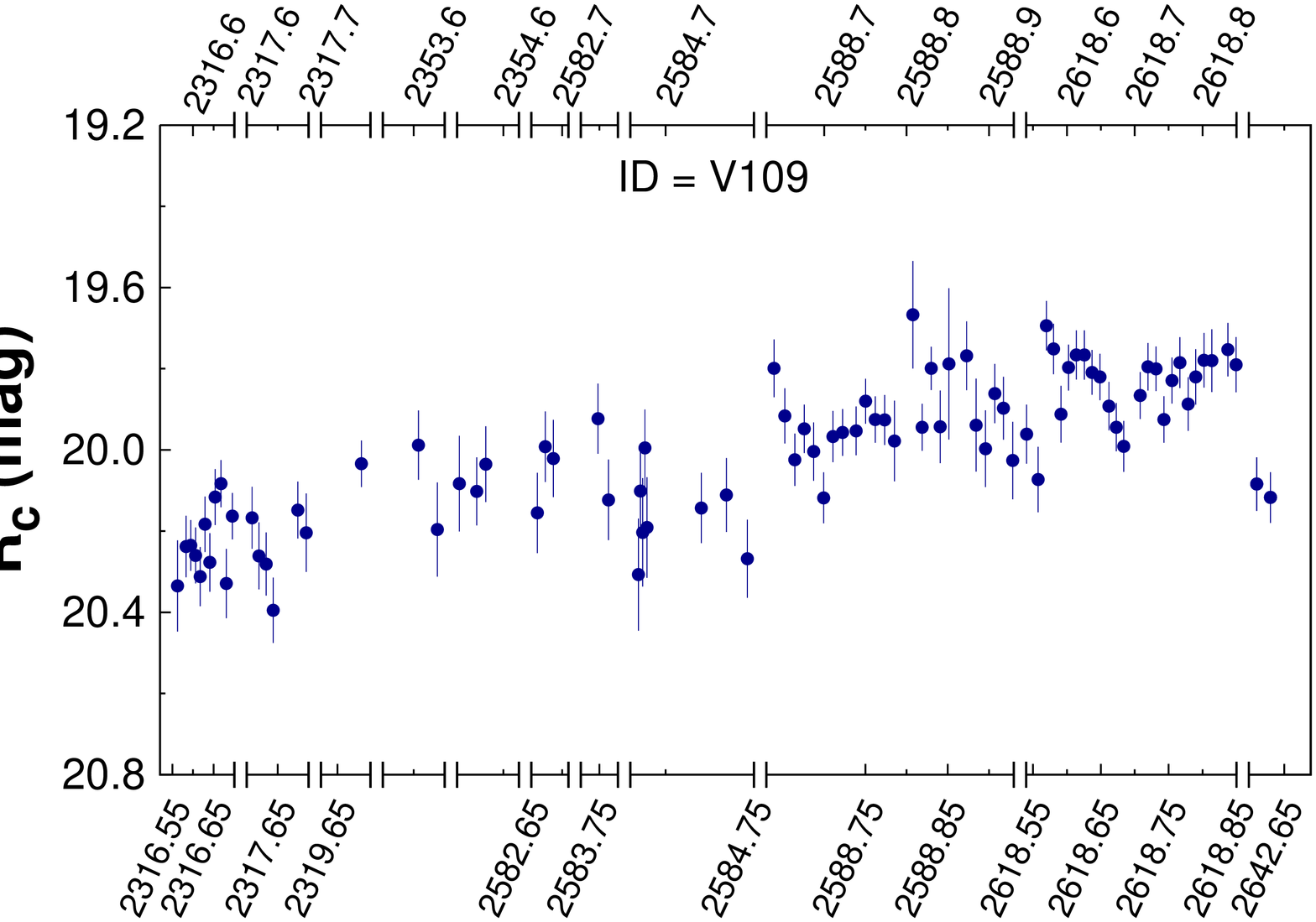}
\hspace{0.1 cm}
\includegraphics[width= 4.0 cm,height = 7.0 cm, angle=270]{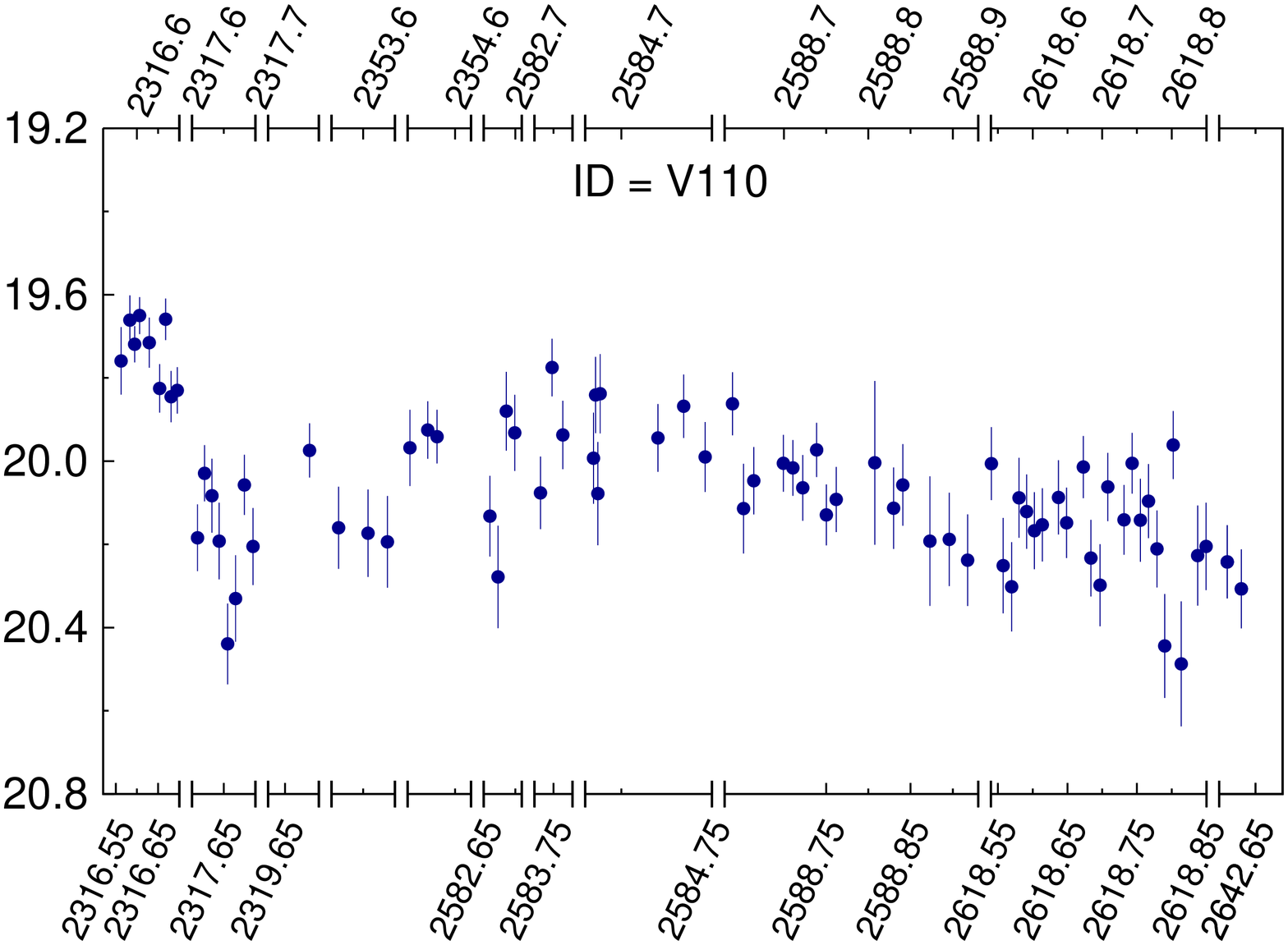} \\
\vspace{0.2 cm}
\includegraphics[width= 4.0 cm,height = 7.0 cm, angle=270]{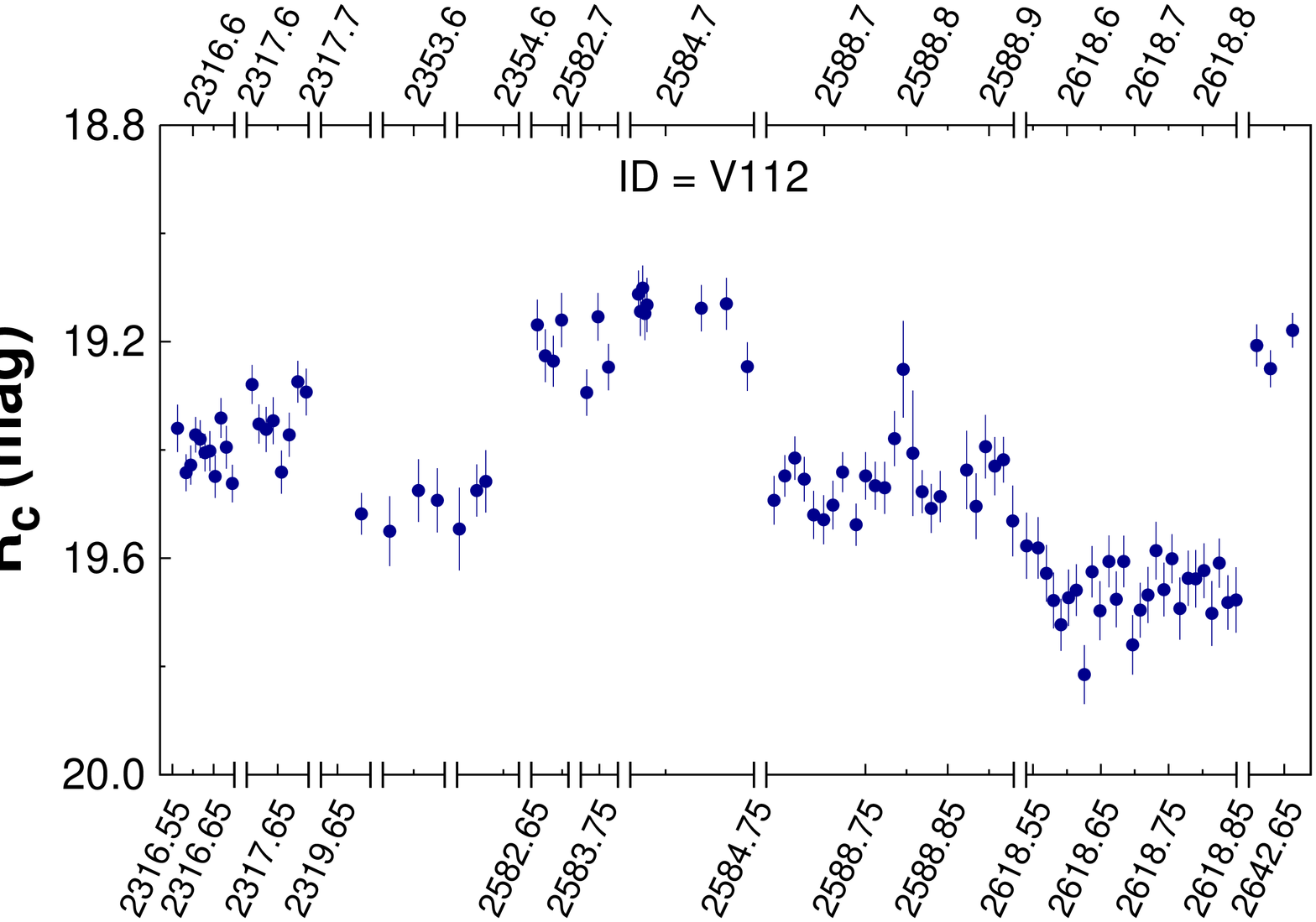}
\hspace{0.1 cm}
\includegraphics[width= 4.0 cm,height = 7.0 cm, angle=270]{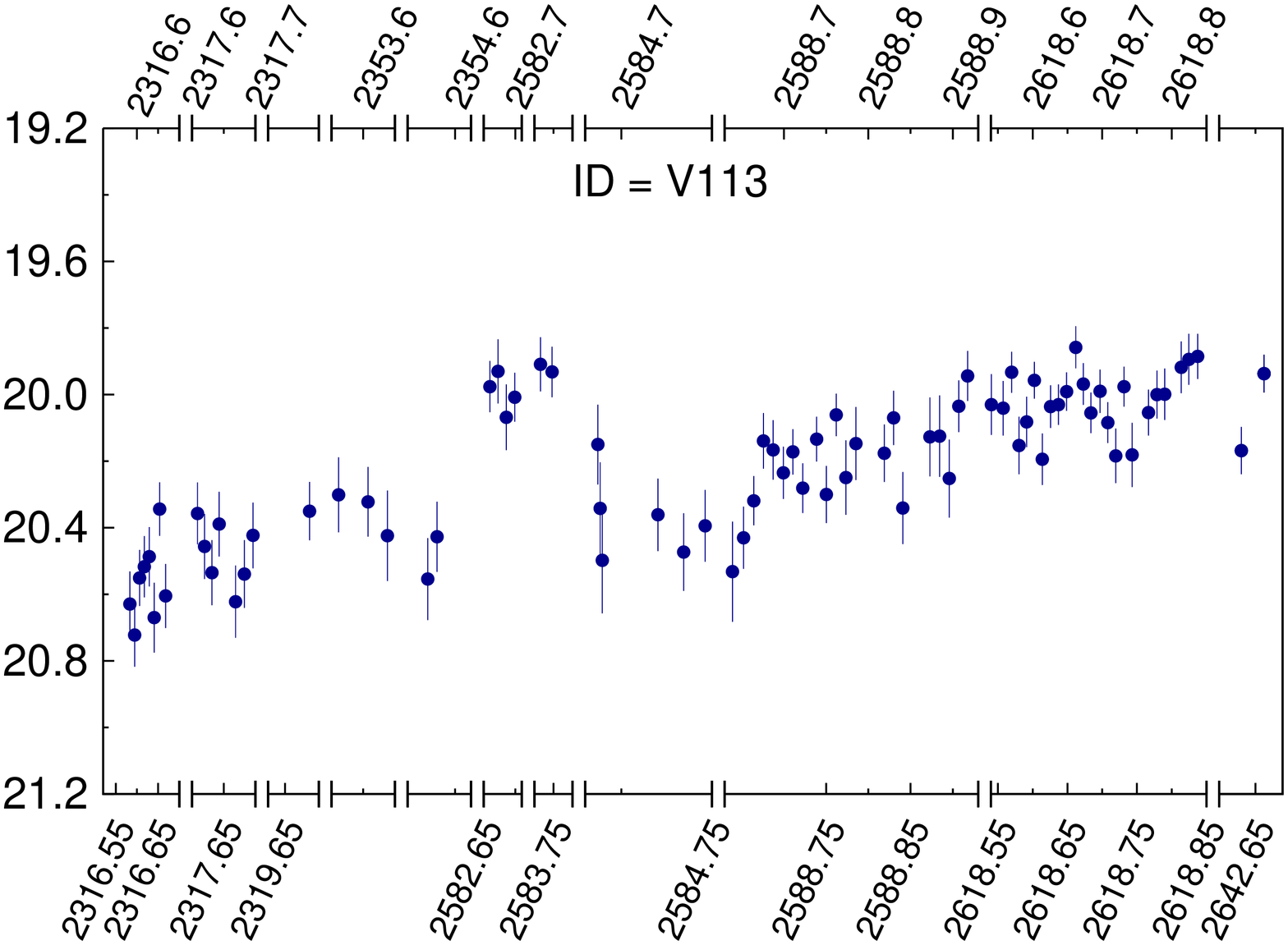} \\
\vspace{0.2 cm}
\includegraphics[width= 4.0 cm,height = 7.0 cm, angle=270]{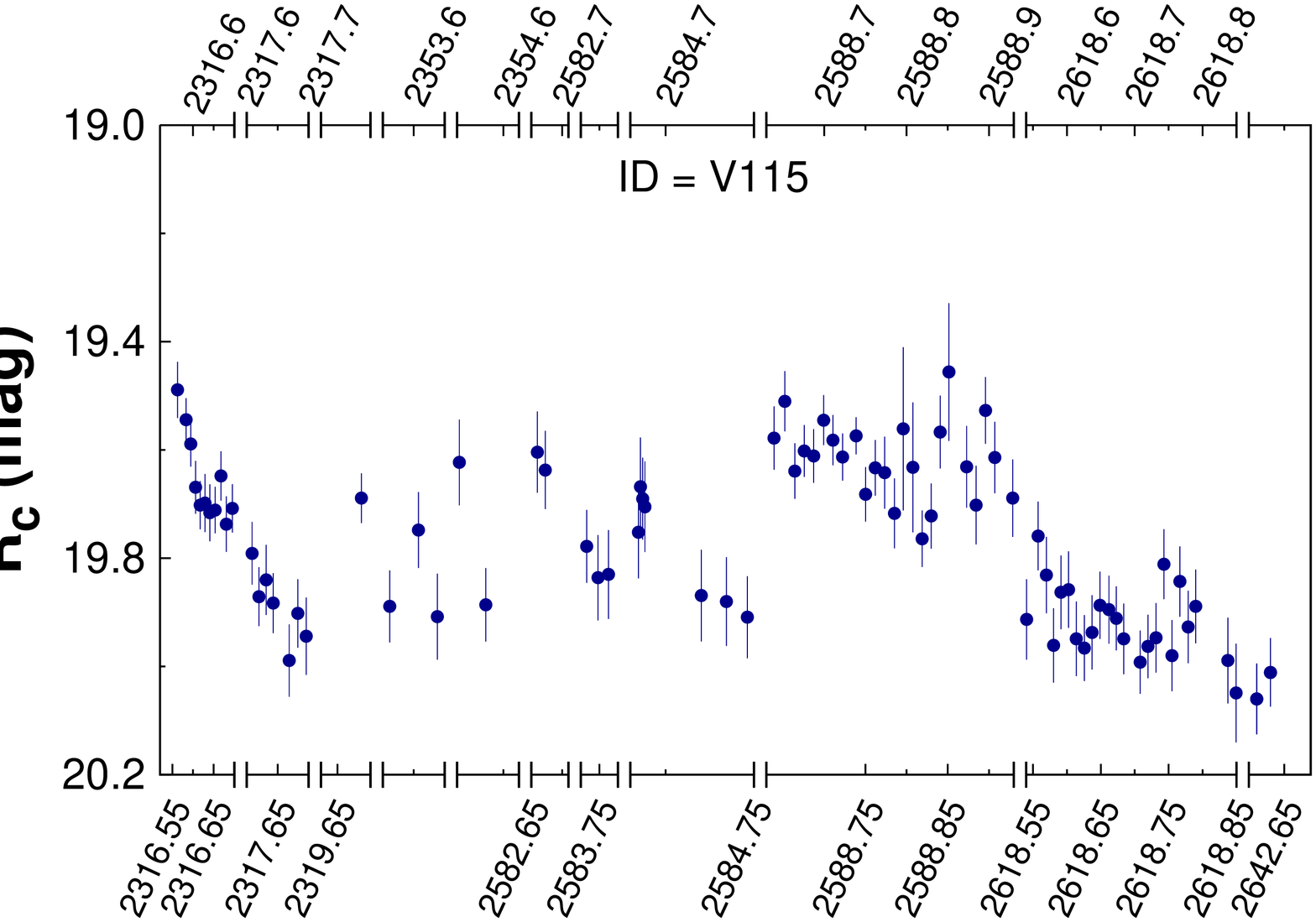}
\hspace{0.1 cm}
\includegraphics[width= 4.0 cm,height = 7.0 cm, angle=270]{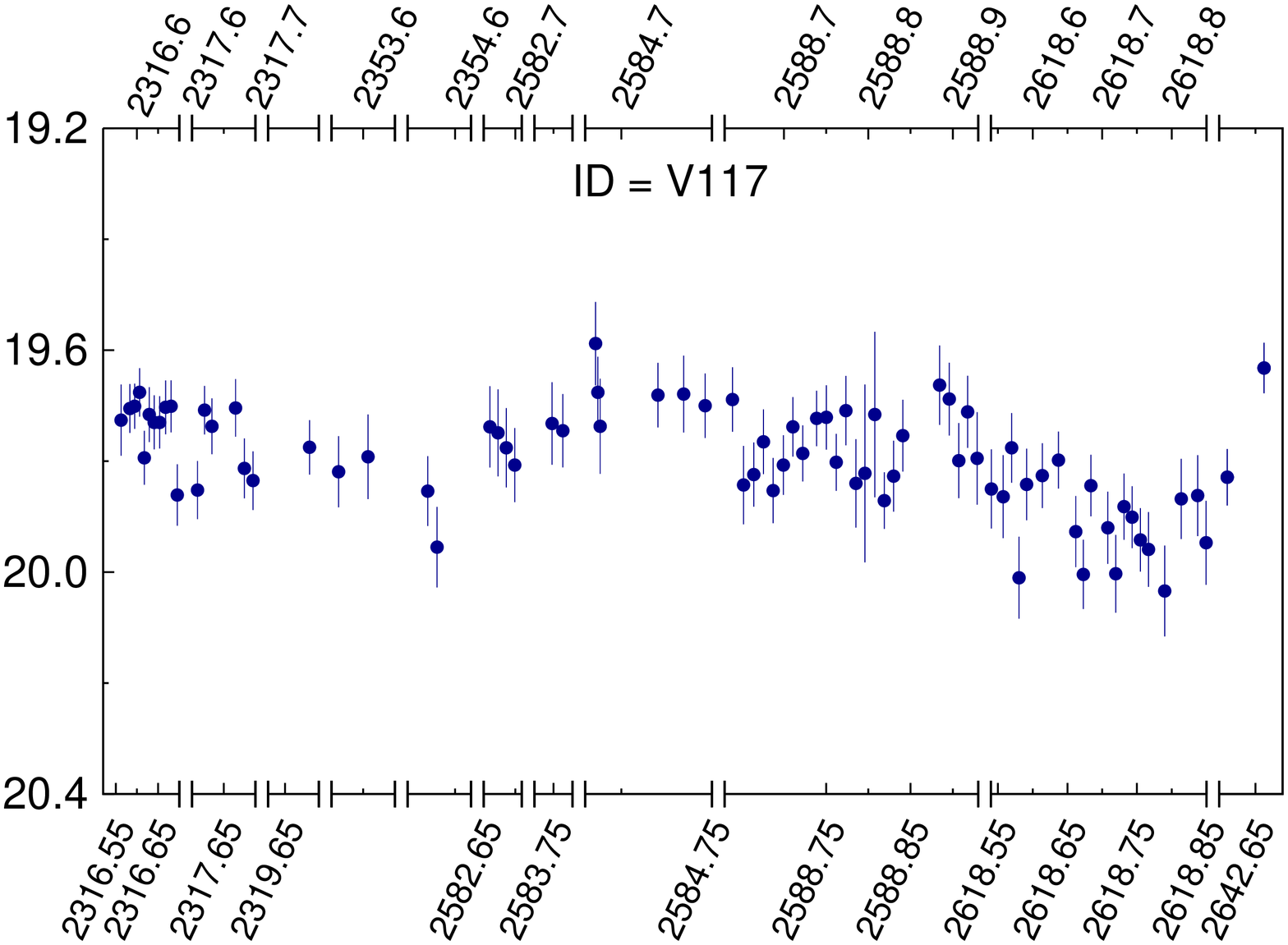} \\
\vspace{0.2 cm}
\includegraphics[width= 4.0 cm,height = 7.0 cm, angle=270]{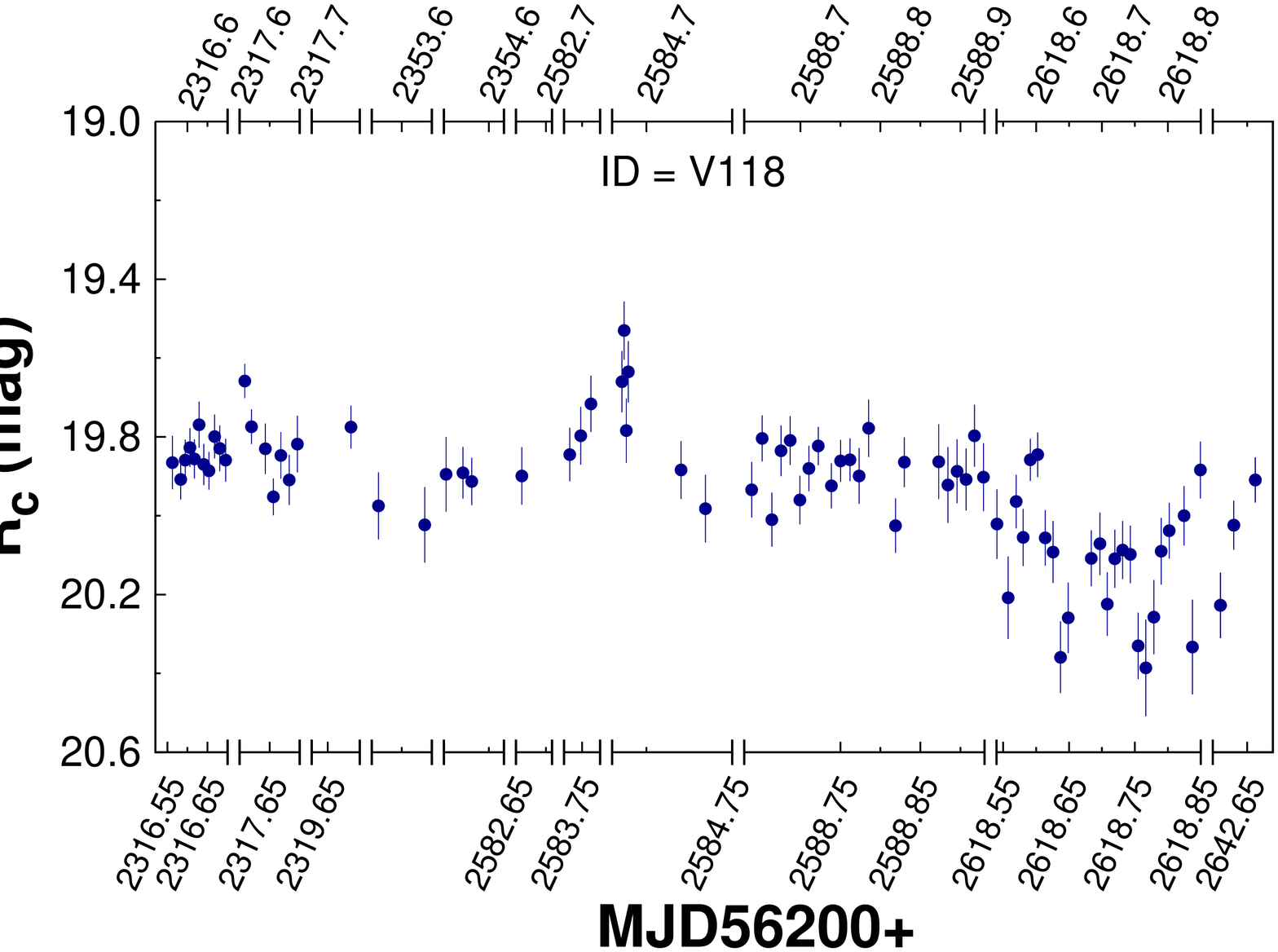}
\hspace{0.1 cm}
\includegraphics[width= 4.0 cm,height = 7.0 cm, angle=270]{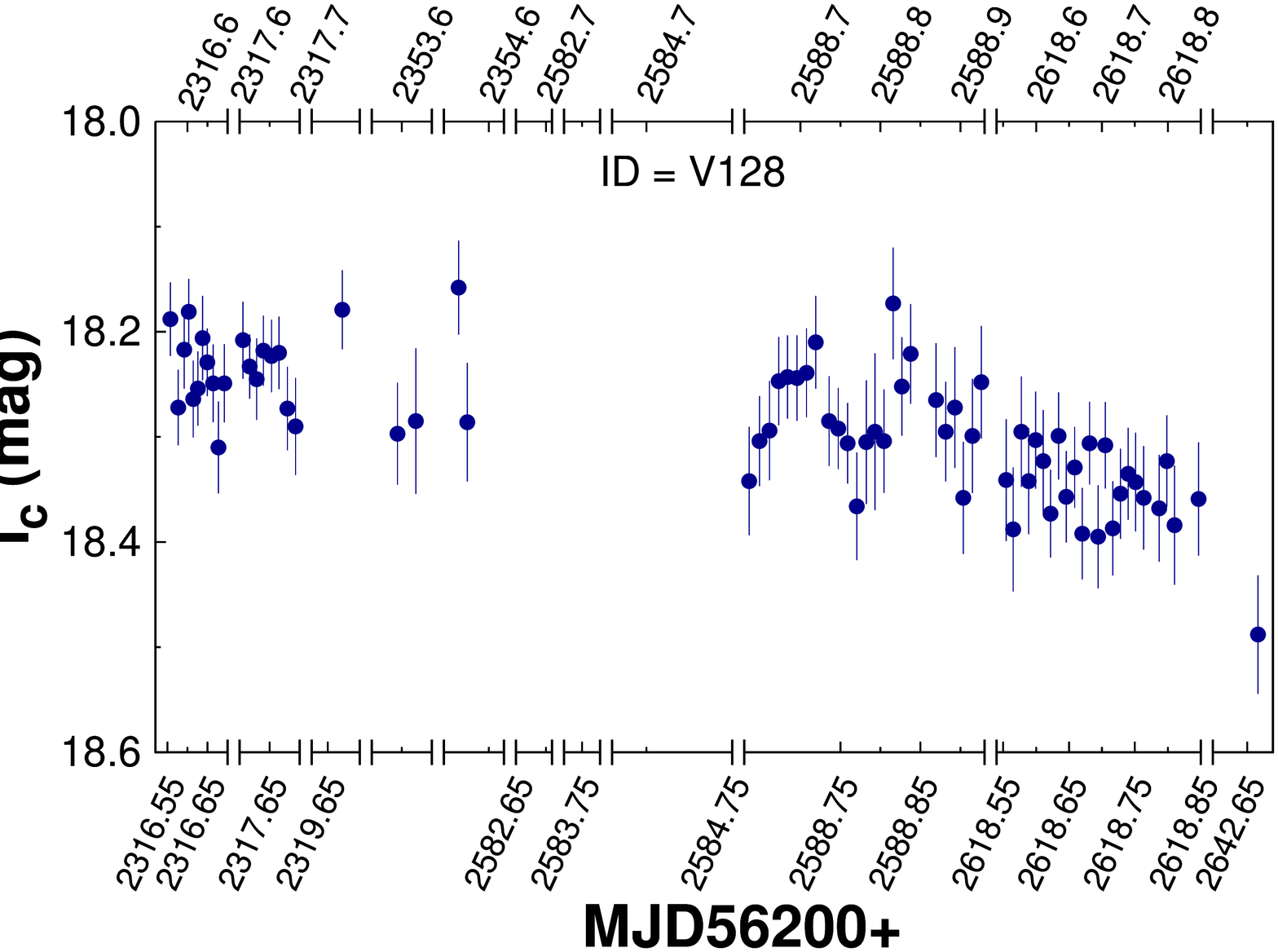} \\

\caption{Contd.}
 
\label{Fig: LC_Class_III_NP}
\end{figure*}

\end{document}